\documentclass[rmp,aps,amsfonts,amsmath,amssymb,nofootinbib,twocolumn,superscriptaddress]{revtex4} 

\usepackage{graphicx}
\usepackage{bm}
\usepackage{IEEEtrantools}
\usepackage{xcolor} 
\usepackage{soul}
\usepackage{todonotes}
\usepackage[normalem]{ulem}
\usepackage{multirow}
\usepackage[T1]{fontenc}

\usepackage{float}
\restylefloat{table}

\usepackage{hyperref}
\usepackage{doi}

\hypersetup{
     colorlinks = true,
     linkcolor = blue,
     anchorcolor = blue,
     citecolor = blue,
     filecolor = blue,
     urlcolor = blue
     }

\begin{document}

	
	
	

\title{The Future of the Correlated Electron Problem}

\author{A. Alexandradinata}
\affiliation{Institute of Condensed Matter Theory, University of Illinois at Urbana-Champaign, Urbana, 61801 IL, USA}\affiliation{Department of Physics, University of Illinois at Urbana-Champaign, Urbana, 61801 IL, USA}

\author{N.P. Armitage}
\email{npa@jhu.edu}
\affiliation{The Institute for Quantum Matter and the Department of Physics and Astronomy, The Johns Hopkins University, Baltimore, MD 21218, USA}

\author{Andrey Baydin}\affiliation{Department of  Electrical and Computer Engineering, Rice  University, Houston, TX 70005, USA}

\author{Wenli Bi}\affiliation{Department of Physics, University of Alabama at Birmingham, Birmingham, AL 35294, USA}

\author{Yue Cao}\affiliation{Materials Science Division, Argonne National Laboratory, Lemont, IL 60439, USA}

\author{Hitesh J. Changlani}
\affiliation{Department of Physics, Florida State University, Tallahassee, FL 32306, USA}
\affiliation{National High Magnetic Field Laboratory, Tallahassee, FL 32304, USA}

\author{Eli Chertkov}\affiliation{Institute of Condensed Matter Theory, University of Illinois at Urbana-Champaign, Urbana, 61801 IL, USA}\affiliation{Department of Physics, University of Illinois at Urbana-Champaign, Urbana, 61801 IL, USA}

\author{Eduardo H. da Silva Neto}
\affiliation{Department of Physics, University of California, Davis, CA 95616, USA}
\affiliation{Department of Physics, Yale University, New Haven, CT 06511, USA}

\author{Luca Delacretaz}\affiliation{Kadanoff Center for Theoretical Physics, University of Chicago, Chicago, IL 60637, USA}

\author{Ismail El Baggari}\affiliation{Department of Physics, Cornell University, Ithaca, NY 14853, USA}

\author{G.M. Ferguson}\affiliation{Department of Physics, Cornell University, Ithaca, NY 14853, USA}

\author{William J. Gannon}\affiliation{Department of Physics and Astronomy, University of Kentucky, Lexington, KY 40506, USA}

\author{Sayed Ali Akbar Ghorashi}\affiliation{Department of Physics, William $\&$ Mary, Williamsburg, VA 23187, USA}

\author{Berit H. Goodge}\affiliation{School of Applied and Engineering Physics, Cornell University, Ithaca, NY 14853, USA}

\author{Olga Goulko}\affiliation{Boise State University, Department of Physics, Boise, ID 83725, USA}
\affiliation{Department of Physics, University of Massachusetts Boston, Boston, MA 02125, USA}

\author{Ga\"el Grissonnanche}\affiliation{Laboratory of Atomic and Solid State Physics, Cornell University, Ithaca, NY 14853, USA}

\author{Alannah Hallas}\affiliation{Department of Physics and Astronomy and Quantum Matter Institute, University of British Columbia, Vancouver, B.C., Canada V6T 1Z1}

\author{Ian M. Hayes}\affiliation{Department of Physics, Maryland Quantum Materials Center, University of Maryland, College Park, MD 20742, USA}

\author{Yu He}
\affiliation{Department of Physics, University of California at Berkeley, Berkeley, CA 94720, USA}
\affiliation{Department of Applied Physics, Yale University, New Haven, CT 06511, USA}

\author{Edwin W. Huang}
\affiliation{Institute of Condensed Matter Theory, University of Illinois at Urbana-Champaign, Urbana, 61801 IL, USA}\affiliation{Department of Physics, University of Illinois at Urbana-Champaign, Urbana, 61801 IL, USA}

\author{Anshul Kogar}\affiliation{Department of Physics and Astronomy, Univ. of California at Los Angeles, Los Angeles, CA 90095, USA}

\author{Divine Kumah}\affiliation{Department of Physics, North Carolina State University, Raleigh, NC 27695, USA}

\author{Jong Yeon Lee}\affiliation{Department of Physics, Harvard University, Cambridge, MA 02138, USA}

\author{Ana\"elle Legros}\affiliation{The Institute for Quantum Matter and the Department of Physics and Astronomy, The Johns Hopkins University, Baltimore, MD 21218, USA}

\author{Fahad Mahmood}
\affiliation{F. Seitz Materials Research Laboratory, University of Illinois at Urbana-Champaign, Urbana, IL 61801, USA}\affiliation{Department of Physics, University of Illinois at Urbana-Champaign, Urbana, 61801 IL, USA}

\author{Yulia Maximenko}
\affiliation{Department of Physics, University of Illinois at Urbana-Champaign, Urbana, 61801 IL, USA}\affiliation{F. Seitz Materials Research Laboratory, University of Illinois at Urbana-Champaign, Urbana, IL 61801, USA}

\author{Nick Pellatz}\affiliation{Department of Physics, University of Colorado, Boulder, CO 80309, USA}

\author{Hryhoriy Polshyn}\affiliation{Department of Physics, University of California, Santa Barbara, CA 93106, USA}

\author{Tarapada Sarkar}\affiliation{Department of Physics, Maryland Quantum Materials Center, University of Maryland, College Park, MD 20742, USA}

\author{Allen Scheie}\affiliation{Neutron Scattering Division, Oak Ridge National Laboratory, Oak Ridge, TN 37831, USA}

\author{Kyle L. Seyler}\affiliation{Department of Physics, California Institute of Technology, Pasadena, CA 91125, USA}

\author{Zhenzhong Shi}\affiliation{Department of Physics, Duke University, Durham, NC 27708, USA}

\author{Brian Skinner}\affiliation{Department of Physics, Ohio State University, Columbus, OH 43210, USA}

\author{Lucia Steinke}\affiliation{Department of Physics, University of Florida, Gainesville, FL 32611, USA}
\affiliation{National High Magnetic Field Laboratory, Tallahassee, FL 32304, USA}

\author{Komalavalli Thirunavukkuarasu}\affiliation{Department of Physics, Florida Agricultural and Mechanical University, Tallahassee, FL 32307, USA}

\author{Tha\'{i}s Victa Trevisan}\affiliation{Ames Laboratory, Ames, IA 50011, USA}

\author{Michael Vogl}\affiliation{Department of Physics, King Fahd University of Petroleum and Minerals, 31261 Dhahran, Saudi Arabia}

\author{Pavel A. Volkov}\affiliation{Department of Physics and Astronomy, Center for Materials Theory, Rutgers University, Piscataway, NJ 08854, USA}

\author{Yao Wang}\affiliation{Department of Physics and Astronomy, Clemson University, Clemson, SC 29631, USA}

\author{Yishu Wang}\affiliation{The Institute for Quantum Matter and the Department of Physics and Astronomy, The Johns Hopkins University, Baltimore, MD 21218, USA}

\author{Di Wei}\affiliation{Geballe Laboratory for Advanced Materials, Stanford University, Stanford, CA 94305, USA}\affiliation{Department of Applied Physics, Stanford University, Stanford, CA 94305, USA}

\author{Kaya Wei}\affiliation{National High Magnetic Field Laboratory, Tallahassee, FL 32304, USA}

\author{Shuolong Yang}\affiliation{Prtizker School of Molecular Engineering, The University of Chicago, Chicago, IL 60637, USA}

\author{Xian Zhang}\affiliation{Department of Mechanical Engineering, Stevens Institute of Technology, Hoboken, NJ 07030, USA}

\author{Ya-Hui Zhang}\affiliation{Department of Physics, Harvard University, Cambridge, MA 02138, USA}

\author{Liuyan Zhao}\affiliation{Department of Physics, University of Michigan, Ann Arbor, MI 48109, USA}

\author{Alfred Zong}\affiliation{Department of Physics, Massachusetts Institute of Technology, Cambridge, MA 02139, USA}\affiliation{Department of Chemistry, University of California Berkeley, Berkeley, CA 94720, USA}

\date{\today}

\begin{abstract}
  A central problem in modern condensed matter physics is the understanding of materials with strong electron correlations.   Despite extensive work, the essential physics of many of these systems is not understood and there is very little ability to make predictions in this class of materials.  In this manuscript we share our personal views on the major open problems in the field of correlated electron systems.  We discuss some possible routes to make progress in this rich and fascinating field. This manuscript is the result of the vigorous discussions and deliberations that took place at Johns Hopkins University during a three-day workshop January 27, 28, and 29, 2020 that brought together six senior scientists and 46 more junior scientists.  Our hope, is that the topics we have presented will provide inspiration for others working in this field and motivation for the idea that significant progress can be made on very hard problems if we focus our collective energies. 
\end{abstract}

\maketitle

\tableofcontents

 \vspace{5 mm}

 The model of non-interacting electrons is well established in solid-state physics.  It is a remarkable fact of nature that, for many materials, the effects of electron-electron interactions can be best captured by {\it ignoring} the correlations they produce.  In other systems, interactions can often be included as a perturbation and manifest through renormalizing parameters such as the effective mass, without altering the qualitative behavior.  Such systems can be adiabatically connected to an interaction-free system.  There are, however, other materials whose properties explicitly manifest strong interactions, which adiabatic connection to an  interaction-free system is not possible, or is not useful.  Such strongly correlated electron systems host a tremendous variety of fascinating macroscopic phenomena including high-temperature superconductivity, quantum spin-liquids, fractionalized topological phases, and strange metals.  Despite many years of intensive work, the essential physics of many of these systems is still not understood, and we do not have an overall perspective on strong electron correlations. Moreover, our predictive power for such systems is lacking. This topic is central to a broader range of scientific disciplines, such as atomic and molecular physics, nuclear and high energy physics, astrophysics, and chemistry, where many-body effects are significant. Despite decades of intensive research, there has been relatively limited progress on an overall picture. Is a unified perspective even possible? Or is the ``Anna Karenina Principle'' in effect\footnote{\url{https://en.wikipedia.org/wiki/Anna_Karenina_principle}} -- all non-interacting systems are alike; each strongly correlated system is strongly correlated in its own way?

In thinking about the future of the correlated electron problem, myriad questions abound.  Is there a general definition of a strongly correlated material?  Is a general framework to understand strong electronic correlations possible?   Are numerical approaches essential?  Can we develop general frameworks to better make predictions?  What new experiments can we design that give essential insight to heretofore unrecognized correlations?  Is “hidden order” ubiquitous?  Can we hope to understand exotic superconductors in the same way we understand conventional superconductors?  What would a ``solution” to the ``problem” even look like?   Is there {\it a problem}?   Or are there {\it many problems}?  What is the future of correlated electrons?   In searching for ``the future” should we come back to the possible avenues not fully explored in the past, or invest in completely new directions, or do both?

On January 27, 28, and 29, 2020, a workshop (organized by NPA) was held at The Johns Hopkins University to try to answer these and other questions\footnote{\url{https://physics-astronomy.jhu.edu/the-future-of-the-correlated-electron-problem-workshop/}}.  Six senior scientists gave lectures on the first day on their ideas to solve parts of the correlated electron problem\footnote{The original lecturers for the workshop were A. Kapitulnik, A.J. Leggett, M.B. Maple, M. Norman, P. Riseborough, and G.A. Sawatzky. MBM was unfortunately unable to travel to Baltimore and so T.M. McQueen generously gave a lecture on materials aspects of the correlated electron problem. However, MBM's slides were used as a reference for the writing of this mansucript.}.   On days 2 and 3, 46 more junior scientists brainstormed, debated, and wrote about their different approaches to understanding correlated electrons.   This manuscript is the result of those vigorous deliberations.   This manuscript was written through collaborative writing software such as Google Docs, Slack, and Overleaf.   Subject topics and the general format was suggested by NPA, but the ultimate topics were chosen by consensus on the morning of the second day.  80$\%$ of the text was written collectively in the first 72 hours after the workshop. All 47 coauthors contributed to the writing and proofing, both at the workshop and afterwards. NPA edited this manuscript.

It is important to note that this manuscript is {\it not} a review and no attempt to be complete has been made.  The topics and opinions expressed are idiosyncratic, and reflect the particular interests and preferences of the people who spoke at and attended the workshop.  It presents their collective vision for the future of the correlated electron problem.   We hope that this document can serve as a starting point for further debate.  And although we do our best to anticipate what directions will be important in the future, we do so with the full expectation (and hope) that much of the below will become irrelevant as some person in some laboratory somewhere in the world will look at some new data coming out of a new experiment on a new material and say, ``That’s funny...''\footnote{The quote ``The most exciting phrase to hear in science, the one that heralds new discoveries, is not `Eureka!' (I found it!) but `That’s funny' '' has been ascribed to Issac Asimov, but various other attributions exist.}.  And we will learn even more about the incredible numbers of ways that electrons can behave in solids.

\section{What is ``the'' problem?}

There is no consensus on the role of strong electron correlations in solids.  Moreover, at present, there is no agreed single definition as to what constitutes {\it the} correlated electron problem. As such, for the purposes of this manuscript, we adopt the following working definition: {\it a correlated electron problem is one in which interactions are so strong or have a character such that theories based on the underlying original ``bare'' particles fail even qualitatively to describe the material properties.}  These original free ``particles'' of a strongly correlated system could be electrons, or spin-flips, or local vibrations.  For instance, exactly solvable models with emergent free quasiparticles ({\it e.g.}  the Kitaev spin liquid~\cite{kitaev2006anyons}) {\it are} strongly correlated by this definition\footnote{Perhaps even this definition is problematic. Are heavy fermion Fermi liquids with masses approximately a 1000 times~\cite{andres19754} the free electron mass strongly correlated by this definition?  Perhaps not, but most would agree that they are strongly correlated.  This leaves us with the only unassailable definition, which is that we know a strongly correlated system ``when we see it.''~\cite{Stewart}}.  The optimistic hope is that a large class of such problems can be understood using a set of similar underlying principles, which are as of yet not understood by us. We believe that such principles should either provide us with a blueprint for a robust predictive power for material realizations {\it or} an understanding of why this is not possible.  While it may be possible to arrive at such principles by general reflection, our judgment is that a general “solution” to the strongly correlated problem will most likely be identified in the common features of candidate solutions to individual strongly correlated ``problems.'' Far from being a mere after-thought to those specific solutions, a general principle arrived at in this way should provide the predictive power needed to find new useful materials with interesting phenomena. In this spirit, we will review some current problems in condensed matter physics that we feel most likely to be fertile in this regard.

\begin{figure}
\includegraphics[width=\columnwidth]{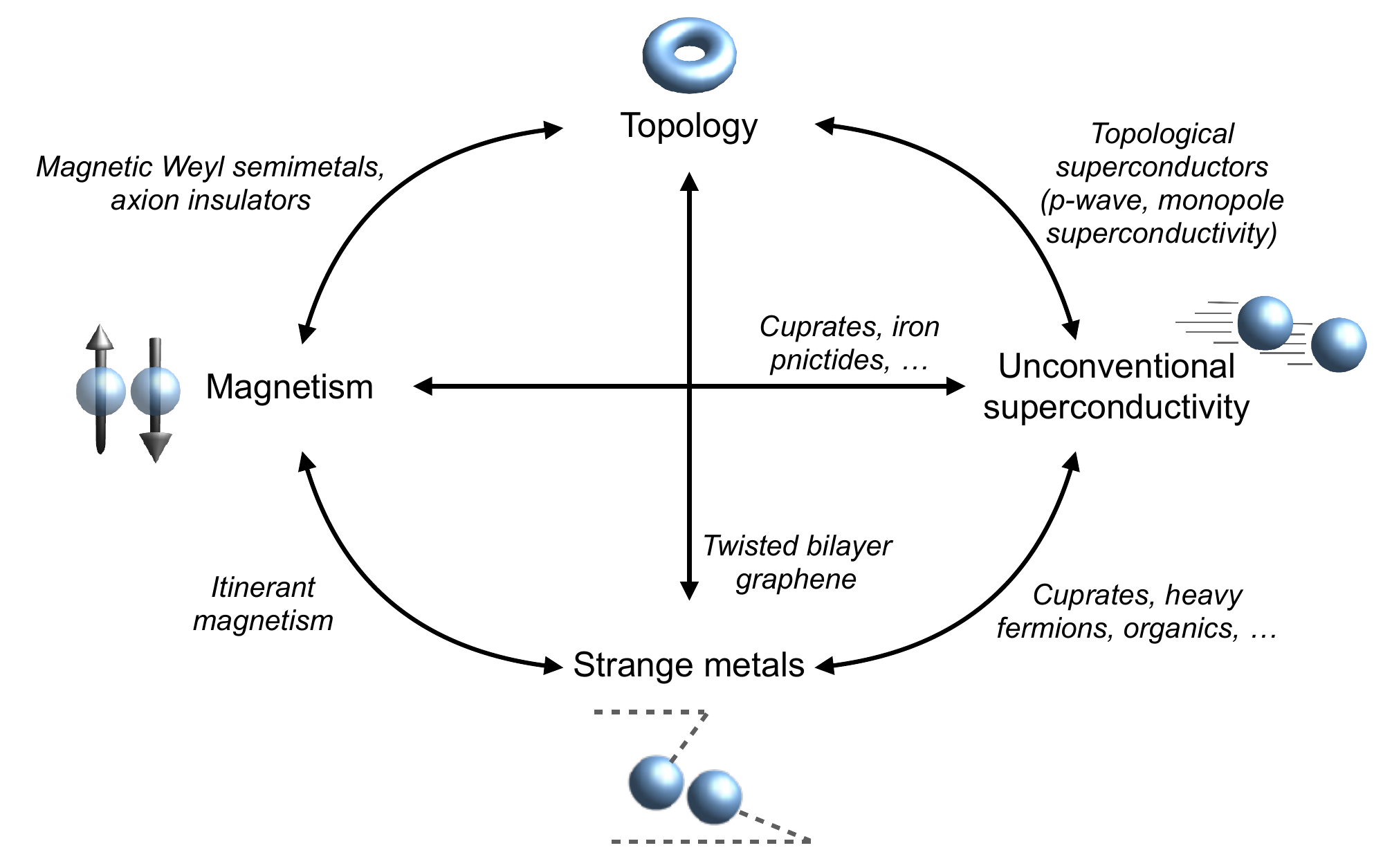}
\caption{Properties such as magnetism, topology, superconductivity and strange metal behavior coexist in some material's phase diagrams. For each binary link connecting these properties, we list a few examples of systems in which both phenomena can be found and which are natural candidates to study the specific relation between these properties.}
\label{Fig.Preliminaries}
\end{figure}

As this workshop was intended to be forward-looking, we present below topics that we see as representative of the {\it future} of the correlated electron problem.  The discovery in 1986 of cuprate high-temperature superconductivity~\cite{bednorz1986possible} was not the start of this field, but gave strong impetus to a vast number of researchers to join it. We thus introduce (A) the field of correlated superconductors at the outset. We then follow with a survey of (B) quantum spin liquids and (C) strange metals, which grew initially from the field of correlated superconductivity, but which currently represent independent thriving fields of study in their own right.  We address (D) quantum criticality and competing orders, which appear to be common among many of correlated systems.  In addition, we include a section on (E) correlated topological materials, which has seen significant outgrowth following the discovery of three-dimensional (3D) symmetry protected topological insulators~\cite{hsieh2008topological}.  Lastly, in contrast to these topical material classes, we also discuss the possibility of (F) returning to ``legacy materials'', which can exhibit similar correlated electron physics and often carry the benefit of accumulated knowledge and perhaps simpler materials synthesis.  Several of these phases or properties can be intertwined in strongly correlated materials~\cite{RafaelARCMP2019} and the link between them is mysterious. Understanding the potential causality or competition between these phenomena could help unveil some universal mechanism between different families of compounds. In Fig.~\ref{Fig.Preliminaries} we show some candidate systems for the study of topology, unconventional superconductivity, magnetism and strange metal behavior. The topics discussed below are undoubtedly not exhaustive and we anticipate an outgrowth of new research areas as completely new subjects emerge and merge.   Indeed, this is what makes this field so particularly dynamic and exciting.

To address the complexity of the correlated electron problem, sophisticated experimental probes, methods for material design and growth, as well as theoretical and numerical tools have been developed in recent years.  We survey a number of these efforts, ranging from the development of spectroscopic and microscopic techniques to efforts to study materials under conditions of ever-increasing extremity.  And we make some suggestions about what we believe is needed experimentally and theoretically to make progress here.  Finally, at the risk of stating the obvious, it must be stressed that in order to make progress, time, energy, and resources must be brought to bear. Funded research, meetings, and publications should have the general treatment of correlated systems as their subject, even if some of these efforts will stray into speculation.

\section{What are the problems?}

\subsection{Correlated Superconductors}
\subsubsection{Definition of the Problem}

Many strongly correlated metals exhibit superconductivity at sufficiently low temperatures.  A comprehensive understanding of these systems is necessary not only for achieving practical applications but also for clarifying many other exotic phenomena of condensed matter. Here, we focus on superconductors that cannot be well-described by the Migdal-Eliashberg theories or their extensions based on electron-phonon coupling as the pairing mechanism. So far, this list includes but is not limited to cuprates~\cite{bednorz1986possible}, iron pnictides~\cite{kamihara2008iron}, iron chalcogenides~\cite{mizuguchi2008superconductivity, Hsu14262}, ruthenates~\cite{maeno1994superconductivity}, nickelates~\cite{li2019superconductivity}, heavy fermion compounds~\cite{steglich1979superconductivity,stewart1984heavy,fisk1988heavy}, and organics~\cite{jerome1980superconductivity}, and perhaps twisted bilayer graphene (TBG)~\cite{cao2018unconventional}.  The discussion presented here is in no way complete, so we refer the interested reader to a number of excellent review articles~\cite{dagotto2005complexity,armitage2010progress,norman2011challenge,si2010heavy,keimer2015quantum,hosono2015iron,mackenzie2017even}.

Although the standard, weak-coupling BCS expression for the superconducting transition temperature places an upper limit on the superconducting critical temperature in many materials (if a moderate Coulomb retardation is assumed)~\cite{cohen1972comments,moussa2006two}, the energy scale of electron-electron interactions in correlated electron systems is generally much higher, suggesting that electronic mechanisms for superconductivity have the potential to produce high-$T_c$. Correlated superconductors generally support a richer set of superconducting ground states, including those with odd-parity and time-reversal symmetry breaking, and of course higher orbital angular momentum wave functions ({\it e.g.}  $d$-wave).

\subsubsection{Possible Structure of a Solution}

A particularly ambitious goal for correlated superconductivity would be to find a theory that would identify a mechanism for the formation of Cooper pairs analogous to the phonon-mediated superconductivity that applies to conventional superconductors. Such a theory might also offer insight into the connection between the exotic normal state properties and the superconducting properties of correlated superconductors. For example, the normal states of many correlated superconductors feature non-Fermi liquid transport (see Sec. \ref{sec:strangemetal}) and quantum critical behavior (see Sec. \ref{sec:qcp}), as well as charge and spin density waves and pseudogap states, but it remains unclear what role these properties play in the formation of superconductivity.

It is possible, however, that it may be difficult to implicate a specific pairing interaction in these systems that would be analogous to the BCS Migdal-Eliashberg phonon mechanism. This is reminiscent of the case of superfluid He$^3$. In this system, even before the discovery of superfluidity, there were proposals for $d$-wave pairing based on van der Waals attraction between atoms~\cite{emery1960possible}. However, based on exchange interactions and the fact that He$^3$ was believed to be almost ferromagnetic, $p$-wave spin-triplet superfluidity was also proposed~\cite{fay1968superfluidity}. Some years later, He$^3$ was indeed found to be a $p$-wave superfluid~\cite{osheroff1972evidence,leggett1975theoretical}, but it was ultimately realized that many interactions contribute to pairing, including density, spin, and transverse current interactions. It was therefore not possible to point to a single pairing mechanism~\cite{leggett1975theoretical,norman2011challenge}, despite the fact that there is a tendency to focus on ``spin-fluctuation exchange.'' ``If one is interested in calculating the actual value of the effective pairing interaction quantitatively, it is by no means obvious that it is a good approximation to limit oneself to the exchange of spin fluctuations only'', as Leggett wrote~\cite{leggett1975theoretical} long ago.  We feel there is a similar lesson here for correlated superconductors.  If one cannot point to a distinct mechanism in a comparatively simple material like He$^3$, it is likely that this issue is even more challenging to resolve in solid-state systems. It may be that electron-phonon mediated superconductivity is a unique case due to disparity between electronic and phononic energy scales and the fact that the lattice and electrons comprise distinctly different sub-systems.

The fact that a specific mechanism may not be able to be implicated in some unconventional superconductors does not mean quantitative questions cannot be asked and answered. One such discussion may revolve around how energy is saved in the formation of a superconductivity~\cite{hirsch1992superconductors,scalapino1998superconducting,demler1998quantitative,leggett1999midinfrared,leggett2006we,leggett2006some}. Such a theory might still provide recipes for constructing superconductors with various properties, and enable control over the critical temperature or symmetry of the superconducting gap. Such an approach would clarify if room temperature superconductivity under normal conditions is a realistic possibility and where to look for exotic superconducting gap symmetries (\textit{e.g.}~odd-parity superconductors or superconductors with multi-component order parameters).  In conventional BCS superconductors for instance, energy is saved through a decrease of the kinetic energy of the {\it ions} and the total potential energy, which outweighs the penalty from the increased electron kinetic energy (see Ref.~\cite{chester1956difference} and note that this calculation was pre-BCS! Also see Ref.~\cite{norman2003electronic}).

\begin{figure}
\includegraphics[width=1\columnwidth]{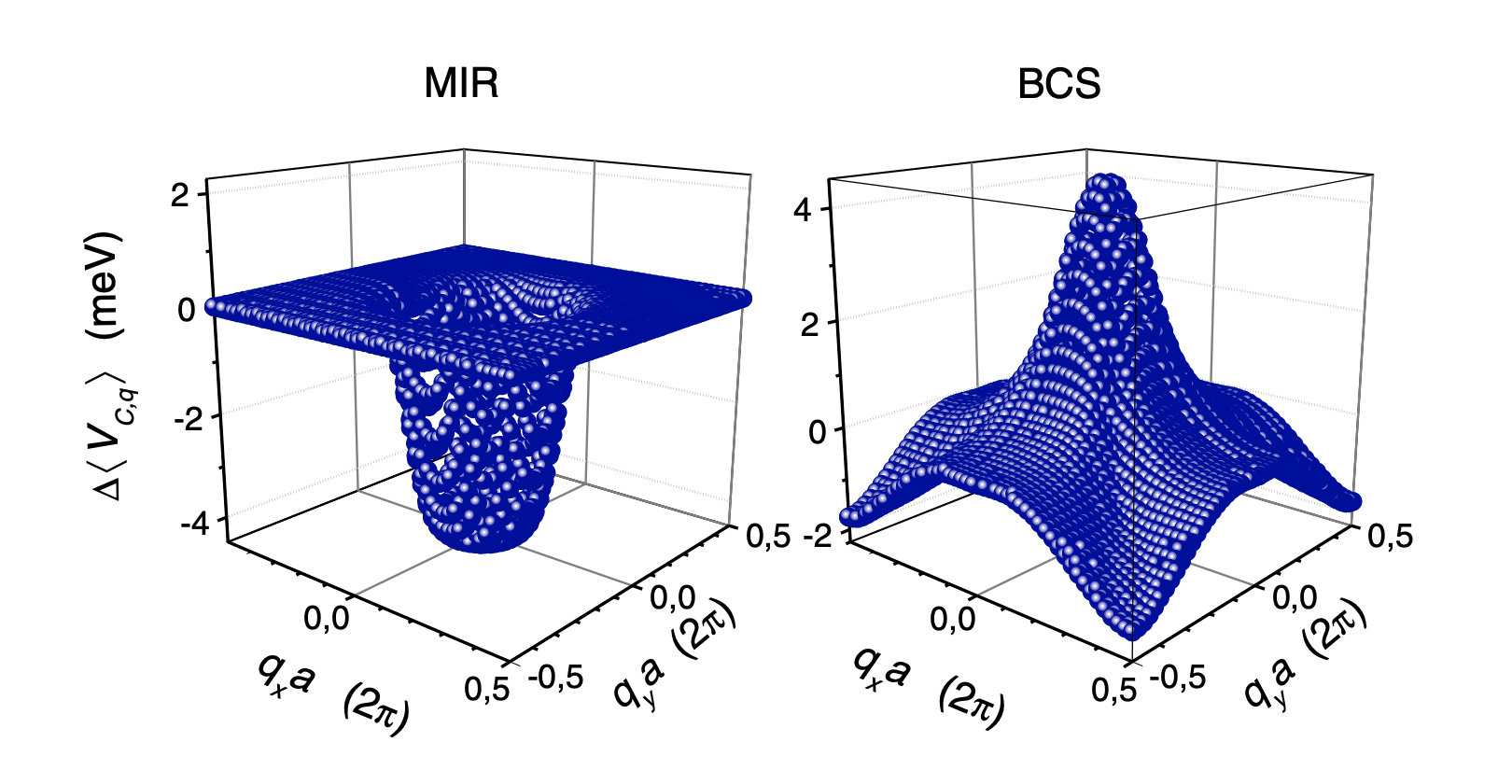}
\includegraphics[width=0.9\columnwidth]{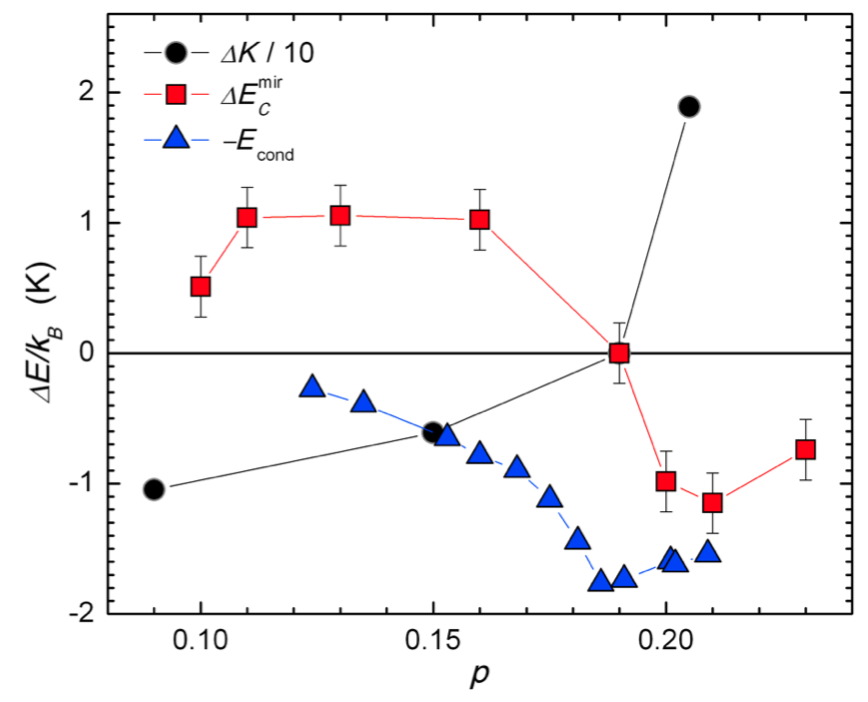}
\caption{(top) The difference between the normal and superconducting state Coulomb interaction energy calculated within the ``midinfrared'' and BCS scenarios~\cite{levallois2016temperature}.   This is one possible model for the source of energy savings in a correlated superconductor without identifying a specific pairing ``glue.'' Here the BCS model calculation is for $d$-wave symmetry, $x=0.16$ hole doping, and with the interaction adjusted such as to give $T_c = 100$ K. (bottom) The estimated difference between normal and superconducting state Coulomb interaction energies ($\Delta E_c^{\text{mir}}$) for Bi$_2$Sr$_2$CaCu$_2$O$_{8-x}$ crystals, together with the total energy difference from heat capacity ($E_{\text{cond}}$) at different doping levels.  Also plotted is the result of a calculation that estimates the changes in kinetic energy ($\Delta K$) when entering the superconducting state.  From Ref.~\cite{levallois2016temperature}.}
\label{MIRscenario}
\end{figure}

There is a significant history of related analyses for exotic superconductors, but still substantial room for progress.   Scalapino and White~\cite{scalapino1998superconducting} suggested that energy lowering in the cuprate's superconducting state could primarily come through the exchange interaction $J\langle \bf{S}_i \cdot \bf{S}_j \rangle $.  In principle one could quantify magnetic energy savings by looking at differences in the neutron scattering structure factor $S(\bf{q},\omega)$ between normal and superconducting state.  Demler and Zhang proposed a particular mechanism along these lines~\cite{demler1998quantitative}.   Experimentally, the exchange energy change was in excess of the condensation energy and it was proposed the difference could be due to the relative cost of kinetic energy in the superconducting state~\cite{demler1998quantitative,dai1999magnetic} as happens in BCS-style superconductors.  This is not necessarily the case however.  For instance, Hirsch has predicted that superconductivity is driven by lowering of electronic kinetic energy due to changes in the electronic mass in the superconducting state~\cite{hirsch1992superconductors}.  Under some scenarios, the kinetic energy changes can be measured in optical conductivity experiments.  Here the experimental situation is inconclusive in the cuprates with possible qualitative differences in kinetic energy savings from underdoped and overdoped regimes~\cite{deutscher2005kinetic,carbone2006doping}.  It was proposed by Leggett that superconductivity in many correlated superconductors is driven by a saving in Coulomb energy that takes place predominantly at long wavelengths and mid-infrared frequencies, which results from the increased screening due to formation of Cooper pairs~\cite{leggett2006we}. This scenario seemingly explained several trends in many known non-BCS high-temperature superconductors, including their quasi-2D nature, their relative insensitivity to other aspects of structure, trends of $T_c$ with the number of CuO$_2$ layers per unit cell in cuprates, and a common prominent mid-infrared absorption. Studies with $q\sim 0$ measurements of the loss function (from optical ellipsometry~\cite{levallois2016temperature}) showed that changes to the Coulomb energy can likely account for a significant portion of the condensation energy for overdoped cuprates (Fig.~\ref{MIRscenario}), but the analysis had to make estimates for the finite $q$ extrapolation of the data. Therefore, further measurements of the loss function performed at finite $q$ with momentum-resolved electron energy loss spectroscopy are needed to test this theory in more detail.   There is also some merit to theoretical analyses or energy savings particularly for numerical works~\cite{maier2004kinetic,gull2012energetics,fratino2016organizing}.  Related analyses, which in some ways is simpler, can also be performed for the single particle spectral function, that is in principle measurable with Angle-Resolved Photoemission Spectroscopy (ARPES)~\cite{norman2000condensation}.   However finite energy and momentum resolution, uncertainties with normalization, and matrix element effects have made such an approach unreliable.  Irrespective of the details, the general point is that questions with quantitative answers {\it can} be asked even without appealing to any paradigms that we associate with BCS, Migdal-Eliashberg theory. 

\begin{figure*}
\begin{center}
\includegraphics[width=0.95\textwidth]{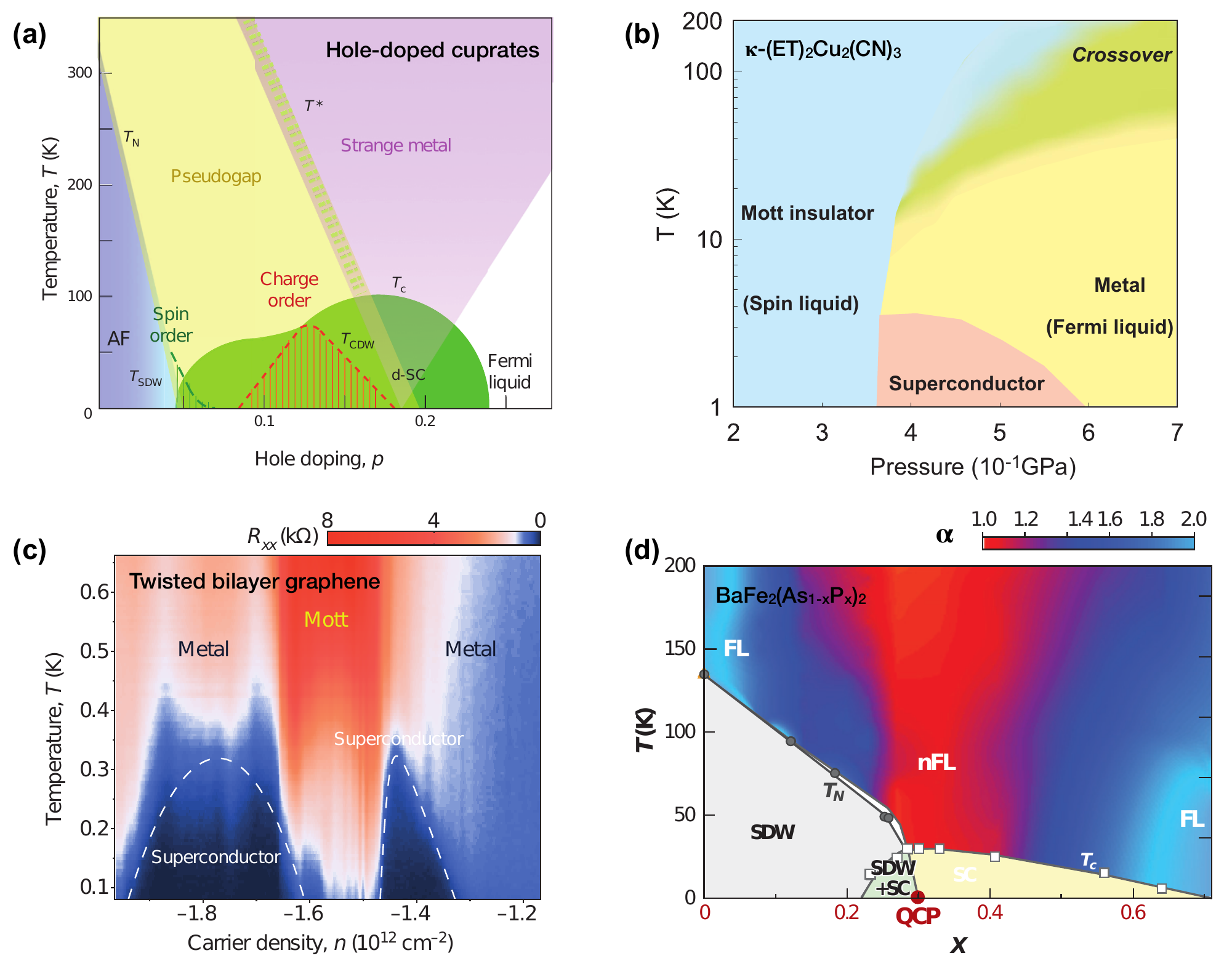}
\caption{Phase diagrams of several unconventional superconductors: (a)~Temperature-hole doping schematic phase diagram of hole-doped cuprates~\cite{keimer2015quantum}. (b)~Temperature-pressure schematic phase diagram of the organic superconductor $\kappa$-(ET)$_2$Cu$_2$(CN)$_3$~\cite{kurosaki2005mott}. (c)~Temperature-carrier density phase diagram of twisted bilayer graphene at the magic angle $\theta=1.16^\circ$~\cite{cao2018unconventional}.  $R_{xx}$ is the longitudinal resistance. (d)~Temperature-phosphorus concentration phase diagram of the iron pnictide BaFe$_2$(As$_{1-x}$P$_x$)$_2$~\cite{shibauchi2014quantum}.  $\alpha$ is the exponent of the temperature dependence of the resistivity. All phase diagrams show multiple competing phases in the vicinity of the superconducting dome.}
\label{Fig.SC_phase_diagrams}
\end{center}
\end{figure*}

\subsubsection{Central Questions}
To organize our thinking, we identify five central questions on the problem of correlated superconductors. 

\begin{enumerate}
    \item  {\bf What is (are) the pairing mechanism(s)? }
    All known superconductors (correlated or otherwise) contain Cooper pairs. The question of the mechanism of superconductivity concerns the effective attractive interaction that is responsible for pair binding. If this interaction arises due to exchange of a soft (low-energy) bosonic excitation, we may speak of the boson as a ``pairing glue'', in analogy to the role of phonons in BCS, Migdal-Eliashberg superconductors.  A fundamental challenge with correlated superconductivity may be that  the same electrons that give rise to the pairing interaction are those that form Cooper pairs.  Furthermore, as seen in the example of He$^3$, multiple types of fluctuations may contribute to the pairing interaction, such that the pairing mechanism cannot be ascribed clearly to a single process.  Moreover, exchange of a pairing boson is not even necessary given that instantaneous (high-energy) interactions can also contribute to pairing in unconventional superconductors~\cite{anderson2007there}.

    \item  {\bf How/where is energy saved?}
    In the absence of a clear answer to the question of the pairing mechanism, it is possible nevertheless to ask reasonably model-independent questions concerning the superconducting state. In particular, what kind of energy ({\it i.e.} kinetic, magnetic exchange, Coulomb) is saved when the system transitions to the superconducting state~\cite{hirsch1992superconductors,scalapino1998superconducting,demler1998quantitative,leggett1999midinfrared,leggett2006we,leggett2006some}? The answer to this question, which can possibly be determined experimentally through optics and momentum-resolved probes such as electron energy loss spectroscopy, inelastic X-ray scattering, inelastic neutron scattering, or ARPES, has significant implications both for developing theoretical models of correlated superconductors and for guiding the experimental search for novel superconductors.

  \item  {\bf What is the order parameter symmetry? }
    The symmetry of the superconducting order parameter is a property that is defined independent of any microscopic mechanism.  However, it can be used to constrain new theories of correlated superconductors.   The most definitive order parameter tests are those that are sensitive to the superconducting order parameter's phase, such as corner SQUID and tri-crystal measurements~\cite{tsuei2000pairing}.  Unfortunately, among correlated superconductors, such studies were -- until recently~\cite{kalenyuk2018phase} -- only performed for the cuprates~\cite{van1995phase, tsuei2000pairing}. It is unknown to what extent the order parameter symmetry varies between different classes of correlated superconductors. While a single order parameter symmetry seems universal for cuprates, this may not be the case in other classes of correlated superconductors~\cite{hirschfeld2011gap}.
    
    It is remarkable how challenging it can be to answer even this simplest of questions for correlated superconductors.   The community has grappled with it recently in the case of Sr$_{2}$RuO$_4$. Largely accepted as the best example of a time-reversal symmetry breaking $p$-wave superconductor~\cite{maeno1994superconductivity,armitage2019superconductivity}, recent nuclear magnetic resonance (NMR) and strain experiments appear to have debunked this conclusion and a reevaluation is underway~\cite{hicks2014strong,pustogow2019constraints}.
    
  \item {\bf Why do very different systems have similar phase diagrams?}
    Many correlated superconductors exhibit striking similarities between their phase diagrams (Fig.~\ref{Fig.SC_phase_diagrams}). These similarities include the existence of additional broken symmetry states, quantum critical points, proximity to magnetism, and superconducting domes~\cite{dagotto2005complexity}. Proximity between superconductivity and magnetism exists, for instance, in cuprates, organic superconductors, and iron pnictides. Similarly, several systems exhibit a Mott insulating state in a parent compound at half filling (often under pressure), including possibly the recently discovered magic-angle TBG. The common features between different phase diagrams suggest that similar phenomena might be responsible for superconductivity in different correlated systems. Although some similarities are highly suggestive, there is no consensus on where to draw the line between essential universal features of the phase diagram and system-specific details.  
    
  \item  {\bf What role does the dimensionality of the electronic structure play?}
    In many (but not all) correlated superconductors, the normal state electronic properties often exhibit quasi-2D behavior (with some exceptions of quasi-1D for organics and 3D for heavy fermions). Is dimensionality an important criterion for these specific unconventional superconductors? If not, are there key mechanistic similarities between systems with the same electronic dimensionality?
\end{enumerate}

\subsubsection{Future Directions}

With the discovery of each new family of correlated superconductors over the past four decades there has been a flurry of experimental activity.  To complement and build upon these studies, we must seek new ways to explore properties of both the normal and superconducting states to begin addressing the questions outlined above.

\begin{figure}
\includegraphics[width=1\columnwidth]{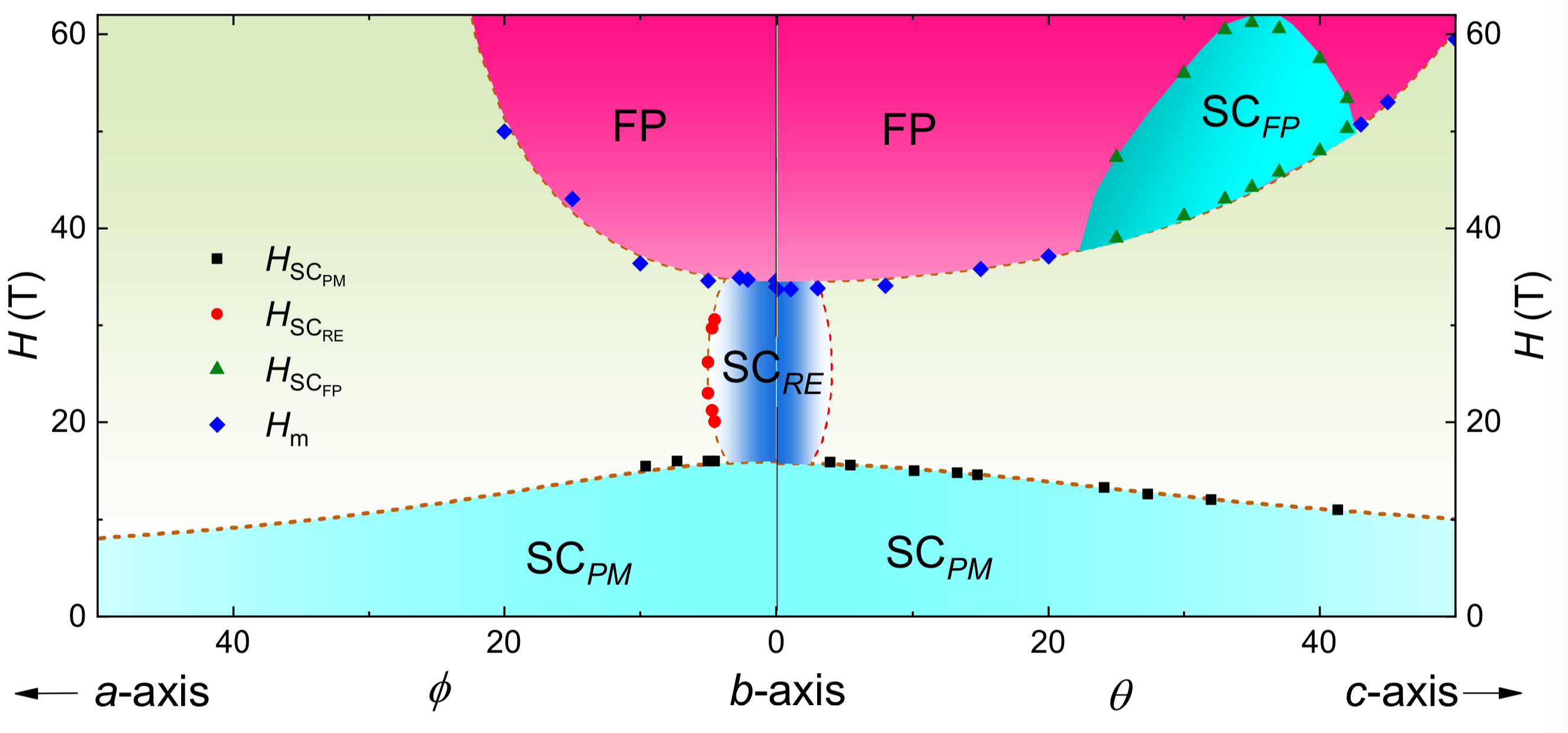}
\caption{Magnetic field - angle phase diagram of UTe$_2$ showing three superconducting phases with, among other aspects, the largest upper critical field in a reentrant superconductor.  The magnetic field is rotated within the $ab$ and $bc$-plane.  Sample temperature is 0.35--0.5~K.   From Ref.~\cite{ran2019extreme}.  FP is field polarized, PM is paramagnetic, RE is reentrant, and SC is superconductor.}
\label{UTe2HighField}
\end{figure}

\begin{enumerate}
\item   Assuming that a pairing mechanism can be identified, new pump-probe style measurements may enable the direct study of the coupling between different degrees of freedom or subsystems. Such measurements include targeted pumping of particular phonon modes combined with time-resolved X-ray and photo emission spectroscopies to separate out the response of the electronic and structural degrees of freedom~\cite{gerber2017femtosecond,Cilento2018,Boschini2018,Mankowsky2014}. In this way phonon mediated pairing (or lack thereof) may be identified.  Similar studies include targeted pumping followed by a broadband measurement of the transient reflection or transmission at frequencies spanning from the THz to the IR region \cite{DalConte2015}. Other advances include time-resolved RIXS which can serve as a momentum-resolved and bulk-sensitive probe of changes in the superconducting gap in response to optical quenches and targeted pumping of particular phonon resonances \cite{cao2019ultrafast}.

\item Failing the identification of a ``pairing glue'', certain spectroscopic techniques can directly address from where in $\omega$, ${\bf q}$, and ${\bf k}$ space the superconducting condensation energy comes~\cite{levallois2016temperature, li2018coherent, husain2019crossover,leggett1999midinfrared,senga2019position}.  It should be possible to determine what region of wave vector and frequency is the Coulomb energy saved (or expended) in the superconducting transition.  New generations of momentum resolved probes (see Fig.~\ref{MIRscenario}) should be able to give insight in this regard.

\item If electronic correlations are important for superconductivity, understanding how the Coulomb interaction is screened by different dielectric environments in layered superconductors could provide insight to the nature of the pairing~\cite{leggett2006some}. Revisiting seldom-studied cuprates for example (such as ones with a large number of CuO$_2$ per unit cell), like the recent work on the five-layer cuprate Ba$_2$Cu$_4$Cu$_5$O$_{10}$(F,O)$_2$ \cite{kunisada2020observation}, gives new perspective on a long-standing problem. Advances in sample synthesis will allow systematic variation of the dielectric environment in correlated superconductors~\cite{logvenov2009high, bovzovic2016dependence, stepanov2019interplay}.

\item Until recently the cuprates were the only correlated superconductors where the order parameter symmetry has been conclusively identified through phase sensitive technique~\cite{van1995phase, tsuei2000pairing}. Although similar studies have been attempted for other materials,~\cite{nelson2004odd, strand2009evidence}, the experimental identification of the order parameter symmetry via such phase sensitive measurements is missing for most correlated superconductors.  A promising development is recent Josephson junction experiments between a conventional $s$-wave Nb and the multiband iron-pnictide superconductor Ba$_{1-x}$Na$_x$Fe$_2$As$_2$ that provide evidence for a sign-reversing $s^\pm$ symmetry of the order parameter in this compound~\cite{kalenyuk2018phase}.  Extending these measurements to new materials is challenging and typically requires either high quality thin films or the ability to integrate bulk crystals into devices, but is an essential direction going forward.   A promising yet relatively unexplored approach to phase-sensitive experiments is conducting mesoscopic imaging experiments on polycrystalline samples~\cite{w2008scanning}. The local energy scale of the experiments allows phase-sensitive measurements to be performed on samples that are unavailable in thin-film form or cannot be interfaced with other superconductors to form Josephson junctions.   Another promising method is Bogoliubov quasiparticle interference imaging that has shown, for instance, that in a number of Fe based superconductors that the gaps have opposite sign on different Fermi surface sheets~\cite{sprau2017discovery,du2018sign}.  Similar experiments have provided supporting evidence for gap changing behavior in cuprate~\cite{gu2019directly}.   These can be applied to more classes of materials.

\item Identifying the relevant common features in the phase diagrams of correlated superconductors is an ongoing challenge. Many of these systems can be tuned by a variety of parameters, including chemical doping, external pressure, strain, and applied field. Combining two or more of these tuning knobs can further access unexplored regions of parameter space~\cite{kim2018uniaxial, gerber2015three} (see Sec. \ref{extreme}).  For example, small uniaxial strains can have as large of an effect on electronic order as a 28~T magnetic field~\cite{kim2018uniaxial}. Through broad and systematic studies across each of these variables in specific systems, it may be possible to construct multi-dimensional phase diagrams that could help identify important commonalities or surprising differences between material classes.   Of course it will be only complicating if such studies only result in the number of tuning variables increasing.  In this regard it is important to establish the causal relation between the driving force and response.  For example, see the work on measures of the nematic susceptibility in iron arsenide superconductors via elastoresistivity~\cite{chu2012divergent}.

 Access to new and more extreme experimental conditions will also be useful in exploring phase space.  For instance, the recently discovered re-entrant superconducting phases in UTe$_2$ under high field and high pressure~\cite{ran2019nearly,ran2019extreme,aoki2019unconventional} call for a search of potential novel phases in other materials to reveal new common features between correlated superconductors.  As shown in Fig.~\ref{UTe2HighField}, this material exhibits by far the highest upper and lower critical fields of any field-induced superconducting phase, more than 40~T and 65~T respectively.

\item It is also important to understand how distinct phases interact and compete at mesoscopic and microscopic scales. Local probes that can identify and measure specific regions or states in isolation will thus provide important insight. Spatially-resolved imaging of magnetism using diamond nitrogen vacancy centers and nano-SQUID could reveal short-range magnetic correlations~\cite{pham2011magnetic,martinez2017nanosquids,vasyukov2013scanning,ceccarelli2019imaging,wolf2015subpicotesla}. This is potentially crucial information for the study of the pseudogap phase of cuprates for example, whose origin has been lengthily debated and could be due to antiferromagnetic spin fluctuations. Localized spectroscopic measurements can distinguish between contributions from mesoscopic inhomogeneities or disorder and other competing phases that would otherwise be averaged by more macroscopic measurements. 
    
\item Advances in focused ion beam (FIB) and other nano-fabrication techniques can be used to realize highly-controlled device geometries for investigating dimensional effects and transport anisotropies~\cite{ronning2017electronic}. One potential pitfall given the quasi-2D nature of many of these superconductors would be the neglect of the $c$-axis properties. As demonstrated by the interlayer tunneling model for cuprate superconductivity, however, understanding the out-of-plane electronic properties should not be neglected~\cite{tsvetkov1998global, leggett1996interlayer, wheatley1988interlayer}.

\item It is useful to identify model systems that can be studied exhaustively and tuned extensively.  It seems unlikely to find a single model compound for unconventional superconductivity that is well-suited for all probes and that can be obtained in good quality over the whole phase diagram.  Nevertheless, recent discoveries of unconventional superconductivity in van der Waals materials and heterostructures such as TBG~\cite{cao2018unconventional} and one unit cell thick cuprate Bi2212~\cite{jiang2014high,sterpetti2017comprehensive,yu2019high} could provide simple enough toy models in which various parameters can be more easily tuned. Growth-integrated techniques for layer-by-layer probes (\textit{e.g.} \textit{in situ} STM~\cite{lv2015mapping} or MBE growth of isolated ``deconstructed” layers~\cite{logvenov2009high}) could help explore the role of single atomic planes or interfaces between different materials.

\end{enumerate}

\subsection{Quantum Spin liquids}

\subsubsection{Challenges in the field}

The essential problem of quantum spin liquids (QSLs)~\cite{balents2010spin,savary2016quantum,knolle2019field,broholm2020quantum} is that despite a multitude of candidate materials, none have been proven as QSLs.  Progress toward positively identifying a QSL faces two roadblocks. The first is the imprecise definition often used for a ``quantum spin liquid''. Spin systems with no long-range order at $T=0$ have often been conflated with spin liquids---but this is not sufficient; other effects, such as disorder~\cite{Zhu_YMGO_2017} or weak interactions~\cite{calder2010neutron}, can also prevent magnetic order. Materials with no long-range magnetic order in the limit of zero temperature are more properly called quantum paramagnets.  Moreover, recent theory suggests that QSLs and long-range ordered states are not necessarily mutually exclusive~\cite{brooks2014magnetic,liu2019competing}. A useful definition for a QSL should not focus on the absence of conventional features, but on the presence of key properties of interest {\it e.g.}  long-range entanglement~\cite{KitaevPreskill2006} and fractionalized excitations~\cite{broholm2020quantum}.  Although there are exactly solvable models that show that such a state in principle can exist~\cite{kitaevspinliquid}, it remains to be shown if such a state exists in real materials, which are subject to disorder, further 
neighbor and ``ring-link'' exchange, and spin-lattice coupling.

This leads to the second roadblock; there are no clear experimental signatures of the long-range entanglement and fractionalized excitations which demonstrate the existence of a QSL ground state. The presence of a diffuse continuum of scattering in neutron spectroscopy experiments is often interpreted as a hallmark for fractionalized excitations but this too is insufficient.  Disorder-induced glassy behavior can also produce a diffuse neutron spectrum~\cite{Zhang_NCNF_2019,Zhu_YMGO_2017}. Furthermore, while the experimental neutron continuum scattering of a 1D spin chain can be accurately related to theoretical models~\cite{mourigal2013fractional}, such straightforward comparisons are not possible in 2D and 3D materials~\cite{knolle2019field,Zhu_YMGO_2017}.

\subsubsection{Routes to Progress}

To make substantial progress in this field, it is necessary to develop a means of directly measuring fractionalized excitations and long-range entanglement. There are many proposed options, most of which require both new theoretical calculations and new experimental techniques. Here we consider some possibilities:

\begin{figure}
\includegraphics[width=0.5\columnwidth]{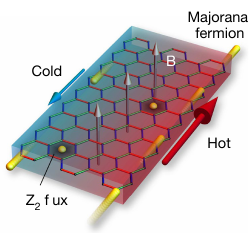}
\includegraphics[width=0.95\columnwidth]{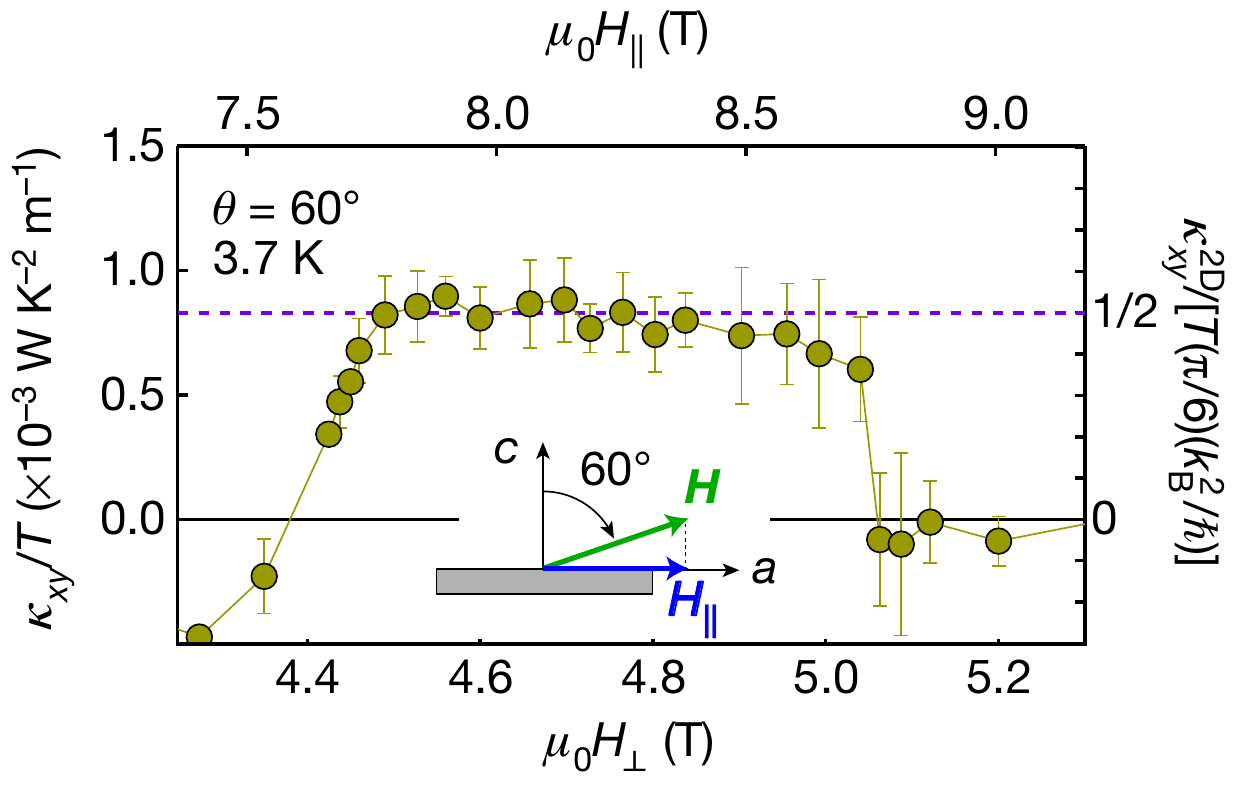}
\caption{(top) Schematic illustration of the thermal Hall conductivity of a Kitaev spin liquid, with a magnetic field perpendicular to the sample plane, resulting in the fractionalization of spins into Majorana fermions (yellow spheres) and $Z_2$ fluxes (hexagons). The charge-neutral Majorana fermions are responsible for the conduction of heat by chiral edge currents. (bottom) Half integer plateau reported in the thermal Hall conductance of $\alpha$-RuCl$_3$. Adapated from Ref.~\cite{kasahara2018majorana}.  }
\label{SpinLiquid}
\end{figure} 

\begin{enumerate}
    \item {\bf New analysis of neutron spectroscopy.} Neutron scattering measures the energy-resolved spin-spin correlations, and this can encode evidence of fractionalization and entanglement~\cite{Morampudi_PRL_2017,Hauke2016}. Rather than focusing on 2D plots of diffuse continua, more sophisticated efforts may extract entanglement bounds~\cite{Hauke2016,scheie2021witnessing,laurell2020dynamics} or fractionalization signatures in the energy-dependence of response functions~\cite{Morampudi_PRL_2017}. 
    \item {\bf Thermal transport.}   Thermal transport is potentially extremely powerful in probing spin degrees of freedom in electrical insulators.   It can be directly sensitive to spin transport or can be sensitive to phonons scattering off of spins.  The technique has a long history but is still underutilized in quantum magnets. A linear term as a function of temperature in thermal conductivity is an expected hallmark of a QSL featuring a spinon Fermi surface~\cite{Xu_YMGO_2016}, and different spin liquids are expected to have distinct, topological thermal Hall conductivity signals~\cite{Katsura_2010,kasahara2018majorana,Zhang_2020}.   A quantized thermal Hall effect is an expected signature of Majorana fermions in a Kitaev spin liquid~\cite{Katsura_2010,kasahara2018majorana,Zhang_2020} and has been reported in $\alpha-$RuCl$_3$ as shown in Fig.~\ref{SpinLiquid}.
    \item {\bf Imaging spin densities around impurities.} It has been theoretically proposed that nonmagnetic impurities will produce a characteristic spin density pattern in a Kitaev spin liquid~\cite{Willans_PRB_2011}. Similar local spin measurements were key to proving Haldane behavior ({\it e.g.}  fractionalized end spins) for $S=1$ chains~\cite{Hagiwara_1990}. Atomic-resolution local spin density measurements may be available in the near future and could be broadly applied to all classes of QSLs (see Sec. \ref{subsection:local probes}).
    \item {\bf Entangled neutron beams.} Recent experiments have shown that an entangled neutron beam can, in principle, probe quantum entanglement between different points on a lattice~\cite{shen2019unveiling}. This technique may provide a direct measure of entanglement in solid-state systems.
    \item {\bf Device fabrication.} The fractionalized quasiparticles of a QSL are predicted to give unique signals if incorporated in microscopic spintronic devices~\cite{Barkeshli2014,Chatterjee2019,Chatterjee2015,aasen2020electrical}.  QSLs sandwiched between metals, superconductors, or ordered magnets, could have their fractionalized excitations directly probed through certain measurements, such as measurements of spin current. Such direct probes would be possible despite the electrically insulating nature of most QSLs.
    \item {\bf Spin noise experiments.} Spin noise measurements are underutilized in quantum magnets. Fractional quasiparticle creation and annihilation are likely to give magnetic noise signal distinct from conventional Boltzmann fluctuations. Such experiments have been done for the classical spin ices~\cite{Dusad2019,watson2019real}, and may yield definitive evidence of quasiparticles in QSL candidates.   As discussed below in Sec.~\ref{Prospects}, there have also been proposals for how to measure entanglement in solid-state systems~\cite{laflorencie2016quantum} with a globally conserved quantity such uninform magnets.  For instance, Song et al. \cite{song2012bipartite} propose that the noise spectrum can be a probe of entanglement in a $O(2)$ quantum magnet that has a magnetic field partially obscured by a superconducting shield.  Related theory (and experiment) could be extended to QSL candidates.
    \item {\bf Out-of-equilibrium relaxation.} If a system is perturbed out of its ground state, relaxation back to equilibrium will generally be different for QSL and non-QSL states (see for example Ref.~\cite{Claassen2017}). For instance, Nasu \textit{et al.} showed that when quenching a Kitaev spin liquid with a magnetic field, the relaxational dynamics are qualitatively different depending on the phase that it is quenched from~\cite{nasu2019nonequilibrium}. In the case of the quench from the high-field classical ferromagnetic (FM) phase, the two Majorana fermions are strongly coupled and the spin dynamics are conventional, originating from the precessional motion of spins.   In a quench from the low-field spin-liquid phase, which is connected to the zero-field Kitaev spin liquid, two Majoranas are observed separately in the time evolutions.  Measuring the relaxation after a quench may allow spin-liquid states to be distinguished from other non-fractionalized states.
    \item {\bf Multidimensional coherent THz spectroscopy.} This newly developed nonlinear optical technique may be able to resolve the difference between a diffuse continuum from local disorder and a diffuse continuum from fractionalized quasiparticles~\cite{wan2019resolving,choi2019theory,lu2017coherent} and measure the lifetime of multi-spinon excited states (see Sec. \ref{subsection:methods-spectroscopies}).  In some cases it allows new information about excitations that can already be seen in linear response~\cite{wan2019resolving}.  In other cases it provides spectroscopic information that is inaccessible at the level of nonlinear response~\cite{lu2017coherent}.
\end{enumerate}

As an overall comment, the effort to distinguish a QSL from glassy or other quantum-disordered phases requires paying close attention to sample quality. Disorder or defects may cause ``false positives'' with regards to creating broad continuum lineshapes even in spin systems that would have well-defined low temperature spin-wave-like excitations~\cite{paddison2017continuous,Zhu_YMGO_2017,zhang2018hierarchy}. Although some theoretical proposals suggest that certain types of disorder can stabilize a QSL~\cite{savary2017disorder,Wu_PRB_2019,Kawamura_2019}, lessons learned from superfluid He$^3$ -- perhaps the cleanest condensed matter physics system in its pure form -- indicate that random disorder tends to destroy quantum coherence.  Rather, disorder must be highly correlated and structured in order to induce quantum effects.  When superfluid He$^3$ is confined in high porosity ($\sim$ 98\% porous) silica aerogels to act as artificial defect scattering centers, the real-space correlations of the gel on length scales comparable to the superfluid correlation length become vitally important.  When the disorder is carefully controlled and thoroughly characterized, a variety of novel superfluid phases can be induced~\cite{thuneberg1998models, pollanen2012new, li2013superfluid}, including a correlated topological phase~\cite{dmitriev2015polar, zhelev2016observation}.  The most important aspect in stabilizing these phases is the presence of particular types of correlated disorder.  Understanding and interpreting these phases and their origins requires a detailed understanding of the underlying disorder itself.

These examples demonstrate that greater caution in claiming experimental evidence for QSL is needed. Finding a QSL state is an exciting possibility in strongly correlated electron physics, but we have to be clear what exactly we are looking for and how to find it.

\subsection{Strange Metals}

\label{sec:strangemetal}
\subsubsection{Definition of Phenomenology}
\begin{figure}
\includegraphics[width=1\columnwidth]{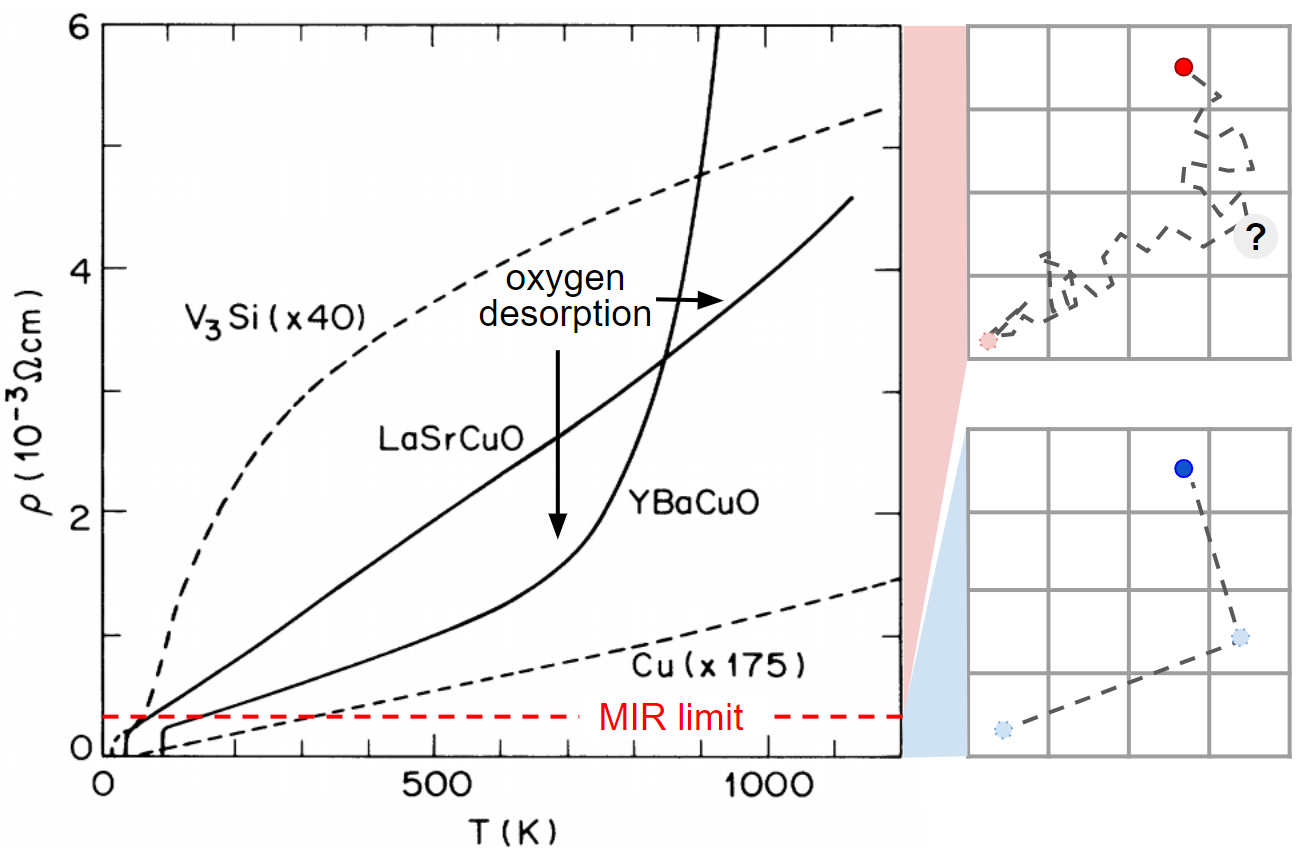}
\caption{Electrical resistivity of normal vs. strange metals. Solid lines are in-plane resistivity of two archetypal cuprate high-temperature superconductors - La$_{2-x}$Sr$_{x}$CuO$_4$ and YBa$_2$Cu$_3$O$_{7-\delta}$ near their respective optimal hole-doping. The upturn at high temperature in YBa$_2$Cu$_3$O$_{7-\delta}$ has been ascribed to oxygen loss. Dashed lines are resistivity curves from normal metals, which either show saturating behavior (V$_3$Si), or maintain a phonon-dominated linear resistivity well below the MIR limit (Cu).  Adapted from Ref.~\cite{gurvitch1987resistivity}. Cartoons on the right represent electron motion with mean free path shorter (top) and longer (bottom) than the primitive unit cell size (gray grids).}
\label{StrangeMetalsFiory}
\end{figure}

The problems presented by ``bad'' and ``strange'' metals are simple to pose, but have withstood an understanding for many decades now.  Early work (see Fig.~\ref{StrangeMetalsFiory}) showed that the normal phase of high-temperature cuprate superconductors exhibits ``bad-metal'' transport signatures -- most notably the violation of the Mott-Ioffe-Regel (MIR) limit $\rho \lesssim \hbar/e^2$ (in 2D) at high temperatures.  The MIR limit corresponds to the notion that metallic transport of electron-like quasiparticles cannot occur with a mean free path much shorter than some microscopic scale of the system ({\it e.g.}  inverse Fermi wavelength, lattice constant)~\cite{gunnarsson2003colloquium}.  Results like these have considerably challenged the Boltzmann or quasiparticle picture of transport in such systems~\cite{gurvitch1987resistivity, hussey2004universality,smreview}. We typically refer to metals as ``bad'' even if the resistivity does not violate the MIR limit, but appears smoothly connected to the regime where it does.  For instance, 19\% doped La$_{2-x}$Sr$_{x}$CuO$_4$ right above $T_c$ has a resistivity that is low enough to not violate MIR, but is smoothly connected to the high temperature regime where MIR is violated~\cite{emery1995superconductivity}.  Furthermore, many of these materials show a ``strange metal'' behavior where the resistivity is linear in temperature and smoothly passes through the MIR value~\cite{emery1995superconductivity}.  The persistence of a linear resistivity at the lowest temperatures when superconductivity is quenched~\cite{daou2009linear, cooper2009anomalous, legros_universal_2019, PhysRevB.80.214531,kuinature} is perhaps one of the most difficult aspects to explain theoretically, due to the lack of a scattering mechanism that gives $\tau\sim 1/T$ at temperatures lower than the Fermi energy, Debye temperature, and other relevant energy scales.  These behaviors contrast with metals whose resistivity saturates near the MIR limit~\cite{gunnarsson2003colloquium,hussey2011dichotomy}.  We note that similar behavior has been recently observed in a cold atom system~\cite{brown2019bad}.  When tuning with magnetic field, possibly related behavior has been found in the cuprates where a linear magnetoresistance is found in magnetic fields up to 80~T, the magnitude of which mirrors the magnitude and doping evolution of the linear in temperature resistivity~\cite{giraldo2018scale}.   Related observations have been made in the pnictides~\cite{hayes2016scaling}.  

Scaling of observables with temperature with unconventional exponents frequently hints at the possible proximity to a quantum critical point (QCP), see also Sec.~\ref{sec:qcp}. Quantum criticality is a recognized framework that features non-quasiparticle transport; however a linear in temperature resistivity evades simple scaling arguments. Indeed, if charge density scales as dimensionality $d$, the Kubo formula for the conductivity leads to $\rho\sim T^{(2-d)/z}$ (where $d$ is the dimension and $z$ is the dynamic critical exponent), at low temperatures~\cite{damle1997nonzero,phillips2005breakdown} (see Ref.~\cite{hartnoll2015scaling} for extensions to scaling theory that accommodate some of the phenomenology of strange metals). Furthermore, although it is tempting to interpret the simple scaling of resistivity with field observed in some strange metals~\cite{hayes2016scaling,sarkarsciadv} as additional signatures of a QCP, recent work has suggested it could be explained classically from macroscopic disorder~\cite{patel2018magnetotransport, boyd2019single}.

While strange metals exhibit their most salient features in transport, their connection to spectral signatures remain less clear. Angle-resolved photoemission spectroscopy (ARPES) experiments in optimally hole-doped cuprates show persistent incoherent spectra near the antinode momenta in the Brillouin zone, with a scaling of widths that is consistent with $\omega/T$.  Such behavior persists even above the pseudogap temperature, indicating its distinction from the pseudogap. A sudden spectroscopic collapse of the strange metal occurs above a critical doping, which also inflicts sudden changes in lower temperature properties such as the pseudogap, the superconductivity, and electron-boson coupling strength~\cite{hashimoto2015direct,he2018rapid,chen2019incoherent}.

An important point is that the transport and single-particle manifestations of strange metals are not obviously equivalent. A good example is the hole-doped cuprate superconductor La$_{\rm 1.6-x}$Nd$_{\rm 0.4}$Sr$_{\rm x}$CuO$_{\rm 4}$ which shows perfect $T$-linear resistivity down to $T = 0$ at a doping right above the pseudogap critical doping $p^*$~\cite{daou2009linear}. Nonetheless, angle-resolved photoemission (ARPES) finds a well defined electron-like Fermi surface~\cite{matt2015electron} and the Wiedemann-Franz law is satisfied~\cite{michon2018wiedemann}.  At the same time, ARPES experiments on overdoped Bi2212 do seem to show a correspondence between the temperatures where sharper features appear in the spectra and an inflection point appears in the resistivity. Therefore, it is not simple to identify which materials should be treated entirely outside of the Fermi liquid framework.

\subsubsection{Outstanding Questions and Perspectives}

In correlated systems in general, and strange metals in particular, there is a general lack of connection between single-particle properties (like ARPES) and multi-particle properties (like electrical transport). When interactions are weak, {\it i.e.} ~non-linearities in the quantum description of the system can be treated perturbatively,  Wick's theorem can be used to reduce multi-body correlation functions into products of single-particle Green's functions, through the use of Feynman diagrams. Development of new multi-particle probes may elucidate the breakdown of Wick's theorem by quantifying non-linearities, or reveal details about interaction effects more generally. Multi-particle photoemission~\cite{haak1978auger,stahl2019noise,su2020coincidence} where the energy and momentum of multiple emitted electrons are measured simultaneously may be very illuminating in this regard (see  Sec. \ref{subsection:methods-spectroscopies}).  The principal issue with such experiments thus far is their relatively poor resolution, mainly limiting the method to getting information on the correlation hole around an electron~\cite{schumann2007mapping}.  If such problems could be overcome, the two-electron removal function would provide direct information, in the cuprates for instance, concerning the development of the superconducting state from the strange metal normal state.  We also believe nonlinear response~\cite{sun2018universal} and direct probes of the dynamic charge susceptibility~\cite{mitrano2018anomalous}) are also promising novel routes to probe the strange metal.

Experimentally, coordinated measurements of the carrier density, resistivity, and other physical parameters (magnetic susceptibility, specific heat, thermal diffusivity, single-particle spectral function) at high temperature near the MIR limit will help elucidate the nature of the dissipation mechanism(s). For example, direct measures of phonon self-energy with inelastic X-ray~\cite{he2018persistent} and neutron scattering~\cite{merritt2019low} can provide a momentum-resolved view of possible lattice dissipation channels. An important issue to manage is the oxygen concentration in cuprates. Because some of the strange metal phenomenology occurs at high temperatures where these systems can be susceptible to oxygen migration, it would be a significant advance for experiments to be able to monitor and control carrier concentration simultaneously. Under a better-defined theoretical framework in the time-domain, ultrafast pump-probe techniques may also help reveal various dissipation mechanisms and even discover new forms of excitations in strange metals.

In addition to the pursuit of a consistent description of the cuprates' strange metal phenomenology seen by different probes, efforts to generalize the description and seek universality in other material systems may provide new perspectives~\cite{stewart2001non, cao2016Hallmarks}. For example, iron-based superconductors exhibit transport phase diagrams that strongly resemble the cuprate strange metal region both with and without magnetic field~\cite{kasahara2010evolution,hayes2016scaling}. Ruddlesden-Popper series materials Sr$_{n+1}$Ru$_{n}$O$_{3n+1}$ offers a platform to study Fermi-liquid-to-strange-metal crossover at a moderate temperature with highly controllable structural motifs~\cite{berg2018theory}.  The recent observation of linear resistivity in magic angle TBG may bring in exceptional tunability and control to the study of electron scattering mechanisms and its relation to superconductivity in this system~\cite{polshyn2019large, cao2020strange}.

Another remarkable feature of strange metals is that they appear to exhibit a transport time scale ($\tau_{\rm tr}$) that is close to saturating conjectured bounds~\cite{sachdev2007quantum,hartnoll2015theory} $\tau_{\rm tr} \geq \alpha \hbar/k_B T$, with $\alpha$ a number of order unity.  While experimental reports of the Planckian limit, including the recently discovered magic-angle TBG, point towards $\alpha \approx 1$~\cite{cao2020strange,bruin_similarity_2013, legros_universal_2019}, it is important to note that no rigorous bound has been established theoretically to date.  A challenge for the coming years is to sharply define $\tau_{\rm tr}$ and derive such a bound, if it exists. In the meantime, it is crucial that experiments attempt to extract transport time scales with as little bias as possible to constrain $\alpha$. One other puzzle is the seemingly similar values of $\alpha$ for a wide range of temperatures -- with resistivities both below and above the MIR value -- despite changes in single-particle observables. This aspect is reproduced in certain models~\cite{mousatov2019bad,cha2019t}, but a deeper understanding remains absent. 

\begin{figure}
\includegraphics[width=1\columnwidth]{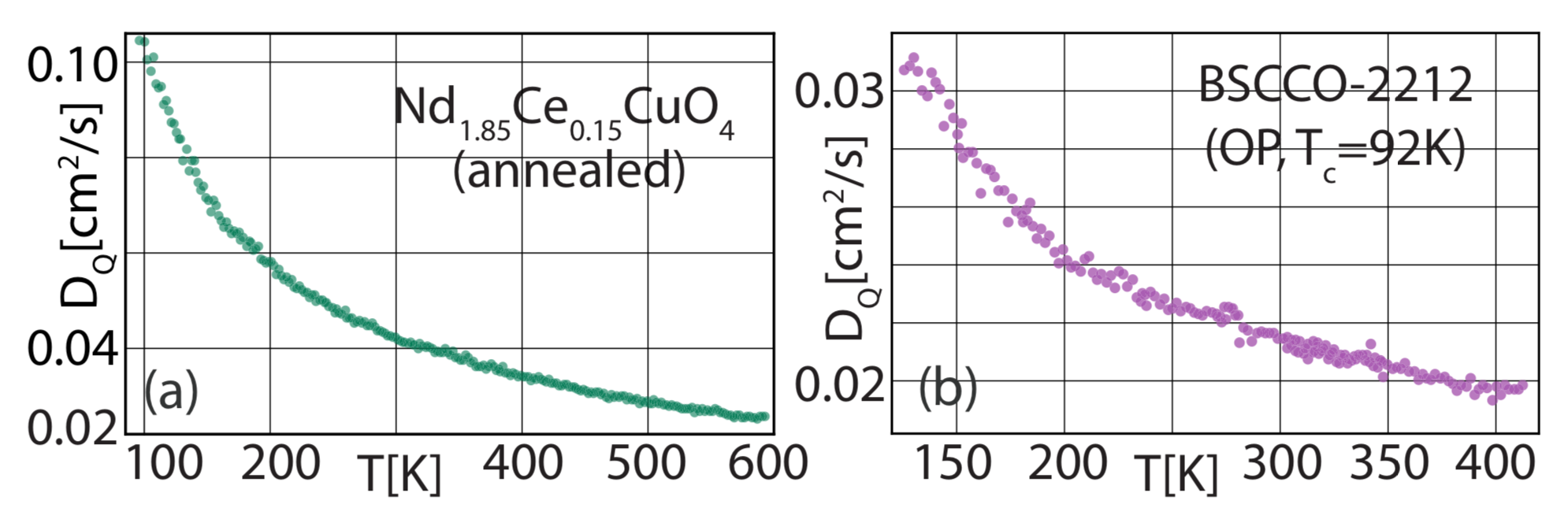}
\caption{Thermal diffusivity in the $ab$ plane as a function of temperature, measured using the optical setup discussed in Ref.~\cite{zhang2017anomalous,zhang_thermal_2019} for Nd$_{1.85}$Ce$_{0.15}$CuO$_4$ (a) and optimally doped Bi$_2$Sr$_2$CaCu$_2$O$_{8+x}$ (b). The data is consistent with a quantum-limited relaxation time that goes as $\hbar/k_B T$.}
\label{ThermalDifussion}
\end{figure}

A promising avenue for further experimental study of $\tau_{\rm tr}$ has been opened up by recent measurements of the thermal diffusivity (Fig.~\ref{ThermalDifussion}) by a novel optical technique wherein a local temperature gradient is established with one laser, and the diffusivity is measured by the reflectivity of a second laser~\cite{zhang_thermal_2019}.  It allows the thermal diffusivity to be obtained directly, without the need to measure the thermal conductivity and specific heat separately.  The diffusivity is connected to the more conventional transport coefficients via the Einstein relations; neglecting thermoelectric effects $\sigma = \chi D_e$ and $\kappa = cD_Q$ (see~\cite{hartnoll2015theory} for a more general relation).  One immediate and striking result of experiments on cuprates has been the observation that, like charge transport, thermal transport also shows a linear dependence on temperature~\cite{zhang_thermal_2019}.  Data at high temperatures is consistent with a thermal transport scenario where entropy is carried by an overdamped diffusive fluid of coupled electrons and phonons characterized by a unique velocity and a quantum-limited relaxation time $\hbar/k_B T$~\cite{zhang2017anomalous}.  The approach to this linear regime is qualitatively consistent with a bound on $\tau_{\rm tr}$, a result that is especially intriguing given that many of the measured cuprates do not show \textit{T}-linear resistivity in the same temperature range.  This apparent wrinkle on the idea of a transport bound highlights the importance of applying novel measurement techniques in the strange metal regime.

As mentioned above, issues related to strange metals are intimately related to those associated with perennial observations of non-Fermi liquid physics.  In some systems, the route to non-Fermi liquid (NFL) behavior seems to involve single ion physics~\cite{Maple10}.  In fact, the first $f$-electron system Y$_{1-x}$U$_x$Pd$_3$ in which NFL behavior was observed exhibits an unconventional Kondo effect in which the Kondo temperature decreases with U concentration due to Fermi level tuning before the system undergoes spin glass ordering at $x\approx0.2$ followed by long-range antiferromagnetic ordering at $x\approx0.4$~\cite{Seaman91,Maple94,Maple95,Maple10}.  The low-T NFL characteristics scale with the Kondo temperature and the U concentration and have anomalous dependencies\footnote{The following dependencies are seen in Y$_{1-x}$U$_x$Pd$_3$.  Electrical resistivity varies as $-T$, the specific heat divided by temperature $C(T)/T$ diverges as $-\log(T)$ with evidence of a residual entropy $S(0) = (1/2)R\log(2)$, and the magnetic susceptibility varies as $T^{1/2}$.}.  Similar weak power law and logarithmic T-dependencies of the NFL characteristics have been found in many $f$-electron systems~\cite{Maple94,Maple95,stewart2001non,Maple10,Loehneysen07}.  Several of these characteristics were consistent with a quadrupolar Kondo effect, a variant of a 2-channel spin-1/2 Kondo effect, which could occur if the U ions were tetravalent and the $f$-electron ground state in the cubic crystal field was a $\Gamma_3$ nonmagnetic doublet which carries a quadrupolar moment~\cite{Cox87,Cox96}.  The quadrupolar Kondo model was first proposed by Cox~\cite{Cox87} to account for the low temperature properties of the heavy fermion compound UBe$_{13}$.  The observation of NFL characteristics in the YUPd$_3$ system came as a surprise and opened a new era of research on NFL behavior in $f$-electron materials.  These observations reinforce the notion that much of NFL behavior and strange metal physics in correlated electron quantum materials arises from the tension between localized and itinerant electron character.  In such materials, the hybridization of localized and itinerant electron states is manifested in complex temperature {\it vs.} composition, pressure and magnetic field phase diagrams consisting of regions in which various correlated electron phenomena and phases reside and, in addition, new phenomena and phases emerge, often in the vicinity of a QCP.   Whether this dichotomy between local and itinerant physics is causing strange metal behavior in general -- is as of yet --  unclear.

Both microscopic and phenomenological approaches will be important in tackling the strange metallic problem theoretically. The central challenge in the phenomenological or effective theory approach is an identification of the appropriate collective excitations and their dynamics, potentially realizing marginal Fermi liquid phenomenology~\cite{varma1989phenomenology}. For example, hydrodynamics assumes that only excitations related to exact or approximate conservation laws survive after fast local thermalization, and thereby directly connects response functions to exact or approximate symmetry assumptions~\cite{davison2014holographic,delacretaz2017bad,lucas2017resistivity}. Constructions based on the Sachdev-Ye Kitaev model have provided novel microscopic approaches to strange metallicity~\cite{song2017strongly,PhysRevX.8.031024,patel2018magnetotransport}. A challenge of any microscopic theory is the emergence of an energy scale $\frac{h}{e^2}T \sigma_{\rm dc}$ that is far below the Fermi energy ({\it e.g.} ~the size of small Fermi pockets~\cite{berg2018theory}) and a mechanism for $T$-linear transport time down to low temperatures (see {\it e.g.} ~Ref.~\cite{patel2019theory} for a recent example in the context of the SYK model). And ultimately, any microscopic theory should give insight into why some metals are strange and some are not.

\subsection{Quantum Criticality}
\label{sec:qcp}

\subsubsection{Status of the field}

Quantum criticality is found in many correlated electron materials and proximity to a quantum critical point has been proposed to be an organizing principle for many strongly interacting systems.   The point of instability between two competing ground states, which each have their own quasiparticle spectrum can have an excitation spectrum that is non-quasiparticle like.   Although quantum criticality has been a focus of the field for many years there are still many open issues theoretically and experimentally.  Conventionally, it has been well-described within the Landau-Ginzburg-Wilson (LGW) paradigm~\cite{sondhi1997continuous,sachdev2007quantum,ma2018modern}, where the renormalization group method has been extended to capture dynamics of order parameters at a continuous transition into a broken symmetry state. However, the LGW framework fails to address the case of a continuous transition between two distinctly ordered phases~\cite{senthil2004deconfined} and provides no description of phases and phase transitions that are not characterized by a symmetry breaking (with the possible exception of the 2D metal-insulator~\cite{abrahams2001metallic} transition in disordered metals where the critical modes may be identified).  And although proximate quantum criticality is frequently invoked to explain anomalous scaling of exponents in strongly interacting metals, there is still no accepted framework for quantum criticality in magnetic heavy fermion metals.  Additionally, the role of disorder in many systems is still unclear.  As proximity to a QCP is frequently used as a ``go-to'' explanation for anomalous system properties, the community would also benefit from having more simple materials to investigate paradigms of quantum criticality.  One could then understand the range of parameter space that is affected by a proximate QCP and get better intuition about prospective signatures of quantum criticality in other systems.  For detailed reviews of the conventional understanding of quantum criticality, we refer readers to Refs.~\cite{sondhi1997continuous,sachdev.2008, ma2018modern, carr2010understanding,gegenwart2008quantum,si2013quantum}.

The limitations of LGW have been demonstrated in model systems and material examples. A prominent example involves itinerant metallic systems where the conducting electrons couple to the critical fluctuations of the order parameter. The gapless nature of particle-hole excitations in this case may dramatically alter the nature of the QCP. For example, such a coupling directly renders the transition first-order in ferromagnetic metals~\cite{belitz1999first} and magnetic Dirac semimetals~\cite{belitz2019magnetic}. On the other hand, quantum phase transitions associated with antiferromagnetic order or nematic order can be continuous~\cite{sondhi1997continuous}.  The critical theory for metallic systems with antiferromagnetic or nematic ordering is relatively well understood in three dimensions~\cite{sachdev1999quantum}.  In contrast, in two dimensions the critical coupling is strongly relevant and a full understanding of the critical behavior is still elusive~\cite{sachdev.2008}. There are attempts to perform controlled calculations introducing additional control parameters, such as the number of electronic species or flavors, which predicts non-Fermi-liquid electronic behavior~\cite{lee2018recent}. The self-energy of electrons is shown to scale as $\Sigma(\omega)\sim \omega^{2/3}$ for the nematic transition and $\sim \sqrt{\omega}$ for the antiferromagnetic one, leading to the breakdown of the quasi-particle picture.  An important frontier here is going beyond these approximations both analytically and numerically. While some progress has been achieved~\cite{lee2018recent,berg2019monte}, it is so far restricted to a few special cases. Developing more general frameworks is thus the work for the future.

Experimentally, strange metal behavior, such as linear-in-temperature resistivity has been observed close to QCPs in several systems including heavy fermions, cuprates and iron-based superconductors~\cite{lohneysen1994non,legros_universal_2019,shibauchi2014quantum}. Although such behavior seems to be associated with a `phase' (see Sec. \ref{sec:strangemetal} above), one reoccurring question is whether such non-Fermi-liquid behaviors can be explained by the framework with a Fermi surface coupled to a critical mode as described above~\cite{schofield1999non}. In particular, for heavy-fermion systems there is frequently a clear critical point between an antiferromagnetic metal with small Fermi surfaces and the heavy Fermi liquid with large Fermi surfaces. Thus naively one may think the critical point is just the onset of the antiferromagnetic order~\cite{si2010heavy}, but the standard Hertz-Millis theory of an antiferromagnetic critical point~\cite{sachdev1999quantum} fails to explain the NFL behavior observed in experiment. Meanwhile, a jump of Fermi surface volume has also been observed, for example in a pressure-tuned critical point in CeRhIn$_5$ through quantum oscillation measurements~\cite{shishido2005drastic}.  This again cannot be explained simply by the onset of magnetic ordering.

Although as mentioned antiferromagnetic transitions in 3D are frequently mean-field-like, exotic scaling has been predicted at the QCP in nodal metals where antiferromagnetic fluctuations and electrons at a nodal point are strongly coupled.  In such cases, the coupling between the electronic and the critical modes can qualitatively change the critical behaviors compared to the pure ordering transition without itinerant electrons. This type of QCP, proposed for the pyrochlore iridates, is beyond the current experimental scope, although there may be indirect evidence from symmetry breaking in a quantum phase transition in Cd$_2$Os$_2$O$_7$~\cite{savary2014new,wang2018strongly}. 

Heavy fermion metals are replete with systems that seem to evade the simplest considerations for criticality.  For instance, in the case of 
the quantum critical point of CeCu$_{6-x}$Au$_x$, which goes from a paramagnetic metal to an antiferromagnetic metal as $x$ increases through a critical value, $x_c \approx 0.1$, inelastic neutron scattering has shown that critical scattering with $\omega/T$ scaling occurs all over the Brillouin zone (instead of just at antiferromagnetic wavevectors)~\cite{schroder1998scaling}.   This is in contrast to the usual notion that when a metal undergoes an antiferromagnetic quantum phase transition, fluctuations induced by quantum criticality are taken to be long-wavelength fluctuations of the order parameter at the ordering wave vector.

New theoretical frameworks may therefore be necessary to understand such physics. One theoretical scenario is that the heavy fermion critical point is associated with ``Kondo breakdown'' instead of the onset of antiferromagnetic order~\cite{senthil2004weak,si.2010,coleman.2010}.  The key idea is that one electron per site gets ``Mott'' localized to form a local spin moment and only the remaining electrons can move coherently, resulting in a sudden drop of carrier density. So far there has been only moderate progress in theories where this Kondo breakdown and the onset of antiferromagnetism can happen simultaneously~\cite{khait2018doped,komijani.2019}.   A separate mechanism has been postulated in the form of spin-charge separation~\cite{coleman2001fermi}. An agreed upon treatment for the realistic 2D or 3D systems is currently lacking. The drop of carrier density from $1+p$ to $p$ ($p$ being the hole doping) has also been observed in hole-doped cuprates below the pseudogap critical point $p^*$ under high magnetic field~\cite{proust2019remarkable}. In this case, no long-range antiferromagnetic order has been observed below the critical point, which suggests a different mechanism of reconstructing the Fermi surface without involving symmetry breaking order parameters. 

In the framework of ``local” quantum criticality~\cite{si2001locally}, the Kondo effect is destroyed because local moments are coupled not just to conduction electrons but also to the fluctuations of the other moments.  The destruction of the Kondo effect leads to the vanishing of quasiparticle weight (and hence a divergent effective mass) on the entire Fermi surface.  In contrast, in an antiferromagnetic QCP, the quasiparticle spectral weight vanishes only near the ``hot spots” ({\it e.g.}  the portions of the Fermi surface that are connected by the antiferromagnetic wave vector).  This does not lead to anomalous transport as the current carried by ``cold'' electrons short circuits the ones in the hot spot.

One may expect related physics underlies the heavy fermion and cuprates critical points. An interesting observation is that both heavy fermions and cuprates are in the ``Mott'' limit where the large Hubbard $U$ induces a constraint on the Hilbert space by forbidding double occupancy. Therefore a framework taking into account Mott physics may be necessary to give an understanding of both the phases and the critical region. For example, a slave boson theory has been proposed to explicitly respect the restriction on the Hilbert space~\cite{lee2006doping}. One specific continuous Kondo breakdown transition can be successfully described using this slave boson framework~\cite{senthil2004weak}. In this theory, carrier density indeed drops across the critical point.  But the theory fails to predict the magnetic ordering onset at the same critical point.  Nevertheless, the partial success of the slave boson theory is encouraging and suggests that a language which captures the Mott physics may be the key to understanding these exotic phases and unconventional critical points found in heavy fermion systems and in cuprates. It is worth noting that several theories were able to reproduce the drop in carrier density observed in hole-doped cuprates at the opening of the pseudogap: models based on an antiferromagnetic QCP \cite{storey2016hall, chatterjee2017thermal, verret2017phenomenological}, and other scenarios for the pseudogap involving a Fermi surface reconstruction in the Yang-Rice-Zhang theory \cite{yang2006phenomenological, verret2017subgap}, the SU(2) fluctuations theory \cite{morice2017evolution} or the FL$^*$ theory \cite{chatterjee2016fractionalized}.

The carrier density drop becomes even more acute in metal-insulator transitions. In a clean system, at integer filling a metal insulator transition can be driven by increasing the interaction strength or decreasing the bandwidth. Due to the lack of a broken symmetry order parameter, such a transition is generically beyond LGW theory and study of this transition may provide more intuition for the more intricate metal-metal transition in heavy fermion systems. A pressure-tuned metal-insulator transition has been found in organic materials~\cite{furukawa2015quantum}, but the role of disorder and inhomogeneity may need to be considered (this is discussed in a somewhat different context below). The recently discovered Moir\'e systems may be a powerful platform to explore this physics.  In several systems (such as ABC trilayer graphene aligned with hexagon boron nitride~\cite{Wang2019Evidence}, TBG~\cite{Cao2019Electric,Shen2019Observation,Liu2019Spin, burg2019correlated} and twisted transition metal dichalcogenides~\cite{wang2019magic}), both the density and the bandwidth can be gate controlled, which makes it much easier to sweep the phase diagram and study phase transitions than in traditional solid state systems.

The 2D superconductor-insulator transition has been considered to be an important model system for quantum phase transitions~\cite{goldman1998superconductor}.  In fact, much of our intuition about what happens in the LGW perspective on quantum criticality has been developed considering bosonic models of this transition~\cite{goldman1998superconductor, wallin1994superconductor}.   Experimentally, one expects that by destroying superconductivity in a 2D thin film with, for example, an applied magnetic field or disorder, a direct transition to an insulating state at $T=0$ occurs. However, recent experiments suggest that this is not the full story.  In fact, most experimental systems actually show a zero temperature transition from the superconducting state to an anomalous metallic phase which has a resistance that is much lower than in the normal state~\cite{kapitulnik2019colloquium}, before ultimately becoming insulating at even higher disorder or fields. Involving at least three phases not readily distinguished by symmetry, this transition will likely require theories beyond the LGW paradigm to explain.

Finally, an additional important point is the possibility of the development of a secondary order near the QCP, the most prominent example of this being superconductivity~\cite{scalapino.2012} that frequently occurs near magnetic critical points in materials like electron-doped cuprates, iron pnictides, and heavy fermions. In this case, the experimental observation of quantum critical scaling would require suppressing the secondary order, which may make, in some cases, experiments under extreme conditions necessary (such as very high magnetic fields in the case of cuprates). This raises the question of whether the observed scaling is affected by these conditions. On the other hand, the emergence of superconductive pairing at the QCP represents an important open problem by itself. While the critical fluctuations may mediate attraction, but on the other hand they can destroy the coherence of the quasiparticles, exemplified by the `strange metal' regime, perhaps rendering the ordinary BCS, Migdal-Eliashberg approach inapplicable. Thus, it is possible to imagine that coherence is lost to such an extent that superconductivity never occurs. Indeed, the ultimate fate of the competition between NFL phenomena and manifestations of an ordered state is a topic of current interest~\cite{raghu2015metallic,wang2016superconductivity}.

\subsubsection{New frameworks}

The above experimental discoveries of unusual metal-insulator and metal-metal transitions clearly indicate that we need new theoretical frameworks.  In the past two decades, one new critical theory beyond the LGW framework that was developed was that of deconfined quantum criticality~\cite{senthil2004deconfined}.  It involves fractionalized degrees of freedom and emergent gauge fields in its description. By employing these degrees of freedoms, one can describe several  continuous transitions which are now allowed in the Landau framework.  Although there has been little experimental evidence for deconfined quantum criticality (DQC), there are some candidate materials~\cite{Zayed2017, LeeQED2018, LeeShastry2019, GuoShastry2019,zou2020field}. In particular, the famous Kitaev material $\alpha$-RuCl$_3$~\cite{kasahara2018majorana,laurell2020dynamical,PhysRevB.100.075110} may host such a deconfined transition between the field-induced N\'{e}el state to putative Ising topological spin liquid phase~\cite{zou2020field}.  It was also recently claimed~\cite{ lee2019signatures} that there was a deconfined quantum phase transition in the pressure tuned transition of the Shastry-Sutherland lattice compound SrCu$_2$(BO$_3$)$_2$~\cite{Zayed2017}.  More investigations in this regard would be interesting.

To make connections with the experimentally relevant metal-metal transitions in heavy fermion systems or in cuprates, a more sophisticated generalization of the existing deconfined QCP is needed. Most theoretically explored realizations of deconfined quantum criticality, however are restricted to insulator-insulator transitions.  There are attempts to describe metal-insulator transitions and metal-metal transitions with carrier density drop using fractionalized degrees of freedom~\cite{senthil2004weak,senthil2008theory}.  It remains to be seen whether these attempts can lead to successful explanation of the mysterious non-Fermi liquid phases in heavy fermion systems, cuprates and other strongly correlated materials.

Although disorder effects broadly exist in correlated systems, they become acutely important near QCPs where susceptibilities tend to diverge~\cite{vojta2019disorder}. For example, they can fundamentally modify the scaling properties in a second order phase transition~\cite{harris1974effect}, or stabilize a QCP by rounding a first-order phase transition~\cite{goswami2008rounding}.  To better connect experimental and theoretical understandings of QCPs and the effects of disorder,  we propose the following workflow:
\begin{enumerate}
\item Identify types and levels of disorders. Typically, experiments only estimate an overall quantitative level of disorder.  The qualitative nature of the disorder, for example the distribution of impurities or defects, the shape and size of extended defects, etc.\ may also play an important role in how it affects the physical observables. In particular, many spectroscopic measurements close to a QCP involve spatial averaging over large scales ({\it e.g.} , X-ray absorption, angle-resolved photoemission). More detailed local experimental probes may be needed ({\it e.g.} , electron energy loss spectroscopy, scanning tunneling spectroscopy), which can identify and characterize different kinds of disorder on smaller length scales.
\item Estimate effective impurity potentials with ab-initio analysis. Based on the experimental findings from local probes, the next step is to derive effective descriptions that capture the specific qualitative nature of disorder in each case of interest. 
\item Perform model calculations (numerical and analytical) that take as input the specific effective impurity potentials derived from experiments in the previous step, and compare them with analogous calculations with or without different types of disorder. This will either reveal a sensitivity of the systems to disorder, which would indicate that disorder is an important factor that affects the physical properties near a QCP, or confirm that the spatial averaging over large scales in spectroscopy measurements does not result in the loss of relevant information. Such calculations could also reveal possibilities to tune correlated materials via disorder. 
\item Identify better model material systems for quantum criticality for which both the canonical theory and extensions to it can be tested.
\end{enumerate}

From the numerical perspective~\cite{Xu_2019}, development of controlled techniques with different implementations of disorder may be important to understand the effects of impurities beyond the DFT level, which may be required in correlated materials. Analytical theory, on the other hand, can move further in the prediction of dynamical and non-equilibrium responses. Nonetheless, predicting dynamical and static quantities in the presence of known impurities allows a more rational comparison with experiments. Further identification of intrinsic behavior and understanding disorder effects can contribute to the ultimate goal of quantum material design in a controllable way.

Further establishing quantum criticality beyond the LGW regime and getting further insight into this physics poses significant challenges for experiments. Currently, a plethora of experimental techniques have been developed to probe order parameters and resolve the phase boundary in the multiple-dimensional phase space of pressure, chemical doping, magnetic field, etc., but more sophisticated and comprehensive techniques to resolve symmetry, topology, dynamic correlations, and scaling laws are needed. The development of these experimental tools is not specific to quantum criticality, and will have broad applications across the study of correlated phenomena.

\subsection{Correlated Topological Matter}\label{sec:corrtopmatter}

\begin{figure}
\includegraphics[]{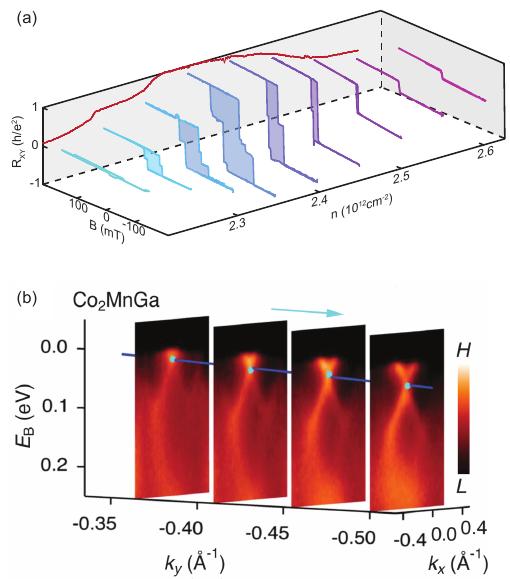}
\caption{(a) Intrinsic quantum anomalous Hall effect in twisted bilayer graphene, as adapted from Ref.~\cite{serlin2020intrinsic}.  R$_{xy}$ as a function of field and density. Hysteresis loop areas are shaded for clarity. The rear wall shows field-training symmetrized values of R$_{xy}$ at B = 0. R$_{xy}(0)$ becomes nonzero when ferromagnetism appears, and it reaches a plateau of $h/e^2$ near a density of n = 2.37 $\times$ 10$^{12}$/cm$^2$. (b) Angle-resolved photoemission spectrum of a nodal-line degeneracy of Co$_2$MnGa, a ferromagnetic Weyl semimetal candidate. From Ref.~\cite{Belopolski1278}}.
\label{CorrelatedTopology}
\end{figure}

The last decade has seen tremendous theoretical and experimental activity at the intersection of band theory and algebraic topology.  Substantial progress has been made in identifying topological insulators and (semi)metals that can in principle be realized by systems of noninteracting (free) fermions~\cite{hasan2010colloquium,hasan2011three,qi2011topological,armitage2018weyl,kitaev2009aip,ryu2010topological,PhysRevB.85.085103,watanabe2018structure}.  It is believe that free-fermion topological insulators can be stable to the perturbative inclusion of many-body interactions~\cite{Fidkowski_2010} {\it e.g.} , the surface states of a 3D $\mathbb{Z}_2$ topological insulator form a 2D Fermi liquid.  The focus has now started to be on topological states of matter, where interactions are important (Here we will not consider topological aspects of systems which can also be characterized as quantum spin liquids.).  There are at least two kinds of these systems.  In the first possibility, interactions drive an ordered state the properties of which determine aspects of the topology, but ultimately a free fermion with little residual interactions description still applies.  Chern insulators (in 2D), axion insulators (in 3D), and magnetic Weyl semimetals (Fig.~\ref{CorrelatedTopology}(b)) are of this category.   In the second possibility, interactions drive systems into a state that has no non-interacting analog~\cite{Wang_2014}.  Here, there are no known examples where this occurs spontaneously, but the fractional quantum Hall effect provides an example for the kinds of effects that could exist.  There may be also be systems that straddle these cases, where the systems are ``like'' those of the non-interacting ones, but perhaps have remaining large residual interactions.  This may be exemplified by the large intra-atomic Coulomb energy characteristic of narrow 4$f$ bands in the heavy-fermion, topological Kondo insulators~\cite{dzero2010topological} or Kondo-Weyl semimetals~\cite{lai2018weyl}.

One ongoing thrust lies in introducing correlations to topological metals. The resultant correlated phase of matter is not necessarily topological but it often has rich physics. As exemplified by (TaSe$_4$)$_2$I, charge-density-wave correlations in a Weyl semimetal result in an insulator whose electromagnetic response mimics axion electrodynamics~\cite{gooth2019axionic}. Introducing pairing correlations to topological metals is also known theoretically to lead to unconventional superconductivity, e.g, on the Fermi surface of magnetic Weyl metals, the nontrivial Chern number necessitates nodes in the superconducting order parameter~\cite{murakami2003berry,li2018topological}.  It would be especially interesting if superconducting correlations are discovered in the Kondo-Weyl semimetals -- this would provide a platform to investigate the interplay of Kondo physics, topology and superconductivity. 

The rigorous classification of interacting topological matter is incomplete. Ongoing and future efforts should focus on constructing physically realistic models (perhaps by beginning with minimal, exactly-soluble models), and identifying many-body topological invariants which are translatable to many-body physical observables. Some of these observables exist on the boundaries of samples, which favor either surface-sensitive experimental probes such as photoemission spectroscopy,~\cite{Belopolski1278,alex2019glideresolved} or mixed-bulk-surface probes such as quantum oscillations from spatially-nonlocal cyclotron orbits~\cite{moll2016transport}. However, not all topological matter has a bulk-boundary correspondence~\cite{SchmidtPRB2012,DellAnnaNJP2017,WangPRL2018,XiongJPC2018,TobiasArxiv2019}. Bulk observables may include thermodynamic quantities such as specific heat, the temperature dependence of which can in principle identify a Kondo-Weyl semimetal~\cite{dzsaber2017kondo}. There has been partial success in exploring the effect of disorder on the stability of many-body topological phases~\cite{PhysRevB.96.201104,PhysRevB.73.045322}. Rigorous general results are still lacking, however, and such a dependence may not exist for a subset of topological phases whose robustness rely on crystallographic spatial symmetries.  A notion of out-of-equilibrium topological matter is also developing~\cite{PhysRevX.3.031005,PhysRevX.6.021013,PhysRevB.100.085138,PhysRevB.97.205402,doi:10.1002/andp.201900336,PhysRevLett.121.036402,PhysRevB.98.041113,rudner2019floquet,Cayssol_2013,RobertPRB2019,lindner2011floquet,WuPRL2018} and promises to be a rich platform to explore the interplay between correlation effects, topology and many-body localization in periodically-driven Floquet systems~\cite{OkaReview2019}. It is known that some out-of-equilibrium phases of matter have no equilibrium counterpart~\cite{RobertPRB2019,LindnerNature2011,WuPRL2018}.

Progress in the theoretical understanding of correlated insulators ({\it e.g.}  TBG) has demonstrated an insightful relation between single-particle band topology and many-body correlations; this relation has been missed in previous formulations of effective Hamiltonians for interacting electrons. Namely, it is increasingly recognized that not all crystallographic spacetime symmetries can be imposed locally on the Wannier functions of a band, owing to a topological obstruction~\cite{PhysRevLett.121.126402}. Despite being exponentially-localized in real space, such Wannier functions cannot be localized to a single atomic site~\cite{alex2019crystallographic}, unlike the traditional atomic orbital. These atypical Wannier functions result in non-standard terms in the (generalized) Hubbard model~\cite{PhysRevX.8.031088}, with exotic correlated ground states such as an SU(4) ferromagnet predicted~\cite{PhysRevLett.122.246401}.  These issues raise interesting questions about constructing tight-binding Hubbard-like model and considerations about what one might consider to be the strong coupling limit of such models.  We speculate that these topics will be important in future studies.

A current bottleneck in the field is that there are very few material candidates of strongly interacting topological matter.  An emerging material platform likely to gain more traction is 2D multi-layer heterostructures with van der Waals inter-layer coupling~\cite{geim2013van,ajayan2016van}. Their advantage over 3D materials lies in enhanced tunablity through gating, stacking, and twisting. The latter results in artificial Moir\'e superlattices that can realize strongly-correlated, fractional Chern insulators in fractionally-filled Hofstadter bands~\cite{spanton2018observation}. Such systems raise the enticing possibility to probe non-abelian quasiparticle statistics. A different class of Moir\'e superlattices in TBG realizes correlated insulating phases that spontaneously break spin- and valley- symmetries, resulting in an intrinsic quantized anomalous Hall effect~\cite{serlin2020intrinsic} (see Fig.~\ref{CorrelatedTopology}(a)). The quick realization of this effect in such a relatively clean system (see also \onlinecite{deng2020quantum}) may be juxtaposed with the same effect realized in (Bi,Sb)$_2$Te$_3$ only after years of optimizing the platform material~\cite{chang2013experimental}.   This suggests that disorder is an experimental barrier to discovering correlated topological materials.

The lack of material candidates is especially acute for topological superconductors~\cite{sato2017topological}.   There are as of yet little convincing evidence for the original proposal of a proximity effect driven 2D superconductor~\cite{fu2008superconducting}.  Thus far topological superconductivity seems to have been best realized in bulk materials like the iron-based superconductor Fe Se$_{1-x}$Te$_x$~\cite{zhang2018observation} or in 1D wires~\cite{mourik2012signatures,zhang2019next}.  However, twisted van der Waals heterostructures are also promising avenues to realize topological superconductivity~\cite{PhysRevLett.121.087001}. Beside establishing concrete material candidates, it will be crucial to further develop experimental techniques to identify topological superconductors, such as local probes ({\it e.g.} , scanning nano-SQUID), which can detect Majorana edge states, as well as bulk probes ({\it e.g.} , nuclear magnetic resonance~\cite{pustogow2019constraints}) to constrain the superconducting order parameter. Future theories would hopefully establish why a particular order parameter is energetically favored; sometimes the reasons can be established without reference to a specific pairing mechanism or a detailed microscopic model.  One particularly interesting example of topological superconductivity is monopole superconductivity in which the Berry phase structure of a magnetic Weyl system ensures an superconducting order parameter with nodes independent of the mechanism~\cite{murakami2003berry,li2018topological}.

An additional issue in this area is that of finding definitive signatures of strongly correlated topological insulators.  It has been proposed that topological insulators are best characterized not as surface conductors, but as bulk magnetoelectrics~\cite{qi2011topological} with a quantized magnetoelectric response coefficient whose size is set by the fine structure constant.  This magnetoelectric effect was observed in the free fermion systems of Bi$_2$Se$_3$ through measurements of a quantized Faraday rotation. As alluded to above, one possibility in the case of the strongly interacting topological phases, is the prospect that 3D analogs of 2D fractional quantum Hall phases could be realized. In the same manner as non-interacting topological insulators are expected to show a magnetoelectric effect quantized in units of the fine structure constant, such fractional topological insulators may be expected to show a magnetoelectric effect that is quantized in rational fractions of the fine structure constant~\cite{swingle2011correlated,maciejko2010fractional,maciejko2015fractionalized}. Such a fractional phase may be uniquely identified by this fractional magnetoelectric effect.   However, a fractional quantized Faraday effect will give a signal smaller that the precision of state-of-the-art THz polarimetry and so new instrumentation may have to be developed.

\subsection{Revisiting old materials in a modern context}

While much of the current research on strongly-correlated electrons focuses on newly developed materials platforms, it is worth considering the value of ``old materials” (or ``legacy materials”) in the context of the strongly-correlated electron problem. Here, by an ``old material” we mean a condensed matter system that was studied for a time by the solid state physics community and then largely abandoned as a research field, perhaps many decades ago.

Many old materials are worth revisiting, as in the intervening decades we have developed both new theoretical ideas and better experimental probes. Bringing these probes to old materials often has the advantage of a greater wealth of expertise about materials synthesis and purification, as compared to more recently-developed materials platforms. At the very least, present-day researchers may find it advantageous to examine old and well-studied materials as test beds for calibrating new experimental methods.  Examples of old materials with interesting correlated electron physics can be found across the spectrum of material types: metals, semimetals, and semiconductors.  What follows is an illustrative, but far from complete, set of examples. The allure of these examples lies either in the materials demonstrating some unique phenomenon, or in the similarity of some of their properties to those of a more complicated material class (such as high-\textit{$T_c$} superconductivity or strange metal behavior).

\subsubsection{Metals} For examples of interesting electron physics, one need not look farther than the elemental metals. Iron is perhaps the simplest example of a magnetic metal, in which magnetism develops in a material with itinerant electrons rather than in an insulator with localized magnetic moments.  The magnetism has both itinerant and localized character~\cite{Stearns1978,iron-review,moriya1984itinerant,ironbook}.  Similar coexistence of magnetism with metallicity can be found in the iron oxides, such as magnetite~\cite{magnetite1, magnetite2, magnetite3}.  This type of magnetism has so far eluded a complete theoretical description (see {\it e.g.} ~\cite{itinerantmagnetismbook}).

Even in the absence of magnetism or superconductivity, one can find unconventional transport properties on display in a number of metals, which may give insight into the correlated electron problem. For example, the phenomenon of linear magnetoresistance has recently attracted significant attention due to its appearance in a variety of topological semimetals~\cite{Mr-HgTe, MR-thinTI, MR-TIs, MR-YPtBi, MR-WTe2, MR-SmB6, MR-Cd3As2, MR-Bi2Se3}, as well as in strange metals~\cite{hayes2016scaling,sarkarsciadv}.  Yet this effect is on display even in pure potassium, which is ostensibly one of the simplest metals, with a nearly perfectly spherical Fermi surface~\cite{PotassiumMR2, PotassiumMR1}. Its postulated origin from the formation of a charge-density-wave~\cite{PotassiumMR1} has not been verified, but linear magnetoresistance indeed has been observed in a broad family of density-wave materials, which can generically arise in a partially gapped Fermi surface with sharp corners~\cite{feng2019linear}. This mechanism, first pointed out by Pippard~\cite{pippard1989magnetoresistance}, could also play a role in magnetoresistance of topological semimetals.

Perhaps even more interesting are the liquid metals, which are good conductors for which there is no crystalline order, and therefore no notions of traditional electron or phonon bands (examples include Hg, Ga, Rb, and various alloys, all of which have a melting temperature near or below room temperature). Liquid metals may therefore constitute excellent test beds for the idea of a ``Planckian'' bound of dissipation (see, {\it e.g.} , Ref.\ \onlinecite{hartnoll2015theory} and Sec. \ref{sec:strangemetal}), where the transport scattering time for charge carriers approaches a maximal value and there may be no Fermi liquid-type quasiparticles.  Recent experiments have shown that liquid metals exhibit large linear magnetoresistance, which is present only in the liquid phase~\cite{liquidmetalMR}.

\subsubsection{ Semimetals} A semimetal is a material in which both an electron and a hole band coincide with or are near to the Fermi level; the existence of semimetals has been understood theoretically since the 1930s (see, {\it e.g.} , \onlinecite{Herringsemimetals}).  Perhaps the most ubiquitously-studied semimetal is graphite, whose band structure has been known since the 1940s~\cite{graphitebands1, graphitebands2}. While graphite seems to exhibit no strongly-correlated physics at zero field, owing to its relatively high band velocities, a magnetic field can quench the electron kinetic energy and greatly increase the role of interactions. A recent investigation using modern pulsed magnetic fields up to 90~T, suggests that graphite may be driven into an excitonic insulator phase~\cite{graphitehighfield}.

\begin{figure}
\includegraphics[width=\columnwidth]{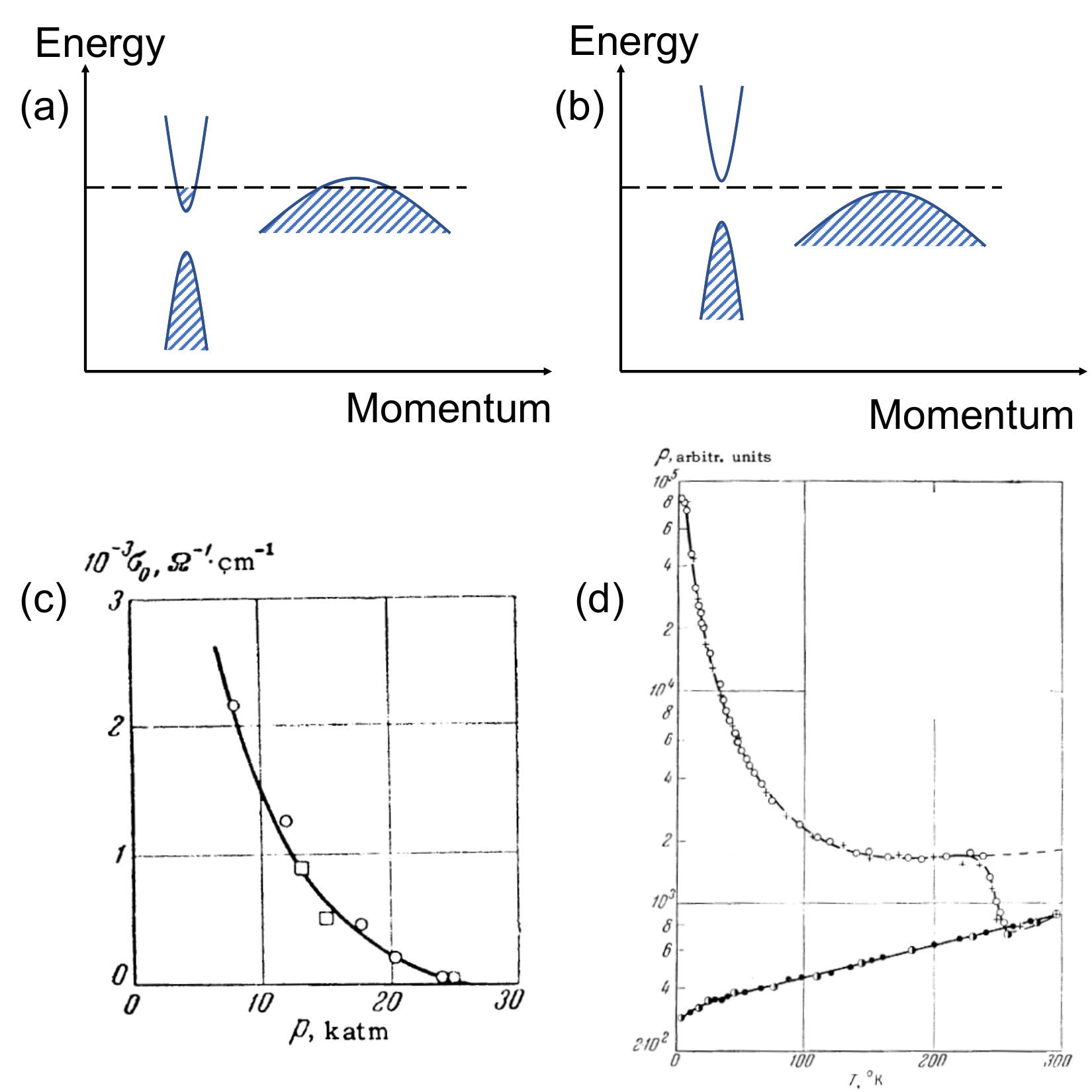}
\caption{The metal-to-insulator transition in elemental bismuth as a function of pressure. (a) Under ambient conditions, the band structure of bismuth is such that both electron and hole pockets intersect the chemical potential (dashed line).  (b) Under high pressure, the electron pockets move up in energy and the hole pockets move down, which produces a metal-insulator transition.
(c) The low-temperature conductivity as a function of pressure (Figure from Ref. \onlinecite{JETPbismuth}), taken at $T \approx 2$\,K).  (d) The transition is evident in the dependence of the resistivity (vertical axis) on temperature (horizontal axis).  The lower curve shows data for a sample at ambient pressure, which has a metallic-like temperature dependence, while the upper curve shows data for a sample under $24,500$\,atm. of pressure, which has an insulating-like temperature dependence. (From Ref. \onlinecite{JETPbismuth}.)
}
\label{fig:boring_bismuth}
\end{figure}

Experimental studies of crystalline bismuth also have a long history.  Large bismuth crystals can be grown with extremely high purity and high electron mobility. In bismuth, light electron bands in three valleys coexist with a heavier hole band, with the electron and hole concentrations being nearly equal and opposite, each on the order of $10^{17}$\,cm$^{-3}$~\cite{bismuthbands}.
This ultralow electron density implies a low Fermi energy in each band, and hence the possibility for electron interactions to play a large role. A magnetic field, in particular, may easily drive crystalline bismuth into a more strongly-interacting phase. For example, one can reach the ``extreme quantum limit'' of magnetic field, in which only a single Landau level in each band is occupied and strongly-correlated states are expected~\cite{halperinhighfieldreview}, using only a few Tesla.  Bismuth has been purported to exhibit valley ferromagnetism at sufficiently high fields~\cite{li2008phase}.  The transport properties (and, in particular, the thermoelectric properties) in a magnetic field remain incompletely understood, despite being studied as early as the 1970s (see, {\it e.g.} , Ref.\ \onlinecite{heremansbismuth}).   Pressure also drives the charge density of bismuth lower and through a metal-insulator transition near 25~kbar~\cite{JETPbismuth, brandt1969pressure}.   As pressure is applied, the electron band moves up in energy and the hole bands move down (as illustrated in Fig.\ \ref{fig:boring_bismuth}), which in this self-compensated material reduces the charge density~\cite{armitage2010infrared}.   The charge density has been inferred to go to zero near 25~kbar, although the nature of the resulting metal-insulator transition is unclear and a simple non-interacting Lifshitz-like transition seems unlikely.   Strong correlations and strongly dressed plasmaron quasiparticles (strongly coupled electron-plasmon composites) were inferred from infrared spectroscopy under pressure near the metal-insulator transition~\cite{tediosi2007charge,armitage2010infrared}.

A poster-child for the value of reconsidering old semimetals is the Kondo insulator SmB$_6$. This material was discovered over 50 years ago, and was the first identified as a ``Kondo insulator'' -- a material for which a heavy electron band (with ``nearly localized'' magnetic moments) hybridizes with another light one to produce a gap at the Fermi level~\cite{SmB61, SmB62, SmB63}. Reconsideration of this material in the 21st century led to the realization that the Kondo band gap might be topological in nature, and the corresponding topological surface states helped to explain a 40-year-old puzzle about the saturation of the electrical resistance at low temperature~\cite{dzero2010topological,TKIreview}.  Recent observation of quantum oscillations in SmB$_6$ increased attention because of their suggestion of a charge-neutral Fermi surface~\cite{SebastianSmB6}, although this remains controversial~\cite{li2020emergent}.  It has also been known that the Sommerfeld coefficient of the specific heat of SmB$_6$ is unusually large, $\gamma\approx10$~mJ/mol$\cdot$K$^2$ \cite{Gabani2001,Phelan2014} -- 10 times larger than \textit{metallic} LaB$_6$.  Such large fermionic specific heat has been shown to be a bulk effect \cite{Wakeham2016}.   Additionally, THz range conductivity experiments of SmB$_6$ have revealed in-gap conduction consistent with a \textit{localized response} with conductivities orders of magnitude larger than the dc value \cite{Travaglini1984,gorshunov1999low,laurita2016anomalous}.  Although impurity band conduction is an obvious culprit, the magnitude of these signals is generally orders of magnitude larger than of corresponding impurity bands in conventional semiconductors.  However, there is some recent evidence that these issues can be understood by realizing that electronic dispersions are quite unlike the parabolic ones of conventional semiconductors and treating the wavefunctions of impurity states appropriately~\cite{skinner2019properties}.

\subsubsection{Semiconductors and insulators} From a practical standpoint, semiconductor physics has been the overwhelming success story of solid state physics, leading to transformative new technologies throughout the second half of the 20th century. It can seem surprising, then, to note that semiconductors continue to harbor surprises and profound mysteries during the 21st century as well.

One prototypical example of an old semiconductor that continues to serve as a fount of new physics is SrTiO$_3$.  This relatively large band-gap material generated significant interest in the 1950s and '60s due to its anomalously large dielectric constant at low temperature, which results from an aborted ferroelectric transition at low temperature~\cite{BarrettSTO, CowleySTO, YamadaSTO, NevilleSTO}. The large dielectric constant enables both metallic and superconducting behavior at anomalously low electron density~\cite{STO-SC1, STO-SC2, STO-metalMR}. At higher temperatures, experiments have shown a metallic conductivity coexisting with an apparently huge violation of the Mott-Ioffe-Regel criterion~\cite{STOIoffeRegel, STOtransportreview}. A naive application of the Drude theory yields a mean-free-path that is shorter than the lattice constant, suggesting that the electron transport in this regime may involve a description beyond the traditional kinetic theory of Fermi liquid quasiparticles.  

Even the world's best-studied semiconductor, silicon, remains a fruitful platform for addressing unsolved problems in strongly correlated electron physics. For example, doped silicon (like essentially all semiconductors) undergoes an insulator-to-metal transition (IMT) with increasing doping. There is a long history of studying this transition in phosphorus-doped silicon (\onlinecite{Si-MIT1, Si-MIT2, Si-MIT3, Si-MIT4, MIT-expreview}), but the nature of the transition was never completely understood~\cite{MIT-theoryreview1, MIT-theoryreview3}.   
While IMTs are commonly discussed from the perspective of the Anderson transition, {\it i.e.} ~a disorder-driven transition for which interactions play no role, this perspective does not adequately describe the IMT in doped semiconductors. The long-ranged Coulomb interactions between electrons localized on discrete dopant atoms play a crucial role in the transition, making the IMT in doped semiconductors a preeminent strongly correlated problem, for which every site energy depends on the occupation of every other site.  As the archetypal example of a doping-induced transition, phosphorus-doped silicon near the IMT remains an excellent platform for testing new experimental probes of temporal and spatial electron correlations (for example, by measuring optical conductivity~\cite{Helgren02a} and in optical pump-probe experiments~\cite{thorsmolle2010ultrafast}). Applying such probes to the doping-induced IMT may provide crucial insight to the correlated electron problem.

Finally, it is worth mentioning that certain semiconductors exhibit anomalous electronic properties in the vicinity of a structural phase transition. For example, Cu$_2$Se undergoes a structural transition at temperature $T \sim 350$\,K. It has been known since 1971 that this transition is accompanied by a spike in thermopower, hinting at the possibility of a strong renormalization of electronic carriers~\cite{Cu2Sethermopower1}. Revisiting the thermopower of Cu$_2$Se in 2018 showed that this spike can produce an enormous thermoelectric figure of merit, $zT \sim 300$, in the immediate vicinity of the transition~\cite{Cu2Sethermopower2}. The nature of the charge and heat transport in the vicinity of this transition remains an open question.

\subsubsection{Superconductivity}

As alluded to elsewhere in this manuscript, the details of the pairing mechanisms of many ``old'' superconductors are still poorly understood.   From the rare earth borocarbides, amorphous and crystalline bismuth, SrTiO$_3$, doped BaBiO$_3$, or even MgB$_2$ many details are unknown~\cite{takagi1997borocarbide,amorphousbismuthSC,gastiasoro2020superconductivity,meregalli1998electron,buzea2001review}. In particular, these systems provide a wealth of test cases to investigate the interaction between superconductivity and magnetism ({\it e.g.} , reentrance in rare earth borocarbides), structure (BaBiO$_3$) and spin-orbit physics (Bi).

A number of doped semiconductors exhibit ``superconducting domes'' akin to the cuprates. For example, in the bismuthates (which are diamagnetic semiconductors), doping with potassium yields a superconducting dome with a maximal $T_c \sim 30$\,K~\cite{bismuthates}.  The electron-phonon coupling calculated by a density functional and Migdal-Eliashberg theory approach is insufficient to account for the high $T_c$ in the bismuthates.  Many of the systems necessitate a non-BCS, Migdal-Eliashberg explanation because the Debye frequency is much larger than the Fermi energy, and therefore electron-phonon coupling of a conventional variety cannot be the mechanism for electron pairing (although it could still be electron-phonon coupling of an unconventional variety).  In this regard, recent work~\cite{yin2013correlation} claims that standard approaches underestimate large nonlocal correlation effects that can enhance the electron-phonon coupling and enhance $T_c$.  The mechanism for superconductivity in doped SrTiO$_3$ also continues to generate intense interest in this regard (see Ref.\ \onlinecite{gastiasoro2020superconductivity} for a review). 

The superconducting properties of bismuth are similarly fascinating. One would generically not expect superconductivity in bismuth owing to its very low density of electron states, but a very recent study has identified superconductivity in crystalline bismuth (with a $T_c \approx 0.5$\,mK)~\cite{bismuthSC}. Such superconductivity at low density cannot be described by the conventional BCS, Migdal-Eliashberg theory, since it requires the Fermi energy to be much larger than the Debye frequency, while in bismuth they are comparable. BCS theory also predicts a ratio between the critical magnetic field and critical temperature that is an order of magnitude larger than the value observed.  Perhaps even more surprising is that \emph{amorphous} bismuth is also a superconductor, with a $T_c \sim 6$\,K that is more than four orders of magnitude larger than in crystalline bismuth~\cite{amorphousbismuthSC}.  As it happens, bismuth is representative of a larger class of materials for which $T_c$ is higher in the amorphous state than in the crystalline state~\cite{amorphousSC1, amorphousSC2}.  There is some evidence that both the electronic density of states and the electron-phonon coupling is larger in amorphous bismuth~\cite{mata2016superconductivity}, but it is fair to say that this is not fully understood.  Finally, the existence of liquid metals (and superconductivity in amorphous metals) suggests the tantalizing (and relatively unexplored) question: can there be a liquid superconductor? 

A superconducting dome ``high-$T_c$''-like phenomenology is also on display in the elemental magnetic metals. In iron the itinerant magnetism in the bcc phase is suppressed with external pressure and enters a superconducting dome in the hcp phase~\cite{Katsuya2001}. As another example, elemental chromium exhibits a superconducting dome much like that of the cuprates (albeit with a much lower $T_c$) when doped with ruthenium, rhodium, or iridium~\cite{dopedCr1, dopedcr2}. One can also produce a relatively high $T_c$ two-dimensional superconductor by growing monolayer or near-monolayer films of lead on appropriate substrates~\cite{PbSC1, PbSC2, PbSC3}.   One would like to understand if superconductivity in these materials is of the unconventional variety.


\section{What can and should we do?}

\subsection{The Role of Materials Synthesis and Discovery}

 \begin{figure}
\includegraphics[width=\columnwidth]{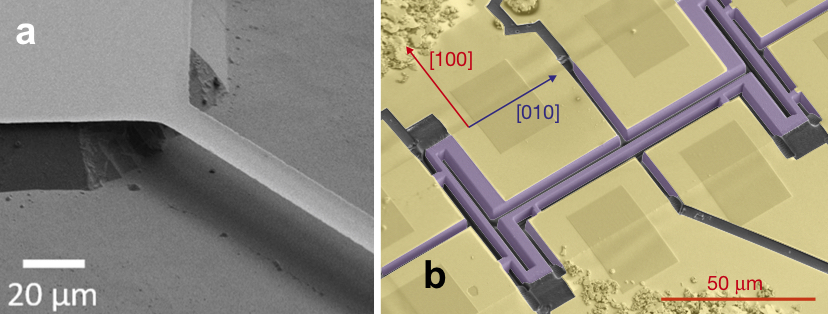}
\caption{Micromachined samples where extreme aspect ratios allow access to increased measurement sensitivity, techniques, or phenomena. (a)~An SEM image of a suspended crystalline sample of BaTiO$_3$ produced using wet-etch techniques.  From Ref.~\cite{lim2020suspended}.  (b)~A microengineered sample of CeRhIn$_5$ machined from a bulk crystal using focused ion beam milling, ideal for measurements at high magnetic fields.  From Ref.~\cite{ronning2017electronic} }
\label{novelsamples}
\end{figure}

Materials play a key and obvious role in the correlated electron problem, as after all, materials actually host the electrons that are correlated.  As McQueen said at the workshop,  ``A materials discovery has occurred when a known or newly created material exhibits phenomenology and behavior not obviously explainable by our current understanding and theories of the universe.''   It is the challenge inherent to the correlated electron problem that it is difficult to know \textit{a priori} where to look for new behavior in these materials, but the past provides guides.   The discovery of superconductivity in the cuprates was driven by the notion that electronic properties of metallic oxides were underexplored and that strong Jahn-Teller coupling might drive superconductivity in new materials~\cite{bednorz1986possible}.   The latter idea is likely not to be the correct explanation of superconductivity in the cuprates, but it was a new physical idea that drove research in a new direction.   The discovery of the fractional quantum Hall effect in 2D electron gases was enabled by the development of ultra-clean heterostructures and a long effort in trying to understand the effects of localization and delocalization in two dimensions.  Localization is not the driver behind the physics of the fractional quantum Hall effect, but again it pushed the community to look in new and unexplored directions.

Materials oriented scientists are vitally important in at least four capacities:  discovering new and interesting materials, improving the quality of existing materials, using aspects like doping to change material properties, and developing heterostructures and new material configurations ({\it e.g.}  ``twisted'' compounds).  Bulk synthesis provides the community with large crystals of both established and new materials, while atomic-scale growth techniques allow for the synthesis of these systems in epitaxial thin film form and in artificial layered heterostructures where structural and chemical degrees of freedom can be systematically controlled.

\begin{figure}
\includegraphics[width=\columnwidth]{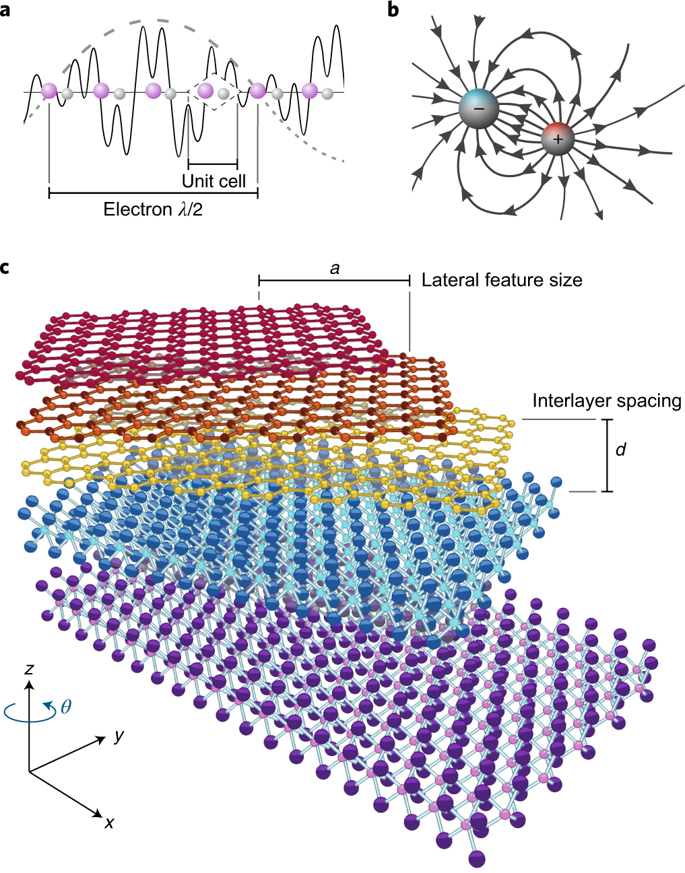}
\caption{2D van der Waals heterostructure ``twistronics" systems provide an incredible opportunity to tune material properties.  (a)~The crystal unit cell introduces a repeating structure that modifies the electronic wavefunctions. (b)~The dielectric environment, which has reduced screening in 2D, modifies the local Coulomb interaction. (c)~2D van der Waal heterostructure quantum metamaterials are composed of individual 2D layers (transition metal dichalcogenide, graphene or boron nitride) and characterized by lateral repeat distance size $a$, interlayer spacing $d$ and atomic-layer twist angle $\theta$.  From Ref.~\cite{song2018electron}.}
\label{twistronics}
\end{figure}

As we move forward in investigating correlated electron materials, there is a need to push well-known existing techniques -- such as flux growth, floating zone, Bridgman, molecular beam epitaxy, pulsed laser deposition, chemical vapor deposition, etc. -- forward by extending them to new frontiers~\cite{Canfield_2019, schmehr2019high} and by incorporating techniques common in other fields such as chemistry and materials engineering. For example, freestanding films developed from wet etch techniques, plasma growth, or exfoliation~\cite{tatarova2017towards,lu2016synthesis,bhaskar2016flexoelectric, lim2020suspended} can serve as novel substrates for extending the range of lattice parameters and crystal symmetries available for growing strongly correlated electron materials. Free standing single atom thin layers have been achieved in the case of graphene~\cite{du2008approaching,tatarova2017towards}, but one may wonder if it is possible for other systems like cuprates.  Topochemical anion exchange allows for the stabilization of oxidation states not available by conventional synthesis techniques~\cite{li2019superconductivity, lefler2019reconfigurable}.  Quite recently, these techniques gave us superconductivity in Ni based oxides~\cite{li2019superconductivity}.  There is the development of hybrid growth techniques with unconventional substrates that can lead to interesting materials~\cite{lichtenberg1992sr2ruo4, yao2019record}.  Modern techniques such as focused ion beam milling (Fig.~\ref{novelsamples}) should continue to be developed for making engineered samples from bulk crystals~\cite{moll2018focused}.   We must also continue to innovate new methods for materials assembly, as evidenced by the remarkable continuing progress in the construction of 2D material heterostructures (see Fig.~\ref{twistronics}) that has culminated in TBG but is not just limited to~\cite{novoselov20162d,rhodes2019disorder,geim2013van, ajayan2016van,cao2018unconventional,kim2016van}.  In the case of TBG, although a magic angle causing relatively flat bands was  anticipated~\cite{bistritzer2011moire}, the remarkable phenomenology exhibited by multiple superconducting domes and insulating regions was a surprise~\cite{cao2018unconventional, yankowitz2019tuning, lu2019superconductors}. This shows that novel combinations of materials can frequently reveal surprising new physics.

A particularly promising direction is the design of materials with novel macromolecular structures.   This is of course relevant in the rise of 2D materials heterostructures and superconducting C$_{60}$, but also in the consideration of other molecular structures.   For instance, there is a series of ``1-2-20'' Pr-based ``cage compounds'' in which the Pr$^{3+}$ and transition metal ion reside in different atomic cages.  The localized Pr$^{3+}$ 4$f$ electron states hybridize with the ligand states of the 16 surrounding X cage ions resulting in a nonmagnetic non-Kramers $\Gamma_3$ ground state in the cubic crystal field.   They have shown evidence or a quadrupolar Kondo effect {\it e.g.}  the two channel Kondo effect.  It has been proposed that magnetostriction could be a very diagnostic test for multipolar orders in this material class~\cite{patri2019unveiling}. The compounds PrTi$_2$Al$_{20}~$\cite{Sakai12} and PrV$_2$Al$_{20}$~\cite{Matsumoto16} have been reported to display unconventional SC with $T_c$’s of 0.2~K and 0.06~K.  The SC coexists with ferroquadrupolar (FQ) order in PrTi$_2$Al$_{20}$ ($T_{\text{FQ}} = 2$~K) and antiferroquadrupolar (AFQ) order in PrV$_2$Al$_{20}$ ($T_{\text{AFQ}} = 0.6$~K).  In another group of Pr-based filled skutterudite ``cage compounds'' the Pr$^{3+}$ ions reside in an atomic cage but have a $\Gamma_1$ singlet ground state in the crystal field.  The PrOs$_4$Sb$_{12}$~\cite{Bauer02,Maple06} and PrPt$_4$Ge$_{12}$~\cite{Maisuradze10} compounds are nonmagnetic and exhibit an unconventional type of SC with $T_c$'s of 1.86~K and 7.9~K, respectively.  The SC appears to have gap nodes and breaks time reversal symmetry.  It has been proposed to be a candidate 3D topological superconductors that could support Majorana fermions~\cite{Kozii16}.  This general idea of using large molecular clusters as a building block for new physics is also relevant for searches for new spin liquid platforms in magnetic cluster compounds like LiZn$_2$Mo$_3$O$_8$~\cite{sheckelton2014local}, and the remarkable configurability of metal-organic frameworks \cite{zhang2017theoretical,yamada2017designing,misumi2020quantum,takenaka2018signature}.   Macromolecular structures represent a whole world of relatively unexplored possibilities.

A key theme in materials synthesis as we consider the future of the correlated electron problem is to understand the role of disorder and defects at both the atomic level and meso-and macroscopic levels.  To that end it would be powerful to incorporate atomic and electronic structural and chemical characterization \textit{in situ} during material synthesis to give information on sample properties in real time~\cite{shen2017situ, he2018combined}.  Moreover, it will be important to more rigorously \textit{ex situ} characterize the interplay between defects and phenomena~\cite{muller2012atomic, cao2017giant}. A particular challenge will be to incorporate materials synthesis with measurements of physical properties under extreme conditions such as low temperatures and high fields where correlated electronic effects and phases are experimentally accessible. 

Future directions need to emphasize the feedback loop between theory which predicts structural motifs for stabilizing novel phases, materials synthesis, and characterization. The expansion of social networks for collaborative synthesis and cloud efforts may accelerate progress in materials development through greater access to advanced techniques and shared understanding of theory and multi-characterization of similar samples.  To enable more efficient exploration and higher throughput in the material synthesis phase space, standardization of protocols and delocalized crowd-sourcing synthesis are arguably as important as methodological innovations. In this regard, we expect future materials innovation to consist of two complimentary modes of operation:
i) individual research labs will continue to lead technological innovations and targeted synthesis of novel materials of interest on a case-by-case basis.
ii) more distributed, either government sponsored or industry invested ``cloud synthesis'' stations will become massively parallel to realize high throughput phase space exploration.

Such organization disentangles the standard part of material discovery from more individual-case based explorations, thereby not only liberating more workforce towards process i), but also improves the comparability, repeatability, and efficiency of process ii) by removing the human preference factor. Moreover, with the rapid developments of robotics and AI assisted material prediction today, process ii) can be executed with such standardization that it readily connects to the industrial scene better, and also enables more efficient material recycling (especially for toxic and rare elements).  When combined with automated, standardized physical property characterizations such as simple Raman spectroscopy and resistivity, such mode of operation provides a basis for the accumulation of big data, and lays the ground for automatic detection of materials with single or combined ``outlier'' properties.  This will obviously require intensive coordination and it will be necessary for researchers, industrial investors and policy makers to gather frequently and discuss standard protocols in light of new instrumentation and scientific developments from the research sector, much like how IEEE operates today.

\subsection{Numerical methods}

In the traditional division between theoretical and experimental physics, numerical (or computational) physics plays a three-fold role: (1) It acts as a bridge between analytical theory and experiment, for example by connecting microscopic or many-body models with complex experimental systems. (2) It facilitates new types of (numerical) experiments. Numerical simulations can be cheaper, faster, and easier to control than physical experiments. (3) It is a theoretical tool to describe complex systems for which an analytical description is unavailable and perhaps even impossible. Strongly correlated electronic materials and models can be examples of such systems.

\begin{figure}
\includegraphics[width=\columnwidth]{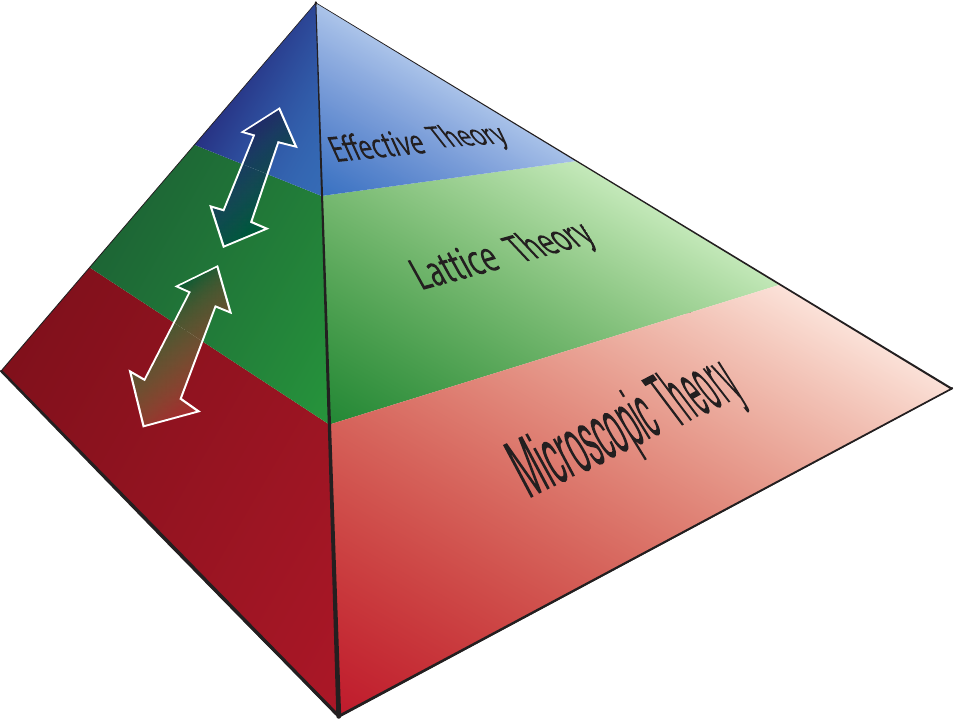}
\caption{At least three classes of theory exist and play an important role in the understanding of correlated systems: microscopic theory, involving \emph{ab initio}  density functional theory based on chemical compositions;  lattice theory, involving many-body wavefunctions with reduced number of degrees of freedom in low-energy lattice models; effective models that attempt to capture essential properties of a system with the minimum additional complexity.}\label{TheoryPyramid}
\end{figure}

For each level of theoretical description in the strongly correlated electron problem, see Fig.~\ref{TheoryPyramid}, there are numerous numerical methods available.  We might hope that the parameters of the upper-level theories can be estimated using the lower-level theories {\it e.g.}  models with fewer degrees of freedom may be parametrized from models with more degrees of freedom.  However, this may not always be possible or efficient.   Some effective theories can only be postulated and their parameters obtained by fit to experiment. Note that the arrows are two-headed as both bottom-up and top-down reasoning are possible. For instance, one can attempt to deduce what materials could give an particular effective Hamiltonian, such as the work that showed the Kitaev model could be possibly realized in strong spin-orbit coupled systems~\cite{jackeli2009mott,kitaevspinliquid}, or what materials could exhibit certain phenomena (such as high-$T_c$ superconductivity). Alternatively, one can attempt to construct lattice models explicitly from microscopic theory using the downfolding techniques discussed below. 

Microscopic model methods, based on directly solving the many-body Schr\"odinger equation, include density functional theory (DFT)~\cite{hohenbergkohn,kohnsham}, \textit{ab-initio} quantum Monte Carlo~\cite{Ceperley80, Wagner_Ceperley_review} and quantum chemistry methods~\cite{Shao_qchem}. There are also many methods to solve effective models, such as Hubbard~\cite{Hirsch_Hubbard, leblanc15}, Heisenberg~\cite{Sandvik_Heisenberg, Yan_Huse_White} and other phenomenological lattice models exist. Within these broad classes, there are many types of both exact and approximate algorithms based on numerical techniques, for example, large scale exact~\cite{Wietek_Lauchli} or selected diagonalization methods~\cite{selci}, numerical renormalization group~\cite{NRG_wilson} and density matrix renormalization group (DMRG)~\cite{white1992,schollwock2005density,2DDMRG}, tensor networks~\cite{TPS_review, vidal_mera,corboz_peps, CPS}, dynamical mean field theory (DMFT)~\cite{Georges_DMFT, Kotliar_DFTDMFT}, density matrix embedding theory (DMET)~\cite{knizia_chan}, and stochastic Monte Carlo methods~\cite{Blankenbecler, Sandvik_SSE, Trivedi_Ceperley, Prokofev1998, zhangcpmc, Gull2011, FCIQMC, SQMC}. There has also been progress in connecting the different levels of theoretical description, {\it i.e.} , building effective models from microscopic models, using techniques known as downfolding~\cite{pavirini,Aryasetiawan_downfolding}, but this is still challenging~\cite{Kent_Kotliar, Honerkamp18}. There are many advantages and limitations of the methods used to study strongly correlated electrons at each level, with properties such as finite system-size, dimensionality, and nature of interactions that determine the range of applicability.

State-of-the-art computing techniques give accessibility to well-controlled approximations of physical quantities that are often unsolvable in the framework of analytical theory.  Strongly correlated electronic systems are composed of a vast, exponential-scaling of many degrees of freedom. Therefore, experimental exploration of all variables is impractical, making simulation-based solutions critical for future success.  Numerical methods can establish a practical connection to experimental observations. Wave-function-based methods at the \textit{ab-initio} or many-body model level give access to some excited-state wavefunctions, and therefore can make predictions about experimental spectra~\cite{Wang2018}. Monte-Carlo-based methods, which utilize rigorous statistical sampling to evaluate the properties of large-scale systems, provide unbiased predictions for finite-temperature observables when applicable~\cite{Blankenbecler,Gull2011}.
Numerical methods can also help guide experiments, for example on symmetry or topology principles, and analytical theory. In some particular cases, numerical methods bring extra mathematical perspectives to describe the physics. For example, in tensor-network-based methods, entanglement entropy, originally introduced in quantum information theory, governs the accuracy of numerical solutions and provides a useful tool for theoretical analysis~\cite{schollwock2005density}.

Numerical methods should not be treated as black boxes. Their limitations need to be clearly stated and understood both by practitioners and users, including experimentalists looking for a theoretical connection to their findings. Microscopic models, by virtue of fully describing the electronic degrees of freedom, are more complex than many-body effective models. Consequently, the methods that directly target microscopic models tend to rely on more approximations than methods for effective models. This trade off, which applies throughout all levels of the pyramid, can be described on a continuum between ``solving an exact model approximately” versus ``solving an approximate model exactly.” These are imperfect, but complementary approaches, that are especially important in the correlated electron problem, where the validity of approximations often lack \textit{a priori} theoretical justification.

A limitation in numerical methods for correlated electron problems is the fermionic sign problem, which restricts the applicability of some exact Monte Carlo techniques to general models of interacting fermions and frustrated spin systems~\cite{Li_Yao}. One school of thought has been to search for, or design, sign problem-free effective models~\cite{berg2019monte,Li_Yao,kaul_sandvik,Alet_sign}. Another approach is diagrammatic Monte Carlo methods, which turn negative signs arising from fermionic statistics into an advantage~\cite{prokofev07}. These techniques are based on the stochastic sampling of Feynman diagram expansions at all relevant orders, without uncontrolled truncation. In this formalism, the alternating fermionic signs result in a faster (exponential) convergence of the diagrammatic series, which can offset the exponential scaling of computational time at a given order \cite{rossi17a,rossi17b}.

While Monte Carlo methods tend to be computationally expensive, it is possible to harness modern computational power and massive parallelism to make progress.  Most methods focus on ground state or equilibrium thermal properties, and obtaining dynamical properties and response functions reliably from them remains a major challenge. For example, traditional quantum Monte Carlo methods are formulated in imaginary time which provides access to thermal properties, but calculating frequency dependent measurements requires numerical analytic continuation \cite{JarrellGubernatis,goulko17nac}. Progress has been made in formulating Monte Carlo methods in real time, providing direct access to the real-frequency axis~\cite{cohen15}, but these face the challenge of the dynamical sign problem~\cite{dynamical_sign}. Tensor network or matrix product state methods~\cite{Daley_2004, White_Feiguin} can be formulated in real-time, but suffer from growing entanglement with time propagation, restricting the accuracy of low-frequency properties~\cite{tMPS}. Exact diagonalization methods can perform exact calculations of dynamical spectra directly on the real frequency axis~\cite{Prelovsek}, but finite-size effects complicate the interpretation of spectra which may exhibit spurious features on small systems. Therefore, using embedding methods and cross-benchmarking multiple numerical methods become important to validating properties in the thermodynamic limit.

There are several difficult-to-compute quantities of interest that are crucial for understanding the underlying physical principles of strongly correlated electron systems, as well as for relating theory and experiment. We need to focus on developing improved methods to study for example, excited states and spectroscopy, finite-temperature systems, dynamical properties, and disorder.  Another key challenge is determining the appropriate parameters and energy scales in effective lattice models for describing real materials. This involves finding a controlled and systematic way to relate, for example, microscopic parameters such as the hopping $t$, Hubbard interaction $U$, and charge transfer energy $\Delta$, to experimentally measurable observables such as $T_c$ (for a concrete example relevant to the cuprates, see Ref.~\cite{Weber_2012}). 

Downfolding techniques connect different levels of description, which will facilitate precise predictions for experiments and enable material design. Equally important is quantifying the accuracy of these low-energy model Hamiltonians -- how do interaction parameters depend on doping, pressure, field? Are higher-body effective interactions (such as ring exchange) important to incorporate~\cite{Paul_2020}?  It must be noted that downfolding methods have been extensively used for deriving tight-binding parameters in the DFT community~\cite{pavirini}, and for DFT+DMFT calculations~\cite{Kotliar_DFTDMFT}, however, the estimation of interaction parameters requires approximations that are not well controlled or understood~\cite{Kent_Kotliar, Honerkamp18}. Many-body formalisms that give equal footing to the kinetic and potential energy parts of the problem, which do not depend on traditional band theory, are being developed~\cite{Rusakov14,Changlani15,Zheng18,Requist19} and ideas from information compression, quantum information, tensor methods, and renormalization groups may have an important role to play. Further discussions and collaborations between the condensed matter physics and quantum chemistry communities is potentially crucial for progress on this front.

Another interesting route related to the interplay between scales is the design of effective models with desired properties. For example, if one desires a superconducting or quantum spin liquid ground state, one can ask what model Hamiltonian realizes such a state. This is an inverse problem that can be approached numerically which may help in the design of future materials and may help develop intuition for where in parameter space to look for desirable strongly correlated phases~\cite{Chertkov18}.

Finally, we foresee that the rapidly developing technologies in artificial intelligence, machine learning, quantum computing, and quantum simulation will play a vital role in how numerical techniques will address the strongly correlated electron problem. Deep neural networks~\cite{LeCun2015} and other machine learning methods, such as computational graphs~\cite{Kochkov2018}, are already actively being used as trial wave functions in variational Monte Carlo~\cite{Carleo17}. Machine learning methods, because of their ability to reveal correlations in large datasets, also hold promise as tools for discovering new strongly correlated material candidates through the analysis of large databases of material properties~\cite{Saal2013}. In addition, the emerging technology of noisy intermediate-scale quantum (NISQ) computers~\cite{Preskill2018} will complement our existing classical computing methodologies. NISQ devices are capable of simulating many-body quantum systems that are difficult to simulate classically, even with the best known supercomputers~\cite{Arute2019}. NISQ devices, and their eventual error-correcting successors, are powerful tools for solving effective models of strongly correlated electrons, such as the Hubbard model, using, for example, hybrid quantum-classical algorithms such as the variational quantum eigensolver~\cite{Peruzzo13}. For these reasons, we believe that machine learning methods and quantum computers will help us make progress in tackling these problems.  There is also great promising in the areas of quantum simulation where ones handles the exponential proliferation of a Hilbert space that characterizes a large system, by ``fighting fire with fire''~\cite{houck2012chip} by simulating a one quantum system by another -- simpler to control -- quantum system.   Possible implementations are in superconducting circuits~\cite{houck2012chip}, quantum dots~\cite{manousakis2002quantum}, and cold atoms~\cite{bloch2012quantum}.

\subsection{Analytical methods}

\subsubsection{New approaches}

The complicated nature of strongly correlated electronic systems severely limits the power of analytical methods that are often based on an expansion with a small parameter around a well-defined ground state. However, considering the difficulties of numerical methods in strongly correlated systems and their limited predictive power compared to the case of single-particle problems, the demand is high for further development of analytical methods to gain unbiased and transparent insight into the physics of strongly electronic systems. Moreover, analytical approaches are, at least in principle, more straightforward to generalize to non-equilibrium problems in a controlled manner, {\it e.g.}  with Keldysh approach, while in their numerical counterparts semi-empirical analytical continuation techniques have to be used.  In spite of the difficulties mentioned, analytical approaches are still the most powerful way to understand correlated systems.  Building on those successes, we discuss below the possible paths for analytical theory of correlated systems to move forward.

One striking example of a solution to a correlated electron problem is the fractional quantum Hall effect (FQHE), where a ground-state wavefunction was explicitly constructed which explained the properties of the system~\cite{laughlin2000fractional}. It would be the ideal to use this strategy to approach some other important states of correlated matter, such as NFL metals and QSLs, and perhaps get new examples of their realization amenable to theoretical studies. In particular, for NFL states one may try to find the wavefunctions from known examples of models having a NFL state (such as fermionic systems at a QCP~\cite{schofield1999non,lohneysen2007fermi} or the Sachdev-Ye-Kitaev (SYK) model~\cite{sachdev1993gapless,Kitaev1,Kitaev2}) and introduce additional parameters to obtain families of NFL wavefunctions with the aim of extracting general properties of NFL states. One may hope to find certain generic forms for the wavefunctions of the excitations in the NFL ground state. Essentially, breakdown of quasiparticle description (or, more precisely, the vanishing of the quasiparticle weight) of excitations is just a sign that single-particle electron-like states are a ``bad basis” for the Hilbert space of excitations in an NFL. One expects the proper eigenfunctions to be a superposition of states with different number of particle-hole excitations, pointing to some kind of generalized many-body coherent state.

The example of the Kitaev model~\cite{kitaev2006anyons,jackeli2009mott,kitaevspinliquid} (discussed above in Sec. \ref{SpinLiquid}) for a spin liquid with anisotropic Ising interactions shows that proposing -- perhaps unrealistic -- toy models with the desired ground state may lead to tremendous progress in both theoretical understanding and future experimental guides.   This is also what transpired with Haldane's honeycomb lattice model in the context of the quantum anomalous Hall effect~\cite{haldane1988model}. In building up theoretical descriptions of many-body systems, exact models -- such as Kitaev's -- are important (despite their frequent artificiality) because they establish the point of principle that a particular phase of matter could exist.   But they can also be motivating to search for new ways to stabilize these states of matter.  In the Kitaev case, it has been shown that despite its contrived nature (Ising interactions with different quantized directions on each bond), its anisotropic interactions can actually arise through the effects of strong spin–orbit coupling~\cite{jackeli2009mott,kitaevspinliquid}.  Thus, constructing new toy models with NFL or QSL ground states while using the known ones as a starting point (cf. SYK~\cite{song2017strongly,PhysRevX.8.031024,patel2019theory} or Kitaev models~\cite{PhysRevB.93.085101}) may be fruitful.

Another strategy for making progress in the correlated electron problem is to exploit general symmetries and properties of quantum mechanical descriptions to derive exact statements that are valid regardless of the correlation strength. In the study of gapped topological phases, such as spin liquid or FQH states, Lieb-Schultz-Mattis (LSM)-type theorems~\cite{LSM_original} have played an important role in theoretical developments. Basically, such theorems state that one can constrain possible macroscopic physics based on microscopic information, such as symmetries or degrees of freedom per unit cell. For gapless systems, an example is the so-called Luttinger's theorem~\cite{luttingerFS} for the volume enclosed by the Fermi surface.  It was shown to be of topological origin by Oshikawa~\cite{oshikawa2000topological}.  This has important repercussions for Kondo lattice systems and possibly the pseudogap state of the cuprates, where in both cases Fermi surface volume differs from simple expectations based on weakly interacting electrons. It might be useful if we can come up with something similar for generic gapless phases of matter, {\it i.e.} , ``liquid'' phases.

Additionally, bounds for certain observable quantities can be deduced analytically from rather general considerations, which makes them also applicable to correlated systems. The examples include the aforementioned bounds on diffusivity and resistivity, which are based on a coarse-grained hydrodynamic description~\cite{PhysRevLett.119.141601,Lucas11344} or the quantum mechanical ``Lieb-Robinson bound''~\cite{mousatov2019planckian}. Energetics of correlated systems may be also better understood using exact relations between the potential and kinetic energy derived from the virial theorem~\cite{LEGGETT19981729,PhysRevX.6.031027}. Studying the consequences of general quantum-mechanical relations/theorems for strongly correlated systems and applying the resulting statements to the analysis of experiments may expand our view of the correlated electron problem and potentially lead to new and useful phenomenologies.

Difficulties regarding strong correlations are shared with other branches of physics as well. One possible way to tackle these is through the development of dual theories between a strongly correlated limit and a weakly correlated regime by breaking down the problem to summation of multiple ``less complicated'' problems.  A successful example in high-energy physics is holographic theories such as anti-de Sitter Space/conformal field theory, which has garnered attention in the condensed matter context~\cite{hartnoll2016holographic}. Within condensed matter physics, there have been many attempts to map interacting systems to single-particle physics, among which bosonization in 1D is probably one of the most successful. Most of the existing approaches, however, have limitations. Further development and generalization of the existing methods to higher dimensions and/or more general correlated systems along the line of some of the ongoing works such as higher-D bosonization~\cite{LUSCHER1989557,PhysRevLett.72.1393,PhysRevB.19.320,PhysRevB.48.7790,kopietz2006bosonization}, or patch theories~\cite{POLCHINSKI1994617}, is therefore highly desirable. Ideally, as in the case of a Fermi liquid, a mapping between a strongly correlated phase to a simple one would enable one to learn physics of the ``difficult'' regime from ``easy'' regime.

In a related fashion, ``analogue theories'' are a research program which investigates analogues of a particular field of physics within other physical systems, with the aim of gaining new insights into the corresponding problems. 
For example, the utilization of analogue theories of gravity and cosmology in various low-energy fields such as ultracold atoms, acoustic and condensed matter systems have lead to many fruitful results~\cite{gravityanalogue}. As a result, they have motivated numerous interesting experimental setups which simulate puzzling problems in gravity such as black holes. However, ``condensed matter analogue theories'' are much less explored~\cite{ghorashi2019criticality,GruzbergQG, Arguello-Luengo2019,doi:10.1080/00018730701223200}. Aside from applications of adS/CFT ideas \cite{hartnoll2016holographic}, particular examples, are recent attempts to map the problem of disordered fermions to gravitational theories with ambitious perspective to use the developments of quantum gravity to gain further understandings in these systems \cite{ghorashi2019criticality,GruzbergQG}. Another recent example, of both analytical and computational importance, is a development made by a group of mathematicians which proposed an interesting method for finding eigenvalues of large random matrices without solving eigenvalue problems \cite{Arnold, Beenakkercommentary}, which develops the idea of ``localization landscape'' proposed in \cite{Filoche14761}. Therefore, considering the tremendous developments of sophisticated analytical and mathematical methods and techniques which are developed in other fields such as high-energy physics and/or mathematical physics, it makes the exploration via ``analogue correlated electronic models'' in these fields a desirable goal.  They could provide an improved toolbox to tackle correlated systems. Moreover, they may also guide further simulation of correlated electronic phases in other physical systems. 

\subsubsection{Non-equilibrium}

Finally, let us discuss an important avenue, where analytical methods do have a certain advantage. Investigating non-equilibrium phenomena in an already quite complicated correlated electronic system requires further developments of an analytical toolbox. One particularly important direction is periodically driven many-body systems. It has been found theoretically that there exist long time scales in which interacting periodically driven quantum many-body systems or Floquet systems can be described by an effective time-independent theory~\cite{PhysRevLett.115.256803,PhysRevB.95.014112,PhysRevLett.116.120401,PhysRevX.7.011026,PhysRevE.93.012130,PhysRevResearch.1.033202,weidinger2017floquet}. Since this makes it possible to use existing equilibrium techniques to understand strongly correlated Floquet systems, a lot of effort has been spent on deriving effective time-independent Hamiltonians that allow such a description~\cite{PhysRevB.95.014112,schweizer2019floquet,PhysRevLett.116.125301,PhysRevA.68.013820,PhysRevX.4.031027,PhysRevLett.115.075301,eckardt2015high,PhysRevB.93.144307,PhysRevB.94.235419,PhysRevB.25.6622}. While much progress has been made in the case of non-interacting systems, progress has been slower in the strongly correlated case (with some results obtained using Keldysh formalism \cite{kandelaki2018}). Most current work focuses on the particular limit of high frequencies. A variety of methods called Floquet Magnus-, van Vleck- or Brillouin-Wigner-type expansions~\cite{PhysRevB.93.144307,PhysRevB.94.235419} have been developed for this regime. However, there has not been much progress for more generic, interesting and experimentally relevant mid- to lower frequency regimes~\cite{Vogl_2020}.

Progress has been limited for two reasons. First, generic interacting systems have an algebra that does not ``close'', which has stymied progress because it leads to complicated effective Hamiltonians.  This means the following.  High frequency expansions include a set of nested commutators of the Hamiltonian at different times $[H(t_1),[H(t_2),...]]$. These generate higher and higher order interaction terms as higher corrections are included and can become quickly uncontrolled -- this is called operator spreading. If new terms keep being generated ad infinitum we say that the algebra is not closed.  Second, for the interacting case, high frequency expansion are at best to be considered asymptotic expansions rather than convergent expansions. This means higher order corrections might improve predictability in the high frequency regime but do not extend the regime to lower frequencies~\cite{PhysRevB.95.014112}. Recent work~\cite{PhysRevX.9.021037} has circumvented part of this problem by using an renormalization group-flow like approach to partially re-sum one of the high frequency expansions.  However, even for this case additional work is needed to improve the method. We anticipate that the flow equation approach~\cite{PhysRevX.9.021037} can be improved by a better choice of generator for the underlying unitary transformations.  Let us motivate this idea.

\begin{figure}
\includegraphics[width=0.85\columnwidth]{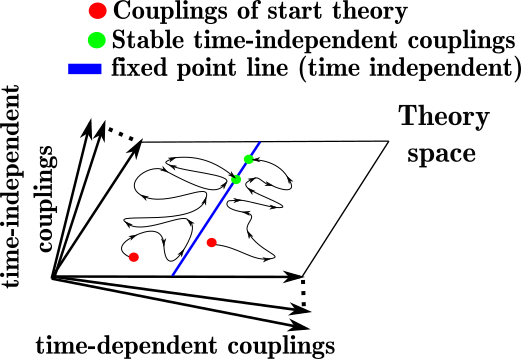}
\caption{One new proposed method to understanding out-of-equilibrium strongly correlated systems exploits infinitesimal unitary transformation steps, from which renormalization-group–like flow equations are derived to produce the effective Hamiltonian.  This graphic shows schematically how couplings in a time-dependent theory flow in the approach of~\cite{PhysRevX.9.021037}.  One finds that as the couplings flow they repeatedly approach a line of fixed points, which ultimately turns out to be unstable until eventually a stable fixed point is reached.}\label{fig:couplig_flow}
\end{figure}

Recent years has seen the development of another similar approach for interacting time independent Hamiltonians -- the so-called Wegner flow approach~\cite{wegner1994flow}. What this approach does is dynamically construct a unitary transformation that diagonalizes an interacting Hamiltonian in an effective non-interacting basis~\cite{kehrein2007flow}. An effective Hamiltonian flows until a non-interacting Hamiltonian is reached at a fixed point. This method also suffers from the issue of operator-spreading - before a fixed point is reached many interaction terms are generated along the flow. However, with a clever choice of unitary transformations the issue can be avoided~\cite{mielke1998flow}.  In the Floquet case the fixed points of the flow equations are time independent Hamiltonians. Along the flow one also suffers from operator spreading. However, it is found that the source of this spreading is that there are many time-independent unstable fixed points that are approached closely as sketched in Fig.~\ref{fig:couplig_flow}.  It might be possible to stabilize these fixed points by a better choice of generator. Finding such a generator could allow reaching lower frequency regimes with less interaction terms being generated.  Success in this regard will lead to progress in understanding of out-of equilibrium strongly correlated systems.

\begin{figure*}
\includegraphics[width=6 in]{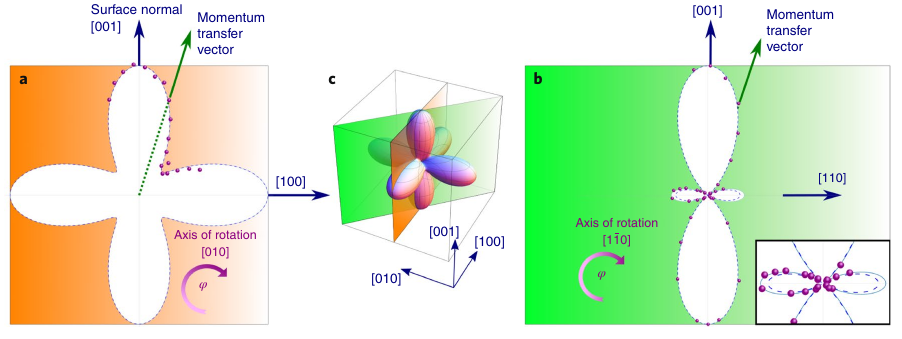}
\caption{New methods for $s$-orbital non-resonant inelastic X-ray scattering using modern synchrotron facilities with high brilliance allows the direct resolution of the orbital occupation.  This the quadrupolar scattering intensity as a function of the momentum transfer direction in the canonical Mott insulator NiO.   It directly shows the three-dimensional (3D) orbital hole density of the Ni high-spin 3$d^8$ configuration in an octahedral coordination, namely the $^3A_2$ $3d(x^2-y^2)3d(3z^2-r^2)$.  Specifically it is the integrated intensities of $M_1$ $(3s \rightarrow 3d)$ X-ray edge spectra plotted on the projections of the orbital shape of the $^3A_2$ $^3A_23d(x^2-y^2)3d(3z^2-r^2)$ hole density.  Here the projections of the 3D orbital shape on two planes are defined by [001] and [100] (a) and [001] and [110] (b). Note that the $3d(x^2-y^2)$ contribution vanishes in the [001]–[110] plane and so only the $3d(3z^2-r^2)$ may be seen. From Ref.~\cite{yavacs2019direct}.}\label{OrbitalImaging}
\end{figure*}

\subsection{Novel Spectroscopic Approaches}\label{subsection:methods-spectroscopies}

A key challenge of the correlated electron problem is that the electron momentum $\bf{k}$ and momentum transfer $\bf{q}$ between electrons in many cases cease to be good variables to describe systems across a wide range of length, energy, and time scales.  If interactions are strong enough, umklapp processes prevent even Bloch states, let alone free-particle states, from forming a good approximate basis to approach the problem.  Moreover, crystal imperfections and disorders on the atomic level often affect electronic properties at much longer wavelengths and even macroscopically~\cite{mesaros2011topological, cao2017giant}. Mesoscopically, the competition between different nearly degenerate broken symmetry states can result in short-ranged and anisotropic correlations. In other cases, correlated electron systems are characterized by patterns of order (hidden-order, non-quasiparticle transport, topological effects, fractionalization) that we have only imperfect tools to characterize.   It is imperative to develop new spectroscopic tools to address the above aspects.

\subsubsection{State-of-the-art spectroscopies}
 
Addressing these challenge calls for a holistic and concerted effort --- not only do we need experimental methods that probe the relevant degrees of freedom (charge, orbital, spin and lattice) in the form of both single-particle spectral and two-particle (and higher) correlation functions, but also ones that combine these probes in multimodal approaches to reveal the cooperation and competition between orders. This requires both utilizing existing experimental tools by pushing their capabilities and resolutions as well as developing new experimental methods.

Over the last few years, multimodal experimental studies combining complementary probes have revealed insights in a wide range of materials~\cite{da2014ubiquitous, comin2014charge, gerber2017femtosecond, zong2019evidence} that cannot be obtained otherwise.  In such studies one hopes that identical sample condition across probes enables more reliable comparison.  In cases such as \textit{in-situ} ARPES/STM studies of epitaxial films grown by MBE, this approach is essential due to inherent limitations of the experimental probes and air sensitivity of their surfaces. Part of this effort requires extending the regions of phase space (temperature, pressure, magnetic fields) over which the combined techniques overlap. Additionally, these efforts encompass the integration of novel tuning parameters with existing techniques.  One such example is the integration of uniaxial strain tuning to spectroscopy methods such as ARPES, photon scattering, STM and NMR~\cite{kissikov2018uniaxial, andrade2018visualizing, kim2018uniaxial, pfau2019detailed}. Techniques that can probe and disentangle contributions from multiple degrees of freedom constitute an important part of the multimodal approach. One such example is resonant elastic and inelastic X-ray scattering, which is sensitive to lattice, charge and spin degrees of freedom and has provided a considerable impact in the study of cuprates and iridates~\cite{kim2012magnetic}. RIXS has now demonstrated sub-30~meV energy resolution at the Cu $L_3$ edge, and sub-10~meV resolution at the Ir $L_3$ edge~\cite{kim2018quartz}. We also believe that there will continue to be exciting developments in momentum resolved electron energy loss spectroscopy that can measure the frequency- and wave-vector-dependent density-density correlation function~\cite{vig2017measurement} and inelastic neutron scattering.

There are also a number of new techniques that can give previously inaccessible information.  Recently, $s$-orbital non-resonant inelastic X-ray scattering using modern synchrotron facilities with high brilliance allows the direct resolution of the orbital occupation~\cite{yavacs2019direct, leedahl2019origin}.  This is an improvement over typical methods of deducing wavefunctions from optical, X-ray and neutron spectroscopy methods in which spectra must be analyzed and interpreted using modeling of spectroscopic information, for example through crystal field excitations~\cite{zhang2014neutron}.  Shown in Fig.~\ref{OrbitalImaging} is the quadrupolar scattering intensity as a function of the momentum transfer direction in the canonical Mott insulator NiO.   It directly shows the three-dimensional (3D) orbital hole density of the Ni high-spin 3$d^8$ configuration in an octahedral coordination, namely the $^3A_2$ $3d(x^2-y^2)3d(3z^2-r^2)$.  As the $3d(x^2-y^2)$ contribution vanishes in the [001]–[110] plane the small lobes of the $3d(3z^2-r^2)$ contribution remain.  This technique can also be used for itinerant systems, and may be invaluable to determine the local orbital in systems where both band formation and electron correlations are important, for example in the entangled spin–orbit states in ruthenium and iridium materials.  It could be interesting to apply it to systems with rare earths where the ground state is often composed of an admixture of complex 4$f$ orbitals.

\subsubsection{Prospects for future developments}
\label{Prospects}

Most spectroscopies focus on the spectral function of quasiparticle excitations or two-particle correlation functions in the limit of linear response. Correlated electron systems may host symmetry protected or symmetry broken phases that do not manifest directly in these spectral and correlation functions~\cite{morimoto2016topological,zhao2017global,zhao2018second}.  In this regard, we believe it will be essential to move beyond the conventional confines of linear response techniques to gain insights about strong correlation effects.  This calls for spectroscopies that explicitly probe higher order susceptibilities ($\chi^{(2)}$, $\chi^{(3)}$…) across a range of energies and wave-vectors. Key examples include 2D coherent spectroscopy (THz to IR) as a probe of fractionalized excitations in quantum spin liquids~\cite{wan2019resolving, choi2019theory, lu2017coherent}. THz emission spectroscopy and optical second harmonic generation (SHG) can be used to probe subtle symmetry-broken and hidden order phases of matter~\cite{fiebig2005second, zhao2017global,zhao2018second}.  In addition to its role in probing symmetries, it has also been proposed that nonlinear response is sensitive to Berry's phase effects in non-interacting systems~\cite{morimoto2016topological,virk2011optical,sipe1993nonlinear}.   It will be interesting to see if such physics can be extended to interacting systems.  One important consideration in designing or conceiving new spectroscopic techniques, is that they should measure a well-defined response function.  Many ultrafast pump-probe experiments using typical 800~nm light (which corresponds to 1.55~eV), which have been applied to correlated systems are interpreted as measuring some general relaxation time without clear perspective of what exactly is relaxing or how. This has limited the impact of such experiments.

\begin{figure}
\includegraphics[width=3 in]{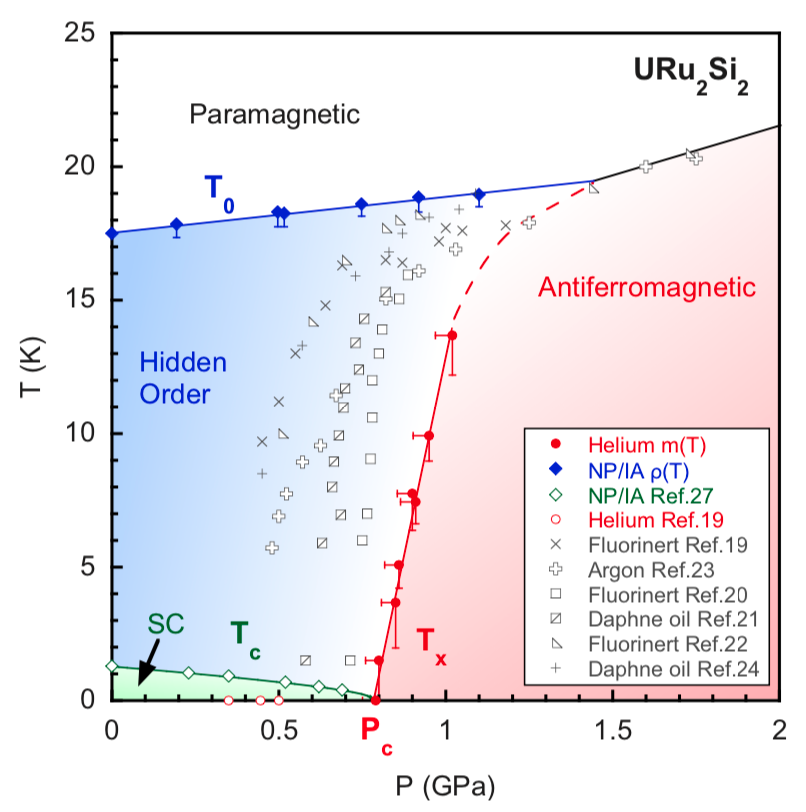}
\caption{(\textit{T,P}) phase diagram of URu$_2$Si$_2$ from resistivity and neutron scattering in the low-pressure hidden-order phase and the high-pressure antiferromagnetic phase.  $T_0$ is defined by the local minimum in the resistivity data; the error bar represents $\Delta T_0$.  $T_x$ is defined where the neutron ordered moment reaches half of its full value, the vertical error bar 90$\%$. The horizontal error bar represents a 5\% uncertainty in pressure. The superconductivity is suppressed at $P_c$ where antiferromagnetism appears. A comparison to other published data shows that the value of the AFM critical pressure $P_c$ is substantially higher under the hydrostatic conditions of Ref.~\cite{butch2010antiferromagnetic} than many previous experiments.  References correspond to those given in Ref.~\cite{butch2010antiferromagnetic}.}\label{HiddenOrder}
\end{figure}

A number of correlated systems show the phenomenon of ``hidden order'' (HO) {\it e.g.}  they may exhibit a clear sign of a phase transition in thermodynamic quantities like heat capacity or signs of a gap developing in spectroscopy, but conventional probes of symmetry breaking give little information on the nature of the ordered state.  Most famously, URu$_2$Si$_2$ shows a large peak in specific heat at $T_o = 17.5$ K, which indicates a classic second-order phase transition~\cite{bourdarot2005hidden,villaume2008signature,tripathi2007sleuthing}.  Although intensive theoretical and experimental studies have been performed, the order parameter of the state below $T_o$ is still undetermined.  For instance, it is difficult to reconcile the small size (if any) of the ordered moment ($< 0.03~\mu_B$) with the large jump of $\Delta C/T_o = 0.3$~J/mol\,{K}$^2$.  It is well established that at low-pressures the HO phase is not simple antiferromagnetism, although there is a transition at a pressure of 0.5~Pa (Fig.~\ref{HiddenOrder}) through a first-order transition to a long-range antiferromagnetic state~\cite{villaume2008signature}.   Theories ranging from orbital currents, singlet-triplet \textit{d}-density wave, hexadecapolar, antiferro-quadrupolar, to ``hastatic'' order have been proposed~\cite{mydosh2014hidden}.  The interest in URu$_2$Si$_2$ is reinforced by the appearance of unconventional superconductivity at $T_{sc} = 1.2$~K under ambient pressure, which disappears at 0.5~GPa.  It is likely that the issue has withstood thirty years of investigation because we do not have the experimental tools that easily couple to the static order or the elementary excitations of this broken symmetry state. For instance, some of the proposed orders are expected to have unconventional excitations with selection rules not easily accessible by conventional electric and magnetic dipole excitations in linear optical response.  In some cases, the excitations can be revealed, but it takes a detailed analysis.  Recently a combination of information from Raman and neutron scattering has been used to understand the nature of the broken symmetries in URu$_2$Si$_2$~\cite{kung2015chirality,buhot2014symmetry}.   A sharp excitation of 1.7~meV with $A_{2g}$ symmetry in the Raman response shows that vertical and diagonal reflection symmetries are broken at the uranium sites.  The appearance of the same excitation in neutron scattering at $(001)$ (corresponding to the inverse the $c$-axis lattice constant) requires the hidden order to be staggered alternating along the $c$ direction. Such order with alternating left- and right-handed states at the uranium sites has no modulation of charge or spin and is hidden to all probes at the zone center except for scatterings of $A_{2g}$ symmetry.  Further development of nonlinear optical techniques that evade the conventional selection rules regarding the linear response of electric and magnetic dipole excitations or enhance the cross-section of unconventional excitation may prove useful to further reveal the nature of these or other hidden ordered states.

It is possible that hidden-order states are very common and give some of the confusing phenomenology of other correlated materials.  For instance, signatures of broken symmetry also exist for the ``pseudogap'' in the cuprates~\cite{xia2008polar,he2011single,zhao2017global}. However, the precise nature of the ordered phases remains unresolved despite intense experimental and theoretical efforts.   Some candidates like the ${\bf q}=0$ and ${\bf q}= (\pi,\pi$) orbital current orders are challenging to verify with conventional scattering technique
~\cite{bourges2011novel,varma1997non,chakravarty2001hidden,croft2017no,huang2012precision} and new techniques that give information on their broken symmetries and possible unconventional excitations will be useful.

In a related fashion, future promising directions include tools that involve using and/or measuring pairs of particles ({\it e.g.}  entangled neutrons, photons, electrons) to extract response functions that are inaccessible to ``conventional'' spectroscopies. One exciting direction is the implementation of ``coincidence'' experiments, such as Auger-photoelectron coincidence spectroscopy (APECS)~\cite{haak1978auger,stefani2002electron}, which should be revisited with the advent of improved photoemission detection technology.  As mentioned above, the principal issue with such experiments is their relatively poor resolution, mainly limiting the method to getting information on the correlation hole around an electron~\cite{schumann2007mapping}.  However if this could be overcome, in addition to allowing some related recent proposals~\cite{stahl2019noise}, such experiments would probe particle-particle correlations \cite{su2020coincidence} at a finite momentum or a given time, in contrast to most other two particle probes that probe particle-hole correlations.  Here one can imagine, for instance, probing Cooper pair correlations by looking at coincidence in $\pm {\bf k}$ emitted electrons. This could give direct access to the anomalous self energy of the pairing interaction (off-diagonal term), instead of via an indirect entry into the diagonal terms in the $2\times2$ Green's function matrix in the Nambu-Gorkov representation.

As an essential property of quantum systems, probing long-range entanglement would be very powerful, but may pose an even bigger challenge. One example is using two neutrons prepared in an entangled state~\cite{shen2019unveiling} to scatter off different areas of a possible spin liquid. Under such conditions, probing the final state of the neutron pair one might be able to obtain the entanglement information of the spins in the material. This type of experiment highlights a future direction in which spectroscopic measurements, in this case beyond conventional neutron scattering, may be able to probe long-range entanglement in strongly correlated systems.  Entanglement entropy has been measured using ultracold bosonic atoms in optical lattices where identical copies of a many-body state are prepared and then interfered~\cite{islam2015measuring}.  Solid-state systems have the obvious problem in this regard that they cannot generally be easily partitioned and interfered.   There have been proposals for how to measure entanglement in solid-state systems~\cite{laflorencie2016quantum}, but they have been mostly limited to measuring systems with a globally conserved quantity, for instance the particle number for Fermi gases or the subsystem magnetization for quantum magnets and are thus far from general.  Klich and Levitov proposed that quantum noise in a quantum point contact can be used as an entanglement meter when driven by a periodic pulse train~\cite{klich2009quantum}.  In a related fashion, Song et al. \cite{song2012bipartite} propose that the noise spectrum can be a probe of entanglement in a $O(2)$ quantum magnet that has a magnetic field partially obscured by a superconducting shield.

\subsection{Local probes}\label{subsection:local probes}

Many strongly correlated systems exhibit a multitude of nearly degenerate phases that either compete or coexist locally even in clean systems. Moreover, given the degree of inhomogeneity inherent to many correlated electron systems~\cite{hamidian2016detection,zhang2016lcmo,liang2018singleshot}, local probes provide crucial tools for the identification and isolated measurement of differing local states or environments. Furthermore, in the presence of nearly degenerate states, or states that either compete or coexist, it is generally difficult to understand the behavior of the whole system as a simple composition of the microscopic parts. Spatially-resolved measurements are therefore crucial for probing individual phases not only in isolation but also for understanding how their interactions contribute to macroscopic behavior (see Fig.~\ref{LocalProbes2}). This idea is illustrated by nanoscale imaging experiments in the colossal magnetoresistive manganites as shown in Fig.~\ref{LocalProbes2}a. In these materials, microscopic competition between the metallic ferromagnetic and insulating antiferromagnetic phases determine the macroscopic transport properties. Beyond the phases themselves, there further exist many intriguing questions to explore regarding the boundaries or walls between such domains which similarly require local visualization. 

\begin{figure*}
\includegraphics[width=\textwidth]{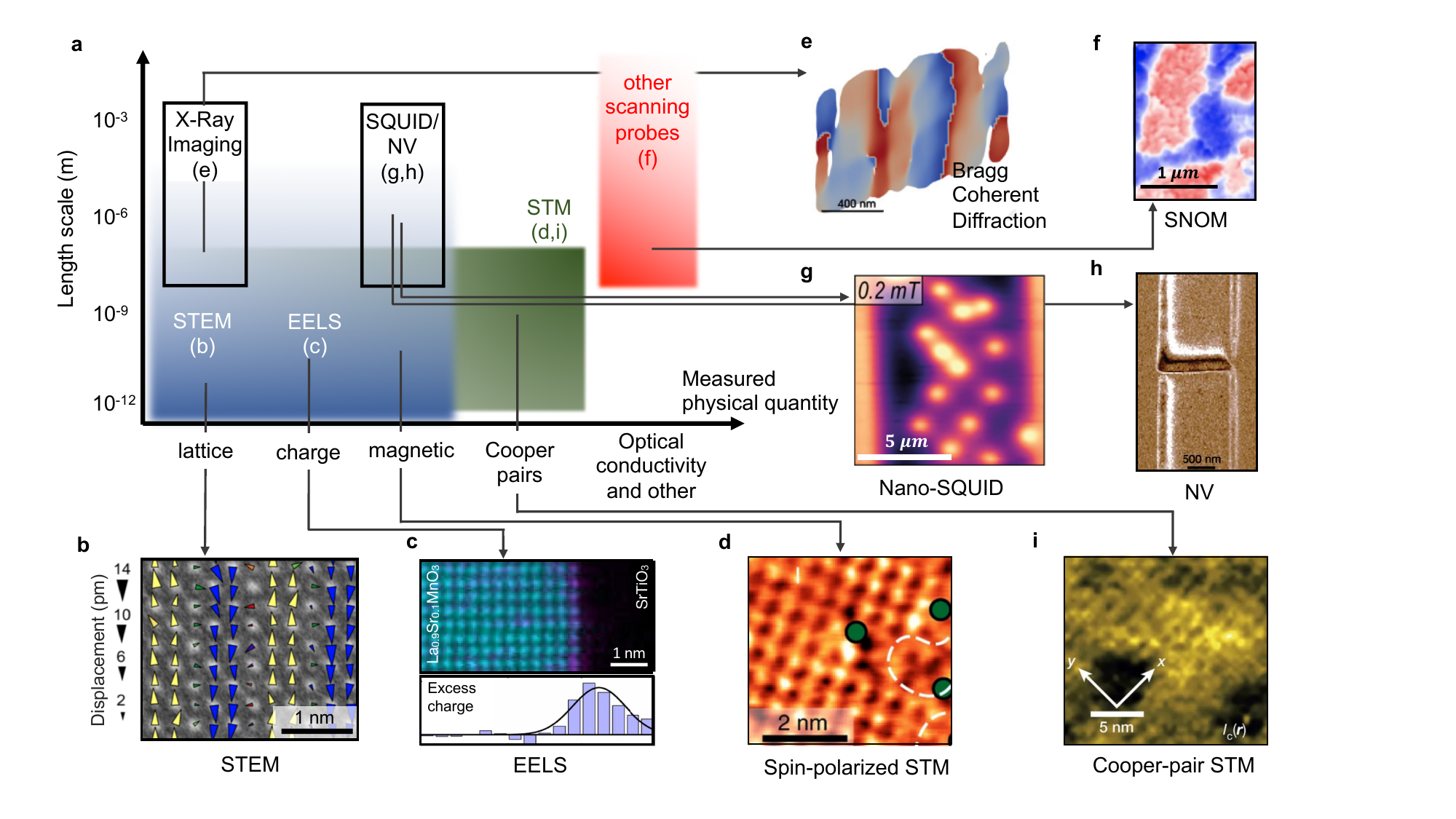}
\caption{Local probes with different sensitivities and spatial resolution. 
(a) Survey of spatially resolved probes with access to information at the picometer to millimeter length scales.
(b) STEM imaging and mapping of picometer lattice displacements in charge-ordered phases. From Ref.~\cite{el2018nature}.
(c) Spatially resolved EELS of valence and charge in an oxide interface. From Ref.~\cite{mundy2014visualizing}.
(d) Spin-polarized STM detection of magnetic moments at the atomic scale. From Ref.~\cite{enayat2014real}.
(e) Coherent Bragg X-ray imaging of structural and charge order domains. From Ref.~\cite{assefa2020scaling}.
(f) SNOM imaging of coexisting metallic and insulating domains. From Ref.~\cite{mcleod2017nanotextured}.
(g) Advanced SQUID microscopy with sub-micron resolution. From Ref.~\cite{ceccarelli2019imaging}. 
(h) NV imaging with high sensitivity to spins. From Ref.~\cite{tetienne2014nanoscale}. 
(i) STM with a superconducting tip enables Cooper pair tunneling and nanoscale imaging of the superconducting condensate. From Ref.~\cite{hamidian2016detection}.
}\label{LocalProbes}
\end{figure*}

In  Fig.~\ref{LocalProbes}, we compile several microscopic methods, the degrees of freedom to which they are sensitive, and their spatial resolutions. Some of these microscopies are relatively well-established, {\it e.g.}  transmission electron microscopy (TEM), and scanning tunneling microscopy (STM), but have been utilized recently in cutting-edge and previously unexpected ways~\cite{el2018nature,mundy2014visualizing,enayat2014real,hamidian2016detection}. Efforts to probe local spins through spin-resolved tunneling and superconducting pairs with superconducting tips have been demonstrated~\cite{enayat2014real,hamidian2016detection} in STM. Recent advances in high-resolution scanning TEM (STEM) enabled in part by the development of new imaging detectors and data processing techniques have expanded the accessible phase space for atomic resolution real-space imaging across a much wider range of temperatures and other \textit{in-situ} conditions~\cite{coll2019towards}. 

New microscopy methods have also been developed in recent years, showing promising sensitivity and spatial resolution despite their infancy. Novel developments in the design and fabrication of nano SQUID devices on a tip has enabled magnetic imaging with single spin sensitivity and 10s of nm spatial resolution~\cite{vasyukov2013scanning,ceccarelli2019imaging}. Diamond nitrogen vacancy (NV) microscopy has recently demonstrated room-temperature field sensitivities as high as $0.9~\times10^{-12}$ T/Hz$^{1/2}$ ~~\cite{wolf2015subpicotesla} with a spatial resolution tens of nanometers and below depending on the microscope design, NV center-to-sample distance, etc.~\cite{levine2019principles,tetienne2014nanoscale}. Utilizing the coherence of X-rays at existing and upcoming diffraction-limited synchrotron facilities, nano X-ray diffraction and coherent X-ray imaging have demonstrated up to $10^{-6}$ strain sensitivity and spatial resolution as high as 1~nm in metals and semiconductors~\cite{robinson2009coherent, hruszkewycz2017high, pfeiffer2018x}, and are expected to be applied to the study of lattice and electronic orders in correlated materials~\cite{chen2016remarkable,robinson2020domain, assefa2020scaling, cao2020complete}.

It should be noted that a host of microscopy approaches have proven to be or will become powerful spectroscopic tools, as in the case of STM and STEM. Few-meV energy resolution and $\sim$\AA~spatial resolution have both been demonstrated by electron energy loss spectroscopy (EELS) in the STEM~\cite{muller_atomic-scale_2008}. Recent experiments have also begun to probe  $\mathbf{q}$ on still small spatial scales with momentum-resolved EELS mapping the dispersion curves in graphene nanostructures~\cite{senga2019position}.  As spectroscopic resolution improves in spatially localized probes, spatial localization is similarly improving for many spectroscopic ``gold standards”.  With tightly focused laser pulses, second harmonic generation (SHG) and magnetic optical Kerr effect (MOKE) could deliver single-digit micron spatial resolution. Micro- and nano- ARPES have recently been realized at synchrotron user facilities across the world~\cite{cattelan2018perspective}. Meanwhile, next generation time-of-flight photoemission ``momentum microscopes'' are also becoming commercially available, enabling simultaneous 2D data collection either in electron momentum or real space~\cite{tusche2015spin}. For more discussion, see Section \ref{subsection:methods-spectroscopies}.

While many well-established techniques routinely probe down to the atomic scale (including STM, AFM, STEM, EELS), physical constraints of the advanced instrumentation required for these techniques often limits their application to specific sample conditions or geometries which in many cases do not extend to the phases of interest for condensed matter systems. For example, electron energy loss spectroscopy (EELS) can be used to probe core electronic structure down to the atomic scale, enabling the direct measurement and visualization of charge at polar interface\cite{mundy2014visualizing} (Fig.~\ref{LocalProbes}c). Compared to other core spectroscopy techniques such as X-ray absorption spectroscopy, however, the stability requirements and subsequent signal limitations of such high resolution EELS experiments are, as yet, generally limited to ambient conditions under vacuum, precluding the detailed study of how such states evolve under temperature, pressure, or other applied stimuli. Thorough exploration of competing and coexisting states in many correlated systems will require improved flexibility of existing imaging experiments as well as the development of new imaging techniques in order to probe local phenomena across a wide range of conditions and systems.   A number of potentially important areas for future work include:

\begin{figure*}
\includegraphics[width=\textwidth]{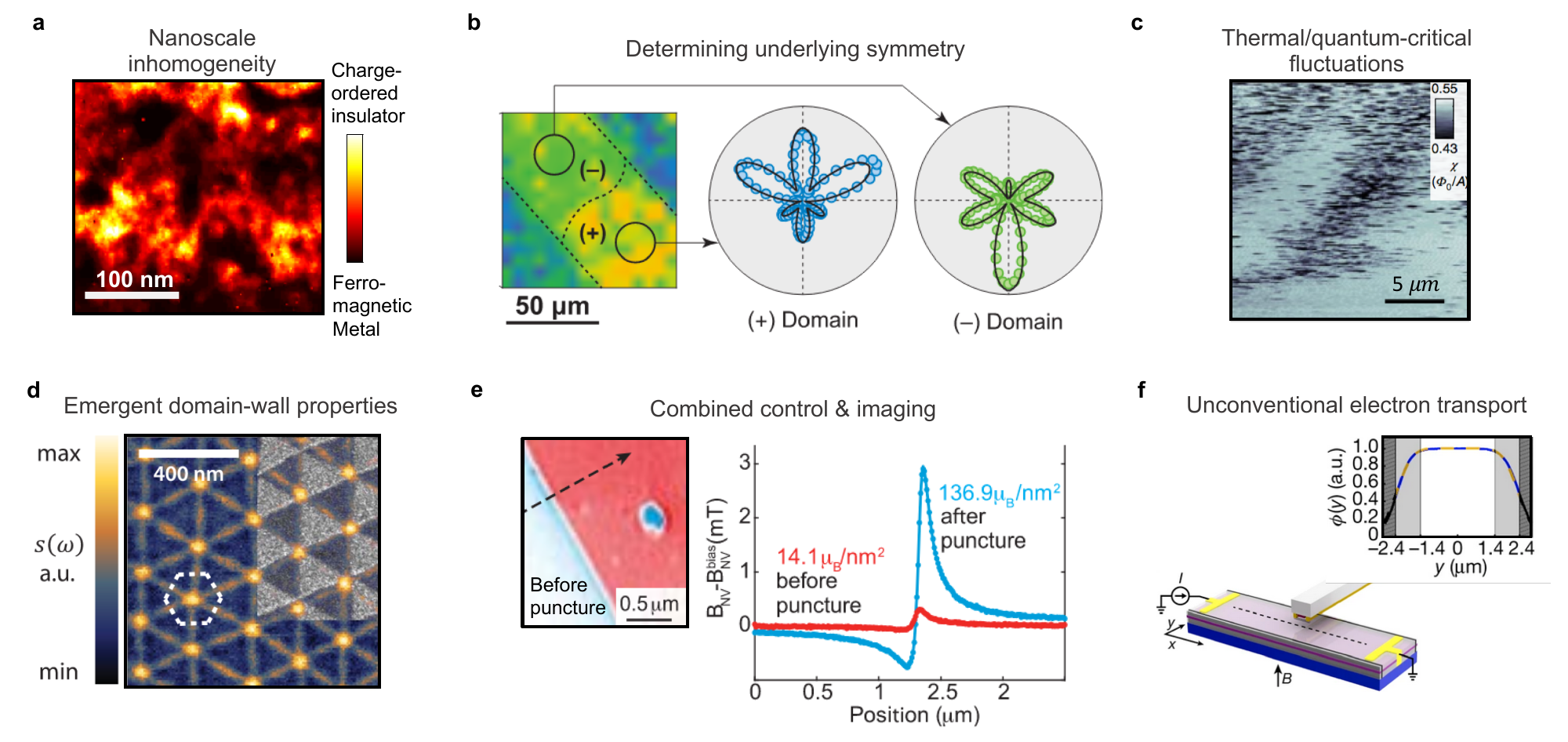}
\caption{Example problems addressed with local measurements. 
(a)~Nanoscale phase coexistence in the (La,Pr,Ca)MnO$_{3}$ manganite.
The competition between charge-ordered insulating patches (bright) and ferromagnetic metallic regions (dark) is visualized using dark-field transmission electron microscopy. 
The contrast reflects the amplitude of the charge order superlattice.
Image adapted from~\cite{uehara1999percolative}.
(b)~Endowing bulk probes such as SHG with spatial resolution can help disentangle the underlying symmetry from its sample-averaged counterpart, especially in the presence of domains or competing states. 
Such capability revealed parity domains in the parity-breaking electronic nematic metal Cd$_2$Re$_2$O$_7$~\cite{harter2017parity}. The 2D false-color map reflects the second harmonic intensity; the polar plots were measured by rotating light polarizations.
(c)~Scanning SQUID imaging of the diamagnetic susceptibility reveals quantum fluctuations in the disordered superconductor NbTiN at mesoscopic scale~\cite{kremen2018imaging}.
(d)~Domain walls often carry exotic properties distinct from the bulk, due to the suppression of a particular order parameter or local change in symmetry. 
IR nano-imaging, for instance, measures enhanced optical conductivity due to plasmons localized at the domain walls in TBG~\cite{sunku2018photonic}.
(e)~Local magnetization in the layered magnetic material CrI$_{3}$ measured using a NV magnetometer.
By adding an \textit{in situ} mechanical stimulus, NV imaging further reveals a local enhancement of the magnetization coupled to structural degrees of freedom~\cite{thiel2019probing}.
(f)~Visualizing electronic transport with enhanced spatial resolution is a promising approach for understanding exotic phenomena such as electron hydrodynamics, strange metals and topological edge modes.
In Ref.~\cite{sulpizio2019visualizing}, a scanning carbon nanotube single-electron transistor, which is sensitive to the potential of flowing electrons, reveals Poiseuille flow in high-mobility graphene devices (see also Ref.~\cite{ku2020imaging}).
}\label{LocalProbes2}
\end{figure*}

\begin{enumerate}
\item 
Sample environments and operation protocols satisfying the unique needs of correlated materials can be further developed. Because interesting electronic properties often emerge at temperatures substantially lower than room temperature, ongoing efforts for realizing stable and compact sample environments are underway for electron, coherent X-ray, and force microscopies~\cite{el2018nature,robinson2020domain,assefa2020scaling}. The additional integration of other \textit{in situ} conditions such as pressure, strain, external fields, etc. will expand these types of characterization to unexplored regions of phase space. The successful integration of such environments into modern microscopes will require innovative and inspired engineering given the space limitations and stability requirements of these techniques.

\item Inputs from theoretical modelling are essential for fully understanding the experimental measurements and for connecting experimental observations with theoretical calculations of electron response and correlation functions. Ultimately, most microscopy measurements provide some sort of contrast which is only physically meaningful if the contrast mechanism can be identified or understood. One example involves scanning near-field optical microscopy (SNOM)~\cite{atkin2012nano,mcleod2017nanotextured,mcleod2019multi}. Simulations of the electromagnetic field distribution around the tip and the sample were crucial in revealing the sensitivity to the plasmon oscillations in graphene~\cite{fei2012gate}. 

\item We encourage establishing a consortium, or a collaboration mechanism for multimodal explorations of correlated materials at the same spatial location under the same condition. This will require developing (1) fiducials that could be used across microscopies, and (2) measurement protocols where air-sensitive, \textit{in situ} experiments precede \textit{ex situ} and/or potentially destructive ones. Similar standard operating procedures are already common in other fields, for example the biological science cryo-EM community.

\item Microscopy studies provide a natural playground for the application of machine learning. The advent of pixelated area detectors across many modern microscopic methods in the last decade fuel the acceleration of data generation. For example, modern coherent X-ray imaging can generate sub-terabytes of data within 24~hours. Its generation, transfer, and storage will require new data infrastructures and management plans not only for user facilities but also individual research labs in the foreseeable future. Moreover, identifying key features in the image, streamlining the data analysis and cross-comparing different microscopic studies will benefit from different approaches of artificial intelligence~\cite{cherukara2018real,laanait2019exascale}.   Machine learning has been recently applied to STM~\cite{cheung2020dictionary}.
\end{enumerate}

\subsection{Spectroscopies and microscopies out of equilibrium}\label{subsubsection: tr}

\begin{figure*}
\includegraphics[width=\textwidth]{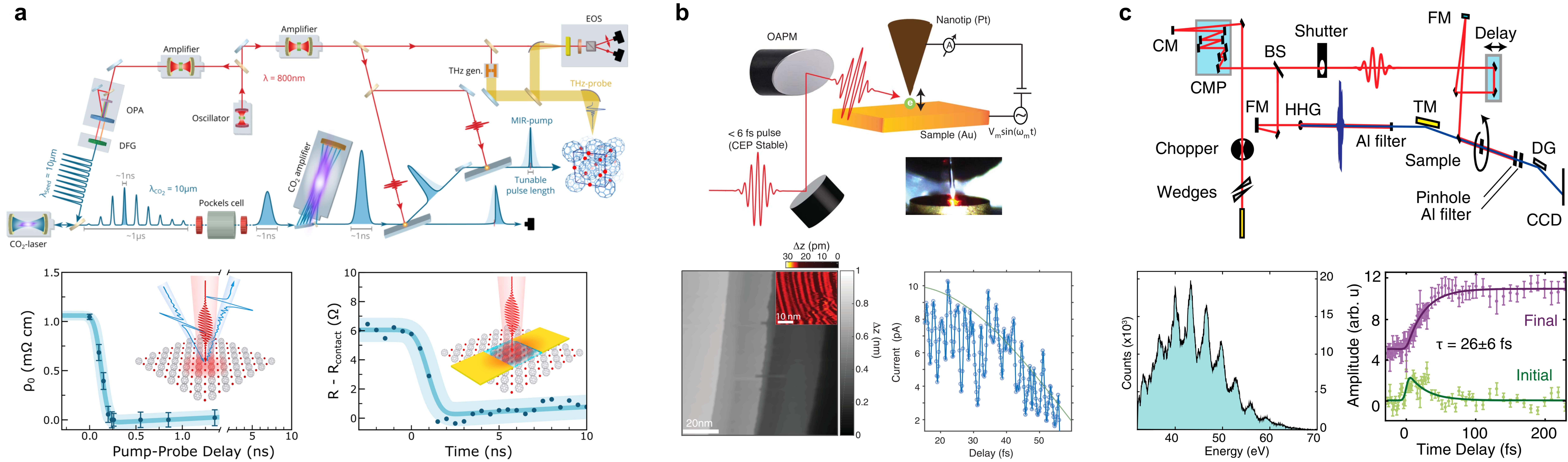}
\caption{Frontiers of nonequilibrium spectroscopies and microscopies. 
(a)~Signatures of MIR pulse-induced superconductivity in K$_3$C$_{60}$ at the nanosecond timescale. \textit{Top}: Schematic of the setup that produces pulses of duration variable between 5~ps and 1.3~ns, centered at 10.6~$\mu$m wavelength, and with a pulse energy of up to 10~mJ. \textit{Bottom}: Transient resistivity obtained from extrapolation of optical conductivity (left) and 2-point transport measurement (right). From Ref.~\onlinecite{budden2021evidence}.
(b)~Attosecond coherent field-driven STM. \textit{Top}: Schematic and photo of the setup where $<6$-fs carrier-envelope phase (CEP)-stable pulses are focused at a Pt/Ir tip. \textit{Bottom left}: Surface topography on a Au surface, generated solely by laser-induced tunneling electrons. Inset shows atomic reconstruction on the Au surface. \textit{Bottom right}: Laser-induced tunneling current for various pump-probe delays, featuring a 1.7-eV plasmon mode in a Au nanorod. From Ref.~\onlinecite{garg2020atto}.
(c)~Table-top attosecond XUV spectroscopy. \textit{Top:} Schematic of the pump-probe setup using few-fs CEP-stable NIR pump and attosecond probe. The probe is produced by high harmonic generation (HHG) and its spectrum spans from 30 to 60~eV (bottom left). \textit{Bottom right}: Characteristic timescale of insulator-metal-transition in VO$_2$ ($26\pm6$~fs) revealed by absorption changes from the vanadium $M_{2,3}$ edge. From Ref.~\onlinecite{jager2017tracking}.
}\label{fig:nonequilibrium}
\end{figure*}

Nonequilibrium spectroscopies provide a new avenue to disentangle different degrees of freedom, and enable the study of collective excitations, metastable and transient states, and fluctuations (Fig.~\ref{fig:nonequilibrium}). Many of the previous time-resolved nonequilibrium studies on strongly correlated materials, such as time-resolved reflectivity and time-resolved photoemission spectroscopy, have heavily utilized photoexcitations at energies around $\sim1.5$~eV~\cite{gedik2005abrupt, demsar1999superconducting,yang2015inequivalence}. This is largely attributed to the commercial development and widespread adoption of 800-nm Ti:sapphire femtosecond lasers. Their photon energy is, however, orders of magnitude larger than the relevant energy scales for collective excitations in correlated material systems. One important future direction is to develop pumps with photon energies targeted in resonance with underlying low-energy excitations, such as phonons, magnons, and other emergent particles. This has been attempted on limited basis in correlated materials: using mid-infrared pumping in resonance with a vibration mode of cuprate superconductors to induce possible nonequilibrium superconductivity~\cite{forst2011nonlinear}; using targeted pumping to establish transient metastable ferroelectric states~\cite{li2019terahertz,nova2019metastable}; and using orbital excitations across the Mott gap to observe the evolution of spin waves in iridates~\cite{dean2016ultrafast}. The development of a continuously tunable, $>100$~kHz repetition rate, $>1~\mu$J pulse energy THz to mid-infrared sources, such as the one at Helmholtz-Zentrum Dresden-Rossendorf~\cite{green2016thz}, will enable future time-resolved spectroscopies to accommodate a wide range of materials with characteristic low-energy excitations in the range of 1--200~meV. Recent effort to achieve intense mid-infrared pulses with variable duration (ps to ns) would further enable sustained optical driving and stabilize transient states~\cite{budden2021evidence} (Fig.~\ref{fig:nonequilibrium}a).

In contrast to the THz/mid-infrared sources mentioned above, sub-fs pulses coming from the high-harmonic generation (HHG) process represent the high end of the spectrum. It is not immediately obvious that the sub-fs and the 10s--100s~eV scales are relevant for most processes in solids, but HHG-based techniques offer two important pieces of information in the time domain.  First, element specificity through XUV/X-ray absorption spectroscopy is afforded through table-top or free-electron laser-based absorption spectroscopies that have been used to track element-specific evolution of local bonding, magnetization, and lattice structure~\cite{geneaux2019transient,jager2017tracking} (see Fig.~\ref{fig:nonequilibrium}c). Without the constraint of the uncertainty principle, both time and energy resolutions can be optimal (as to fs in time and $\sim10$~meV in energy).  Second, phase sensitivity through a holographic detection for photoelectrons is allowed by leveraging the interference between different quantum paths of photoemission among successive harmonics in an HHG pulse train.  The phase of a complex wave in atomic orbitals has been recently imaged~\cite{villeneuve2017coherent,huismans2011time}. It is our hope that a similar holographic detection can be applied to materials, which may enable phase-sensitive photoelectron spectroscopy and, for instance, allow the investigation into the sign of the superconducting gap in correlated superconductors.

We envision that novel nonequilibrium spectroscopies will facilitate addressing some specific pertinent issues that are discussed above as well as driving new phenomena. For instance, theoretical calculations have predicted that circularly polarized light can provide a knob to break time-reversal symmetry and drive frustrated Mott insulators into a chiral spin liquid~\cite{claassen2017dynamical,quito2020floquet}. This can be realized using spectroscopies such as mid-infrared pumped time-resolved second harmonic generation. In a Kondo breakdown QCP, electron lifetime diverges following a multiscale temperature scaling law, which was proposed to be addressed by time-resolved optical reflectivity and photoemission spectroscopies~\cite{paul2008multiscale}. The challenge here is to have sufficient sensitivity when approaching a zero-excitation limit to minimize transient heating. In correlated superconductors, the amplitude mode of superconductivity, which is often termed the condensed matter analogue of the Higgs boson, has been reported by THz pump-optical probe reflectivity measurements~\cite{katsumi2018higgs}. The pump photon energy has to be smaller than twice the superconducting gap (or the antinodal gap in the case of a $d$-wave superconductor). Future developments which combine THz pumping and other spectroscopic probes such as photoemission or scanning tunneling spectroscopy can further detail the momentum- and spatial dependence of such collective modes. Notably, the collective mode is a direct manifestation of the order parameter~\cite{baldini2020verwey}. Last but not least, studies of order parameters by driving competing phases may provide insight into the phase competition at their fundamental interaction timescale~\cite{wandel2020light,kogar2020light} and may help find the roots of node-antinode dichotomy in cuprate superconductors~\cite{hashimoto2014energy}.

In addition, time-resolved STM (Fig.~\ref{fig:nonequilibrium}b), time-resolved scanning SQUID~\cite{cui2017scanning}, time-resolved neutron diffraction, and time-resolved RIXS are tangible future directions for nonequilibrium spectroscopies~\cite{dean2016ultrafast,cao2019ultrafast}. The pump excitation here can be a pulsed electric or magnetic field which facilitates transitions between different correlated states. Near-field imaging techniques in combination with femtosecond laser excitations may enable the imaging of hidden ordered phases in correlated materials~\cite{mcleod2019multi}. A further extension of time-resolved near-field imaging is to employ second harmonics as a pathway to reveal symmetry-breaking phases~\cite{neacsu2009second}. In such novel nonequilibrium imaging techniques, a method to acquire a broad field-of-view image in a single shot can be a paradigm-shifting development in studying spatially inhomogeneous correlated phases~\cite{zhang2016lcmo,liang2018singleshot}.

\subsection{Experimental probes in extreme environments}
\label{extreme}

\begin{table*}[t!]
 \small
\begin{center}
\begin{tabular}{||c|p{2.8cm}|p{2.8cm}|p{2.8cm}|p{2.8cm}|p{2.8cm}||} 
\hline
\hline
{Technique} & {Low temperature} & {Pressure} & {dc B field} & {Pulsed B field} & {Ultrafast E field$^\mathsection$} \\
\hline
\hline
Electrical transport & 6~mK~\cite{pan2008experimental} & 200~GPa~\cite{drozdov2015conventional} & 45~T~\cite{fang2022fermi} & 95~T~\cite{ramshaw2018quantum} & 400~kV/cm~\cite{mciver2020ahe} \\ \cline{2-6}
Thermal transport & 50~mK~\cite{toews2013thermal} & 50~GPa~\cite{hohensee2015thermal} & 45~T~\cite{grissonnanche2014direct} & & \\ \hline
Heat capacity & 0.6~mK~\cite{greywall1986he} & 4.4~GPa~\cite{zheng2014high} & 45~T~\cite{riggs2011heat} & 60~T~\cite{terashima2018magnetic} & \\ \cline{2-6}
Magnetic properties & 0.2~mK~\cite{bismuthSC} & 20~GPa~\cite{jackson2005high} & 45~T~\cite{Jaime2012} & 75~T~\cite{zuo2015magnetic} & 9~MV/cm~\cite{schlauderer2019temporal} \\ \hline
Broadband FTIR & 0.15~K$^{\mathparagraph}$  &  16.5~GPa~\cite{challener1986far}  & 35~T~\cite{brinzari2013electron} & &  \\ \cline{2-6}
Broadband NIR &   & 400~GPa~\cite{loubeyre2020highp} & 35~T~\cite{brinzari2013electron} & 74~T~\cite{zaric2006excitons} &  \\ \cline{2-6}
Raman and PL & 20~mK~(PL)~\cite{Hayne1999} & 1~TPa~(Raman)~\cite{Dubrovinskaiae2016} & 45~T (Raman)~\cite{kim2013measurement} & 89~T (PL)~\cite{crooker2007tuning} &  \\ \cline{2-6}
Time-domain THz & 0.4~K~\cite{curtis2016cyclotron} & 34.4 MPa \cite{zhang2017high} & 25~T~\cite{baydin2020time} & $\sim$30~T~\cite{baydin2020time} & 70~MV/cm~\cite{schubert2014thz} \\ \cline{2-6}
X-ray & 220~mK~\cite{Suzuki2002,suzuki2004} & 1~TPa~\cite{Dubrovinskaiae2016} & 10~T~\cite{paolasini2007id20} & 43~T~\cite{narumi2012x} & 1~MV/cm~\cite{kozina2019sto} \\ \cline{2-6}
Neutron & 30~mK~\cite{ross2011qsi}  & 94~GPa~\cite{Boehler2017} & 15~T (inelastic)~\cite{NAP18355} $^*$ & 40~T (diffraction)~\cite{duc2018neutron} &  \\ \cline{2-6}
EPR/ESR & 1.4~K~\cite{takahashi2005rotating} & 2.5~GPa~\cite{sakurai2015development} & 45~T~\cite{takahashi2005rotating} & 63~T~\cite{zvyagin2011field} & \\ \cline{2-6}
NMR & 20~mK~\cite{pustogow2019constraints} & 90~GPa~\cite{meier2018nmr}. & 45~T~\cite{frachet2020hidden} & 56~T~\cite{tokunaga2019high} &  \\ \cline{2-6}
ARPES & 1~K~\cite{zabolotnyy2012sro} &   & 10$^6 \mathrm{A/cm}^2$ ~\cite{kaminski2016arpes}$^\dagger$ &  & 25~kV/cm~\cite{reimann2018arpes}\\ \cline{2-6}
EELS & 10~K~\cite{zhao2018direct} & &  & & \\ \hline
STM & 10~mK~\cite{song2010stm} &  & 34~T~\cite{tao2017stm} & & 100~MV/cm~\cite{garg2020atto}\\ \cline{2-6}
STEM & 4.5~K~\cite{Behler1993} &  & &  & \\ \cline{2-6}
SNOM & 20~K~\cite{yang2013snom} &  & 7~T~\cite{yang2013snom} &  &  \\ \cline{2-6}
MIM & 450~mK~\cite{allen2019mim} &  & 9~T~\cite{ma2015mim} & & \\
\hline
\hline
\end{tabular}
\end{center}
\caption{Present limits of extreme environments used in the study of strongly correlated systems. The values listed are the highest or lowest implemented to the best of our knowledge. The list is not meant to be exhaustive, but to highlight areas of recent activity or avenues for future improvement. Empty cells indicate that the environment is either incompatible with the technique or no reports were found. Here we have not included the considerable research into destructive fields that can reach up to 1200 T (for a controlled explosion)~\cite{nakamura2018record}, that enables even higher pulsed magnetic fields for electrical transport and magnetic property measurements. To explore current capabilities of high field measurements in U.S. national laboratories, one can also refer to the NHMFL website \url{https://nationalmaglab.org/}. $^ \mathsection$dc or quasi-dc electrostatic gating on 2D materials which can attain a field exceeding $\geq1$~MV/cm, is compatible with most environments listed. Here, numbers and references are limited to E fields generated by an ultrashort light pulses.  $^\mathparagraph$\url{https://kbfi.ee/chemical-physics/research-facilities/?lang=en}.  $^*$HZB did have 26~T steady field~\cite{prokevs2017magnetic}, but that facility is now closed.  The highest field inelastic facility in the world is now at ILL at the modest 15 T.  $^\dagger$Magnetic field cannot be applied in traditional ARPES, but the large currents that were applied in the referenced experiment perhaps induce a related effect.
\label{tab:extremecondition}}
\end{table*}

Probing correlated electron phenomena and accessing energy scales relevant to underlying interactions often require extreme experimental environments, such as ultra low temperatures, high pressure, high magnetic fields, and high electric fields.   Extreme environmental conditions can induce new correlated states or may be used to probe the energy scale of correlations involved in an existing phase.  In Table~\ref{tab:extremecondition}, we list the present limits in extreme environments discussed below.

\subsubsection{Ultralow temperature}

Ultralow temperature allows for the observation of new ground states and quantum effects that may be masked or destroyed by thermal excitations. Experiments in this extreme limit have led to notable discoveries, most recently establishing superconductivity near a magnetic QCP as a relatively universal phenomenon~\cite{schuberth2016emergence}, demonstrating a superconducting phase at extremely low electron density in crystalline bismuth~\cite{bismuthSC}, or finding delicate phases in the 2D electron gasses under high field and very low temperature ({\it i.e.}, even-denominator fractional quantum Hall phases~\cite{willett1987observation} are found in some cases at temperatures only as low as 5~mK~\cite{xia2004electron}). Advancing experimental capabilities at ultralow temperature could lead to more significant advances. For example, thermal conductivity at ultralow temperature, {\it i.e.}, below the tens of millikelvins that are typically achievable in a dilution refrigerator, may help clarify the ground state in quantum spin liquid candidate materials like herbertsmithite if samples of sufficient quality become available.

\subsubsection{High electric/magnetic field}

Strong electric fields, up to $\sim$1~V/\AA, constitute another knob for tuning phase transitions in strongly correlated systems. In the dc limit, electrostatic gating of ultrathin materials has enabled the precision control of carrier concentration and band structure~\cite{goldman2014gating}. In the ac limit, strong fields in mid-infrared or terahertz laser pulses have led to new dynamical states of matter, such as symmetry-breaking or topologically nontrivial phases, some of which do not exist at equilibrium~\cite{salen2019thz,mciver2020ahe}. Furthermore, nonlinear response associated with extreme fields offers a sensitive probe of symmetry~\cite{torchinsky2017shg}, topology~\cite{sodemann2015nhe}, electron correlation~\cite{silva2018hhg}, and perhaps spin fractionalization~\cite{wan2019resolving}. Therefore, access to high electric fields is instrumental in both manipulating and measuring properties of many correlated systems.

\begin{figure}
\includegraphics[width=0.6\columnwidth]{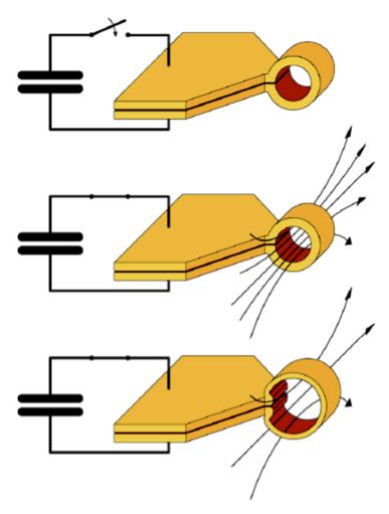}
\caption{Sketch of a single-turn coil used in Megagauss facilities to reach magnetic fields significantly higher than 100~T in pulsed field (up to $\sim 300$~T in a 5-mm-diameter bore) for a few microseconds. A current pulse of $\sim 3$~MA from a capacitor bank heatss and expands the coil as it generates the field (a semi-destructive technique). From Ref.~\cite{battesti2018high}.}
\label{Fig.Megagauss_fields}
\end{figure}

With 100~T reached in pulsed field at Los Alamos National Laboratory in 2012 (and now available in other high field laboratories) as the highest non-destructive magnetic field ever realized, these last decades have witnessed great advances in magnetic field technologies~\cite{battesti2018high}. The development of Megagauss magnets (semi-destructive fields where the coil is destroyed at each pulse but the sample space preserved, see Fig.~\ref{Fig.Megagauss_fields}), which are able to reach as high as $\sim$ 300~T~\cite{Portugall_1999}, foresees brand new kinds of experiments in condensed matter. High magnetic fields have already proved themselves to be an effective tool in disentangling competing and coexisting states of matter. For example, the normal state of several unconventional superconductors, like high-$T_c$ cuprates, has been extensively studied down to low temperatures by suppressing the superconductivity with magnetic field~\cite{proust2019remarkable,shi2020vortex}. High magnetic fields can also be extremely useful to observe quantum oscillations that can give a direct measurement of the fermiology and the quasiparticles behavior in quantum materials~\cite{sebastian2015quantum}.  High magnetic fields allow the stabilization of a remarkable field induced unconventional superconducting phase in UTe$_2$ (Fig.~\ref{UTe2HighField}) that has the highest upper and lower critical fields of any field-induced superconducting phase (more than 40~T and 65~T respectively)~\cite{ran2019nearly,ran2019extreme}.  High magnetic fields are also essential to probe integer and fractional quantum Hall effects in two-dimensional electron systems, and are expected to continue to play a critical role as the field of 2D materials is developed further~\cite{dean2011multicomponent}. Moreover, they allow the study of the ``quantum limit'', in which all charge carriers are confined to the lowest Landau level~\cite{moll2016magnetic}, a particularly pertinent state to explore in correlated topological materials.  Unfortunately, magnetic field cannot be applied in traditional ARPES experiments, although photoemission experiments with large currents (10$^6 \mathrm{A/cm}^2$) have been performed~\cite{kaminski2016arpes}.  For at least that application on superconductors, they perhaps induce a related effect. Magnetic fields can be applied in another novel momentum- and energy-resolved tunneling spectroscopy technique that also probes single-particle spectral functions~\cite{jang2017full}.

There are of course many other interesting applications of large fields not listed here, especially in the domain of magnetism. We need new technological developments to access even higher fields for longer periods of time, for example to study the normal state of some cuprates with a large upper critical field (YBCO, Hg1201, Bi2212, etc.). These could be obtained thanks to advances in the megagauss technology that currently reaches around 150--200~T, but are not widely developed to date. Improvements in the pulse duration and the cooling time for standard pulsed field magnets would also help, respectively by expanding the types of measurements that can be achieved in pulsed field (thermal transport for example is more challenging than electrical transport, see Table~\ref{tab:extremecondition}) and by increasing the number of data points taken. Finally, advances in all-superconducting dc magnets, as the 32~T magnet in service at NHMFL since 2017, will avoid the vibrations due to water cooling used for resistive coils, allowing new types of vibration-sensitive experiments, and reduces operating costs. These types of magnets though remain challenging to implement because of the use of high-$T_c$ superconducting coils that are delicate to manufacture.

\subsubsection{High pressure/strain}

Application of hydrostatic pressure is one of the cleanest ways to continuously tune the interplay between spin, charge, lattice and orbital degrees of freedom. Due to the often competing interactions in correlated systems, novel electronic and magnetic phenomena can be quickly masked by the presence of disorder induced by chemical substitution. This is a well-known problem in investigations that attempt to tune correlations using chemical doping. The best examples of materials where experiments under high pressure have contributed a wealth of information for understanding quantum phase transitions are heavy-fermions~\cite{si2010heavy, chen2016}. Not only can 3D interactions may be fine-tuned with pressure, but also the role of inter-planar coupling, {\it e.g.}  for organic superconductors, can be studied via pressure-induced dimensional crossover, such as quasi-1D to quasi-2D, quasi-2D to 3D~\cite{Nagata1998,Valla2002,Pashkin2010,Zhang2019}. Pressure has also proven to be powerful in tuning quantum critical points and quantum spin liquid states in magnetically frustrated systems~\cite{Mirebeau2002,Dressel2011,Powell_2011,Klein2018}. However due to the stringent requirements such as small sample volume, space for pressure cells, etc., there is a constant need for development to improve high-pressure generation technologies.  One successful example is the combination of high pressure diamond anvil cell technology with synchrotron X-ray techniques.  Here improved source brilliance and small focus sizes allows diffraction, scattering, and spectroscopy in ways not allowed previously~\cite{Shen2017,wang2019x}. IR spectroscopy at synchrotron beamlines also allows advantages for high pressures~\cite{kimura2012infrared,piccinini2005far}.  There have been recent successes with combining NMR experiments with pressures as high as 90~GPa~\cite{meier2018nmr} and neutron diffraction experiment above 90~GPa~\cite{Boehler2017}. Although earth sciences and other areas of condensed matter physics have mastered the art of general high pressures using diamond anvil cells, combining very high pressure with many of the techniques used for quantum materials has been prohibitive. For example, combining high pressure with highly surface sensitive techniques such as ARPES has remained completely unaccomplished for obvious reasons.   

As most correlated electron systems are anisotropic, hydrostatic pressure is often not the ideal tuning parameter. Therefore, uniaxial strain is becoming one of the most commonly used technique to study electronic and magnetic properties in correlated systems~\cite{Tokunaga2008,Tarantini2011,Masaki2012}. Furthermore, uniaxial and/or biaxial strain/pressure allow us to overcome some of the limitations of hydrostatic pressure generation methods. In some cases such as superconducting heavy fermions, it is even desired to increase the structural anisotropy to tune $T_c$~\cite{PhysRevLett.91.076402}. Uniaxial strain has been successfully combined with experimental probes for which the application of hydrostatic pressure is currently not possible. For example, experimental probes such as STM, SEM and ARPES can use uniaxial strain on single crystals or thin films to tune the system as a substitute for hydrostatic pressure~\cite{trainer2019stm, ricco2018cro, flototto2018bi2se3}. Using the substrate of thin films to create unaxial or biaxial strain that manipulates material properties is a well established technique in the field of correlated electron systems such as cuprates~\cite{Abrecht2003}, manganites~\cite{Liao2014} and titanates~\cite{Zhang2013}, but the recent focus has been on using strain as a continuous in situ probe of materials with piezoelectric stacks. This has been proven to be extremely powerful when probing nematic correlations in the pnictides~\cite{chu2012divergent}. For instance, Ref. \onlinecite{chu2012divergent} showed how measurement of the divergent nematic susceptibility of the iron pnictide superconductor Ba(Fe$_{1-x}$Co$_x$)$_2$As$_2$ can distinguish an electronic nematic phase transition from a simple ferroelastic distortion. In situ strain and STM have been combined to show how nematic fluctuations and nematic order in an iron-based superconductor change across the phase diagram~\cite{andrade2018visualizing}.  It was shown that sizable nematic correlations persist to high temperatures and that there is strong {\it nonlinear} coupling between structure and electronic nematicity even at temperatures above the tetragonal to orthorhombic transition.

\subsubsection{Challenges and outlooks}

Despite the rapid progress in pushing various experimental limits, probes in \textit{multiple} extreme environments are uncommon due to the high level of experimental difficulty. We believe that efforts should be undertaken to push boundaries in these experimental techniques by both integrating setups with new extreme environments, and by combining multiple environments. Due to the complexity of many-body interaction in correlated electron material, it is often necessary to leverage more than one extreme environment.
For instance, the combination of high magnetic fields up to 60~T and hydrostatic pressure up to 4~GPa in URu$_2$Si$_2$ has recently shed new light on the subtle competition between the hidden-order state and neighboring magnetically ordered states \cite{knafo2020destabilization}.
In some situations, one extreme environment may be employed to help access a phase boundary within the experimental limits of a second environment.  For example, on YBCO, the extremely high critical fields of 150~T at optimal doping hindered investigations of the normal state at low temperature. With the application of high pressure, however, critical field values can in principle be lowered to a more accessible field regime, thereby allowing studies of the quantum phases below the superconducting transition. The use of this extreme environment, acting similarly as doping, enabled for instance the access to the entire overdoped regime in pristine YBCO by lowering $T_c$ and moving the end of the superconducting dome~\cite{alireza2017accessing}.  This approach also minimizes effects from varying the sample quality and the environments in different set-ups or laboratories, as multiple experiments are performed simultaneously on the same sample.

The experimental complexity of many advanced scattering, spectroscopy, and microscopy measurements also poses great challenges for accessing extreme sample conditions. For example, static magnetic fields for inelastic neutron scattering are still nowadays limited to a modest field of 15~T at the Institut Laue-Langevin\footnote{The Helmholtz-Zentrum-Berlin facility that used to have a 26~T steady field magnet~\cite{prokevs2017magnetic}, but it is now shut down. \\  \url{https://www.helmholtz-berlin.de/pubbin/news_seite?nid=14076;sprache=en;seitenid=74699} \\ \url{https://www.helmholtz-berlin.de/projects/rueckbau/ber/index_en.html}}. Implementing extreme conditions in these setups necessitates a collaborative effort to surmount numerous engineering challenges, but we expect that such endeavor is of great interest to the community and will pay off in the long run.

\section{Concluding remarks}

In this manuscript, we have attempted to lay out the results of our discussions on the Future of the Correlated Electron Problem.  These are hard problems and progress on them takes time.  But progress has cycles and going forward it will make sense to stubbornly come back to stubborn old problems with new ideas.

We have so far stayed away from sociological and philosophical aspects surrounding the Future of the Correlated Electron Problem, but such discussions frequently surfaced during the workshop. Issues such as questionable reproducibility, over-analysis/interpretation, overabundance of jargon, pressure to produce high-impact publications, and a pursuit of ``novelty'' and therefore a lack of systematic studies, are definitely not unique to the correlated electron community. Nonetheless, both senior and junior scientists in this field should make a concerted effort to address these issues, which only complicate the already complex problems. As a community, we can take simple, concrete steps, and one example would be giving more credits to reporting conflicting or null results. Indeed, recently we have seen several high-profile cases concerning reproducibility of experimental data in the literature {\it e.g.}  triplet-pairing in superconducting Sr$_2$RuO$_4$~\cite{pustogow2019constraints,ishida1998spin}, giant current-induced diamagnetism in Ca$_2$RuO$_4$~\cite{zhao2019nonequilibrium}, and chiral Majorana fermions in a quantum anomalous Hall-superconductor device~\cite{kayyalha2020absence}. These examples serve to remind us of the high standards and open debate are important when pushing forward in this field.

We have made the above forecasts, predictions, and recommendations not from an expectation that we will be ultimately be proven correct.  It is of course ``difficult to make predictions, especially about the future''\footnote{This famous expression has been variously attributed to a collection of individuals as diverse as N. Bohr, S. Goldwyn, K.K. Steincke, {\it and} Yogi Berra. \url{https://quoteinvestigator.com/2013/10/20\/no-predict/}}. Our hope, however, is that the topics we have presented will provide inspiration for others working in this field and motivation for the idea that significant progress can be made on very hard problems if we focus our collective energies.   Irrespective of the particular path taken, it is clear that the Future of the Correlated Electron Problem will be full of fascinating physics and unexpected twists and turns that will challenge us for years to come.

\section{Acknowledgements}

We thank NSF CMP program for suggestions regarding the topic and general structure of the workshop.  This project was supported by the NSF DMR-2002329 and The Gordon and Betty Moore Foundation (GBMF) EPiQS initiative.  We would like to sincerely thank A. Kapitulnik, A. J. Leggett, M.B. Maple, T.M. McQueen, M. Norman, P. S. Riseborough, and G. A. Sawatzky for their lectures at the workshop and advice on the writing of this manuscript.  We would also like to thank G. Blumberg, C. Broholm, S. Crooker, N. Drichko, and A. Patel for helpful consultation on topics discussed herein.  A number of individuals also had independent support:  (AA, EH; GBMF-4305 and GBMF-8691), (IMH; GBMF-9071), (HJC; NHMFL is supported by the NSF DMR-1644779 and the state of Florida), (YH, AZ; Miller Institute for Basic Research in Science), (YC; US DOE-BES DE-AC02-06CH11357), (AS; Spallation Neutron Source, a DOE Office of Science User Facility operated by ORNL),  (SAAG; ARO-W911NF-18-1-0290, NSF DMR-1455233), (YW; DOE-BES DE-SC0019331, GBMF-4532).

\bigskip

\textbf{Competing interests:} The Authors declare no Competing Financial or Non-Financial Interests.

\bigskip

 \textbf{Data availability:  }
No data sets were generated or analysed during the current study.

\bigskip

 \textbf{Author contributions:  }
All authors contributed to the writing of the manuscript.   NPA organized the workshop and edited the manuscript.

\bibliographystyle{plainnat}
\bibliography{CorrelatedReferencesFull}

\begin{thebibliography}{746}
\providecommand{\natexlab}[1]{#1}
\providecommand{\url}[1]{\texttt{#1}}
\expandafter\ifx\csname urlstyle\endcsname\relax
  \providecommand{\doi}[1]{doi: #1}\else
  \providecommand{\doi}{doi: \begingroup \urlstyle{rm}\Url}\fi

\bibitem[lic()]{lichtenberg1992sr2ruo4}
{Sr$_2$RuO$_4$: A metallic substrate for the epitaxial growth of
  {YBa}$_2${Cu}$_3${O}$_{7-\delta}$}.
\newblock \doi{10.1063/1.106432}.

\bibitem[Aasen et~al.(2020)Aasen, Mong, Hunt, Mandrus, and
  Alicea]{aasen2020electrical}
David Aasen, Roger S.~K. Mong, Benjamin~M. Hunt, David Mandrus, and Jason
  Alicea.
\newblock {Electrical Probes of the Non-Abelian Spin Liquid in Kitaev
  Materials}.
\newblock \emph{Phys. Rev. X}, 10:\penalty0 031014, 2020.
\newblock \doi{10.1103/PhysRevX.10.031014}.
\newblock URL \url{https://link.aps.org/doi/10.1103/PhysRevX.10.031014}.

\bibitem[Abanin et~al.(2015)Abanin, De~Roeck, and
  Huveneers]{PhysRevLett.115.256803}
Dmitry~A. Abanin, Wojciech De~Roeck, and Fran\c{c}ois Huveneers.
\newblock Exponentially slow heating in periodically driven many-body systems.
\newblock \emph{Phys. Rev. Lett.}, 115:\penalty0 256803, 2015.
\newblock \doi{10.1103/PhysRevLett.115.256803}.
\newblock URL \url{https://link.aps.org/doi/10.1103/PhysRevLett.115.256803}.

\bibitem[Abanin et~al.(2017)Abanin, De~Roeck, Ho, and
  Huveneers]{PhysRevB.95.014112}
Dmitry~A. Abanin, Wojciech De~Roeck, Wen~Wei Ho, and Fran\c{c}ois Huveneers.
\newblock Effective hamiltonians, prethermalization, and slow energy absorption
  in periodically driven many-body systems.
\newblock \emph{Phys. Rev. B}, 95:\penalty0 014112, 2017.
\newblock \doi{10.1103/PhysRevB.95.014112}.
\newblock URL \url{https://link.aps.org/doi/10.1103/PhysRevB.95.014112}.

\bibitem[Abrahams et~al.(2001)Abrahams, Kravchenko, and
  Sarachik]{abrahams2001metallic}
Elihu Abrahams, Sergey~V Kravchenko, and Myriam~P Sarachik.
\newblock Metallic behavior and related phenomena in two dimensions.
\newblock \emph{Reviews of Modern Physics}, 73\penalty0 (2):\penalty0 251,
  2001.
\newblock \doi{10.1103/RevModPhys.73.251}.

\bibitem[Abrecht et~al.(2003)Abrecht, Ariosa, Cloetta, Mitrovic, Onellion, Xi,
  Margaritondo, and Pavuna]{Abrecht2003}
M.~Abrecht, D.~Ariosa, D.~Cloetta, S.~Mitrovic, M.~Onellion, X.~X. Xi,
  G.~Margaritondo, and D.~Pavuna.
\newblock Strain and high temperature superconductivity: Unexpected results
  from direct electronic structure measurements in thin films.
\newblock \emph{Phys. Rev. Lett.}, 91:\penalty0 057002, 2003.
\newblock \doi{10.1103/PhysRevLett.91.057002}.
\newblock URL \url{https://link.aps.org/doi/10.1103/PhysRevLett.91.057002}.

\bibitem[Ajayan et~al.(2016)Ajayan, Kim, and Banerjee]{ajayan2016van}
Pulickel Ajayan, Philip Kim, and Kaustav Banerjee.
\newblock {van der Waals materials}.
\newblock \emph{Physics Today}, 69:\penalty0 9--38, 2016.
\newblock \doi{10.1063/PT.3.3297}.

\bibitem[Alet et~al.(2016)Alet, Damle, and Pujari]{Alet_sign}
Fabien Alet, Kedar Damle, and Sumiran Pujari.
\newblock Sign-problem-free {Monte} {Carlo} simulation of certain frustrated
  quantum magnets.
\newblock \emph{Phys. Rev. Lett.}, 117:\penalty0 197203, 2016.
\newblock \doi{10.1103/PhysRevLett.117.197203}.
\newblock URL \url{https://link.aps.org/doi/10.1103/PhysRevLett.117.197203}.

\bibitem[Alexandradinata et~al.(2020{\natexlab{a}})Alexandradinata, H{\"o}ller,
  Wang, Cheng, and Lu]{alex2019crystallographic}
A~Alexandradinata, J~H{\"o}ller, Chong Wang, Hengbin Cheng, and Ling Lu.
\newblock Crystallographic splitting theorem for band representations and
  fragile topological photonic crystals.
\newblock \emph{Physical Review B}, 102\penalty0 (11):\penalty0 115117,
  2020{\natexlab{a}}.
\newblock \doi{10.1103/PhysRevB.102.115117}.

\bibitem[Alexandradinata et~al.(2020{\natexlab{b}})Alexandradinata, Wang,
  Bernevig, and Zaletel]{alex2019glideresolved}
A~Alexandradinata, Zhijun Wang, B~Andrei Bernevig, and Michael Zaletel.
\newblock Glide-resolved photoemission spectroscopy: Measuring topological
  invariants in nonsymmorphic space groups.
\newblock \emph{Physical Review B}, 101\penalty0 (23):\penalty0 235166,
  2020{\natexlab{b}}.
\newblock \doi{10.1103/PhysRevB.101.235166}.

\bibitem[Alireza et~al.(2017)Alireza, Zhang, Guo, Porras, Loew, Hsu, Lonzarich,
  Le~Tacon, Keimer, and Sebastian]{alireza2017accessing}
P.~L. Alireza, G.~H. Zhang, W.~Guo, J.~Porras, T.~Loew, Y.-T. Hsu, G.~G.
  Lonzarich, M.~Le~Tacon, B.~Keimer, and Suchitra~E. Sebastian.
\newblock {Accessing the entire overdoped regime in pristine YBa$_2$Cu$_3$O$_6+
  x$ by application of pressure}.
\newblock \emph{Physical Review B}, 95\penalty0 (10):\penalty0 100505, 2017.
\newblock \doi{10.1103/PhysRevB.95.100505}.

\bibitem[Allen et~al.(1979)Allen, Batlogg, and Wachter]{SmB63}
J.~W. Allen, B.~Batlogg, and P.~Wachter.
\newblock {Large low-temperature Hall effect and resistivity in mixed-valent
  Sm${\mathrm{B}}_{6}$}.
\newblock \emph{Phys. Rev. B}, 20:\penalty0 4807--4813, 1979.
\newblock \doi{10.1103/PhysRevB.20.4807}.
\newblock URL \url{https://link.aps.org/doi/10.1103/PhysRevB.20.4807}.

\bibitem[Allen et~al.(2019)Allen, Cui, {Yue Ma}, Mogi, Kawamura, Fulga,
  Goldhaber-Gordon, Tokura, and Shen]{allen2019mim}
Monica Allen, Yongtao Cui, Eric {Yue Ma}, Masataka Mogi, Minoru Kawamura,
  Ion~Cosma Fulga, David Goldhaber-Gordon, Yoshinori Tokura, and Zhi-Xun Shen.
\newblock {Visualization of an axion insulating state at the transition between
  2 chiral quantum anomalous Hall states}.
\newblock \emph{Proceedings of the National Academy of Sciences}, 116\penalty0
  (29):\penalty0 14511--14515, 2019.
\newblock \doi{10.1073/pnas.1818255116}.

\bibitem[Anderson(2007)]{anderson2007there}
Philip~W Anderson.
\newblock Is there glue in cuprate superconductors?
\newblock \emph{Science}, pages 1705--1707, 2007.
\newblock \doi{10.1126/science.1141598}.

\bibitem[Andrade et~al.(2018)Andrade, Berger, Rosenthal, Wang, Xing, Wang, Jin,
  Fernandes, Millis, and Pasupathy]{andrade2018visualizing}
Erick~F Andrade, Ayelet~Notis Berger, Ethan~P Rosenthal, Xiaoyu Wang, Lingyi
  Xing, Xiancheng Wang, Changqing Jin, Rafael~M Fernandes, Andrew~J Millis, and
  Abhay~N Pasupathy.
\newblock Visualizing the nonlinear coupling between strain and electronic
  nematicity in the iron pnictides by elasto-scanning tunneling spectroscopy.
\newblock \emph{arXiv preprint arXiv:1812.05287}, 2018.
\newblock \doi{10.48550/arXiv.1812.05287}.

\bibitem[Andres et~al.(1975)Andres, Graebner, and Ott]{andres19754}
K~Andres, JE~Graebner, and HR~Ott.
\newblock {4 f-Virtual-Bound-State Formation in CeAl$_3$ at Low Temperatures}.
\newblock \emph{Phys. Rev. Lett.}, 35\penalty0 (26):\penalty0 1779, 1975.
\newblock \doi{10.1103/PhysRevLett.35.1779}.

\bibitem[Aoki et~al.(2019)Aoki, Nakamura, Honda, Li, Homma, Shimizu, Sato,
  Knebel, Brison, Pourret, et~al.]{aoki2019unconventional}
Dai Aoki, Ai~Nakamura, Fuminori Honda, DeXin Li, Yoshiya Homma, Yusei Shimizu,
  Yoshiki~J Sato, Georg Knebel, Jean-Pascal Brison, Alexandre Pourret, et~al.
\newblock {Unconventional superconductivity in heavy fermion UTe$_2$}.
\newblock \emph{Journal of the Physical Society of Japan}, 88\penalty0
  (4):\penalty0 043702, 2019.
\newblock \doi{10.7566/JPSJ.88.043702}.

\bibitem[Argüello-Luengo et~al.(2019)Argüello-Luengo, González-Tudela, Shi,
  Zoller, and Cirac]{Arguello-Luengo2019}
Javier Argüello-Luengo, Alejandro González-Tudela, Tao Shi, Peter Zoller, and
  J.~Ignacio Cirac.
\newblock Analogue quantum chemistry simulation.
\newblock \emph{Nature}, 574:\penalty0 215--218, 2019.
\newblock \doi{10.1038/s41586-019-1614-4}.

\bibitem[Armitage(2019)]{armitage2019superconductivity}
N~Peter Armitage.
\newblock Superconductivity mystery turns 25.
\newblock \emph{Nature}, 576:\penalty0 386, 2019.
\newblock \doi{10.1038/d41586-019-03734-7}.

\bibitem[Armitage et~al.(2010{\natexlab{a}})Armitage, Fournier, and
  Greene]{armitage2010progress}
NP~Armitage, P~Fournier, and RL~Greene.
\newblock Progress and perspectives on electron-doped cuprates.
\newblock \emph{Reviews of Modern Physics}, 82\penalty0 (3):\penalty0 2421,
  2010{\natexlab{a}}.
\newblock \doi{10.1103/RevModPhys.82.2421}.

\bibitem[Armitage et~al.(2010{\natexlab{b}})Armitage, Tediosi, L{\'e}vy,
  Giannini, Forro, and Van Der~Marel]{armitage2010infrared}
NP~Armitage, Riccardo Tediosi, Florence L{\'e}vy, Enrico Giannini, L~Forro, and
  Dirk Van Der~Marel.
\newblock {Infrared conductivity of elemental bismuth under pressure: Evidence
  for an avoided Lifshitz-type semimetal-semiconductor transition}.
\newblock \emph{Phys. Rev. Lett.}, 104\penalty0 (23):\penalty0 237401,
  2010{\natexlab{b}}.
\newblock \doi{10.1103/PhysRevLett.104.237401}.

\bibitem[Armitage et~al.(2018)Armitage, Mele, and Vishwanath]{armitage2018weyl}
NP~Armitage, EJ~Mele, and Ashvin Vishwanath.
\newblock {Weyl and {Dirac} semimetals in three-dimensional solids}.
\newblock \emph{Reviews of Modern Physics}, 90\penalty0 (1):\penalty0 015001,
  2018.
\newblock \doi{10.1103/RevModPhys.90.015001}.

\bibitem[Arnold et~al.(2019)Arnold, David, Filoche, Jerison, and
  Mayboroda]{Arnold}
Douglas~N. Arnold, Guy David, Marcel Filoche, David Jerison, and Svitlana
  Mayboroda.
\newblock Computing spectra without solving eigenvalue problems.
\newblock \emph{SIAM Journal on Scientific Computing}, 41\penalty0
  (1):\penalty0 B69--B92, 2019.
\newblock \doi{10.1137/17M1156721}.
\newblock URL \url{, https://doi.org/10.1137/17M1156721}.

\bibitem[Arute et~al.(2019)Arute, Arya, Bab {BCS}~h, Bacon, Bardin, Barends,
  Biswas, Boixo, Brandao, Buell, Burkett, Chen, Chen, Chiaro, Collins,
  Courtney, Dunsworth, Farhi, Foxen, and Martinis]{Arute2019}
Frank Arute, Kunal Arya, Ryan Bab {BCS}~h, Dave Bacon, Joseph Bardin, Rami
  Barends, Rupak Biswas, Sergio Boixo, Fernando Brandao, David Buell, Brian
  Burkett, Yu~Chen, Zijun Chen, Ben Chiaro, Roberto Collins, William Courtney,
  Andrew Dunsworth, Edward Farhi, Brooks Foxen, and John Martinis.
\newblock Quantum supremacy using a programmable superconducting processor.
\newblock \emph{Nature}, 574:\penalty0 505--510, 2019.
\newblock \doi{10.1038/s41586-019-1666-5}.

\bibitem[Aryasetiawan et~al.(2004)Aryasetiawan, Imada, Georges, Kotliar,
  Biermann, and Lichtenstein]{Aryasetiawan_downfolding}
F.~Aryasetiawan, M.~Imada, A.~Georges, G.~Kotliar, S.~Biermann, and A.~I.
  Lichtenstein.
\newblock Frequency-dependent local interactions and low-energy effective
  models from electronic structure calculations.
\newblock \emph{Phys. Rev. B}, 70:\penalty0 195104, 2004.
\newblock \doi{10.1103/PhysRevB.70.195104}.
\newblock URL \url{https://link.aps.org/doi/10.1103/PhysRevB.70.195104}.

\bibitem[Assaf et~al.(2013)Assaf, Cardinal, Wei, Katmis, Moodera, and
  Heiman]{MR-thinTI}
B.~A. Assaf, T.~Cardinal, P.~Wei, F.~Katmis, J.~S. Moodera, and D.~Heiman.
\newblock Linear magnetoresistance in topological insulator thin films: Quantum
  phase coherence effects at high temperatures.
\newblock \emph{Applied Physics Letters}, 102\penalty0 (1):\penalty0 012102,
  2013.
\newblock \doi{10.1063/1.4773207}.
\newblock URL \url{, https://doi.org/10.1063/1.4773207}.

\bibitem[Assefa et~al.(2020)Assefa, Cao, Diao, Harder, Cha, Kisslinger, Gu,
  Tranquada, Dean, and Robinson]{assefa2020scaling}
Tadesse~A Assefa, Y~Cao, J~Diao, RJ~Harder, W~Cha, K~Kisslinger, GD~Gu,
  JM~Tranquada, MPM Dean, and IK~Robinson.
\newblock {Scaling behavior of low-temperature orthorhombic domains in the
  prototypical high-temperature superconductor
  La$_{1.875}$Ba$_{0.125}$CuO$_4$}.
\newblock \emph{Phys. Rev. B}, 101\penalty0 (5):\penalty0 054104, 2020.
\newblock \doi{10.1103/PhysRevB.101.054104}.

\bibitem[Atkin et~al.(2012)Atkin, Berweger, Jones, and Raschke]{atkin2012nano}
Joanna~M. Atkin, Samuel Berweger, Andrew~C. Jones, and Markus~B. Raschke.
\newblock Nano-optical imaging and spectroscopy of order, phases, and domains
  in complex solids.
\newblock \emph{Advances in Physics}, 61\penalty0 (6):\penalty0 745--842, 2012.
\newblock \doi{10.1080/00018732.2012.737982}.

\bibitem[Baldini et~al.(2020)Baldini, Belvin, Rodriguez-Vega, Ozel, Legut,
  Koz{\l}owski, Ole{\'{s}}, Parlinski, Piekarz, Lorenzana, Fiete, and
  Gedik]{baldini2020verwey}
Edoardo Baldini, Carina~A. Belvin, Martin Rodriguez-Vega, Ilkem~Ozge Ozel,
  Dominik Legut, Andrzej Koz{\l}owski, Andrzej~M. Ole{\'{s}}, Krzysztof
  Parlinski, Przemys{\l}aw Piekarz, Jos{\'{e}} Lorenzana, Gregory~A. Fiete, and
  Nuh Gedik.
\newblock {Discovery of the soft electronic modes of the trimeron order in
  magnetite}.
\newblock \emph{Nature Physics}, 2020.
\newblock \doi{10.1038/s41567-020-0823-y}.

\bibitem[Balents(2010)]{balents2010spin}
Leon Balents.
\newblock Spin liquids in frustrated magnets.
\newblock \emph{Nature}, 464\penalty0 (7286):\penalty0 199--208, 2010.
\newblock \doi{10.1038/nature08917}.

\bibitem[Balla and Brandt(1965)]{JETPbismuth}
D~Balla and NB~Brandt.
\newblock Investigation of the effect of uniform compression on the temperature
  dependence of the electrical conductivity of bismuth.
\newblock \emph{Soviet Physics JETP}, 20\penalty0 (5):\penalty0 1111, 1965.
\newblock \doi{not available}.

\bibitem[Barcel{\'{o}} et~al.(2011)Barcel{\'{o}}, Liberati, and
  Visser]{gravityanalogue}
C.~Barcel{\'{o}}, Liberati, and M~Visser.
\newblock Analogue gravity.
\newblock \emph{Living Rev. Relativ.}, 14, 2011.
\newblock \doi{10.12942/lrr-2011-3}.

\bibitem[Barkeshli et~al.(2014)Barkeshli, Berg, and Kivelson]{Barkeshli2014}
Maissam Barkeshli, Erez Berg, and Steven Kivelson.
\newblock Coherent transmutation of electrons into fractionalized anyons.
\newblock \emph{Science}, 346\penalty0 (6210):\penalty0 722--725, 2014.
\newblock \doi{10.1126/science.1253251}.
\newblock URL \url{https://science.sciencemag.org/content/346/6210/722}.

\bibitem[Barrett(1952)]{BarrettSTO}
John~H Barrett.
\newblock Dielectric constant in perovskite type crystals.
\newblock \emph{Physical Review}, 86\penalty0 (1):\penalty0 118, 1952.
\newblock \doi{10.1103/PhysRev.86.118}.

\bibitem[Battesti et~al.(2018)Battesti, Beard, B{\"o}ser, Bruyant, Budker,
  Crooker, Daw, Flambaum, Inada, Irastorza, et~al.]{battesti2018high}
R{\'e}my Battesti, Jerome Beard, Sebastian B{\"o}ser, Nicolas Bruyant, Dmitry
  Budker, Scott~A Crooker, Edward~J Daw, Victor~V Flambaum, Toshiaki Inada,
  Igor~G Irastorza, et~al.
\newblock High magnetic fields for fundamental physics.
\newblock \emph{Physics Reports}, 765:\penalty0 1--39, 2018.
\newblock \doi{10.1016/j.physrep.2018.07.005}.

\bibitem[Bauer et~al.(2002)Bauer, Frederick, Ho, Zapf, and Maple]{Bauer02}
ED~Bauer, NA~Frederick, P-C Ho, VS~Zapf, and MB~Maple.
\newblock Superconductivity and heavy fermion behavior in pros 4 sb 12.
\newblock \emph{Phys. Rev. B}, 65\penalty0 (10):\penalty0 100506, 2002.
\newblock \doi{10.1103/PhysRevB.65.100506}.

\bibitem[Baydin et~al.(2021)Baydin, Makihara, Peraca, and Kono]{baydin2020time}
Andrey Baydin, Takuma Makihara, Nicolas~Marquez Peraca, and Junichiro Kono.
\newblock {Time-domain terahertz spectroscopy in high magnetic fields}.
\newblock \emph{Frontiers of Optoelectronics}, 14:\penalty0 110--129, 2021.
\newblock \doi{10.1007/s12200-020-1101-4}.

\bibitem[Bednorz and M{\"u}ller(1986)]{bednorz1986possible}
J~George Bednorz and K~Alex M{\"u}ller.
\newblock {Possible high T$_c$ superconductivity in the Ba-La- Cu- O system}.
\newblock \emph{Zeitschrift f{\"u}r Physik B Condensed Matter}, 64\penalty0
  (2):\penalty0 189--193, 1986.
\newblock \doi{10.1007/BF01303701}.

\bibitem[Beenakker(2019)]{Beenakkercommentary}
Carlo Beenakker.
\newblock Hidden landscape of an anderson insulator.
\newblock \emph{Journal Club of Condensed Matter Physics}, 2019.
\newblock \doi{10.36471/JCCM_August_2019_01}.

\bibitem[Behler et~al.(1993)Behler, Lang, Pan, Thommen-Geiser, and
  Güntherodt]{Behler1993}
S.~Behler, H.~P. Lang, S.~H. Pan, V.~Thommen-Geiser, and H.~J. Güntherodt.
\newblock {Imaging C$_{60}$ fullerite at 4.5 K by scanning tunneling
  microscopy}.
\newblock \emph{Zeitschrift für Physik B Condensed Matter}, 91\penalty0
  (1):\penalty0 1--2, 1993.
\newblock URL \url{https://doi.org/10.1007/BF01316700}.

\bibitem[Belitz and Kirkpatrick(2019)]{belitz2019magnetic}
D~Belitz and TR~Kirkpatrick.
\newblock Magnetic quantum phase transitions in a clean {Dirac} metal.
\newblock \emph{Phys. Rev. B}, 100\penalty0 (17):\penalty0 174433, 2019.
\newblock \doi{10.1103/PhysRevB.100.174433}.

\bibitem[Belitz et~al.(1999)Belitz, Kirkpatrick, and Vojta]{belitz1999first}
Dietrich Belitz, Theodore~R Kirkpatrick, and Thomas Vojta.
\newblock First order transitions and multicritical points in weak itinerant
  ferromagnets.
\newblock \emph{Phys. Rev. Lett.}, 82\penalty0 (23):\penalty0 4707, 1999.
\newblock \doi{10.1103/PhysRevLett.82.4707}.

\bibitem[Belopolski et~al.(2019)Belopolski, Manna, Sanchez, Chang, Ernst, Yin,
  Zhang, Cochran, Shumiya, Zheng, Singh, Bian, Multer, Litskevich, Zhou, Huang,
  Wang, Chang, Xu, Bansil, Felser, Lin, and Hasan]{Belopolski1278}
Ilya Belopolski, Kaustuv Manna, Daniel~S. Sanchez, Guoqing Chang, Benedikt
  Ernst, Jiaxin Yin, Songtian~S. Zhang, Tyler Cochran, Nana Shumiya, Hao Zheng,
  Bahadur Singh, Guang Bian, Daniel Multer, Maksim Litskevich, Xiaoting Zhou,
  Shin-Ming Huang, Baokai Wang, Tay-Rong Chang, Su-Yang Xu, Arun Bansil,
  Claudia Felser, Hsin Lin, and M.~Zahid Hasan.
\newblock {Discovery of topological {Weyl} fermion lines and drumhead surface
  states in a room temperature magnet}.
\newblock \emph{Science}, 365\penalty0 (6459):\penalty0 1278--1281, 2019.
\newblock \doi{10.1126/science.aav2327}.
\newblock URL \url{https://science.sciencemag.org/content/365/6459/1278}.

\bibitem[Berg et~al.(2019)Berg, Lederer, Schattner, and Trebst]{berg2019monte}
Erez Berg, Samuel Lederer, Yoni Schattner, and Simon Trebst.
\newblock {Monte {Carlo} studies of quantum critical metals}.
\newblock \emph{Annual Review of Condensed Matter Physics}, 10:\penalty0
  63--84, 2019.
\newblock \doi{10.1146/annurev-conmatphys-031218-013339}.

\bibitem[Bhaskar et~al.(2016)Bhaskar, Banerjee, Abdollahi, Wang, Schlom,
  Rijnders, and Catalan]{bhaskar2016flexoelectric}
Umesh~Kumar Bhaskar, Nirupam Banerjee, Amir Abdollahi, Zhe Wang, Darrell~G
  Schlom, Guus Rijnders, and Gustau Catalan.
\newblock A flexoelectric microelectromechanical system on silicon.
\newblock \emph{Nature nanotechnology}, 11\penalty0 (3):\penalty0 263--266,
  2016.
\newblock \doi{10.1038/nnano.2015.260}.

\bibitem[Bhattacharya et~al.(2016)Bhattacharya, Skinner, Khalsa, and
  Suslov]{STO-metalMR}
Anand Bhattacharya, Brian Skinner, Guru Khalsa, and Alexey~V Suslov.
\newblock Spatially inhomogeneous electron state deep in the extreme quantum
  limit of strontium titanate.
\newblock \emph{Nature Communications}, 7\penalty0 (1):\penalty0 1--9, 2016.
\newblock \doi{10.1038/ncomms12974}.

\bibitem[Bistritzer and MacDonald(2011)]{bistritzer2011moire}
Rafi Bistritzer and Allan~H MacDonald.
\newblock Moir{\'e} bands in twisted double-layer graphene.
\newblock \emph{Proceedings of the National Academy of Sciences}, 108\penalty0
  (30):\penalty0 12233--12237, 2011.
\newblock \doi{10.1073/pnas.1108174108}.

\bibitem[Blankenbecler et~al.(1981)Blankenbecler, Scalapino, and
  Sugar]{Blankenbecler}
R.~Blankenbecler, D.~J. Scalapino, and R.~L. Sugar.
\newblock {Monte {Carlo} calculations of coupled boson-fermion systems. I}.
\newblock \emph{Phys. Rev. D}, 24:\penalty0 2278--2286, 1981.
\newblock \doi{10.1103/PhysRevD.24.2278}.

\bibitem[Bloch et~al.(2012)Bloch, Dalibard, and Nascimbene]{bloch2012quantum}
Immanuel Bloch, Jean Dalibard, and Sylvain Nascimbene.
\newblock Quantum simulations with ultracold quantum gases.
\newblock \emph{Nature Physics}, 8\penalty0 (4):\penalty0 267--276, 2012.
\newblock \doi{10.1038/nphys2259}.

\bibitem[Boehler et~al.(2013)Boehler, Guthrie, Molaison, dos Santos,
  Sinogeikin, Machida, Pradhan, and Tulk]{Boehler2017}
R~Boehler, M~Guthrie, J.J. Molaison, A.M. dos Santos, S~Sinogeikin, S.~Machida,
  N~Pradhan, and C.A. Tulk.
\newblock {Large-volume diamond cells for neutron diffraction above 90 GPa}.
\newblock \emph{High Pressure Research}, 33\penalty0 (3):\penalty0 546--554,
  2013.
\newblock \doi{10.1080/08957959.2013.823197}.
\newblock URL
  \url{http://www.tandfonline.com/doi/abs/10.1080/08957959.2013.823197}.

\bibitem[Booth et~al.(2009)Booth, Thom, and Alavi]{FCIQMC}
George~H Booth, Alex J~W Thom, and Ali Alavi.
\newblock {Fermion {Monte} {Carlo} without fixed nodes: a game of life, death,
  and annihilation in Slater determinant space.}
\newblock \emph{J. Chem. Phys.}, 131\penalty0 (5):\penalty0 054106, 2009.
\newblock \doi{10.1063/1.3193710}.

\bibitem[Boschini et~al.(2018)Boschini, da~Silva~Neto, Razzoli, Zonno, Peli,
  Day, Michiardi, Schneider, Zwartsenberg, Nigge, Zhong, Schneeloch, Gu,
  Zhdanovich, Mills, Levy, Jones, Giannetti, and Damascelli]{Boschini2018}
F.~Boschini, E.~H. da~Silva~Neto, E.~Razzoli, M.~Zonno, S.~Peli, R.~P. Day,
  M.~Michiardi, M.~Schneider, B.~Zwartsenberg, P.~Nigge, R.~D. Zhong,
  J.~Schneeloch, G.~D. Gu, S.~Zhdanovich, A.~K. Mills, G.~Levy, D.~J. Jones,
  C.~Giannetti, and A.~Damascelli.
\newblock Collapse of superconductivity in cuprates via ultrafast quenching of
  phase coherence.
\newblock \emph{Nature Materials}, 17\penalty0 (5):\penalty0 416--420, 2018.
\newblock \doi{10.1038/s41563-018-0045-1}.
\newblock URL \url{https://doi.org/10.1038/s41563-018-0045-1}.

\bibitem[Bourdarot et~al.(2005)Bourdarot, Bombardi, Burlet, Enderle, Flouquet,
  Lejay, Kernavanois, Mineev, Paolasini, Zhitomirsky,
  et~al.]{bourdarot2005hidden}
F~Bourdarot, A~Bombardi, P~Burlet, M~Enderle, J~Flouquet, P~Lejay,
  N~Kernavanois, VP~Mineev, L~Paolasini, ME~Zhitomirsky, et~al.
\newblock Hidden order in uru2si2.
\newblock \emph{Physica B: Condensed Matter}, 359:\penalty0 986--993, 2005.
\newblock \doi{10.1016/j.physb.2005.01.318}.

\bibitem[Bourges and Sidis(2011)]{bourges2011novel}
Philippe Bourges and Yvan Sidis.
\newblock {Novel magnetic order in the pseudogap state of high-$T_c$ copper
  oxides superconductors}.
\newblock \emph{Comptes Rendus Physique}, 12\penalty0 (5-6):\penalty0 461--479,
  2011.
\newblock \doi{10.1016/j.crhy.2011.04.006}.

\bibitem[Boyd and Phillips(2019)]{boyd2019single}
Christian Boyd and Philip~W Phillips.
\newblock {Single-parameter scaling in the magnetoresistance of optimally doped
  La$_{2- x}$Sr$_x$CuO$_4$}.
\newblock \emph{Phys. Rev. B}, 100\penalty0 (15):\penalty0 155139, 2019.
\newblock \doi{10.1103/PhysRevB.100.155139}.

\bibitem[Bo{\v{z}}ovi{\'c} et~al.(2016)Bo{\v{z}}ovi{\'c}, He, Wu, and
  Bollinger]{bovzovic2016dependence}
I~Bo{\v{z}}ovi{\'c}, X~He, J~Wu, and AT~Bollinger.
\newblock Dependence of the critical temperature in overdoped copper oxides on
  superfluid density.
\newblock \emph{Nature}, 536\penalty0 (7616):\penalty0 309--311, 2016.
\newblock \doi{10.1038/nature19061}.

\bibitem[Brandt and Ponomarev(1969)]{brandt1969pressure}
NB~Brandt and Ya~G Ponomarev.
\newblock Pressure-induced electron transitions in bismuth- tin, bismuth-lead,
  bismuth-antimony, and bismuth-antimony-lead alloys.
\newblock \emph{SOV PHYS JETP}, 28\penalty0 (4):\penalty0 635--646, 1969.
\newblock \doi{not available}.

\bibitem[Brinzari et~al.(2013)Brinzari, Haraldsen, Chen, Sun, Kim, Tung,
  Litvinchuk, Schlueter, Smirnov, Manson, et~al.]{brinzari2013electron}
Tatiana~V Brinzari, Jason~T Haraldsen, Peng Chen, Q-C Sun, Younghee Kim, L-C
  Tung, Alexander~P Litvinchuk, John~A Schlueter, Dmitry Smirnov, Jamie~L
  Manson, et~al.
\newblock Electron-phonon and magnetoelastic interactions in ferromagnetic
  {Co[N(CN)$_2$]$_2$}.
\newblock \emph{Phys. Rev. Lett.}, 111\penalty0 (4):\penalty0 047202, 2013.
\newblock \doi{10.1103/PhysRevLett.111.047202}.

\bibitem[Broholm et~al.(2020)Broholm, Cava, Kivelson, Nocera, Norman, and
  Senthil]{broholm2020quantum}
C~Broholm, RJ~Cava, SA~Kivelson, DG~Nocera, MR~Norman, and T~Senthil.
\newblock Quantum spin liquids.
\newblock \emph{Science}, 367\penalty0 (6475), 2020.
\newblock \doi{10.1126/science.aay0668}.

\bibitem[Brooks-Bartlett et~al.(2014)Brooks-Bartlett, Banks, Jaubert,
  Harman-Clarke, and Holdsworth]{brooks2014magnetic}
ME~Brooks-Bartlett, Simon~T Banks, Ludovic~DC Jaubert, Adam Harman-Clarke, and
  Peter~CW Holdsworth.
\newblock Magnetic-moment fragmentation and monopole crystallization.
\newblock \emph{Physical Review X}, 4\penalty0 (1):\penalty0 011007, 2014.
\newblock \doi{10.1103/PhysRevX.4.011007}.

\bibitem[Brown et~al.(2019)Brown, Mitra, Guardado-Sanchez, Nourafkan, Reymbaut,
  H{\'e}bert, Bergeron, Tremblay, Kokalj, Huse, et~al.]{brown2019bad}
Peter~T Brown, Debayan Mitra, Elmer Guardado-Sanchez, Reza Nourafkan, Alexis
  Reymbaut, Charles-David H{\'e}bert, Simon Bergeron, A-MS Tremblay, Jure
  Kokalj, David~A Huse, et~al.
\newblock {Bad metallic transport in a cold atom Fermi-Hubbard system}.
\newblock \emph{Science}, 363\penalty0 (6425):\penalty0 379--382, 2019.
\newblock \doi{10.1126/science.aat4134}.

\bibitem[Bruin et~al.(2013)Bruin, Sakai, Perry, and
  Mackenzie]{bruin_similarity_2013}
J.~a.~N. Bruin, H.~Sakai, R.~S. Perry, and A.~P. Mackenzie.
\newblock Similarity of {Scattering} {Rates} in {Metals} {Showing} {T}-{Linear}
  {Resistivity}.
\newblock \emph{Science}, 339\penalty0 (6121):\penalty0 804--807, 2013.
\newblock \doi{10.1126/science.1227612}.
\newblock URL \url{https://science.sciencemag.org/content/339/6121/804}.

\bibitem[Brun et~al.(2016)Brun, Cren, and Roditchev]{PbSC3}
Christophe Brun, Tristan Cren, and Dimitri Roditchev.
\newblock Review of 2{D} superconductivity: the ultimate case of epitaxial
  monolayers.
\newblock \emph{Superconductor Science and Technology}, 30\penalty0
  (1):\penalty0 013003, 2016.
\newblock \doi{10.1088/0953-2048/30/1/013003}.

\bibitem[Budden et~al.(2021)Budden, Gebert, Buzzi, Jotzu, Wang, Matsuyama,
  Meier, Laplace, Pontiroli, Ricc{\`o}, et~al.]{budden2021evidence}
M~Budden, T~Gebert, M~Buzzi, G~Jotzu, E~Wang, T~Matsuyama, G~Meier, Y~Laplace,
  D~Pontiroli, M~Ricc{\`o}, et~al.
\newblock {Evidence for metastable photo-induced superconductivity in
  K$_3$C$_60$}.
\newblock \emph{Nature Physics}, 17\penalty0 (5):\penalty0 611--618, 2021.
\newblock \doi{10.1038/s41567-020-01148-11}.

\bibitem[Buhot et~al.(2014)Buhot, M{\'e}asson, Gallais, Cazayous, Sacuto,
  Lapertot, and Aoki]{buhot2014symmetry}
J~Buhot, M-A M{\'e}asson, Y~Gallais, M~Cazayous, A~Sacuto, G~Lapertot, and
  D~Aoki.
\newblock Symmetry of the excitations in the hidden order state of uru 2 si 2.
\newblock \emph{Phys. Rev. Lett.}, 113\penalty0 (26):\penalty0 266405, 2014.
\newblock \doi{10.1103/PhysRevLett.113.266405}.

\bibitem[Bukov et~al.(2016)Bukov, Kolodrubetz, and
  Polkovnikov]{PhysRevLett.116.125301}
Marin Bukov, Michael Kolodrubetz, and Anatoli Polkovnikov.
\newblock {Schrieffer-Wolff Transformation for Periodically Driven Systems:
  Strongly Correlated Systems with Artificial Gauge Fields}.
\newblock \emph{Phys. Rev. Lett.}, 116:\penalty0 125301, 2016.
\newblock \doi{10.1103/PhysRevLett.116.125301}.
\newblock URL \url{https://link.aps.org/doi/10.1103/PhysRevLett.116.125301}.

\bibitem[Burg et~al.(2019)Burg, Zhu, Taniguchi, Watanabe, MacDonald, and
  Tutuc]{burg2019correlated}
G~William Burg, Jihang Zhu, Takashi Taniguchi, Kenji Watanabe, Allan~H
  MacDonald, and Emanuel Tutuc.
\newblock Correlated insulating states in twisted double bilayer graphene.
\newblock \emph{Phys. Rev. Lett.}, 123\penalty0 (19):\penalty0 197702, 2019.
\newblock \doi{10.1103/PhysRevLett.123.197702}.

\bibitem[Butch et~al.(2011)Butch, Syers, Kirshenbaum, Hope, and
  Paglione]{MR-YPtBi}
N.~P. Butch, P.~Syers, K.~Kirshenbaum, A.~P. Hope, and J.~Paglione.
\newblock {Superconductivity in the topological semimetal YPtBi}.
\newblock \emph{Phys. Rev. B}, 84:\penalty0 220504, 2011.
\newblock \doi{10.1103/PhysRevB.84.220504}.
\newblock URL \url{https://link.aps.org/doi/10.1103/PhysRevB.84.220504}.

\bibitem[Butch et~al.(2010)Butch, Jeffries, Chi, Le{\~a}o, Lynn, and
  Maple]{butch2010antiferromagnetic}
Nicholas~P Butch, Jason~R Jeffries, Songxue Chi, Juscelino~Batista Le{\~a}o,
  Jeffrey~W Lynn, and M~Brian Maple.
\newblock {Antiferromagnetic critical pressure in URu$_2$Si$_2$ under
  hydrostatic conditions}.
\newblock \emph{Physical Review B}, 82\penalty0 (6):\penalty0 060408, 2010.
\newblock \doi{10.1103/PhysRevB.82.060408}.

\bibitem[Buzea and Yamashita(2001)]{buzea2001review}
Cristina Buzea and Tsutomu Yamashita.
\newblock {Review of the superconducting properties of MgB$_2$}.
\newblock \emph{Superconductor Science and Technology}, 14\penalty0
  (11):\penalty0 R115, 2001.
\newblock \doi{10.1088/0953-2048/14/11/201}.

\bibitem[Byeon et~al.(2019)Byeon, Sobota, Delime-Codrin, Choi, Hirata, Adachi,
  Kiyama, Matsuura, Yamamoto, Matsunami, et~al.]{Cu2Sethermopower2}
Dogyun Byeon, Robert Sobota, K{\'e}vin Delime-Codrin, Seongho Choi, Keisuke
  Hirata, Masahiro Adachi, Makoto Kiyama, Takashi Matsuura, Yoshiyuki Yamamoto,
  Masaharu Matsunami, et~al.
\newblock {Discovery of colossal Seebeck effect in metallic Cu$_2$Se}.
\newblock \emph{Nature Communications}, 10\penalty0 (1):\penalty0 1--7, 2019.
\newblock \doi{10.1038/s41467-018-07877-5}.

\bibitem[Calder et~al.(2010)Calder, Fennell, Kockelmann, Lau, Cava, and
  Bramwell]{calder2010neutron}
S~Calder, T~Fennell, W~Kockelmann, GC~Lau, R~J Cava, and S~T Bramwell.
\newblock Neutron scattering and crystal field studies of the rare earth double
  perovskite ba$_2$ersbo$_6$.
\newblock \emph{Journal of Physics: Condensed Matter}, 22\penalty0
  (11):\penalty0 116007, 2010.
\newblock \doi{10.1088/0953-8984/22/11/116007}.

\bibitem[Canfield(2019)]{Canfield_2019}
Paul~C Canfield.
\newblock New materials physics.
\newblock \emph{Reports on Progress in Physics}, 83\penalty0 (1):\penalty0
  016501, 2019.
\newblock \doi{10.1088/1361-6633/ab514b}.
\newblock URL \url{http://dx.doi.org/10.1088/1361-6633/ab514b}.

\bibitem[Canovi et~al.(2016)Canovi, Kollar, and Eckstein]{PhysRevE.93.012130}
Elena Canovi, Marcus Kollar, and Martin Eckstein.
\newblock Stroboscopic prethermalization in weakly interacting periodically
  driven systems.
\newblock \emph{Phys. Rev. E}, 93:\penalty0 012130, 2016.
\newblock \doi{10.1103/PhysRevE.93.012130}.
\newblock URL \url{https://link.aps.org/doi/10.1103/PhysRevE.93.012130}.

\bibitem[Cao et~al.(2019)Cao, Mazzone, Meyers, Hill, Liu, Wall, and
  Dean]{cao2019ultrafast}
Y.~Cao, D.~G. Mazzone, D.~Meyers, J.~P. Hill, X.~Liu, S.~Wall, and M.~P.~M.
  Dean.
\newblock Ultrafast dynamics of spin and orbital correlations in quantum
  materials: an energy-and momentum-resolved perspective.
\newblock \emph{Philosophical Transactions of the Royal Society A},
  377\penalty0 (2145):\penalty0 20170480, 2019.
\newblock \doi{10.1098/rsta.2017.0480}.

\bibitem[Cao et~al.(2018)Cao, Fatemi, Fang, Watanabe, Taniguchi, Kaxiras, and
  Jarillo-Herrero]{cao2018unconventional}
Yuan Cao, Valla Fatemi, Shiang Fang, Kenji Watanabe, Takashi Taniguchi,
  Efthimios Kaxiras, and Pablo Jarillo-Herrero.
\newblock Unconventional superconductivity in magic-angle graphene
  superlattices.
\newblock \emph{Nature}, 556\penalty0 (7699):\penalty0 43--50, 2018.
\newblock \doi{10.1038/nature26160}.

\bibitem[Cao et~al.(2020{\natexlab{a}})Cao, Chowdhury, Rodan-Legrain,
  Rubies-Bigorda, Watanabe, Taniguchi, Senthil, and
  Jarillo-Herrero]{cao2020strange}
Yuan Cao, Debanjan Chowdhury, Daniel Rodan-Legrain, Oriol Rubies-Bigorda, Kenji
  Watanabe, Takashi Taniguchi, T~Senthil, and Pablo Jarillo-Herrero.
\newblock Strange metal in magic-angle graphene with near planckian
  dissipation.
\newblock \emph{Phys. Rev. Lett.}, 124\penalty0 (7):\penalty0 076801,
  2020{\natexlab{a}}.
\newblock \doi{10.1103/PhysRevLett.124.076801}.

\bibitem[Cao et~al.(2020{\natexlab{b}})Cao, Rodan-Legrain, Rubies-Bigorda,
  Park, Watanabe, Taniguchi, and Jarillo-Herrero]{Cao2019Electric}
Yuan Cao, Daniel Rodan-Legrain, Oriol Rubies-Bigorda, Jeong~Min Park, Kenji
  Watanabe, Takashi Taniguchi, and Pablo Jarillo-Herrero.
\newblock Tunable correlated states and spin-polarized phases in twisted
  bilayer--bilayer graphene.
\newblock \emph{Nature}, 583:\penalty0 215, 2020{\natexlab{b}}.
\newblock \doi{10.1038/s41586-020-2260-6}.

\bibitem[Cao et~al.(2016)Cao, Wang, Waugh, Reber, Li, Zhou, Parham, Park,
  Plumb, Rotenberg, et~al.]{cao2016Hallmarks}
Yue Cao, Qiang Wang, Justin~A. Waugh, Theodore~J. Reber, Haoxiang Li, Xiaoqing
  Zhou, Stephen Parham, S.-R. Park, Nicholas~C. Plumb, Eli Rotenberg, et~al.
\newblock {Hallmarks of the Mott-metal crossover in the hole-doped
  pseudospin-1/2 Mott insulator Sr$_2$IrO$_4$}.
\newblock \emph{Nature Communications}, 7\penalty0 (1):\penalty0 11367, 2016.
\newblock \doi{10.1038/ncomms11367}.

\bibitem[Cao et~al.(2017)Cao, Liu, Xu, Yin, Meyers, Kim, Casa, Upton, Gog,
  Berlijn, et~al.]{cao2017giant}
Yue Cao, Xuerong Liu, Wenhu Xu, Wei-Guo Yin, Derek Meyers, Jungho Kim, Diego
  Casa, MH~Upton, Thomas Gog, Tom Berlijn, et~al.
\newblock {Giant spin gap and magnon localization in the disordered
  {Heisenberg} antiferromagnet Sr$_2$Ir$_{1-x}$Ru$_x$O$_4$}.
\newblock \emph{Phys. Rev. B}, 95\penalty0 (12):\penalty0 121103 (R), 2017.
\newblock \doi{10.1103/PhysRevB.95.121103}.

\bibitem[Cao et~al.(2020{\natexlab{c}})Cao, Assefa, Banerjee, Wieteska, Wang,
  Pasupathy, Tong, Liu, Lu, Sun, et~al.]{cao2020complete}
Yue Cao, Tadesse Assefa, Soham Banerjee, Andrew Wieteska, Dennis Wang, Abhay
  Pasupathy, Xiao Tong, Yu~Liu, Wenjian Lu, Yu-Ping Sun, et~al.
\newblock {Complete Strain Mapping of Nanosheets of Tantalum Disulfide}.
\newblock \emph{ACS Appl. Mater. Interfaces}, 12\penalty0 (38):\penalty0
  43173--43179, 2020{\natexlab{c}}.
\newblock \doi{10.1021/acsami.0c06517}.

\bibitem[Capellmann(1982)]{iron-review}
H~Capellmann.
\newblock The magnetism of iron and other 3-d transition metals.
\newblock \emph{Journal of Magnetism and Magnetic Materials}, 28\penalty0
  (3):\penalty0 250--260, 1982.
\newblock \doi{10.1016/0304-8853(82)90057-9}.

\bibitem[Carbone et~al.(2006)Carbone, Kuzmenko, Molegraaf, Van~Heumen, Lukovac,
  Marsiglio, Van Der~Marel, Haule, Kotliar, Berger, et~al.]{carbone2006doping}
Fabrizio Carbone, AB~Kuzmenko, HJA Molegraaf, Erik Van~Heumen, Vladimir
  Lukovac, Frank Marsiglio, Dirk Van Der~Marel, Kristjan Haule, G~Kotliar,
  H~Berger, et~al.
\newblock {Doping dependence of the redistribution of optical spectral weight
  in {Bi}$_2${Sr}$_2${CaCu}$_2${O}$_{8+\delta}$}.
\newblock \emph{Phys. Rev. B}, 74\penalty0 (6):\penalty0 064510, 2006.
\newblock \doi{10.1103/PhysRevB.74.064510}.

\bibitem[{Carleo, Giuseppe and Troyer, Matthias}(2017)]{Carleo17}
{Carleo, Giuseppe and Troyer, Matthias}.
\newblock Solving the quantum many-body problem with artificial neural
  networks.
\newblock \emph{Science}, 355\penalty0 (6325):\penalty0 602--606, 2017.
\newblock \doi{10.1126/science.aag2302}.
\newblock URL \url{https://science.sciencemag.org/content/355/6325/602}.

\bibitem[Carr(2010)]{carr2010understanding}
Lincoln Carr.
\newblock \emph{Understanding Quantum Phase Transitions}.
\newblock CRC press, 2010.
\newblock \doi{10.1201/b10273}.

\bibitem[Castro~Neto and Fradkin(1994)]{PhysRevLett.72.1393}
A.~H. Castro~Neto and Eduardo Fradkin.
\newblock {Bosonization of the low energy excitations of {Fermi} liquids}.
\newblock \emph{Phys. Rev. Lett.}, 72:\penalty0 1393--1397, 1994.
\newblock \doi{10.1103/PhysRevLett.72.1393}.
\newblock URL \url{https://link.aps.org/doi/10.1103/PhysRevLett.72.1393}.

\bibitem[Cattelan and Fox(2018)]{cattelan2018perspective}
Mattia Cattelan and Neil~A Fox.
\newblock A perspective on the application of spatially resolved {ARPES} for
  2{D} materials.
\newblock \emph{Nanomaterials}, 8\penalty0 (5):\penalty0 284, 2018.
\newblock \doi{10.3390/nano8050284}.

\bibitem[Cayssol et~al.(2013)Cayssol, Dóra, Simon, and Moessner]{Cayssol_2013}
Jérôme Cayssol, Balázs Dóra, Ferenc Simon, and Roderich Moessner.
\newblock Floquet topological insulators.
\newblock \emph{physica status solidi (RRL) - Rapid Research Letters},
  7\penalty0 (1-2):\penalty0 101–108, 2013.
\newblock \doi{10.1002/pssr.201206451}.
\newblock URL \url{http://dx.doi.org/10.1002/pssr.201206451}.

\bibitem[Ceccarelli et~al.(2019)Ceccarelli, Vasyukov, Wyss, Romagnoli, Rossi,
  Moser, and Poggio]{ceccarelli2019imaging}
L~Ceccarelli, D~Vasyukov, M~Wyss, G~Romagnoli, N~Rossi, L~Moser, and M~Poggio.
\newblock Imaging pinning and expulsion of individual superconducting vortices
  in amorphous mosi thin films.
\newblock \emph{Phys. Rev. B}, 100\penalty0 (10):\penalty0 104504, 2019.
\newblock \doi{10.1103/PhysRevB.100.104504}.

\bibitem[Ceperley and Alder(1980)]{Ceperley80}
D.~M. Ceperley and B.~J. Alder.
\newblock Ground state of the electron gas by a stochastic method.
\newblock \emph{Phys. Rev. Lett.}, 45:\penalty0 566--569, 1980.
\newblock \doi{10.1103/PhysRevLett.45.566}.
\newblock URL \url{https://link.aps.org/doi/10.1103/PhysRevLett.45.566}.

\bibitem[Cha et~al.(2020)Cha, Patel, Gull, and Kim]{cha2019t}
Peter Cha, Aavishkar~A Patel, Emanuel Gull, and Eun-Ah Kim.
\newblock {Slope invariant $T$-linear resistivity from local self-energy}.
\newblock \emph{Physical Review Research}, 2\penalty0 (3):\penalty0 033434,
  2020.
\newblock \doi{10.1103/PhysRevResearch.2.033434}.

\bibitem[Chakravarty et~al.(2001)Chakravarty, Laughlin, Morr, and
  Nayak]{chakravarty2001hidden}
Sudip Chakravarty, RB~Laughlin, Dirk~K Morr, and Chetan Nayak.
\newblock Hidden order in the cuprates.
\newblock \emph{Phys. Rev. B}, 63\penalty0 (9):\penalty0 094503, 2001.
\newblock \doi{10.1103/PhysRevB.63.094503}.

\bibitem[Challener and Thompson(1986)]{challener1986far}
WA~Challener and JD~Thompson.
\newblock Far-infrared spectroscopy in diamond anvil cells.
\newblock \emph{Applied spectroscopy}, 40\penalty0 (3):\penalty0 298--303,
  1986.
\newblock \doi{10.1366/0003702864509303}.

\bibitem[Chang et~al.(2013)Chang, Zhang, Feng, Shen, Zhang, Guo, Li, Ou, Wei,
  Wang, et~al.]{chang2013experimental}
Cui-Zu Chang, Jinsong Zhang, Xiao Feng, Jie Shen, Zuocheng Zhang, Minghua Guo,
  Kang Li, Yunbo Ou, Pang Wei, Li-Li Wang, et~al.
\newblock {Experimental observation of the quantum anomalous Hall effect in a
  magnetic topological insulator}.
\newblock \emph{Science}, 340\penalty0 (6129):\penalty0 167--170, 2013.
\newblock \doi{10.1126/science.1234414}.

\bibitem[Changlani et~al.(2009)Changlani, Kinder, Umrigar, and Chan]{CPS}
Hitesh~J. Changlani, Jesse~M. Kinder, C.~J. Umrigar, and Garnet Kin-Lic Chan.
\newblock Approximating strongly correlated wave functions with correlator
  product states.
\newblock \emph{Phys. Rev. B}, 80:\penalty0 245116, 2009.
\newblock \doi{10.1103/PhysRevB.80.245116}.
\newblock URL \url{https://link.aps.org/doi/10.1103/PhysRevB.80.245116}.

\bibitem[Changlani et~al.(2015)Changlani, Zheng, and Wagner]{Changlani15}
Hitesh~J. Changlani, Huihuo Zheng, and Lucas~K. Wagner.
\newblock Density-matrix based determination of low-energy model hamiltonians
  from ab initio wavefunctions.
\newblock \emph{The Journal of Chemical Physics}, 143\penalty0 (10):\penalty0
  102814, 2015.
\newblock \doi{10.1063/1.4927664}.
\newblock URL \url{, https://doi.org/10.1063/1.4927664}.

\bibitem[Chatterjee and Sachdev(2015)]{Chatterjee2015}
Shubhayu Chatterjee and Subir Sachdev.
\newblock Probing excitations in insulators via injection of spin currents.
\newblock \emph{Phys. Rev. B}, 92:\penalty0 165113, 2015.
\newblock \doi{10.1103/PhysRevB.92.165113}.
\newblock URL \url{https://link.aps.org/doi/10.1103/PhysRevB.92.165113}.

\bibitem[Chatterjee and Sachdev(2016)]{chatterjee2016fractionalized}
Shubhayu Chatterjee and Subir Sachdev.
\newblock {Fractionalized {Fermi} liquid with bosonic chargons as a candidate
  for the pseudogap metal}.
\newblock \emph{Phys. Rev. B}, 94\penalty0 (20):\penalty0 205117, 2016.
\newblock \doi{10.1103/PhysRevB.94.205117}.

\bibitem[Chatterjee et~al.(2017)Chatterjee, Sachdev, and
  Eberlein]{chatterjee2017thermal}
Shubhayu Chatterjee, Subir Sachdev, and Andreas Eberlein.
\newblock Thermal and electrical transport in metals and superconductors across
  antiferromagnetic and topological quantum transitions.
\newblock \emph{Phys. Rev. B}, 96\penalty0 (7):\penalty0 075103, 2017.
\newblock \doi{10.1103/PhysRevB.96.075103}.

\bibitem[Chatterjee et~al.(2019)Chatterjee, Rodriguez-Nieva, and
  Demler]{Chatterjee2019}
Shubhayu Chatterjee, Joaquin~F. Rodriguez-Nieva, and Eugene Demler.
\newblock Diagnosing phases of magnetic insulators via noise magnetometry with
  spin qubits.
\newblock \emph{Phys. Rev. B}, 99:\penalty0 104425, 2019.
\newblock \doi{10.1103/PhysRevB.99.104425}.
\newblock URL \url{https://link.aps.org/doi/10.1103/PhysRevB.99.104425}.

\bibitem[Chen et~al.(2019{\natexlab{a}})Chen, Jiang, Wu, Lyu, Li, Chittari,
  Watanabe, Taniguchi, Shi, Jung, Zhang, and Wang]{Wang2019Evidence}
Guorui Chen, Lili Jiang, Shuang Wu, Bosai Lyu, Hongyuan Li, Bheema~Lingam
  Chittari, Kenji Watanabe, Takashi Taniguchi, Zhiwen Shi, Jeil Jung, Yuanbo
  Zhang, and Feng Wang.
\newblock {Evidence of a gate-tunable Mott insulator in a trilayer graphene
  moir{\'e }superlattice}.
\newblock \emph{Nature Physics}, 15\penalty0 (3):\penalty0 237--241,
  2019{\natexlab{a}}.
\newblock \doi{10.1038/s41567-018-0387-2}.
\newblock URL \url{https://doi.org/10.1038/s41567-018-0387-2}.

\bibitem[Chen et~al.(2019{\natexlab{b}})Chen, Hashimoto, He, Song, Xu, He,
  Devereaux, Eisaki, Lu, Zaanen, et~al.]{chen2019incoherent}
Su-Di Chen, Makoto Hashimoto, Yu~He, Dongjoon Song, Ke-Jun Xu, Jun-Feng He,
  Thomas~P Devereaux, Hiroshi Eisaki, Dong-Hui Lu, Jan Zaanen, et~al.
\newblock {Incoherent strange metal sharply bounded by a critical doping in
  Bi2212}.
\newblock \emph{Science}, 366\penalty0 (6469):\penalty0 1099--1102,
  2019{\natexlab{b}}.
\newblock \doi{10.1126/science.aaw8850}.

\bibitem[Chen et~al.(2016{\natexlab{a}})Chen, Thampy, Mazzoli, Barbour, Miao,
  Gu, Cao, Tranquada, Dean, and Wilkins]{chen2016remarkable}
X.~M. Chen, V.~Thampy, C.~Mazzoli, A.~M. Barbour, H.~Miao, G.~D. Gu, Y.~Cao,
  J.~M. Tranquada, M.~P.~M. Dean, and S.~B. Wilkins.
\newblock {Remarkable stability of charge density wave order in
  La$_{1.875}$Ba$_{0.125}$CuO$_4$}.
\newblock \emph{Phys. Rev. Lett.}, 117\penalty0 (16):\penalty0 167001,
  2016{\natexlab{a}}.
\newblock \doi{10.1103/PhysRevLett.117.167001}.

\bibitem[Chen et~al.(2016{\natexlab{b}})Chen, Weng, Michael, Lu, and
  Yuan]{chen2016}
Ye~Chen, Zongfa Weng, Smidman Michael, Xin Lu, and Huiqiu Yuan.
\newblock High-pressure studies on heavy fermion systems.
\newblock \emph{Chinese Physics B}, 25\penalty0 (7):\penalty0 077401,
  2016{\natexlab{b}}.
\newblock \doi{10.1088/1674-1056/25/7/077401}.
\newblock URL \url{http://dx.doi.org/10.1088/1674-1056/25/7/077401}.

\bibitem[Chertkov and Clark(2018)]{Chertkov18}
Eli Chertkov and Bryan~K. Clark.
\newblock Computational inverse method for constructing spaces of quantum
  models from wave functions.
\newblock \emph{Phys. Rev. X}, 8:\penalty0 031029, 2018.
\newblock \doi{10.1103/PhysRevX.8.031029}.
\newblock URL \url{https://link.aps.org/doi/10.1103/PhysRevX.8.031029}.

\bibitem[Cherukara et~al.(2018)Cherukara, Nashed, and
  Harder]{cherukara2018real}
Mathew~J Cherukara, Youssef~SG Nashed, and Ross~J Harder.
\newblock Real-time coherent diffraction inversion using deep generative
  networks.
\newblock \emph{Scientific Reports}, 8:\penalty0 16520, 2018.
\newblock \doi{10.1038/s41598-018-34525-1}.

\bibitem[Chester(1956)]{chester1956difference}
GV~Chester.
\newblock Difference between normal and superconducting states of a metal.
\newblock \emph{Physical Review}, 103\penalty0 (6):\penalty0 1693, 1956.
\newblock \doi{10.1103/PhysRev.103.1693}.

\bibitem[Cheung et~al.(2020)Cheung, Shin, Lau, Chen, Sun, Zhang, M{\"u}ller,
  Eremin, Wright, and Pasupathy]{cheung2020dictionary}
Sky~C Cheung, John~Y Shin, Yenson Lau, Zhengyu Chen, Ju~Sun, Yuqian Zhang,
  Marvin~A M{\"u}ller, Ilya~M Eremin, John~N Wright, and Abhay~N Pasupathy.
\newblock {Dictionary learning in Fourier-transform scanning tunneling
  spectroscopy}.
\newblock \emph{Nature Communications}, 11:\penalty0 1081, 2020.
\newblock \doi{10.1038/s41467-020-14633-1}.

\bibitem[Choi et~al.(2020)Choi, Lee, and Kim]{choi2019theory}
Wonjune Choi, Ki~Hoon Lee, and Yong~Baek Kim.
\newblock {Theory of two-dimensional nonlinear spectroscopy for the Kitaev spin
  liquid}.
\newblock \emph{Physical Review Letters}, 124\penalty0 (11):\penalty0 117205,
  2020.
\newblock \doi{10.1103/PhysRevLett.124.117205}.

\bibitem[Chowdhury et~al.(2018)Chowdhury, Werman, Berg, and
  Senthil]{PhysRevX.8.031024}
Debanjan Chowdhury, Yochai Werman, Erez Berg, and T.~Senthil.
\newblock Translationally invariant non-fermi-liquid metals with critical
  {Fermi} surfaces: Solvable models.
\newblock \emph{Phys. Rev. X}, 8:\penalty0 031024, 2018.
\newblock \doi{10.1103/PhysRevX.8.031024}.
\newblock URL \url{https://link.aps.org/doi/10.1103/PhysRevX.8.031024}.

\bibitem[Chu et~al.(2012)Chu, Kuo, Analytis, and Fisher]{chu2012divergent}
Jiun-Haw Chu, Hsueh-Hui Kuo, James~G Analytis, and Ian~R Fisher.
\newblock Divergent nematic susceptibility in an iron arsenide superconductor.
\newblock \emph{Science}, 337\penalty0 (6095):\penalty0 710--712, 2012.
\newblock \doi{10.1126/science.1221713}.

\bibitem[Cilento et~al.(2018)Cilento, Manzoni, Sterzi, Peli, Ronchi, Crepaldi,
  Boschini, Cacho, Chapman, Springate, Eisaki, Greven, Berciu, Kemper,
  Damascelli, Capone, Giannetti, and Parmigiani]{Cilento2018}
Federico Cilento, Giulia Manzoni, Andrea Sterzi, Simone Peli, Andrea Ronchi,
  Alberto Crepaldi, Fabio Boschini, Cephise Cacho, Richard Chapman, Emma
  Springate, Hiroshi Eisaki, Martin Greven, Mona Berciu, Alexander~F. Kemper,
  Andrea Damascelli, Massimo Capone, Claudio Giannetti, and Fulvio Parmigiani.
\newblock Dynamics of correlation-frozen antinodal quasiparticles in
  superconducting cuprates.
\newblock \emph{Science Advances}, 4\penalty0 (2):\penalty0 eaar1998, 2018.
\newblock \doi{10.1126/sciadv.aar1998}.
\newblock URL \url{https://advances.sciencemag.org/content/4/2/eaar1998}.

\bibitem[Claassen et~al.(2017{\natexlab{a}})Claassen, Jiang, Moritz, and
  Devereaux]{Claassen2017}
Martin Claassen, Hong-Chen Jiang, Brian Moritz, and Thomas~P. Devereaux.
\newblock Dynamical time-reversal symmetry breaking and photo-induced chiral
  spin liquids in frustrated mott insulators.
\newblock \emph{Nature Communications}, 8\penalty0 (1):\penalty0 1192,
  2017{\natexlab{a}}.
\newblock \doi{10.1038/s41467-017-00876-y}.
\newblock URL \url{https://doi.org/10.1038/s41467-017-00876-y}.

\bibitem[Claassen et~al.(2017{\natexlab{b}})Claassen, Jiang, Moritz, and
  Devereaux]{claassen2017dynamical}
Martin Claassen, Hong-Chen Jiang, Brian Moritz, and Thomas~P Devereaux.
\newblock Dynamical time-reversal symmetry breaking and photo-induced chiral
  spin liquids in frustrated mott insulators.
\newblock \emph{Nature Communications}, 8\penalty0 (1):\penalty0 1192,
  2017{\natexlab{b}}.
\newblock \doi{10.1038/s41467-017-00876-y}.

\bibitem[Cohen et~al.(2015)Cohen, Gull, Reichman, and Millis]{cohen15}
Guy Cohen, Emanuel Gull, David~R. Reichman, and Andrew~J. Millis.
\newblock Taming the dynamical sign problem in real-time evolution of quantum
  many-body problems.
\newblock \emph{Phys. Rev. Lett.}, 115:\penalty0 266802, 2015.
\newblock \doi{10.1103/PhysRevLett.115.266802}.
\newblock URL \url{https://link.aps.org/doi/10.1103/PhysRevLett.115.266802}.

\bibitem[Cohen and Anderson(1972)]{cohen1972comments}
Marvin~L Cohen and PW~Anderson.
\newblock Comments on the maximum superconducting transition temperature.
\newblock In \emph{AIP Conference Proceedings}, volume~4, pages 17--27.
  American Institute of Physics, 1972.
\newblock \doi{10.1063/1.2946185}.

\bibitem[Coleman and Nevidomskyy(2010)]{coleman.2010}
Piers Coleman and Andriy~H. Nevidomskyy.
\newblock {Frustration and the {Kondo} Effect in Heavy Fermion Materials}.
\newblock \emph{Journal of Low Temperature Physics}, 161\penalty0 (1):\penalty0
  182--202, 2010.
\newblock \doi{10.1007/s10909-010-0213-4}.
\newblock URL \url{https://doi.org/10.1007/s10909-010-0213-4}.

\bibitem[Coleman et~al.(2001)Coleman, P{\'e}pin, Si, and
  Ramazashvili]{coleman2001fermi}
Piers Coleman, C~P{\'e}pin, Qimiao Si, and Revaz Ramazashvili.
\newblock {How do {Fermi} liquids get heavy and die?}
\newblock \emph{Journal of Physics: Condensed Matter}, 13\penalty0
  (35):\penalty0 R723, 2001.
\newblock \doi{10.1088/0953-8984/13/35/202}.

\bibitem[Coll et~al.(2019)Coll, Fontcuberta, Althammer, Bibes, Boschker,
  Calleja, Cheng, Cuoco, Dittmann, Dkhil, et~al.]{coll2019towards}
Mariona Coll, Josep Fontcuberta, M~Althammer, Manuel Bibes, H~Boschker, Albert
  Calleja, G~Cheng, M~Cuoco, R~Dittmann, B~Dkhil, et~al.
\newblock Towards oxide electronics: a roadmap.
\newblock \emph{Applied Surface Science}, 482:\penalty0 1, 2019.
\newblock \doi{10.1016/j.apsusc.2019.03.312}.

\bibitem[Collignon et~al.(2019)Collignon, Lin, Rischau, Fauqué, and
  Behnia]{STOtransportreview}
Clément Collignon, Xiao Lin, Carl~Willem Rischau, Benoît Fauqué, and Kamran
  Behnia.
\newblock Metallicity and superconductivity in doped strontium titanate.
\newblock \emph{Annual Review of Condensed Matter Physics}, 10\penalty0
  (1):\penalty0 25--44, 2019.
\newblock \doi{10.1146/annurev-conmatphys-031218-013144}.
\newblock URL \url{, https://doi.org/10.1146/annurev-conmatphys-031218-013144}.

\bibitem[Comin et~al.(2014)Comin, Frano, Yee, Yoshida, Eisaki, Schierle,
  Weschke, Sutarto, He, Soumyanarayanan, et~al.]{comin2014charge}
R~Comin, A~Frano, Michael~Manchun Yee, Y~Yoshida, H~Eisaki, E~Schierle,
  E~Weschke, R~Sutarto, F~He, Anjan Soumyanarayanan, et~al.
\newblock Charge order driven by {Fermi}-arc instability in
  {Bi}$_2${Sr}$_{2-x}${La}$_x${Cu}$_2${O}$_{6+\delta}$.
\newblock \emph{Science}, 343\penalty0 (6169):\penalty0 390--392, 2014.
\newblock \doi{10.1126/science.1242996}.

\bibitem[Conte et~al.(2015)Conte, Vidmar, Gole{\v{z}}, Mierzejewski, Soavi,
  Peli, Banfi, Ferrini, Comin, Ludbrook, Chauviere, Zhigadlo, Eisaki, Greven,
  Lupi, Damascelli, Brida, Capone, Bon{\v{c}}a, Cerullo, and
  Giannetti]{DalConte2015}
S.~Dal Conte, L.~Vidmar, D.~Gole{\v{z}}, M.~Mierzejewski, G.~Soavi, S.~Peli,
  F.~Banfi, G.~Ferrini, R.~Comin, B.~M. Ludbrook, L.~Chauviere, N.~D. Zhigadlo,
  H.~Eisaki, M.~Greven, S.~Lupi, A.~Damascelli, D.~Brida, M.~Capone,
  J.~Bon{\v{c}}a, G.~Cerullo, and C.~Giannetti.
\newblock Snapshots of the retarded interaction of charge carriers with
  ultrafast fluctuations in cuprates.
\newblock \emph{Nature Physics}, 11\penalty0 (5):\penalty0 421--426, 2015.
\newblock \doi{10.1038/nphys3265}.
\newblock URL \url{https://doi.org/10.1038/nphys3265}.

\bibitem[Cooper et~al.(2009)Cooper, Wang, Vignolle, Lipscombe, Hayden, Tanabe,
  Adachi, Koike, Nohara, Takagi, et~al.]{cooper2009anomalous}
RA~Cooper, Y~Wang, B~Vignolle, OJ~Lipscombe, SM~Hayden, Y~Tanabe, T~Adachi,
  Y~Koike, Minoru Nohara, H~Takagi, et~al.
\newblock {Anomalous criticality in the electrical resistivity of
  La$_{2-x}$Sr$_x$CuO$_4$}.
\newblock \emph{Science}, 323\penalty0 (5914):\penalty0 603--607, 2009.
\newblock \doi{10.1126/science.1165015}.

\bibitem[Corboz et~al.(2014)Corboz, Rice, and Troyer]{corboz_peps}
Philippe Corboz, T.~M. Rice, and Matthias Troyer.
\newblock {Competing States in the $t$-$J$ Model: Uniform $d$-Wave State versus
  Stripe State}.
\newblock \emph{Phys. Rev. Lett.}, 113:\penalty0 046402, 2014.
\newblock \doi{10.1103/PhysRevLett.113.046402}.
\newblock URL \url{https://link.aps.org/doi/10.1103/PhysRevLett.113.046402}.

\bibitem[Council(2013)]{NAP18355}
National~Research Council.
\newblock \emph{High Magnetic Field Science and Its Application in the United
  States: Current Status and Future Directions}.
\newblock The National Academies Press, Washington, DC, 2013.
\newblock ISBN 978-0-309-28634-3.
\newblock \doi{10.17226/18355}.

\bibitem[Cowley(1964)]{CowleySTO}
RA~Cowley.
\newblock Lattice dynamics and phase transitions of strontium titanate.
\newblock \emph{Physical Review}, 134\penalty0 (4A):\penalty0 A981, 1964.
\newblock \doi{10.1103/PhysRev.134.A981}.

\bibitem[Cox(1987)]{Cox87}
DL~Cox.
\newblock {Quadrupolar {Kondo} effect in uranium heavy-electron materials?}
\newblock \emph{Phys. Rev. Lett.}, 59\penalty0 (11):\penalty0 1240, 1987.
\newblock \doi{10.1103/PhysRevLett.59.1240}.

\bibitem[Cox and Jarrell(1996)]{Cox96}
DL~Cox and M~Jarrell.
\newblock {The two-channel {Kondo} route to non-Fermi-liquid metals}.
\newblock \emph{Journal of Physics: Condensed Matter}, 8\penalty0
  (48):\penalty0 9825, 1996.
\newblock \doi{10.1088/0953-8984/8/48/012}.

\bibitem[Croft et~al.(2017)Croft, Blackburn, Kulda, Liang, Bonn, Hardy, and
  Hayden]{croft2017no}
Thomas~P Croft, Elizabeth Blackburn, Jiri Kulda, Ruixing Liang, Doug~A Bonn,
  WN~Hardy, and SM~Hayden.
\newblock {No evidence for orbital loop currents in charge-ordered
  YBa$_2$Cu$_3$O$_{6+x}$ from polarized neutron diffraction}.
\newblock \emph{Phys. Rev. B}, 96\penalty0 (21):\penalty0 214504, 2017.
\newblock \doi{10.1103/PhysRevB.96.214504}.

\bibitem[Crooker and Samarth(2007)]{crooker2007tuning}
SA~Crooker and Nitin Samarth.
\newblock {Tuning alloy disorder in diluted magnetic semiconductors in high
  fields to 89 T}.
\newblock \emph{Applied Physics Letters}, 90\penalty0 (10):\penalty0 102109,
  2007.
\newblock \doi{10.1063/1.2711370}.

\bibitem[Cui et~al.(2017)Cui, Kirtley, Wang, Kratz, Rosenberg, Watson,
  Gibson~Jr, Ketchen, and Moler]{cui2017scanning}
Zheng Cui, John~R Kirtley, Yihua Wang, Philip~A Kratz, Aaron~J Rosenberg,
  Christopher~A Watson, Gerald~W Gibson~Jr, Mark~B Ketchen, and Kathryn~A
  Moler.
\newblock Scanning squid sampler with 40-ps time resolution.
\newblock \emph{Review of Scientific Instruments}, 88\penalty0 (8):\penalty0
  083703, 2017.
\newblock \doi{10.1063/1.4986525}.

\bibitem[Curtis et~al.(2016)Curtis, Tokumoto, Hatke, Cherian, Reno, McGill,
  Karaiskaj, and Hilton]{curtis2016cyclotron}
Jeremy~A Curtis, Takahisa Tokumoto, AT~Hatke, Judy~G Cherian, John~L Reno,
  Stephen~A McGill, Denis Karaiskaj, and David~J Hilton.
\newblock Cyclotron decay time of a two-dimensional electron gas from 0.4 to
  100 k.
\newblock \emph{Physical Review B}, 93\penalty0 (15):\penalty0 155437, 2016.
\newblock \doi{10.1103/PhysRevB.93.155437}.

\bibitem[da~Silva~Neto et~al.(2014)da~Silva~Neto, Aynajian, Frano, Comin,
  Schierle, Weschke, Gyenis, Wen, Schneeloch, Xu, et~al.]{da2014ubiquitous}
Eduardo~H da~Silva~Neto, Pegor Aynajian, Alex Frano, Riccardo Comin, Enrico
  Schierle, Eugen Weschke, Andr{\'a}s Gyenis, Jinsheng Wen, John Schneeloch,
  Zhijun Xu, et~al.
\newblock Ubiquitous interplay between charge ordering and high-temperature
  superconductivity in cuprates.
\newblock \emph{Science}, 343\penalty0 (6169):\penalty0 393--396, 2014.
\newblock \doi{10.1126/science.1243479}.

\bibitem[Dagotto(2005)]{dagotto2005complexity}
Elbio Dagotto.
\newblock Complexity in strongly correlated electronic systems.
\newblock \emph{Science}, 309\penalty0 (5732):\penalty0 257--262, 2005.
\newblock \doi{10.1126/science.1107559}.

\bibitem[Dai et~al.(1999)Dai, Mook, Hayden, Aeppli, Perring, Hunt, and
  Do{\u{g}}an]{dai1999magnetic}
Pengcheng Dai, Herbert~A Mook, Stephen~M Hayden, Gabriel Aeppli, Toby~G
  Perring, Rodney~Dale Hunt, and F~Do{\u{g}}an.
\newblock The magnetic excitation spectrum and thermodynamics of high-$t_c$
  superconductors.
\newblock \emph{Science}, 284\penalty0 (5418):\penalty0 1344--1347, 1999.
\newblock \doi{10.1126/science.284.5418.1344}.

\bibitem[Daley et~al.(2004)Daley, Kollath, Schollwöck, and Vidal]{Daley_2004}
A~J Daley, C~Kollath, U~Schollwöck, and G~Vidal.
\newblock Time-dependent density-matrix renormalization-group using adaptive
  effective hilbert spaces.
\newblock \emph{Journal of Statistical Mechanics: Theory and Experiment},
  2004\penalty0 (04):\penalty0 P04005, 2004.
\newblock \doi{10.1088/1742-5468/2004/04/p04005}.
\newblock URL \url{https://doi.org/10.1088%2F1742-5468%2F2004%2F04%2Fp04005}.

\bibitem[Damle and Sachdev(1997)]{damle1997nonzero}
Kedar Damle and Subir Sachdev.
\newblock Nonzero-temperature transport near quantum critical points.
\newblock \emph{Phys. Rev. B}, 56\penalty0 (14):\penalty0 8714, 1997.
\newblock \doi{10.1103/PhysRevB.56.8714}.

\bibitem[Daou et~al.(2009)Daou, Doiron-Leyraud, LeBoeuf, Li, Lalibert{\'e},
  Cyr-Choiniere, Jo, Balicas, Yan, Zhou, et~al.]{daou2009linear}
Ramzy Daou, Nicolas Doiron-Leyraud, David LeBoeuf, SY~Li, Francis
  Lalibert{\'e}, Olivier Cyr-Choiniere, YJ~Jo, Luis Balicas, J-Q Yan, J-S Zhou,
  et~al.
\newblock {Linear temperature dependence of resistivity and change in the
  {Fermi} surface at the pseudogap critical point of a high-$T_c$
  superconductor}.
\newblock \emph{Nature Physics}, 5\penalty0 (1):\penalty0 31--34, 2009.
\newblock \doi{10.1038/nphys1109}.

\bibitem[Davison et~al.(2014)Davison, Schalm, and
  Zaanen]{davison2014holographic}
Richard~A Davison, Koenraad Schalm, and Jan Zaanen.
\newblock Holographic duality and the resistivity of strange metals.
\newblock \emph{Phys. Rev. B}, 89\penalty0 (24):\penalty0 245116, 2014.
\newblock \doi{10.1103/PhysRevB.89.245112}.

\bibitem[Dean et~al.(2011)Dean, Young, Cadden-Zimansky, Wang, Ren, Watanabe,
  Taniguchi, Kim, Hone, and Shepard]{dean2011multicomponent}
Cory~R Dean, Andrea~F Young, Pet Cadden-Zimansky, L~Wang, H~Ren, Kenji
  Watanabe, T~Taniguchi, P~Kim, J~Hone, and KL~Shepard.
\newblock Multicomponent fractional quantum hall effect in graphene.
\newblock \emph{Nature Physics}, 7\penalty0 (9):\penalty0 693--696, 2011.
\newblock \doi{10.1038/nphys2007}.

\bibitem[Dean et~al.(2016)Dean, Cao, Liu, Wall, Zhu, Mankowsky, Thampy, Chen,
  Vale, Casa, et~al.]{dean2016ultrafast}
Mark~PM Dean, Yue Cao, Xuerong Liu, Simon Wall, Dilling Zhu, Roman Mankowsky,
  V~Thampy, XM~Chen, James~G Vale, Diego Casa, et~al.
\newblock {Ultrafast energy-and momentum-resolved dynamics of magnetic
  correlations in the photo-doped Mott insulator Sr$_2$IrO$_4$}.
\newblock \emph{Nature materials}, 15\penalty0 (6):\penalty0 601--605, 2016.
\newblock \doi{10.1038/nmat4641}.

\bibitem[Delacr{\'e}taz et~al.(2017)Delacr{\'e}taz, Gout{\'e}raux, Hartnoll,
  and Karlsson]{delacretaz2017bad}
Luca~V Delacr{\'e}taz, Blaise Gout{\'e}raux, Sean~A Hartnoll, and Anna
  Karlsson.
\newblock Bad metals from fluctuating density waves.
\newblock \emph{SciPost Phys}, 3\penalty0 (025):\penalty0 1612--04381, 2017.
\newblock \doi{10.21468/SciPostPhys.3.3.024}.

\bibitem[Demler and Zhang(1998)]{demler1998quantitative}
Eugene Demler and Shou-Cheng Zhang.
\newblock Quantitative test of a microscopic mechanism of high-temperature
  superconductivity.
\newblock \emph{Nature}, 396\penalty0 (6713):\penalty0 733--735, 1998.
\newblock \doi{10.1038/25482}.

\bibitem[Demsar et~al.(1999)Demsar, Podobnik, Kabanov, Wolf, and
  Mihailovic]{demsar1999superconducting}
J~Demsar, B~Podobnik, VV~Kabanov, Th~Wolf, and D~Mihailovic.
\newblock {Superconducting gap $\Delta_c$, the pseudogap $\Delta_p$, and pair
  fluctuations above $T_c$ in overdoped {Y}$_{1-
  x}${Ca}$_x${Ba}$_2${Cu}$_3${O}$_{7-\delta}$ from femtosecond time-domain
  spectroscopy}.
\newblock \emph{Phys. Rev. Lett.}, 82\penalty0 (24):\penalty0 4918, 1999.
\newblock \doi{10.1103/PhysRevLett.82.4918}.

\bibitem[Deng et~al.(2020)Deng, Yu, Shi, Guo, Xu, Wang, Chen, and
  Zhang]{deng2020quantum}
Yujun Deng, Yijun Yu, Meng~Zhu Shi, Zhongxun Guo, Zihan Xu, Jing Wang, Xian~Hui
  Chen, and Yuanbo Zhang.
\newblock {Quantum anomalous Hall effect in intrinsic magnetic topological
  insulator MnBi$_2$Te$_4$}.
\newblock \emph{Science}, 2020.
\newblock \doi{10.1126/science.aax8156}.

\bibitem[Deutscher et~al.(2005)Deutscher, Santander-Syro, and
  Bontemps]{deutscher2005kinetic}
Guy Deutscher, Andr{\'e}s~Felipe Santander-Syro, and Nicole Bontemps.
\newblock Kinetic energy change with doping upon superfluid condensation in
  high-temperature superconductors.
\newblock \emph{Phys. Rev. B}, 72\penalty0 (9):\penalty0 092504, 2005.
\newblock \doi{10.1103/PhysRevB.72.092504}.

\bibitem[Dmitriev et~al.(2015)Dmitriev, Senin, Soldatov, and
  Yudin]{dmitriev2015polar}
VV~Dmitriev, AA~Senin, AA~Soldatov, and AN~Yudin.
\newblock {Polar phase of superfluid $^3$He in anisotropic aerogel}.
\newblock \emph{Phys. Rev. Lett.}, 115\penalty0 (16):\penalty0 165304, 2015.
\newblock \doi{10.1103/PhysRevLett.115.165304}.

\bibitem[Dobrosavljevic et~al.(2012)Dobrosavljevic, Trivedi, and
  Valles~Jr]{MIT-theoryreview3}
Vladimir Dobrosavljevic, Nandini Trivedi, and James~M Valles~Jr.
\newblock \emph{Conductor Insulator Quantum Phase Transitions}.
\newblock Oxford University Press, 2012.
\newblock \doi{10.1093/acprof:oso/9780199592593.001.0001}.

\bibitem[Doiron-Leyraud et~al.(2009)Doiron-Leyraud, Auban-Senzier, Ren\'e~de
  Cotret, Bourbonnais, J\'erome, Bechgaard, and Taillefer]{PhysRevB.80.214531}
Nicolas Doiron-Leyraud, Pascale Auban-Senzier, Samuel Ren\'e~de Cotret, Claude
  Bourbonnais, Denis J\'erome, Klaus Bechgaard, and Louis Taillefer.
\newblock {Correlation between linear resistivity and ${T}_{c}$ in the
  Bechgaard salts and the pnictide superconductor
  $\text{Ba}{({\text{Fe}}_{1\ensuremath{-}x}{\text{Co}}_{x})}_{2}{\text{As}}_{2}$}.
\newblock \emph{Phys. Rev. B}, 80:\penalty0 214531, 2009.
\newblock \doi{10.1103/PhysRevB.80.214531}.
\newblock URL \url{https://link.aps.org/doi/10.1103/PhysRevB.80.214531}.

\bibitem[Dressel(2011)]{Dressel2011}
Martin Dressel.
\newblock {Quantum criticality in organic conductors? {Fermi} liquid versus
  non-Fermi-liquid behaviour}.
\newblock \emph{Journal of Physics: Condensed Matter}, 23\penalty0
  (29):\penalty0 293201, 2011.
\newblock \doi{10.1088/0953-8984/23/29/293201}.
\newblock URL \url{https://doi.org/10.1088%2F0953-8984%2F23%2F29%2F293201}.

\bibitem[Drozdov et~al.(2015)Drozdov, Eremets, Troyan, Ksenofontov, and
  Shylin]{drozdov2015conventional}
AP~Drozdov, MI~Eremets, IA~Troyan, Vadim Ksenofontov, and SI~Shylin.
\newblock Conventional superconductivity at 203 kelvin at high pressures in the
  sulfur hydride system.
\newblock \emph{Nature}, 525\penalty0 (7567):\penalty0 73--76, 2015.
\newblock \doi{10.1038/nature14964}.

\bibitem[Du et~al.(2008)Du, Skachko, Barker, and Andrei]{du2008approaching}
Xu~Du, Ivan Skachko, Anthony Barker, and Eva~Y Andrei.
\newblock Approaching ballistic transport in suspended graphene.
\newblock \emph{Nature nanotechnology}, 3\penalty0 (8):\penalty0 491--495,
  2008.
\newblock \doi{10.1038/nnano.2008.199}.

\bibitem[Du et~al.(2018)Du, Yang, Altenfeld, Gu, Yang, Eremin, Hirschfeld,
  Mazin, Lin, Zhu, et~al.]{du2018sign}
Zengyi Du, Xiong Yang, Dustin Altenfeld, Qiangqiang Gu, Huan Yang, Ilya Eremin,
  Peter~J Hirschfeld, Igor~I Mazin, Hai Lin, Xiyu Zhu, et~al.
\newblock {Sign reversal of the order parameter in (Li$_{1- x}$Fe$_ x$)
  OHFe$_{1- y}$Zn$_ y$Se}.
\newblock \emph{Nature Physics}, 14\penalty0 (2):\penalty0 134--139, 2018.
\newblock \doi{10.1038/nphys4299}.

\bibitem[Dubrovinskaia et~al.(2016)Dubrovinskaia, Dubrovinsky, Solopova,
  Abakumov, Turner, Hanfland, Bykova, Bykov, Prescher, Prakapenka, Petitgirard,
  Chuvashova, Gasharova, Mathis, Ershov, Snigireva, and
  Snigirev]{Dubrovinskaiae2016}
Natalia Dubrovinskaia, Leonid Dubrovinsky, Natalia~A. Solopova, Artem Abakumov,
  Stuart Turner, Michael Hanfland, Elena Bykova, Maxim Bykov, Clemens Prescher,
  Vitali~B. Prakapenka, Sylvain Petitgirard, Irina Chuvashova, Biliana
  Gasharova, Yves-Laurent Mathis, Petr Ershov, Irina Snigireva, and Anatoly
  Snigirev.
\newblock Terapascal static pressure generation with ultrahigh yield strength
  nanodiamond.
\newblock \emph{Science Advances}, 2\penalty0 (7), 2016.
\newblock \doi{10.1126/sciadv.1600341}.
\newblock URL \url{https://advances.sciencemag.org/content/2/7/e1600341}.

\bibitem[Duc et~al.(2018)Duc, Tonon, Billette, Rollet, Knafo, Bourdarot,
  B{\'{e}}ard, Mantegazza, Longuet, Lorenzo, Leli{\`{e}}vre-Berna, Frings, and
  Regnault]{duc2018neutron}
F.~Duc, X.~Tonon, J.~Billette, B.~Rollet, W.~Knafo, F.~Bourdarot,
  J.~B{\'{e}}ard, F.~Mantegazza, B.~Longuet, J.~E. Lorenzo,
  E.~Leli{\`{e}}vre-Berna, P.~Frings, and L.-P. Regnault.
\newblock {40-Tesla pulsed-field cryomagnet for single crystal neutron
  diffraction}.
\newblock \emph{Review of Scientific Instruments}, 89\penalty0 (5):\penalty0
  053905, 2018.
\newblock \doi{10.1063/1.5028487}.

\bibitem[Dusad et~al.(2019)Dusad, Kirschner, Hoke, Roberts, Eyal, Flicker,
  Luke, Blundell, and Davis]{Dusad2019}
Ritika Dusad, Franziska K.~K. Kirschner, Jesse~C. Hoke, Benjamin~R. Roberts,
  Anna Eyal, Felix Flicker, Graeme~M. Luke, Stephen~J. Blundell, and
  J.~C.~S{\'e}amus Davis.
\newblock Magnetic monopole noise.
\newblock \emph{Nature}, 571\penalty0 (7764):\penalty0 234--239, 2019.
\newblock \doi{10.1038/s41586-019-1358-1}.
\newblock URL \url{https://doi.org/10.1038/s41586-019-1358-1}.

\bibitem[Dzero et~al.(2010)Dzero, Sun, Galitski, and
  Coleman]{dzero2010topological}
Maxim Dzero, Kai Sun, Victor Galitski, and Piers Coleman.
\newblock Topological {Kondo} insulators.
\newblock \emph{Phys. Rev. Lett.}, 104\penalty0 (10):\penalty0 106408, 2010.
\newblock \doi{10.1103/PhysRevLett.104.106408}.

\bibitem[Dzero et~al.(2016)Dzero, Xia, Galitski, and Coleman]{TKIreview}
Maxim Dzero, Jing Xia, Victor Galitski, and Piers Coleman.
\newblock Topological {Kondo} insulators.
\newblock \emph{Annual Review of Condensed Matter Physics}, 7\penalty0
  (1):\penalty0 249--280, 2016.
\newblock \doi{10.1146/annurev-conmatphys-031214-014749}.
\newblock URL \url{, https://doi.org/10.1146/annurev-conmatphys-031214-014749}.

\bibitem[Dzsaber et~al.(2017)Dzsaber, Prochaska, Sidorenko, Eguchi, Svagera,
  Waas, Prokofiev, Si, and Paschen]{dzsaber2017kondo}
S~Dzsaber, L~Prochaska, A~Sidorenko, G~Eguchi, R~Svagera, M~Waas, A~Prokofiev,
  Q~Si, and S~Paschen.
\newblock Kondo insulator to semimetal transformation tuned by spin-orbit
  coupling.
\newblock \emph{Phys. Rev. Lett.}, 118\penalty0 (24):\penalty0 246601, 2017.
\newblock \doi{10.1103/PhysRevLett.118.246601}.

\bibitem[Eckardt and Anisimovas(2015)]{eckardt2015high}
Andr{\'e} Eckardt and Egidijus Anisimovas.
\newblock {High-frequency approximation for periodically driven quantum systems
  from a {Floquet} -space perspective}.
\newblock \emph{New journal of physics}, 17\penalty0 (9):\penalty0 093039,
  2015.
\newblock \doi{10.1088/1367-2630/17/9/093039}.

\bibitem[Eichstaedt et~al.(2019)Eichstaedt, Zhang, Laurell, Okamoto, Eguiluz,
  and Berlijn]{PhysRevB.100.075110}
Casey Eichstaedt, Yi~Zhang, Pontus Laurell, Satoshi Okamoto, Adolfo~G. Eguiluz,
  and Tom Berlijn.
\newblock {Deriving models for the Kitaev spin-liquid candidate material
  $\ensuremath{\alpha}\text{\ensuremath{-}}{\mathrm{RuCl}}_{3}$ from first
  principles}.
\newblock \emph{Phys. Rev. B}, 100:\penalty0 075110, 2019.
\newblock \doi{10.1103/PhysRevB.100.075110}.
\newblock URL \url{https://link.aps.org/doi/10.1103/PhysRevB.100.075110}.

\bibitem[El~Baggari et~al.(2018)El~Baggari, Savitzky, Admasu, Kim, Cheong,
  Hovden, and Kourkoutis]{el2018nature}
Ismail El~Baggari, Benjamin~H Savitzky, Alemayehu~S Admasu, Jaewook Kim,
  Sang-Wook Cheong, Robert Hovden, and Lena~F Kourkoutis.
\newblock Nature and evolution of incommensurate charge order in manganites
  visualized with cryogenic scanning transmission electron microscopy.
\newblock \emph{Proceedings of the National Academy of Sciences}, 115\penalty0
  (7):\penalty0 1445--1450, 2018.
\newblock \doi{10.1073/pnas.1714901115}.

\bibitem[Else et~al.(2017)Else, Bauer, and Nayak]{PhysRevX.7.011026}
Dominic~V. Else, Bela Bauer, and Chetan Nayak.
\newblock Prethermal phases of matter protected by time-translation symmetry.
\newblock \emph{Phys. Rev. X}, 7:\penalty0 011026, 2017.
\newblock \doi{10.1103/PhysRevX.7.011026}.
\newblock URL \url{https://link.aps.org/doi/10.1103/PhysRevX.7.011026}.

\bibitem[Emery and Kivelson(1995)]{emery1995superconductivity}
11~VJ Emery and SA~Kivelson.
\newblock Superconductivity in bad metals.
\newblock \emph{Phys. Rev. Lett.}, 74\penalty0 (16):\penalty0 3253, 1995.
\newblock \doi{10.1038/374434a0}.

\bibitem[Emery and Sessler(1960)]{emery1960possible}
VJ~Emery and AM~Sessler.
\newblock Possible phase transition in liquid he$^3$.
\newblock \emph{Physical Review}, 119\penalty0 (1):\penalty0 43, 1960.
\newblock \doi{10.1103/PhysRev.119.43}.

\bibitem[Enayat et~al.(2014)Enayat, Sun, Singh, Aluru, Schmaus, Yaresko, Liu,
  Lin, Tsurkan, Loidl, et~al.]{enayat2014real}
Mostafa Enayat, Zhixiang Sun, Udai~Raj Singh, Ramakrishna Aluru, Stefan
  Schmaus, Alexander Yaresko, Yong Liu, Chengtian Lin, Vladimir Tsurkan, Alois
  Loidl, et~al.
\newblock {Real-space imaging of the atomic-scale magnetic structure of
  Fe$_{1+y}$Te}.
\newblock \emph{Science}, 345\penalty0 (6197):\penalty0 653--656, 2014.
\newblock \doi{10.1126/science.1251682}.

\bibitem[Fang et~al.(2022)Fang, Grissonnanche, Legros, Verret, Lalibert{\'e},
  Collignon, Ataei, Dion, Zhou, Graf, et~al.]{fang2022fermi}
Yawen Fang, Ga{\"e}l Grissonnanche, Ana{\"e}lle Legros, Simon Verret, Francis
  Lalibert{\'e}, Cl{\'e}ment Collignon, Amirreza Ataei, Maxime Dion, Jianshi
  Zhou, David Graf, et~al.
\newblock Fermi surface transformation at the pseudogap critical point of a
  cuprate superconductor.
\newblock \emph{Nature Physics}, 18\penalty0 (5):\penalty0 558--564, 2022.
\newblock \doi{10.1038/s41567-022-01514-1}.

\bibitem[Fay and Layzer(1968)]{fay1968superfluidity}
D~Fay and A~Layzer.
\newblock Superfluidity of low-density fermion systems.
\newblock \emph{Phys. Rev. Lett.}, 20\penalty0 (5):\penalty0 187, 1968.
\newblock \doi{10.1103/PhysRevLett.20.187}.

\bibitem[Fei et~al.(2012)Fei, Rodin, Andreev, Bao, McLeod, Wagner, Zhang, Zhao,
  Thiemens, Dominguez, et~al.]{fei2012gate}
Zhe Fei, A.~S. Rodin, Gregory~O. Andreev, W.~Bao, A.~S. McLeod, M.~Wagner,
  L.~M. Zhang, Z.~Zhao, M.~Thiemens, G.~Dominguez, et~al.
\newblock Gate-tuning of graphene plasmons revealed by infrared nano-imaging.
\newblock \emph{Nature}, 487\penalty0 (7405):\penalty0 82--85, 2012.
\newblock \doi{10.1038/nature11253}.

\bibitem[Feng et~al.(2015)Feng, Pang, Wu, Wang, Weng, Li, Dai, Fang, Shi, and
  Lu]{MR-Cd3As2}
Junya Feng, Yuan Pang, Desheng Wu, Zhijun Wang, Hongming Weng, Jianqi Li,
  Xi~Dai, Zhong Fang, Youguo Shi, and Li~Lu.
\newblock {Large linear magnetoresistance in {Dirac} semimetal
  ${\mathrm{Cd}}_{3}{\mathrm{As}}_{2}$ with {Fermi} surfaces close to the
  {Dirac} points}.
\newblock \emph{Phys. Rev. B}, 92:\penalty0 081306, 2015.
\newblock \doi{10.1103/PhysRevB.92.081306}.
\newblock URL \url{https://link.aps.org/doi/10.1103/PhysRevB.92.081306}.

\bibitem[Feng et~al.(2019)Feng, Wang, Silevitch, Yan, Kobayashi, Hedo, Nakama,
  {\=O}nuki, Suslov, Mihaila, et~al.]{feng2019linear}
Yejun Feng, Yishu Wang, DM~Silevitch, J-Q Yan, Riki Kobayashi, Masato Hedo,
  Takao Nakama, Yoshichika {\=O}nuki, AV~Suslov, B~Mihaila, et~al.
\newblock Linear magnetoresistance in the low-field limit in density-wave
  materials.
\newblock \emph{Proceedings of the National Academy of Sciences}, 116\penalty0
  (23):\penalty0 11201--11206, 2019.
\newblock \doi{10.1073/pnas.1820092116}.

\bibitem[Fernandes et~al.(2019)Fernandes, Orth, and Schmalian]{RafaelARCMP2019}
Rafael~M. Fernandes, Peter~P. Orth, and Jörg Schmalian.
\newblock Intertwined vestigial order in quantum materials: Nematicity and
  beyond.
\newblock \emph{Annual Review of Condensed Matter Physics}, 10\penalty0
  (1):\penalty0 133--154, 2019.
\newblock \doi{10.1146/annurev-conmatphys-031218-013200}.

\bibitem[Fidkowski and Kitaev(2010)]{Fidkowski_2010}
Lukasz Fidkowski and Alexei Kitaev.
\newblock Effects of interactions on the topological classification of free
  fermion systems.
\newblock \emph{Phys. Rev. B}, 81\penalty0 (13), 2010.
\newblock \doi{10.1103/physrevb.81.134509}.
\newblock URL \url{http://dx.doi.org/10.1103/PhysRevB.81.134509}.

\bibitem[Fiebig et~al.(2005)Fiebig, Pavlov, and Pisarev]{fiebig2005second}
Manfred Fiebig, Victor~V Pavlov, and Roman~V Pisarev.
\newblock Second-harmonic generation as a tool for studying electronic and
  magnetic structures of crystals.
\newblock \emph{JOSA B}, 22\penalty0 (1):\penalty0 96--118, 2005.
\newblock \doi{10.1103/PhysRevLett.87.137202}.

\bibitem[Filoche and Mayboroda(2012)]{Filoche14761}
Marcel Filoche and Svitlana Mayboroda.
\newblock Universal mechanism for anderson and weak localization.
\newblock \emph{Proceedings of the National Academy of Sciences}, 109\penalty0
  (37):\penalty0 14761--14766, 2012.
\newblock \doi{10.1073/pnas.1120432109}.
\newblock URL \url{https://www.pnas.org/content/109/37/14761}.

\bibitem[Fisk et~al.(1988)Fisk, Hess, Pethick, Pines, Smith, Thompson, and
  Willis]{fisk1988heavy}
Z~Fisk, DW~Hess, CJ~Pethick, D~Pines, JL~Smith, JD~Thompson, and JO~Willis.
\newblock Heavy-electron metals: New highly correlated states of matter.
\newblock \emph{Science}, 239\penalty0 (4835):\penalty0 33--42, 1988.
\newblock \doi{10.1126/science.239.4835.33}.

\bibitem[Fl{\"{o}}totto et~al.(2018)Fl{\"{o}}totto, Bai, Chan, Chen, Wang,
  Rossi, Xu, Zhang, Hlevyack, Denlinger, Hong, Chou, Mittemeijer, Eckstein, and
  Chiang]{flototto2018bi2se3}
David Fl{\"{o}}totto, Yang Bai, Yang-Hao Chan, Peng Chen, Xiaoxiong Wang, Paul
  Rossi, Cai-Zhi Xu, Can Zhang, Joseph~A. Hlevyack, Jonathan~D. Denlinger,
  Hawoong Hong, Mei-Yin Chou, Eric~J. Mittemeijer, James~N. Eckstein, and
  Tai-Chang Chiang.
\newblock {In Situ Strain Tuning of the {Dirac} Surface States in Bi$_2$Se$_3$
  Films}.
\newblock \emph{Nano Letters}, 18\penalty0 (9):\penalty0 5628--5632, 2018.
\newblock \doi{10.1021/acs.nanolett.8b02105}.

\bibitem[F{\"o}rst et~al.(2011)F{\"o}rst, Manzoni, Kaiser, Tomioka, Tokura,
  Merlin, and Cavalleri]{forst2011nonlinear}
Michael F{\"o}rst, Cristian Manzoni, Stefan Kaiser, Yasuhide Tomioka, Yoshinori
  Tokura, Roberto Merlin, and Andrea Cavalleri.
\newblock Nonlinear phononics as an ultrafast route to lattice control.
\newblock \emph{Nature Physics}, 7\penalty0 (11):\penalty0 854, 2011.
\newblock \doi{10.1038/nphys2055}.

\bibitem[Frachet et~al.(2020)Frachet, Vinograd, Zhou, Benhabib, Wu, Mayaffre,
  Kr{\"a}mer, Ramakrishna, Reyes, Debray, et~al.]{frachet2020hidden}
Mehdi Frachet, Igor Vinograd, Rui Zhou, Siham Benhabib, Shangfei Wu, Hadrien
  Mayaffre, Steffen Kr{\"a}mer, Sanath~K Ramakrishna, Arneil~P Reyes,
  J{\'e}r{\^o}me Debray, et~al.
\newblock Hidden magnetism at the pseudogap critical point of a cuprate
  superconductor.
\newblock \emph{Nature Physics}, page~1, 2020.
\newblock \doi{10.1038/s41567-020-0950-5}.

\bibitem[Fratino et~al.(2016)Fratino, S{\'e}mon, Sordi, and
  Tremblay]{fratino2016organizing}
L~Fratino, P~S{\'e}mon, Giovanni Sordi, and A-MS Tremblay.
\newblock An organizing principle for two-dimensional strongly correlated
  superconductivity.
\newblock \emph{Scientific Reports}, 6\penalty0 (1):\penalty0 1, 2016.
\newblock \doi{10.1038/srep22715}.

\bibitem[Fu and Kane(2008)]{fu2008superconducting}
Liang Fu and Charles~L Kane.
\newblock Superconducting proximity effect and majorana fermions at the surface
  of a topological insulator.
\newblock \emph{Phys. Rev. Lett.}, 100\penalty0 (9):\penalty0 096407, 2008.
\newblock \doi{10.1103/PhysRevLett.100.096407}.

\bibitem[Furukawa et~al.(2015)Furukawa, Miyagawa, Taniguchi, Kato, and
  Kanoda]{furukawa2015quantum}
Tetsuya Furukawa, Kazuya Miyagawa, Hiromi Taniguchi, Reizo Kato, and Kazushi
  Kanoda.
\newblock Quantum criticality of mott transition in organic materials.
\newblock \emph{Nature Physics}, 11\penalty0 (3):\penalty0 221--224, 2015.
\newblock \doi{10.1038/nphys3235}.

\bibitem[Gabani et~al.(2001)Gabani, Flachbart, Konovalova, Orendac, Paderno,
  Pavlik, and Sebek]{Gabani2001}
S.~Gabani, K.~Flachbart, E.~Konovalova, M.~Orendac, Y.~Paderno, V.~Pavlik, and
  J.~Sebek.
\newblock Properties of the in-gap states in smb6.
\newblock \emph{Solid State Communications}, 117\penalty0 (11):\penalty0 641 --
  644, 2001.
\newblock \doi{10.1016/S0038-1098(01)00004-7}.

\bibitem[Garg and Kern(2020)]{garg2020atto}
M~Garg and K~Kern.
\newblock {Attosecond coherent manipulation of electrons in tunneling
  microscopy}.
\newblock \emph{Science}, 367\penalty0 (6476):\penalty0 411--415, 2020.
\newblock \doi{10.1126/science.aaz1098}.

\bibitem[Gastiasoro et~al.(2020)Gastiasoro, Ruhman, and
  Fernandes]{gastiasoro2020superconductivity}
Maria~N Gastiasoro, Jonathan Ruhman, and Rafael~M Fernandes.
\newblock {Superconductivity in dilute SrTiO$_3$: A review}.
\newblock \emph{Annals of Physics}, page 168107, 2020.
\newblock \doi{10.1016/j.aop.2020.168107}.

\bibitem[Gedik et~al.(2005)Gedik, Langner, Orenstein, Ono, Abe, and
  Ando]{gedik2005abrupt}
N~Gedik, M~Langner, J~Orenstein, S~Ono, Yasushi Abe, and Yoichi Ando.
\newblock Abrupt transition in quasiparticle dynamics at optimal doping in a
  cuprate superconductor system.
\newblock \emph{Phys. Rev. Lett.}, 95\penalty0 (11):\penalty0 117005, 2005.
\newblock \doi{10.1103/PhysRevLett.95.117005}.

\bibitem[Gegenwart et~al.(2008)Gegenwart, Si, and
  Steglich]{gegenwart2008quantum}
Philipp Gegenwart, Qimiao Si, and Frank Steglich.
\newblock Quantum criticality in heavy-fermion metals.
\newblock \emph{Nature Physics}, 4\penalty0 (3):\penalty0 186--197, 2008.
\newblock \doi{10.1038/nphys892}.

\bibitem[Geim and Grigorieva(2013)]{geim2013van}
Andre~K Geim and Irina~V Grigorieva.
\newblock {Van der Waals heterostructures}.
\newblock \emph{Nature}, 499\penalty0 (7459):\penalty0 419--425, 2013.
\newblock \doi{10.1038/nature12385}.

\bibitem[Geneaux et~al.(2019)Geneaux, Marroux, Guggenmos, Neumark, and
  Leone]{geneaux2019transient}
Romain Geneaux, Hugo J.~B. Marroux, Alexander Guggenmos, Daniel~M. Neumark, and
  Stephen~R. Leone.
\newblock Transient absorption spectroscopy using high harmonic generation: a
  review of ultrafast x-ray dynamics in molecules and solids.
\newblock \emph{Philosophical Transactions of the Royal Society A:
  Mathematical, Physical and Engineering Sciences}, 377\penalty0
  (2145):\penalty0 20170463, 2019.
\newblock \doi{10.1098/rsta.2017.0463}.

\bibitem[Georges et~al.(1996)Georges, Kotliar, Krauth, and
  Rozenberg]{Georges_DMFT}
Antoine Georges, Gabriel Kotliar, Werner Krauth, and Marcelo~J. Rozenberg.
\newblock Dynamical mean-field theory of strongly correlated fermion systems
  and the limit of infinite dimensions.
\newblock \emph{Rev. Mod. Phys.}, 68:\penalty0 13--125, 1996.
\newblock \doi{10.1103/RevModPhys.68.13}.
\newblock URL \url{https://link.aps.org/doi/10.1103/RevModPhys.68.13}.

\bibitem[Gerber et~al.(2017)Gerber, Yang, Zhu, Soifer, Sobota, Rebec, Lee, Jia,
  Moritz, Jia, et~al.]{gerber2017femtosecond}
S~Gerber, S-L Yang, D~Zhu, H~Soifer, JA~Sobota, S~Rebec, JJ~Lee, T~Jia,
  B~Moritz, C~Jia, et~al.
\newblock Femtosecond electron-phonon lock-in by photoemission and x-ray
  free-electron laser.
\newblock \emph{Science}, 357\penalty0 (6346):\penalty0 71--75, 2017.
\newblock \doi{10.1126/science.aak9946}.

\bibitem[Gerber et~al.(2015)Gerber, Jang, Nojiri, Matsuzawa, Yasumura, Bonn,
  Liang, Hardy, Islam, Mehta, et~al.]{gerber2015three}
Simon Gerber, H~Jang, H~Nojiri, S~Matsuzawa, H~Yasumura, DA~Bonn, R~Liang,
  WN~Hardy, Z~Islam, A~Mehta, et~al.
\newblock {Three-dimensional charge density wave order in YBa$_2$Cu$_3$O$_6.67$
  at high magnetic fields}.
\newblock \emph{Science}, 350\penalty0 (6263):\penalty0 949--952, 2015.
\newblock \doi{10.1126/science.aac6257}.

\bibitem[Ghorashi(2020)]{doi:10.1002/andp.201900336}
Sayed Ali~Akbar Ghorashi.
\newblock {Hybrid Dispersion {Dirac} Semimetal and Hybrid {Weyl} Phases in
  Luttinger Semimetals: A Dynamical Approach}.
\newblock \emph{Annalen der Physik}, 532\penalty0 (2):\penalty0 1900336, 2020.
\newblock \doi{10.1002/andp.201900336}.
\newblock URL
  \url{https://onlinelibrary.wiley.com/doi/abs/10.1002/andp.201900336}.

\bibitem[Ghorashi et~al.(2018)Ghorashi, Hosur, and Ting]{PhysRevB.97.205402}
Sayed Ali~Akbar Ghorashi, Pavan Hosur, and Chin-Sen Ting.
\newblock {Irradiated three-dimensional Luttinger semimetal: A factory for
  engineering {Weyl} semimetals}.
\newblock \emph{Phys. Rev. B}, 97:\penalty0 205402, 2018.
\newblock \doi{10.1103/PhysRevB.97.205402}.
\newblock URL \url{https://link.aps.org/doi/10.1103/PhysRevB.97.205402}.

\bibitem[Ghorashi et~al.(2020)Ghorashi, Karcher, Davis, and
  Foster]{ghorashi2019criticality}
Sayed Ali~Akbar Ghorashi, Jonas~F. Karcher, Seth~M. Davis, and Matthew~S.
  Foster.
\newblock {Criticality across the energy spectrum from random artificial
  gravitational lensing in two-dimensional {Dirac} superconductors}.
\newblock \emph{Phys. Rev. B}, 101:\penalty0 214521, 2020.
\newblock \doi{10.1103/PhysRevB.101.214521}.
\newblock URL \url{https://link.aps.org/doi/10.1103/PhysRevB.101.214521}.

\bibitem[Giraldo-Gallo et~al.(2018)Giraldo-Gallo, Galvis, Stegen, Modic,
  Balakirev, Betts, Lian, Moir, Riggs, Wu, et~al.]{giraldo2018scale}
Paula Giraldo-Gallo, JA~Galvis, Zachary Stegen, KA~Modic, FF~Balakirev,
  JB~Betts, Xiujun Lian, Camilla Moir, SC~Riggs, J~Wu, et~al.
\newblock Scale-invariant magnetoresistance in a cuprate superconductor.
\newblock \emph{Science}, 361\penalty0 (6401):\penalty0 479--481, 2018.
\newblock \doi{10.1038/s41586-018-0270-9}.

\bibitem[Goldman and Markovic(1998)]{goldman1998superconductor}
Allen~M Goldman and Nina Markovic.
\newblock Superconductor-insulator transitions in the two-dimensional limit.
\newblock \emph{Physics Today}, 51:\penalty0 39--44, 1998.
\newblock \doi{10.1016/S1386-9477(02)00932-3}.

\bibitem[Goldman(2014)]{goldman2014gating}
A.M. Goldman.
\newblock Electrostatic gating of ultrathin films.
\newblock \emph{Annual Review of Materials Research}, 44\penalty0 (1):\penalty0
  45--63, 2014.
\newblock \doi{10.1146/annurev-matsci-070813-113407}.
\newblock URL \url{https://doi.org/10.1146/annurev-matsci-070813-113407}.

\bibitem[Goldman and Dalibard(2014)]{PhysRevX.4.031027}
N.~Goldman and J.~Dalibard.
\newblock Periodically driven quantum systems: Effective hamiltonians and
  engineered gauge fields.
\newblock \emph{Phys. Rev. X}, 4:\penalty0 031027, 2014.
\newblock \doi{10.1103/PhysRevX.4.031027}.
\newblock URL \url{https://link.aps.org/doi/10.1103/PhysRevX.4.031027}.

\bibitem[Gooth et~al.(2019)Gooth, Bradlyn, Honnali, Schindler, Kumar, Noky, Qi,
  Shekhar, Sun, Wang, et~al.]{gooth2019axionic}
J~Gooth, B~Bradlyn, S~Honnali, C~Schindler, N~Kumar, J~Noky, Y~Qi, C~Shekhar,
  Y~Sun, Z~Wang, et~al.
\newblock {Axionic charge-density wave in the {Weyl} semimetal
  (TaSe$_4$)$_2$I}.
\newblock \emph{Nature}, 575\penalty0 (7782):\penalty0 315--319, 2019.
\newblock \doi{10.1038/s41586-019-1630-4}.

\bibitem[Gorshunov et~al.(1999)Gorshunov, Sluchanko, Volkov, Dressel, Knebel,
  Loidl, and Kunii]{gorshunov1999low}
B~Gorshunov, N~Sluchanko, A~Volkov, Martin Dressel, Georg Knebel, Alois Loidl,
  and S~Kunii.
\newblock Low-energy electrodynamics of smb 6.
\newblock \emph{Phys. Rev. B}, 59\penalty0 (3):\penalty0 1808, 1999.
\newblock \doi{10.1103/PhysRevB.59.1808}.

\bibitem[Goswami et~al.(2008)Goswami, Schwab, and
  Chakravarty]{goswami2008rounding}
Pallab Goswami, David Schwab, and Sudip Chakravarty.
\newblock Rounding by disorder of first-order quantum phase transitions:
  emergence of quantum critical points.
\newblock \emph{Phys. Rev. Lett.}, 100\penalty0 (1):\penalty0 015703, 2008.
\newblock \doi{10.1103/PhysRevLett.100.015703}.

\bibitem[{Goulko} et~al.(2017){Goulko}, {Mishchenko}, {Pollet}, {Prokof'ev},
  and {Svistunov}]{goulko17nac}
O.~{Goulko}, A.~S. {Mishchenko}, L.~{Pollet}, N.~{Prokof'ev}, and
  B.~{Svistunov}.
\newblock {Numerical analytic continuation: Answers to well-posed questions}.
\newblock \emph{\prb}, 95\penalty0 (1):\penalty0 014102, 2017.
\newblock \doi{10.1103/PhysRevB.95.014102}.

\bibitem[Green et~al.(2016)Green, Kovalev, Asgekar, Geloni, Lehnert, Golz,
  Kuntzsch, Bauer, Hauser, Voigtlaender, Wustmann, Koesterke, Schwarz, Freitag,
  Arnold, Teichert, Justus, Seidel, Ilgner, Awari, Nicoletti, Kaiser, Laplace,
  Rajasekaran, Zhang, Winnerl, Schneider, Schay, Lorincz, Rauscher, Radu,
  M{\"{a}}hrlein, Kim, Lee, Kampfrath, Wall, Heberle, Malnasi-Csizmadia,
  Steiger, M{\"{u}}ller, Helm, Schramm, Cowan, Michel, Cavalleri, Fisher,
  Stojanovic, and Gensch]{green2016thz}
B~Green, S~Kovalev, V~Asgekar, G~Geloni, U~Lehnert, T~Golz, M~Kuntzsch,
  C~Bauer, J~Hauser, J~Voigtlaender, B~Wustmann, I~Koesterke, M~Schwarz,
  M~Freitag, A~Arnold, J~Teichert, M~Justus, W~Seidel, C~Ilgner, N~Awari,
  D~Nicoletti, S~Kaiser, Y~Laplace, S~Rajasekaran, L~Zhang, S~Winnerl,
  H~Schneider, G~Schay, I~Lorincz, A~A Rauscher, I~Radu, S~M{\"{a}}hrlein, T~H
  Kim, J~S Lee, T~Kampfrath, S~Wall, J~Heberle, A~Malnasi-Csizmadia, A~Steiger,
  A~S M{\"{u}}ller, M~Helm, U~Schramm, T~Cowan, P~Michel, A~Cavalleri, A~S
  Fisher, N~Stojanovic, and M~Gensch.
\newblock {High-Field High-Repetition-Rate Sources for the Coherent {THz}
  Control of Matter}.
\newblock \emph{Scientific Reports}, 6\penalty0 (1):\penalty0 22256, 2016.
\newblock \doi{10.1038/srep22256}.

\bibitem[Greene et~al.(2020)Greene, Mandal, Poniatowski, and Sarkar]{smreview}
Richard~L Greene, Pampa~R Mandal, Nicholas~R Poniatowski, and Tarapada Sarkar.
\newblock The strange metal state of the electron-doped cuprates.
\newblock \emph{Annual Review of Condensed Matter Physics}, 11:\penalty0
  213--229, 2020.
\newblock \doi{10.1146/annurev-conmatphys-031119-050558}.

\bibitem[Greywall(1986)]{greywall1986he}
Dennis~S Greywall.
\newblock He 3 specific heat and thermometry at millikelvin temperatures.
\newblock \emph{Phys. Rev. B}, 33\penalty0 (11):\penalty0 7520, 1986.
\newblock \doi{10.1103/PhysRevB.33.7520}.

\bibitem[Grissonnanche et~al.(2014)Grissonnanche, Cyr-Choini{\`e}re,
  Lalibert{\'e}, De~Cotret, Juneau-Fecteau, Dufour-Beaus{\'e}jour, Delage,
  LeBoeuf, Chang, Ramshaw, et~al.]{grissonnanche2014direct}
G~Grissonnanche, O~Cyr-Choini{\`e}re, F~Lalibert{\'e}, S~Ren{\'e} De~Cotret,
  A~Juneau-Fecteau, S~Dufour-Beaus{\'e}jour, M-E Delage, D~LeBoeuf, J~Chang,
  BJ~Ramshaw, et~al.
\newblock Direct measurement of the upper critical field in cuprate
  superconductors.
\newblock \emph{Nature Communications}, 5:\penalty0 3280, 2014.
\newblock \doi{10.1038/ncomms4280}.

\bibitem[Gruzberg et~al.(2017)Gruzberg, Kl\"umper, Nuding, and
  Sedrakyan]{GruzbergQG}
I.~A. Gruzberg, A.~Kl\"umper, W.~Nuding, and A.~Sedrakyan.
\newblock {Geometrically disordered network models, quenched quantum gravity,
  and critical behavior at quantum Hall plateau transitions}.
\newblock \emph{Phys. Rev. B}, 95:\penalty0 125414, 2017.
\newblock \doi{10.1103/PhysRevB.95.125414}.
\newblock URL \url{https://link.aps.org/doi/10.1103/PhysRevB.95.125414}.

\bibitem[Gu et~al.(2019)Gu, Wan, Tang, Du, Yang, Wang, Zhong, Wen, Gu, and
  Wen]{gu2019directly}
Qiangqiang Gu, Siyuan Wan, Qingkun Tang, Zengyi Du, Huan Yang, Qiang-Hua Wang,
  Ruidan Zhong, Jinsheng Wen, Genda~D Gu, and Hai-Hu Wen.
\newblock {Directly visualizing the sign change of $d$-wave superconducting gap
  in {Bi}$_2${Sr}$_2${CaCu}$_2${O}$_{8+\delta}$ by phase-referenced
  quasiparticle interference}.
\newblock \emph{Nature Communications}, 10\penalty0 (1):\penalty0 1--10, 2019.
\newblock \doi{10.1038/s41467-019-09340-5}.

\bibitem[Gull and Millis(2012)]{gull2012energetics}
Emanuel Gull and Andrew~J Millis.
\newblock Energetics of superconductivity in the two-dimensional hubbard model.
\newblock \emph{Phys. Rev. B}, 86\penalty0 (24):\penalty0 241106, 2012.
\newblock \doi{10.1103/PhysRevB.86.241106}.

\bibitem[Gull et~al.(2011)Gull, Millis, Lichtenstein, Rubtsov, Troyer, and
  Werner]{Gull2011}
Emanuel Gull, Andrew~J Millis, Alexander~I Lichtenstein, Alexey~N Rubtsov,
  Matthias Troyer, and Philipp Werner.
\newblock {Continuous-time {Monte} {Carlo} methods for quantum impurity
  models}.
\newblock \emph{Reviews of Modern Physics}, 83\penalty0 (2):\penalty0 349,
  2011.
\newblock \doi{10.1103/RevModPhys.83.349}.

\bibitem[Gunnarsson et~al.(2003)Gunnarsson, Calandra, and
  Han]{gunnarsson2003colloquium}
O~Gunnarsson, M~Calandra, and JE~Han.
\newblock Colloquium: Saturation of electrical resistivity.
\newblock \emph{Reviews of Modern Physics}, 75\penalty0 (4):\penalty0 1085,
  2003.
\newblock \doi{10.1103/RevModPhys.75.1085}.

\bibitem[Guo et~al.(2020)Guo, Sun, Zhao, Wang, Hong, Sidorov, Ma, Wu, Li, Meng,
  et~al.]{GuoShastry2019}
Jing Guo, Guangyu Sun, Bowen Zhao, Ling Wang, Wenshan Hong, Vladimir~A Sidorov,
  Nvsen Ma, Qi~Wu, Shiliang Li, Zi~Yang Meng, et~al.
\newblock {Quantum Phases of SrCu$_2$(BO$_3$)$_2$ from High-Pressure
  Thermodynamics}.
\newblock \emph{Physical Review Letters}, 124\penalty0 (20):\penalty0 206602,
  2020.
\newblock \doi{10.1103/PhysRevLett.124.206602}.

\bibitem[Gurvitch and Fiory(1987)]{gurvitch1987resistivity}
M~Gurvitch and AT~Fiory.
\newblock Resistivity of la$_{1.825}$sr$_{0.175}$cuo$_4$ and yba$_2$cu$_3$o$_7$
  to 1100 k: absence of saturation and its implications.
\newblock \emph{Phys. Rev. Lett.}, 59\penalty0 (12):\penalty0 1337, 1987.
\newblock \doi{10.1103/PhysRevLett.59.1337}.

\bibitem[Gusev et~al.(2013)Gusev, Olshanetsky, Kvon, Mikhailov, and
  Dvoretsky]{Mr-HgTe}
G.~M. Gusev, E.~B. Olshanetsky, Z.~D. Kvon, N.~N. Mikhailov, and S.~A.
  Dvoretsky.
\newblock {Linear magnetoresistance in HgTe quantum wells}.
\newblock \emph{Phys. Rev. B}, 87:\penalty0 081311, 2013.
\newblock \doi{10.1103/PhysRevB.87.081311}.
\newblock URL \url{https://link.aps.org/doi/10.1103/PhysRevB.87.081311}.

\bibitem[Haak et~al.(1978)Haak, Sawatzky, and Thomas]{haak1978auger}
HW~Haak, GA~Sawatzky, and TD~Thomas.
\newblock Auger-photoelectron coincidence measurements in copper.
\newblock \emph{Phys. Rev. Lett.}, 41\penalty0 (26):\penalty0 1825, 1978.
\newblock \doi{10.1103/PhysRevLett.41.1825}.

\bibitem[Hagiwara et~al.(1990)Hagiwara, Katsumata, Affleck, Halperin, and
  Renard]{Hagiwara_1990}
M.~Hagiwara, K.~Katsumata, Ian Affleck, B.~I. Halperin, and J.~P. Renard.
\newblock {Observation of S=1/2 degrees of freedom in an S=1 linear-chain
  {Heisenberg} antiferromagnet}.
\newblock \emph{Phys. Rev. Lett.}, 65:\penalty0 3181--3184, 1990.
\newblock \doi{10.1103/PhysRevLett.65.3181}.
\newblock URL \url{https://link.aps.org/doi/10.1103/PhysRevLett.65.3181}.

\bibitem[Haldane(1988)]{haldane1988model}
F~Duncan~M Haldane.
\newblock Model for a quantum hall effect without landau levels:
  Condensed-matter realization of the "parity anomaly".
\newblock \emph{Phys. Rev. Lett.}, 61\penalty0 (18):\penalty0 2015, 1988.
\newblock \doi{10.1103/PhysRevLett.61.2015}.

\bibitem[Halperin(1987)]{halperinhighfieldreview}
Bertrand~I Halperin.
\newblock Possible states for a three-dimensional electron gas in a strong
  magnetic field.
\newblock \emph{Japanese Journal of Applied Physics}, 26\penalty0
  (S3-3):\penalty0 1913, 1987.
\newblock \doi{10.7567/JJAPS.26S3.1913}.

\bibitem[Hamidian et~al.(2016)Hamidian, Edkins, Joo, Kostin, Eisaki, Uchida,
  Lawler, Kim, Mackenzie, Fujita, et~al.]{hamidian2016detection}
MH~Hamidian, SD~Edkins, Sang~Hyun Joo, A~Kostin, H~Eisaki, S~Uchida, MJ~Lawler,
  E-A Kim, AP~Mackenzie, K~Fujita, et~al.
\newblock {Detection of a Cooper-pair density wave in
  Bi$_2$Sr$_2$CaCu$_2$O$_{8+x}$}.
\newblock \emph{Nature}, 532\penalty0 (7599):\penalty0 343--347, 2016.
\newblock \doi{10.1038/nature17411}.

\bibitem[Harris(1974)]{harris1974effect}
A~Brooks Harris.
\newblock Effect of random defects on the critical behaviour of ising models.
\newblock \emph{Journal of Physics C: Solid State Physics}, 7\penalty0
  (9):\penalty0 1671, 1974.
\newblock \doi{10.1088/0022-3719/7/9/009}.

\bibitem[Harter et~al.(2017)Harter, Zhao, Yan, Mandrus, and
  Hsieh]{harter2017parity}
JW~Harter, ZY~Zhao, J-Q Yan, DG~Mandrus, and David Hsieh.
\newblock {A parity-breaking electronic nematic phase transition in the
  spin-orbit coupled metal Cd$_2$Re$_2$O$_7$}.
\newblock \emph{Science}, 356\penalty0 (6335):\penalty0 295--299, 2017.
\newblock \doi{10.1126/science.aad1188}.

\bibitem[Hartman et~al.(2017)Hartman, Hartnoll, and
  Mahajan]{PhysRevLett.119.141601}
Thomas Hartman, Sean~A. Hartnoll, and Raghu Mahajan.
\newblock Upper bound on diffusivity.
\newblock \emph{Phys. Rev. Lett.}, 119:\penalty0 141601, 2017.
\newblock \doi{10.1103/PhysRevLett.119.141601}.
\newblock URL \url{https://link.aps.org/doi/10.1103/PhysRevLett.119.141601}.

\bibitem[Hartnoll(2015)]{hartnoll2015theory}
Sean~A Hartnoll.
\newblock Theory of universal incoherent metallic transport.
\newblock \emph{Nature Physics}, 11\penalty0 (1):\penalty0 54--61, 2015.
\newblock \doi{10.1038/nphys3174}.

\bibitem[Hartnoll and Karch(2015)]{hartnoll2015scaling}
Sean~A Hartnoll and Andreas Karch.
\newblock Scaling theory of the cuprate strange metals.
\newblock \emph{Phys. Rev. B}, 91\penalty0 (15):\penalty0 155126, 2015.
\newblock \doi{10.1103/PhysRevB.91.155126}.

\bibitem[Hartnoll et~al.(2018)Hartnoll, Lucas, and
  Sachdev]{hartnoll2016holographic}
Sean~A Hartnoll, Andrew Lucas, and Subir Sachdev.
\newblock \emph{Holographic Quantum Matter}.
\newblock MIT press, 2018.
\newblock \doi{10.48550/arXiv.1612.07324}.

\bibitem[Hasan and Kane(2010)]{hasan2010colloquium}
M~Zahid Hasan and Charles~L Kane.
\newblock Colloquium: topological insulators.
\newblock \emph{Reviews of Modern Physics}, 82\penalty0 (4):\penalty0 3045,
  2010.
\newblock \doi{10.1103/RevModPhys.82.3045}.

\bibitem[Hasan and Moore(2011)]{hasan2011three}
M~Zahid Hasan and Joel~E Moore.
\newblock Three-dimensional topological insulators.
\newblock \emph{Annu. Rev. Condens. Matter Phys.}, 2\penalty0 (1):\penalty0
  55--78, 2011.
\newblock \doi{10.1146/annurev-conmatphys-062910-140432}.

\bibitem[Hashimoto et~al.(2014)Hashimoto, Vishik, He, Devereaux, and
  Shen]{hashimoto2014energy}
Makoto Hashimoto, Inna~M Vishik, Rui-Hua He, Thomas~P Devereaux, and Zhi-Xun
  Shen.
\newblock Energy gaps in high-transition-temperature cuprate superconductors.
\newblock \emph{Nature Physics}, 10\penalty0 (7):\penalty0 483--495, 2014.
\newblock \doi{10.1038/nphys3009}.

\bibitem[Hashimoto et~al.(2015)Hashimoto, Nowadnick, He, Vishik, Moritz, He,
  Tanaka, Moore, Lu, Yoshida, et~al.]{hashimoto2015direct}
Makoto Hashimoto, Elizabeth~A Nowadnick, Rui-Hua He, Inna~M Vishik, Brian
  Moritz, Yu~He, Kiyohisa Tanaka, Robert~G Moore, Donghui Lu, Yoshiyuki
  Yoshida, et~al.
\newblock {Direct spectroscopic evidence for phase competition between the
  pseudogap and superconductivity in
  {Bi}$_2${Sr}$_2${CaCu}$_2${O}$_{8+\delta}$}.
\newblock \emph{Nature Materials}, 14\penalty0 (1):\penalty0 37--42, 2015.
\newblock \doi{10.1038/nmat4116}.

\bibitem[Hauke et~al.(2016)Hauke, Heyl, Tagliacozzo, and Zoller]{Hauke2016}
Philipp Hauke, Markus Heyl, Luca Tagliacozzo, and Peter Zoller.
\newblock Measuring multipartite entanglement through dynamic susceptibilities.
\newblock \emph{Nature Physics}, 12\penalty0 (8):\penalty0 778--782, 2016.
\newblock \doi{10.1038/nphys3700}.
\newblock URL \url{https://doi.org/10.1038/nphys3700}.

\bibitem[Hayes et~al.(2016)Hayes, McDonald, Breznay, Helm, Moll, Wartenbe,
  Shekhter, and Analytis]{hayes2016scaling}
Ian~M Hayes, Ross~D McDonald, Nicholas~P Breznay, Toni Helm, Philip~JW Moll,
  Mark Wartenbe, Arkady Shekhter, and James~G Analytis.
\newblock {Scaling between magnetic field and temperature in the
  high-temperature superconductor BaFe$_2$(As$_{1-x}$P$_x$)$_2$}.
\newblock \emph{Nature Physics}, 12\penalty0 (10):\penalty0 916--919, 2016.
\newblock \doi{10.1038/nphys3773}.

\bibitem[Hayne et~al.(1999)Hayne, Jones, Bogaerts, Riva, Usher, Peeters,
  Herlach, Moshchalkov, and Henini]{Hayne1999}
M.~Hayne, C.~L. Jones, R.~Bogaerts, C.~Riva, A.~Usher, F.~M. Peeters,
  F.~Herlach, V.~V. Moshchalkov, and M.~Henini.
\newblock Photoluminescence of negatively charged excitons in high magnetic
  fields.
\newblock \emph{Phys. Rev. B}, 59:\penalty0 2927--2931, 1999.
\newblock \doi{10.1103/PhysRevB.59.2927}.

\bibitem[He et~al.(2018{\natexlab{a}})He, Yee, McNally, Simonson, Zellman,
  Klemm, Kamenov, Geschwind, Zebro, Ghose, et~al.]{he2018combined}
Hua He, Chuck-Hou Yee, Daniel~E McNally, Jack~W Simonson, Shelby Zellman, Mason
  Klemm, Plamen Kamenov, Gayle Geschwind, Ashley Zebro, Sanjit Ghose, et~al.
\newblock {Combined computational and experimental investigation of the
  La$_2$CuO$_{4-x}$S$_x$ (0$\leq$ x$\leq$ 4) quaternary system}.
\newblock \emph{Proceedings of the National Academy of Sciences}, 115\penalty0
  (31):\penalty0 7890--7895, 2018{\natexlab{a}}.
\newblock \doi{10.1073/pnas.1800284115}.

\bibitem[He et~al.(2011)He, Hashimoto, Karapetyan, Koralek, Hinton, Testaud,
  Nathan, Yoshida, Yao, Tanaka, et~al.]{he2011single}
Rui-Hua He, M~Hashimoto, H~Karapetyan, JD~Koralek, JP~Hinton, JP~Testaud,
  V~Nathan, Y~Yoshida, Hong Yao, K~Tanaka, et~al.
\newblock From a single-band metal to a high-temperature superconductor via two
  thermal phase transitions.
\newblock \emph{Science}, 331\penalty0 (6024):\penalty0 1579--1583, 2011.
\newblock \doi{10.1126/science.1198415}.

\bibitem[He et~al.(2018{\natexlab{b}})He, Hashimoto, Song, Chen, He, Vishik,
  Moritz, Lee, Nagaosa, Zaanen, et~al.]{he2018rapid}
Y~He, M~Hashimoto, D~Song, S-D Chen, J~He, IM~Vishik, B~Moritz, D-H Lee,
  N~Nagaosa, J~Zaanen, et~al.
\newblock {Rapid change of superconductivity and electron-phonon coupling
  through critical doping in {Bi}-2212}.
\newblock \emph{Science}, 362\penalty0 (6410):\penalty0 62--65,
  2018{\natexlab{b}}.
\newblock \doi{10.1126/science.aar3394}.

\bibitem[He et~al.(2018{\natexlab{c}})He, Wu, Song, Lee, Said, Alatas, Bosak,
  Girard, Souliou, Ruiz, et~al.]{he2018persistent}
Yu~He, Shan Wu, Yu~Song, W-S Lee, Ayman~H Said, Ahmet Alatas, Alexei Bosak,
  Adrien Girard, Sofia-Michaela Souliou, Alejandro Ruiz, et~al.
\newblock {Persistent low-energy phonon broadening near the charge-order $q$
  vector in the bilayer cuprate {Bi}$_2${Sr}$_2${CaCu}$_2${O}$_{8+\delta}$}.
\newblock \emph{Phys. Rev. B}, 98\penalty0 (3):\penalty0 035102,
  2018{\natexlab{c}}.
\newblock \doi{10.1103/PhysRevB.98.035102}.

\bibitem[Helbig et~al.(2020)Helbig, Hofmann, Imhof, Abdelghany, Kiessling,
  Molenkamp, Lee, Szameit, Greiter, and Thomale]{TobiasArxiv2019}
T~Helbig, T~Hofmann, S~Imhof, M~Abdelghany, T~Kiessling, LW~Molenkamp, CH~Lee,
  A~Szameit, M~Greiter, and R~Thomale.
\newblock Generalized bulk-boundary correspondence in non-hermitian
  topolectrical circuits.
\newblock \emph{Nature Physics}, pages 1--4, 2020.
\newblock \doi{10.1038/s41567-020-0922-9}.

\bibitem[Helgren et~al.(2002)Helgren, Armitage, and Gr\"uner]{Helgren02a}
E.~Helgren, N.~P. Armitage, and G.~Gr\"uner.
\newblock {Electrodynamics of a Coulomb Glass in $n$-Type Silicon}.
\newblock \emph{Phys. Rev. Lett.}, 89\penalty0 (24):\penalty0 246601, 2002.
\newblock \doi{10.1103/PhysRevLett.89.246601}.

\bibitem[Herring(1937)]{Herringsemimetals}
Conyers Herring.
\newblock Accidental degeneracy in the energy bands of crystals.
\newblock \emph{Phys. Rev.}, 52:\penalty0 365--373, 1937.
\newblock \doi{10.1103/PhysRev.52.365}.
\newblock URL \url{https://link.aps.org/doi/10.1103/PhysRev.52.365}.

\bibitem[Hicks et~al.(2014)Hicks, Brodsky, Yelland, Gibbs, Bruin, Barber,
  Edkins, Nishimura, Yonezawa, Maeno, et~al.]{hicks2014strong}
Clifford~W Hicks, Daniel~O Brodsky, Edward~A Yelland, Alexandra~S Gibbs, Jan~AN
  Bruin, Mark~E Barber, Stephen~D Edkins, Keigo Nishimura, Shingo Yonezawa,
  Yoshiteru Maeno, et~al.
\newblock {Strong increase of T$_c$ of Sr$_2$RuO$_4$ under both tensile and
  compressive strain}.
\newblock \emph{Science}, 344\penalty0 (6181):\penalty0 283--285, 2014.
\newblock \doi{10.1126/science.1248292}.

\bibitem[Hirsch(1985)]{Hirsch_Hubbard}
J.~E. Hirsch.
\newblock Two-dimensional hubbard model: Numerical simulation study.
\newblock \emph{Phys. Rev. B}, 31:\penalty0 4403--4419, 1985.
\newblock \doi{10.1103/PhysRevB.31.4403}.
\newblock URL \url{https://link.aps.org/doi/10.1103/PhysRevB.31.4403}.

\bibitem[Hirsch(1992)]{hirsch1992superconductors}
JE~Hirsch.
\newblock Superconductors that change color when they become superconducting.
\newblock \emph{Physica C: Superconductivity}, 201\penalty0 (3-4):\penalty0
  347--361, 1992.
\newblock \doi{10.1016/0921-4534(92)90483-S}.

\bibitem[Hirschfeld et~al.(2011)Hirschfeld, Korshunov, and
  Mazin]{hirschfeld2011gap}
PJ~Hirschfeld, MM~Korshunov, and II~Mazin.
\newblock {Gap symmetry and structure of Fe-based superconductors}.
\newblock \emph{Reports on Progress in Physics}, 74\penalty0 (12):\penalty0
  124508, 2011.
\newblock \doi{10.1088/0034-4885/74/12/124508}.

\bibitem[Hohenberg and Kohn(1964)]{hohenbergkohn}
P.~Hohenberg and W.~Kohn.
\newblock Inhomogeneous electron gas.
\newblock \emph{Phys. Rev.}, 136:\penalty0 B864--B871, 1964.
\newblock \doi{10.1103/PhysRev.136.B864}.

\bibitem[Hohensee et~al.(2015)Hohensee, Wilson, and
  Cahill]{hohensee2015thermal}
Gregory~T Hohensee, RB~Wilson, and David~G Cahill.
\newblock Thermal conductance of metal--diamond interfaces at high pressure.
\newblock \emph{Nature Communications}, 6\penalty0 (1):\penalty0 1--9, 2015.
\newblock \doi{10.1038/ncomms7578}.

\bibitem[Holmes et~al.(2016)Holmes, Tubman, and Umrigar]{selci}
Adam~A. Holmes, Norm~M. Tubman, and C.~J. Umrigar.
\newblock Heat-bath configuration interaction: An efficient selected
  configuration interaction algorithm inspired by heat-bath sampling.
\newblock \emph{Journal of Chemical Theory and Computation}, 12\penalty0
  (8):\penalty0 3674--3680, 2016.
\newblock \doi{10.1021/acs.jctc.6b00407}.

\bibitem[Honerkamp et~al.(2018)Honerkamp, Shinaoka, Assaad, and
  Werner]{Honerkamp18}
Carsten Honerkamp, Hiroshi Shinaoka, Fakher~F. Assaad, and Philipp Werner.
\newblock Limitations of constrained random phase approximation downfolding.
\newblock \emph{Phys. Rev. B}, 98:\penalty0 235151, 2018.
\newblock \doi{10.1103/PhysRevB.98.235151}.
\newblock URL \url{https://link.aps.org/doi/10.1103/PhysRevB.98.235151}.

\bibitem[Hosono and Kuroki(2015)]{hosono2015iron}
Hideo Hosono and Kazuhiko Kuroki.
\newblock Iron-based superconductors: Current status of materials and pairing
  mechanism.
\newblock \emph{Physica C: Superconductivity and its Applications},
  514:\penalty0 399--422, 2015.
\newblock \doi{10.1016/j.physc.2015.02.020}.

\bibitem[Houck et~al.(2012)Houck, T{\"u}reci, and Koch]{houck2012chip}
Andrew~A Houck, Hakan~E T{\"u}reci, and Jens Koch.
\newblock On-chip quantum simulation with superconducting circuits.
\newblock \emph{Nature Physics}, 8\penalty0 (4):\penalty0 292--299, 2012.
\newblock \doi{10.1038/nphys2251}.

\bibitem[Houghton and Marston(1993)]{PhysRevB.48.7790}
A.~Houghton and J.~B. Marston.
\newblock Bosonization and fermion liquids in dimensions greater than one.
\newblock \emph{Phys. Rev. B}, 48:\penalty0 7790--7808, 1993.
\newblock \doi{10.1103/PhysRevB.48.7790}.
\newblock URL \url{https://link.aps.org/doi/10.1103/PhysRevB.48.7790}.

\bibitem[Hruszkewycz et~al.(2017)Hruszkewycz, Allain, Holt, Murray, Holt,
  Fuoss, and Chamard]{hruszkewycz2017high}
SO~Hruszkewycz, M~Allain, MV~Holt, CE~Murray, JR~Holt, PH~Fuoss, and V~Chamard.
\newblock {High-resolution three-dimensional structural microscopy by
  single-angle Bragg ptychography}.
\newblock \emph{Nature Materials}, 16\penalty0 (2):\penalty0 244--251, 2017.
\newblock \doi{10.1038/nmat4798}.

\bibitem[Hsieh et~al.(2008)Hsieh, Qian, Wray, Xia, Hor, Cava, and
  Hasan]{hsieh2008topological}
David Hsieh, Dong Qian, Lewis Wray, Yiman Xia, Yew~San Hor, Robert~Joseph Cava,
  and M~Zahid Hasan.
\newblock {A topological {Dirac} insulator in a quantum spin Hall phase}.
\newblock \emph{Nature}, 452\penalty0 (7190):\penalty0 970--974, 2008.
\newblock \doi{10.1038/nature06843}.

\bibitem[Hsu et~al.(2008)Hsu, Luo, Yeh, Chen, Huang, Wu, Lee, Huang, Chu, Yan,
  and Wu]{Hsu14262}
Fong-Chi Hsu, Jiu-Yong Luo, Kuo-Wei Yeh, Ta-Kun Chen, Tzu-Wen Huang, Phillip~M.
  Wu, Yong-Chi Lee, Yi-Lin Huang, Yan-Yi Chu, Der-Chung Yan, and Maw-Kuen Wu.
\newblock {Superconductivity in the PbO-type structure $\alpha$-FeSe}.
\newblock \emph{Proceedings of the National Academy of Sciences}, 105\penalty0
  (38):\penalty0 14262--14264, 2008.
\newblock \doi{10.1073/pnas.0807325105}.
\newblock URL \url{https://www.pnas.org/content/105/38/14262}.

\bibitem[Huang et~al.(2012)Huang, Pacradouni, Kennett, Komiya, and
  Sonier]{huang2012precision}
W~Huang, V~Pacradouni, MP~Kennett, S~Komiya, and JE~Sonier.
\newblock {Precision search for magnetic order in the pseudogap regime of
  La$_{2-x}$Sr$_x$CuO$_4$ by muon spin relaxation}.
\newblock \emph{Phys. Rev. B}, 85\penalty0 (10):\penalty0 104527, 2012.
\newblock \doi{10.1103/PhysRevB.85.104527}.

\bibitem[Huismans et~al.(2011)Huismans, Rouzee, Gijsbertsen, Jungmann,
  Smolkowska, Logman, Lepine, Cauchy, Zamith, Marchenko, Bakker, Berden,
  Redlich, van~der Meer, Muller, Vermin, Schafer, Spanner, Ivanov, Smirnova,
  Bauer, Popruzhenko, and Vrakking]{huismans2011time}
Y.~Huismans, A.~Rouzee, A.~Gijsbertsen, J.~H. Jungmann, A.~S. Smolkowska, P.~S.
  W.~M. Logman, F.~Lepine, C.~Cauchy, S.~Zamith, T.~Marchenko, J.~M. Bakker,
  G.~Berden, B.~Redlich, A.~F.~G. van~der Meer, H.~G. Muller, W.~Vermin, K.~J.
  Schafer, M.~Spanner, M.~Y. Ivanov, O.~Smirnova, D.~Bauer, S.~V. Popruzhenko,
  and M.~J.~J. Vrakking.
\newblock {Time-Resolved Holography with Photoelectrons}.
\newblock \emph{Science}, 331\penalty0 (6013):\penalty0 61--64, 2011.
\newblock \doi{10.1126/science.1198450}.

\bibitem[Husain et~al.(2019)Husain, Mitrano, Rak, Rubeck, Uchoa, March, Dwyer,
  Schneeloch, Zhong, Gu, et~al.]{husain2019crossover}
Ali~A Husain, Matteo Mitrano, Melinda~S Rak, Samantha Rubeck, Bruno Uchoa,
  Katia March, Christian Dwyer, John Schneeloch, Ruidan Zhong, Genda~D Gu,
  et~al.
\newblock Crossover of charge fluctuations across the strange metal phase
  diagram.
\newblock \emph{Physical Review X}, 9\penalty0 (4):\penalty0 041062, 2019.
\newblock \doi{10.1103/PhysRevX.9.041062}.

\bibitem[Hussey et~al.(2004)Hussey, Takenaka, and
  Takagi]{hussey2004universality}
NE~Hussey, K~Takenaka, and H~Takagi.
\newblock {Universality of the Mott--Ioffe--Regel limit in metals}.
\newblock \emph{Philosophical Magazine}, 84\penalty0 (27):\penalty0 2847--2864,
  2004.
\newblock \doi{10.1038/nature02070}.

\bibitem[Hussey et~al.(2011)Hussey, Cooper, Xu, Wang, Mouzopoulou, Vignolle,
  and Proust]{hussey2011dichotomy}
NE~Hussey, RA~Cooper, Xiaofeng Xu, Y~Wang, I~Mouzopoulou, B~Vignolle, and Cyril
  Proust.
\newblock {Dichotomy in the $T$-linear resistivity in hole-doped cuprates}.
\newblock \emph{Philosophical Transactions of the Royal Society A:
  Mathematical, Physical and Engineering Sciences}, 369\penalty0
  (1941):\penalty0 1626--1639, 2011.
\newblock \doi{10.1098/rsta.2010.0333}.

\bibitem[Ishida et~al.(1998)Ishida, Mukuda, Kitaoka, Asayama, Mao, Mori, and
  Maeno]{ishida1998spin}
K.~Ishida, H.~Mukuda, Y.~Kitaoka, K.~Asayama, Z.~Q. Mao, Y.~Mori, and Y.~Maeno.
\newblock {Spin-triplet superconductivity in Sr$_2$RuO$_4$ identified by
  $^{17}$O Knight shift}.
\newblock \emph{Nature}, 396\penalty0 (6712):\penalty0 658--660, 1998.
\newblock \doi{10.1038/25315}.

\bibitem[Islam et~al.(2015)Islam, Ma, Preiss, Tai, Lukin, Rispoli, and
  Greiner]{islam2015measuring}
Rajibul Islam, Ruichao Ma, Philipp~M Preiss, M~Eric Tai, Alexander Lukin,
  Matthew Rispoli, and Markus Greiner.
\newblock Measuring entanglement entropy in a quantum many-body system.
\newblock \emph{Nature}, 528\penalty0 (7580):\penalty0 77--83, 2015.
\newblock \doi{10.1038/nature15750}.

\bibitem[Issi et~al.(1976)Issi, Michenaud, and Heremans]{heremansbismuth}
J-P Issi, J-P Michenaud, and J~Heremans.
\newblock Electron scattering in compensated bismuth.
\newblock \emph{Phys. Rev. B}, 14\penalty0 (12):\penalty0 5156, 1976.
\newblock \doi{10.1103/PhysRevB.14.5156}.

\bibitem[Itin and Katsnelson(2015)]{PhysRevLett.115.075301}
A.~P. Itin and M.~I. Katsnelson.
\newblock Effective hamiltonians for rapidly driven many-body lattice systems:
  Induced exchange interactions and density-dependent hoppings.
\newblock \emph{Phys. Rev. Lett.}, 115:\penalty0 075301, 2015.
\newblock \doi{10.1103/PhysRevLett.115.075301}.
\newblock URL \url{https://link.aps.org/doi/10.1103/PhysRevLett.115.075301}.

\bibitem[Jackeli and Khaliullin(2009)]{jackeli2009mott}
G~Jackeli and G~Khaliullin.
\newblock {Mott insulators in the strong spin-orbit coupling limit: from
  {Heisenberg} to a quantum compass and Kitaev models}.
\newblock \emph{Phys. Rev. Lett.}, 102\penalty0 (1):\penalty0 017205, 2009.
\newblock \doi{10.1103/PhysRevLett.102.017205}.

\bibitem[Jackson et~al.(2005)Jackson, Malba, Weir, Baker, and
  Vohra]{jackson2005high}
DD~Jackson, V~Malba, ST~Weir, PA~Baker, and YK~Vohra.
\newblock {High-pressure magnetic susceptibility experiments on the heavy
  lanthanides Gd, Tb, Dy, Ho, Er, and Tm}.
\newblock \emph{Phys. Rev. B}, 71\penalty0 (18):\penalty0 184416, 2005.
\newblock \doi{10.1103/PhysRevB.71.184416}.

\bibitem[Jager et~al.(2017)Jager, Ott, Kraus, Kaplan, Pouse, Marvel, Haglund,
  Neumark, and Leone]{jager2017tracking}
Marieke~F. Jager, Christian Ott, Peter~M. Kraus, Christopher~J. Kaplan, Winston
  Pouse, Robert~E. Marvel, Richard~F. Haglund, Daniel~M. Neumark, and
  Stephen~R. Leone.
\newblock {Tracking the insulator-to-metal phase transition in VO$_2$ with
  few-femtosecond extreme UV transient absorption spectroscopy}.
\newblock \emph{Proceedings of the National Academy of Sciences}, 114\penalty0
  (36):\penalty0 9558--9563, 2017.
\newblock \doi{10.1073/pnas.1707602114}.

\bibitem[Jaime et~al.(2012)Jaime, Daou, Crooker, Weickert, Uchida, A.E., C.D.,
  H.A., and B.D.]{Jaime2012}
M.~Jaime, R.~Daou, S.A. Crooker, F.~Weickert, A.~Uchida, Feiguin A.E., Batista
  C.D., Dabkowska H.A., and Gaulin B.D.
\newblock Magnetostriction and magnetic texture to 100.75 tesla in frustrated
  {SrCu$_2$(BO$_3$)$_2$}.
\newblock \emph{Proceedings of the National Academy of Sciences}, 109\penalty0
  (31):\penalty0 12404--12407, 2012.
\newblock \doi{10.1073/pnas.1200743109}.

\bibitem[Jakli\ifmmode~\check{c}\else \v{c}\fi{} and
  Prelov\ifmmode~\check{s}\else \v{s}\fi{}ek(1994)]{Prelovsek}
J.~Jakli\ifmmode~\check{c}\else \v{c}\fi{} and P.~Prelov\ifmmode~\check{s}\else
  \v{s}\fi{}ek.
\newblock Lanczos method for the calculation of finite-temperature quantities
  in correlated systems.
\newblock \emph{Phys. Rev. B}, 49:\penalty0 5065--5068, 1994.
\newblock \doi{10.1103/PhysRevB.49.5065}.
\newblock URL \url{https://link.aps.org/doi/10.1103/PhysRevB.49.5065}.

\bibitem[Jang et~al.(2017)Jang, Yoo, Pfeiffer, West, Baldwin, and
  Ashoori]{jang2017full}
Joonho Jang, Heun~Mo Yoo, LN~Pfeiffer, KW~West, KW~Baldwin, and Raymond~C
  Ashoori.
\newblock Full momentum-and energy-resolved spectral function of a 2{D}
  electronic system.
\newblock \emph{Science}, 358\penalty0 (6365):\penalty0 901--906, 2017.
\newblock \doi{10.1126/science.aam7073}.

\bibitem[Jarrell and Gubernatis(1996)]{JarrellGubernatis}
Mark Jarrell and J.E. Gubernatis.
\newblock Bayesian inference and the analytic continuation of imaginary-time
  quantum {Monte} {Carlo} data.
\newblock \emph{Physics Reports}, 269\penalty0 (3):\penalty0 133 -- 195, 1996.
\newblock \doi{10.1016/0370-1573(95)00074-7}.

\bibitem[J{\'e}rome et~al.(1980)J{\'e}rome, Mazaud, Ribault, and
  Bechgaard]{jerome1980superconductivity}
D~J{\'e}rome, A~Mazaud, M~Ribault, and K~Bechgaard.
\newblock Superconductivity in a synthetic organic conductor (tmtsf)$_2$pf$_6$.
\newblock \emph{Journal de Physique Lettres}, 41\penalty0 (4):\penalty0 95--98,
  1980.
\newblock \doi{10.1051/jphyslet:0198000410409500}.

\bibitem[Jiang et~al.(2014)Jiang, Hu, You, Li, Li, Wang, Mu, Chen, Zhang, Yu,
  et~al.]{jiang2014high}
Da~Jiang, Tao Hu, Lixing You, Qiao Li, Ang Li, Haomin Wang, Gang Mu, Zhiying
  Chen, Haoran Zhang, Guanghui Yu, et~al.
\newblock {High-$T_c$ superconductivity in ultrathin
  Bi$_2$Sr$_2$CaCu$_2$O$_{8+x}$ down to half-unit-cell thickness by protection
  with graphene}.
\newblock \emph{Nature Communications}, 5\penalty0 (1):\penalty0 1--6, 2014.
\newblock \doi{10.1038/ncomms6708}.

\bibitem[Jin et~al.(2015)Jin, Wiendlocha, and Heremans]{bismuthbands}
Hyungyu Jin, Bartlomiej Wiendlocha, and Joseph~P Heremans.
\newblock P-type doping of elemental bismuth with indium, gallium and tin: a
  novel doping mechanism in solids.
\newblock \emph{Energy \& Environmental Science}, 8\penalty0 (7):\penalty0
  2027--2040, 2015.
\newblock \doi{10.1039/C5EE01309G}.

\bibitem[Jin et~al.(2011)Jin, Butch, Kirshenbaum, Paglione, and
  Greene]{kuinature}
K.~Jin, N.~P. Butch, K.~Kirshenbaum, J.~Paglione, and R.~L. Greene.
\newblock Link between spin fluctuations and electron pairing in copper oxide
  superconductors.
\newblock \emph{Nature}, 476\penalty0 (7358):\penalty0 73--75, 2011.
\newblock \doi{10.1038/nature10308}.
\newblock URL \url{https://doi.org/10.1038/nature10308}.

\bibitem[Kalenyuk et~al.(2018)Kalenyuk, Pagliero, Borodianskyi, Kordyuk, and
  Krasnov]{kalenyuk2018phase}
Aleksey~A Kalenyuk, Alessandro Pagliero, Evgenii~A Borodianskyi, AA~Kordyuk,
  and Vladimir~M Krasnov.
\newblock {Phase-Sensitive Evidence for the Sign-Reversal s$\pm$ Symmetry of
  the Order Parameter in an Iron-Pnictide Superconductor Using Nb/Ba$_{1-
  x}$Na$_x$Fe$_2$As$_2$ Josephson Junctions}.
\newblock \emph{Phys. Rev. Lett.}, 120\penalty0 (6):\penalty0 067001, 2018.
\newblock \doi{10.1103/PhysRevLett.120.067001}.

\bibitem[Kamihara et~al.(2008)Kamihara, Watanabe, Hirano, and
  Hosono]{kamihara2008iron}
Yoichi Kamihara, Takumi Watanabe, Masahiro Hirano, and Hideo Hosono.
\newblock {Iron-based layered superconductor La[O$_{1-x}$F$_x$]FeAs ($x$=0.05-
  0.12) with $T_c$= 26 K}.
\newblock \emph{Journal of the American Chemical Society}, 130\penalty0
  (11):\penalty0 3296--3297, 2008.
\newblock \doi{10.1021/ja800073m}.

\bibitem[Kaminski et~al.(2016)Kaminski, Rosenkranz, Norman, Randeria, Li,
  Raffy, and Campuzano]{kaminski2016arpes}
A.~Kaminski, S.~Rosenkranz, M.~R. Norman, M.~Randeria, Z.~Z. Li, H.~Raffy, and
  J.~C. Campuzano.
\newblock {Destroying Coherence in High-Temperature Superconductors with
  Current Flow}.
\newblock \emph{Physical Review X}, 6\penalty0 (3):\penalty0 031040, 2016.
\newblock \doi{10.1103/PhysRevX.6.031040}.

\bibitem[Kandelaki and Rudner(2018)]{kandelaki2018}
Ervand Kandelaki and Mark~S. Rudner.
\newblock Many-body dynamics and gap opening in interacting periodically driven
  systems.
\newblock \emph{Phys. Rev. Lett.}, 121:\penalty0 036801, 2018.
\newblock \doi{10.1103/PhysRevLett.121.036801}.
\newblock URL \url{https://link.aps.org/doi/10.1103/PhysRevLett.121.036801}.

\bibitem[Kang and Vafek(2018)]{PhysRevX.8.031088}
Jian Kang and Oskar Vafek.
\newblock Symmetry, maximally localized {Wannier} states, and a low-energy
  model for twisted bilayer graphene narrow bands.
\newblock \emph{Phys. Rev. X}, 8:\penalty0 031088, 2018.
\newblock \doi{10.1103/PhysRevX.8.031088}.
\newblock URL \url{https://link.aps.org/doi/10.1103/PhysRevX.8.031088}.

\bibitem[Kang and Vafek(2019)]{PhysRevLett.122.246401}
Jian Kang and Oskar Vafek.
\newblock Strong coupling phases of partially filled twisted bilayer graphene
  narrow bands.
\newblock \emph{Phys. Rev. Lett.}, 122:\penalty0 246401, 2019.
\newblock \doi{10.1103/PhysRevLett.122.246401}.
\newblock URL \url{https://link.aps.org/doi/10.1103/PhysRevLett.122.246401}.

\bibitem[Kapitulnik et~al.(2019)Kapitulnik, Kivelson, and
  Spivak]{kapitulnik2019colloquium}
Aharon Kapitulnik, Steven~A Kivelson, and Boris Spivak.
\newblock Colloquium: anomalous metals: failed superconductors.
\newblock \emph{Reviews of Modern Physics}, 91\penalty0 (1):\penalty0 011002,
  2019.
\newblock \doi{10.1103/RevModPhys.91.011002}.

\bibitem[Kar et~al.(2003)Kar, Raychaudhuri, Ghosh, L\"ohneysen, and
  Weiss]{Si-MIT3}
Swastik Kar, A.~K. Raychaudhuri, Arindam Ghosh, H.~v. L\"ohneysen, and
  G.~Weiss.
\newblock {Observation of Non-Gaussian Conductance Fluctuations at Low
  Temperatures in Si:P(B) at the Metal-Insulator Transition}.
\newblock \emph{Phys. Rev. Lett.}, 91:\penalty0 216603, 2003.
\newblock \doi{10.1103/PhysRevLett.91.216603}.
\newblock URL \url{https://link.aps.org/doi/10.1103/PhysRevLett.91.216603}.

\bibitem[Kasahara et~al.(2010)Kasahara, Shibauchi, Hashimoto, Ikada, Tonegawa,
  Okazaki, Shishido, Ikeda, Takeya, Hirata, et~al.]{kasahara2010evolution}
S~Kasahara, T~Shibauchi, K~Hashimoto, K~Ikada, S~Tonegawa, Ryuji Okazaki,
  H~Shishido, H~Ikeda, H~Takeya, K~Hirata, et~al.
\newblock {Evolution from non-Fermi-to Fermi-liquid transport via isovalent
  doping in BaFe$_2$(As$_{1-x}$P$_x$)$_2$ superconductors}.
\newblock \emph{Phys. Rev. B}, 81\penalty0 (18):\penalty0 184519, 2010.
\newblock \doi{10.1103/PhysRevB.81.184519}.

\bibitem[Kasahara et~al.(2018)Kasahara, Ohnishi, Mizukami, Tanaka, Ma, Sugii,
  Kurita, Tanaka, Nasu, Motome, et~al.]{kasahara2018majorana}
Y~Kasahara, T~Ohnishi, Y~Mizukami, O~Tanaka, Sixiao Ma, K~Sugii, N~Kurita,
  H~Tanaka, J~Nasu, Y~Motome, et~al.
\newblock {Majorana quantization and half-integer thermal quantum Hall effect
  in a Kitaev spin liquid}.
\newblock \emph{Nature}, 559\penalty0 (7713):\penalty0 227--231, 2018.
\newblock \doi{10.1038/s41586-018-0274-0}.

\bibitem[Katsumi et~al.(2018)Katsumi, Tsuji, Hamada, Matsunaga, Schneeloch,
  Zhong, Gu, Aoki, Gallais, and Shimano]{katsumi2018higgs}
Kota Katsumi, Naoto Tsuji, Yuki~I Hamada, Ryusuke Matsunaga, John Schneeloch,
  Ruidan~D Zhong, Genda~D Gu, Hideo Aoki, Yann Gallais, and Ryo Shimano.
\newblock {Higgs Mode in the $d$-Wave Superconductor
  {Bi}$_2${Sr}$_2${CaCu}$_2${O}$_{8+x}$ Driven by an Intense Terahertz Pulse}.
\newblock \emph{Phys. Rev. Lett.}, 120\penalty0 (11):\penalty0 117001, 2018.
\newblock \doi{10.1103/PhysRevLett.120.117001}.

\bibitem[Katsura et~al.(2010)Katsura, Nagaosa, and Lee]{Katsura_2010}
Hosho Katsura, Naoto Nagaosa, and Patrick~A. Lee.
\newblock Theory of the thermal hall effect in quantum magnets.
\newblock \emph{Phys. Rev. Lett.}, 104:\penalty0 066403, 2010.
\newblock \doi{10.1103/PhysRevLett.104.066403}.
\newblock URL \url{https://link.aps.org/doi/10.1103/PhysRevLett.104.066403}.

\bibitem[Kaul et~al.(2013)Kaul, Melko, and Sandvik]{kaul_sandvik}
Ribhu~K. Kaul, Roger~G. Melko, and Anders~W. Sandvik.
\newblock Bridging lattice-scale physics and continuum field theory with
  quantum {Monte} {Carlo} simulations.
\newblock \emph{Annual Review of Condensed Matter Physics}, 4\penalty0
  (1):\penalty0 179--215, 2013.
\newblock \doi{10.1146/annurev-conmatphys-030212-184215}.
\newblock URL \url{, https://doi.org/10.1146/annurev-conmatphys-030212-184215}.

\bibitem[Kawamura and Uematsu(2019)]{Kawamura_2019}
Hikaru Kawamura and Kazuki Uematsu.
\newblock Nature of the randomness-induced quantum spin liquids in two
  dimensions.
\newblock \emph{Journal of Physics: Condensed Matter}, 31\penalty0
  (50):\penalty0 504003, 2019.
\newblock \doi{10.1088/1361-648x/ab400c}.
\newblock URL \url{https://doi.org/10.1088%2F1361-648x%2Fab400c}.

\bibitem[Kayyalha et~al.(2020)Kayyalha, Xiao, Zhang, Shin, Jiang, Wang, Zhao,
  Xiao, Zhang, Fijalkowski, Mandal, Winnerlein, Gould, Li, Molenkamp, Chan,
  Samarth, and Chang]{kayyalha2020absence}
Morteza Kayyalha, Di~Xiao, Ruoxi Zhang, Jaeho Shin, Jue Jiang, Fei Wang, Yi-Fan
  Zhao, Run Xiao, Ling Zhang, Kajetan~M. Fijalkowski, Pankaj Mandal, Martin
  Winnerlein, Charles Gould, Qi~Li, Laurens~W. Molenkamp, Moses H.~W. Chan,
  Nitin Samarth, and Cui-Zu Chang.
\newblock {Absence of evidence for chiral Majorana modes in quantum anomalous
  Hall-superconductor devices}.
\newblock \emph{Science}, 367\penalty0 (6473):\penalty0 64--67, 2020.
\newblock \doi{10.1126/science.aax6361}.

\bibitem[Kehrein(2007)]{kehrein2007flow}
Stefan Kehrein.
\newblock \emph{The Flow Equation Approach to Many-Particle Systems}, volume
  217.
\newblock Springer, 2007.
\newblock \doi{10.1007/3-540-34068-8}.

\bibitem[Keimer et~al.(2015)Keimer, Kivelson, Norman, Uchida, and
  Zaanen]{keimer2015quantum}
Bernhard Keimer, Steven~A Kivelson, Michael~R Norman, Shinichi Uchida, and
  J~Zaanen.
\newblock From quantum matter to high-temperature superconductivity in copper
  oxides.
\newblock \emph{Nature}, 518\penalty0 (7538):\penalty0 179--186, 2015.
\newblock \doi{10.1038/nature14165}.

\bibitem[Kent and Kotliar(2018)]{Kent_Kotliar}
Paul R.~C. Kent and Gabriel Kotliar.
\newblock Toward a predictive theory of correlated materials.
\newblock \emph{Science}, 361\penalty0 (6400):\penalty0 348--354, 2018.
\newblock \doi{10.1126/science.aat5975}.
\newblock URL \url{https://science.sciencemag.org/content/361/6400/348}.

\bibitem[Khait et~al.(2018)Khait, Azaria, Hubig, Schollw{\"o}ck, and
  Auerbach]{khait2018doped}
Ilia Khait, Patrick Azaria, Claudius Hubig, Ulrich Schollw{\"o}ck, and Assa
  Auerbach.
\newblock {Doped {Kondo} chain, a heavy Luttinger liquid}.
\newblock \emph{Proceedings of the National Academy of Sciences}, 115\penalty0
  (20):\penalty0 5140--5144, 2018.
\newblock \doi{10.1073/pnas.1719374115}.

\bibitem[Kim et~al.(2018{\natexlab{a}})Kim, Souliou, Barber, Lefran{\c{c}}ois,
  Minola, Tortora, Heid, Nandi, Borzi, Garbarino, et~al.]{kim2018uniaxial}
H-H Kim, SM~Souliou, ME~Barber, E~Lefran{\c{c}}ois, M~Minola, M~Tortora,
  R~Heid, N~Nandi, Rodolfo~Alberto Borzi, G~Garbarino, et~al.
\newblock Uniaxial pressure control of competing orders in a high-temperature
  superconductor.
\newblock \emph{Science}, 362\penalty0 (6418):\penalty0 1040--1044,
  2018{\natexlab{a}}.
\newblock \doi{10.1126/science.aat4708}.

\bibitem[Kim et~al.(2012)Kim, Casa, Upton, Gog, Kim, Mitchell, Van~Veenendaal,
  Daghofer, van Den~Brink, Khaliullin, et~al.]{kim2012magnetic}
Jungho Kim, D~Casa, MH~Upton, T~Gog, Young-June Kim, JF~Mitchell,
  M~Van~Veenendaal, M~Daghofer, J~van Den~Brink, G~Khaliullin, et~al.
\newblock {Magnetic excitation spectra of Sr$_2$IrO$_4$ probed by resonant
  inelastic X-ray scattering: establishing links to cuprate superconductors}.
\newblock \emph{Phys. Rev. Lett.}, 108\penalty0 (17):\penalty0 177003, 2012.
\newblock \doi{10.1103/PhysRevLett.108.177003}.

\bibitem[Kim et~al.(2018{\natexlab{b}})Kim, Casa, Said, Krakora, Kim, Kasman,
  Huang, and Gog]{kim2018quartz}
Jungho Kim, D~Casa, Ayman Said, Rich Krakora, Bum~Jun Kim, Elina Kasman,
  Xianrong Huang, and T~Gog.
\newblock {Quartz-based flat-crystal resonant inelastic x-ray scattering
  spectrometer with sub-10 meV energy resolution}.
\newblock \emph{Scientific Reports}, 8\penalty0 (1):\penalty0 1--9,
  2018{\natexlab{b}}.
\newblock \doi{10.1038/s41598-018-20396-z}.

\bibitem[Kim et~al.(2016)Kim, Yankowitz, Fallahazad, Kang, Movva, Huang,
  Larentis, Corbet, Taniguchi, Watanabe, et~al.]{kim2016van}
Kyounghwan Kim, Matthew Yankowitz, Babak Fallahazad, Sangwoo Kang, Hema~CP
  Movva, Shengqiang Huang, Stefano Larentis, Chris~M Corbet, Takashi Taniguchi,
  Kenji Watanabe, et~al.
\newblock {van der Waals heterostructures with high accuracy rotational
  alignment}.
\newblock \emph{Nano letters}, 16\penalty0 (3):\penalty0 1989--1995, 2016.
\newblock \doi{10.1021/acs.nanolett.5b05263}.

\bibitem[Kim et~al.(2013)Kim, Poumirol, Lombardo, Kalugin, Georgiou, Kim,
  Novoselov, Ferrari, Kono, Kashuba, et~al.]{kim2013measurement}
Y~Kim, JM~Poumirol, Antonio Lombardo, NG~Kalugin, T~Georgiou, YJ~Kim,
  KS~Novoselov, AC~Ferrari, J~Kono, O~Kashuba, et~al.
\newblock Measurement of filling-factor-dependent magnetophonon resonances in
  graphene using raman spectroscopy.
\newblock \emph{Phys. Rev. Lett.}, 110\penalty0 (22):\penalty0 227402, 2013.
\newblock \doi{10.1103/PhysRevLett.110.227402}.

\bibitem[Kimura and Okamura(2012)]{kimura2012infrared}
Shin-ichi Kimura and Hidekazu Okamura.
\newblock Infrared and terahertz spectroscopy of strongly correlated electron
  systems under extreme conditions.
\newblock \emph{Journal of the Physical Society of Japan}, 82\penalty0
  (2):\penalty0 021004, 2012.
\newblock \doi{10.7566/JPSJ.82.021004}.

\bibitem[Kissikov et~al.(2018)Kissikov, Sarkar, Lawson, {BCS}~h, Timmons,
  Tanatar, Prozorov, Bud’ko, Canfield, Fernandes,
  et~al.]{kissikov2018uniaxial}
T~Kissikov, R~Sarkar, M~Lawson, BT~{BCS}~h, Erik~I Timmons, Makariy~A Tanatar,
  Ruslan Prozorov, SL~Bud’ko, Paul~C Canfield, Rafael~M Fernandes, et~al.
\newblock \emph{Nature Communications}, 9\penalty0 (1):\penalty0 1--6, 2018.
\newblock \doi{10.1038/s41467-018-03377-8}.

\bibitem[Kitaev(2009)]{kitaev2009aip}
A~Kitaev.
\newblock Periodic table for topological insulators and superconductors.
\newblock \emph{AIP Conference Proceedings}, 1134, 2009.
\newblock \doi{10.1063/1.3149495}.

\bibitem[Kitaev(2006)]{kitaev2006anyons}
Alexei Kitaev.
\newblock Anyons in an exactly solved model and beyond.
\newblock \emph{Annals of Physics}, 321\penalty0 (1):\penalty0 2--111, 2006.
\newblock \doi{10.1016/j.aop.2005.10.005}.

\bibitem[Kitaev(2015{\natexlab{a}})]{Kitaev1}
Alexei Kitaev.
\newblock A simple model of quantum holography (part 1).
\newblock \url{http://online.kitp.ucsb.edu/online/entangled15/kitaev/},
  2015{\natexlab{a}}.

\bibitem[Kitaev(2015{\natexlab{b}})]{Kitaev2}
Alexei Kitaev.
\newblock A simple model of quantum holography (part 2).
\newblock \url{http://online.kitp.ucsb.edu/online/entangled15/kitaev2/},
  2015{\natexlab{b}}.

\bibitem[Kitaev and Preskill(2006)]{KitaevPreskill2006}
Alexei Kitaev and John Preskill.
\newblock Topological entanglement entropy.
\newblock \emph{Phys. Rev. Lett.}, 96:\penalty0 110404, 2006.
\newblock \doi{10.1103/PhysRevLett.96.110404}.
\newblock URL \url{https://link.aps.org/doi/10.1103/PhysRevLett.96.110404}.

\bibitem[Klein et~al.(2018)Klein, Walsh, Clarke, Guo, Bi, Fabbris, Meng,
  Haskel, Alp, {Van Duyne}, Jacobsen, and Freedman]{Klein2018}
Ryan~A Klein, James P.~S. Walsh, Samantha~M Clarke, Yinsheng Guo, Wenli Bi,
  Gilberto Fabbris, Yue Meng, Daniel Haskel, E~Ercan Alp, Richard~P. {Van
  Duyne}, Steven~D Jacobsen, and Danna~E Freedman.
\newblock {Impact of Pressure on Magnetic Order in Jarosite}.
\newblock \emph{Journal of the American Chemical Society}, 140\penalty0
  (38):\penalty0 12001--12009, 2018.
\newblock \doi{10.1021/jacs.8b05601}.
\newblock URL \url{https://pubs.acs.org/doi/abs/10.1021/jacs.8b05601
  https://pubs.acs.org/doi/10.1021/jacs.8b05601}.

\bibitem[Klich and Levitov(2009)]{klich2009quantum}
Israel Klich and Leonid Levitov.
\newblock Quantum noise as an entanglement meter.
\newblock \emph{Phys. Rev. Lett.}, 102\penalty0 (10):\penalty0 100502, 2009.
\newblock \doi{10.1103/PhysRevLett.102.100502}.

\bibitem[Knafo et~al.(2020)Knafo, Araki, Lapertot, Aoki, Knebel, and
  Braithwaite]{knafo2020destabilization}
W~Knafo, S~Araki, G~Lapertot, D~Aoki, G~Knebel, and D~Braithwaite.
\newblock Destabilization of hidden order in {URu$_2$Si$_2$} under magnetic
  field and pressure.
\newblock \emph{Nature Physics}, pages 1--7, 2020.
\newblock \doi{10.1038/s41567-020-0927-4}.

\bibitem[Knizia and Chan(2012)]{knizia_chan}
Gerald Knizia and Garnet Kin-Lic Chan.
\newblock Density matrix embedding: A simple alternative to dynamical
  mean-field theory.
\newblock \emph{Phys. Rev. Lett.}, 109:\penalty0 186404, 2012.
\newblock \doi{10.1103/PhysRevLett.109.186404}.
\newblock URL \url{https://link.aps.org/doi/10.1103/PhysRevLett.109.186404}.

\bibitem[Knolle and Moessner(2019)]{knolle2019field}
Johannes Knolle and Roderich Moessner.
\newblock A field guide to spin liquids.
\newblock \emph{Annual Review of Condensed Matter Physics}, 10:\penalty0
  451--472, 2019.
\newblock \doi{10.1146/annurev-conmatphys-031218-013401}.

\bibitem[{Kochkov} and {Clark}(2018)]{Kochkov2018}
Dmitrii {Kochkov} and Bryan~K. {Clark}.
\newblock {Variational optimization in the AI era: Computational Graph States
  and Supervised Wave-function Optimization}.
\newblock \emph{arXiv}, page arXiv:1811.12423, 2018.
\newblock \doi{10.48550/arXiv.1811.12423}.

\bibitem[Kogar et~al.(2020)Kogar, Zong, Dolgirev, Shen, Straquadine, Bie, Wang,
  Rohwer, Tung, Yang, Li, Yang, Weathersby, Park, Kozina, Sie, Wen,
  Jarillo-Herrero, Fisher, Wang, and Gedik]{kogar2020light}
Anshul Kogar, Alfred Zong, Pavel~E. Dolgirev, Xiaozhe Shen, Joshua Straquadine,
  Ya-Qing Bie, Xirui Wang, Timm Rohwer, I-Cheng Tung, Yafang Yang, Renkai Li,
  Jie Yang, Stephen Weathersby, Suji Park, Michael~E. Kozina, Edbert~J. Sie,
  Haidan Wen, Pablo Jarillo-Herrero, Ian~R. Fisher, Xijie Wang, and Nuh Gedik.
\newblock {Light-induced charge density wave in LaTe$_3$}.
\newblock \emph{Nature Physics}, 16\penalty0 (2):\penalty0 159--163, 2020.
\newblock \doi{10.1038/s41567-019-0705-3}.

\bibitem[Kohn and Sham(1965)]{kohnsham}
W.~Kohn and L.~J. Sham.
\newblock Self-consistent equations including exchange and correlation effects.
\newblock \emph{Phys. Rev.}, 140:\penalty0 A1133--A1138, 1965.
\newblock \doi{10.1103/PhysRev.140.A1133}.

\bibitem[Komijani and Coleman(2019)]{komijani.2019}
Yashar Komijani and Piers Coleman.
\newblock Emergent critical charge fluctuations at the {Kondo} breakdown of
  heavy fermions.
\newblock \emph{Phys. Rev. Lett.}, 122:\penalty0 217001, 2019.
\newblock \doi{10.1103/PhysRevLett.122.217001}.
\newblock URL \url{https://link.aps.org/doi/10.1103/PhysRevLett.122.217001}.

\bibitem[Koonce et~al.(1967)Koonce, Cohen, Schooley, Hosler, and
  Pfeiffer]{STO-SC1}
C.~S. Koonce, Marvin~L. Cohen, J.~F. Schooley, W.~R. Hosler, and E.~R.
  Pfeiffer.
\newblock {Superconducting Transition Temperatures of Semiconducting
  SrTi${\mathrm{O}}_{3}$}.
\newblock \emph{Phys. Rev.}, 163:\penalty0 380--390, 1967.
\newblock \doi{10.1103/PhysRev.163.380}.
\newblock URL \url{https://link.aps.org/doi/10.1103/PhysRev.163.380}.

\bibitem[Kopietz(2008)]{kopietz2006bosonization}
Peter Kopietz.
\newblock \emph{Bosonization of Interacting Fermions in Arbitrary Dimensions},
  volume~48.
\newblock Springer Science \& {BCS} iness Media, 2008.
\newblock \doi{10.1007/978-3-540-68495-4}.

\bibitem[Kotliar et~al.(2006)Kotliar, Savrasov, Haule, Oudovenko, Parcollet,
  and Marianetti]{Kotliar_DFTDMFT}
G.~Kotliar, S.~Y. Savrasov, K.~Haule, V.~S. Oudovenko, O.~Parcollet, and C.~A.
  Marianetti.
\newblock Electronic structure calculations with dynamical mean-field theory.
\newblock \emph{Rev. Mod. Phys.}, 78:\penalty0 865--951, 2006.
\newblock \doi{10.1103/RevModPhys.78.865}.

\bibitem[Kozii et~al.(2016)Kozii, Venderbos, and Fu]{Kozii16}
Vladyslav Kozii, J{\"o}rn~WF Venderbos, and Liang Fu.
\newblock Three-dimensional majorana fermions in chiral superconductors.
\newblock \emph{Science Advances}, 2\penalty0 (12):\penalty0 e1601835, 2016.
\newblock \doi{10.1126/sciadv.1601835}.

\bibitem[Kozina et~al.(2019)Kozina, Fechner, Marsik, van Driel, Glownia,
  Bernhard, Radovic, Zhu, Bonetti, Staub, and Hoffmann]{kozina2019sto}
M.~Kozina, M.~Fechner, P.~Marsik, T.~van Driel, J.~M. Glownia, C.~Bernhard,
  M.~Radovic, D.~Zhu, S.~Bonetti, U.~Staub, and M.~C. Hoffmann.
\newblock {Terahertz-driven phonon upconversion in SrTiO$_3$}.
\newblock \emph{Nature Physics}, 15\penalty0 (4):\penalty0 387--392, 2019.
\newblock \doi{10.1038/s41567-018-0408-1}.

\bibitem[Kravchenko and Sarachik(2003)]{MIT-expreview}
SV~Kravchenko and MP~Sarachik.
\newblock Metal--insulator transition in two-dimensional electron systems.
\newblock \emph{Reports on Progress in Physics}, 67\penalty0 (1):\penalty0 1,
  2003.
\newblock \doi{10.1088/0034-4885/67/1/R01}.

\bibitem[Kremen et~al.(2018)Kremen, Khan, Loh, Baturina, Trivedi, Frydman, and
  Kalisky]{kremen2018imaging}
A~Kremen, H~Khan, YL~Loh, TI~Baturina, N~Trivedi, A~Frydman, and B~Kalisky.
\newblock Imaging quantum fluctuations near criticality.
\newblock \emph{Nature physics}, 14\penalty0 (12):\penalty0 1205--1210, 2018.
\newblock \doi{10.1038/s41567-018-0264-z}.

\bibitem[Ku et~al.(2020)Ku, Zhou, Li, Shin, Shi, Burch, Anderson, Pierce, Xie,
  Hamo, Vool, Zhang, Casola, Taniguchi, Watanabe, Fogler, Kim, Yacoby, and
  Walsworth]{ku2020imaging}
Mark J.~H. Ku, Tony~X. Zhou, Qing Li, Young~J. Shin, Jing~K. Shi, Claire Burch,
  Laurel~E. Anderson, Andrew~T. Pierce, Yonglong Xie, Assaf Hamo, Uri Vool,
  Huiliang Zhang, Francesco Casola, Takashi Taniguchi, Kenji Watanabe,
  Michael~M. Fogler, Philip Kim, Amir Yacoby, and Ronald~L. Walsworth.
\newblock {Imaging viscous flow of the {Dirac} fluid in graphene}.
\newblock \emph{Nature}, 583\penalty0 (7817):\penalty0 537--541, 2020.
\newblock \doi{10.1038/s41586-020-2507-2}.

\bibitem[K{\"u}bler(2017)]{itinerantmagnetismbook}
J{\"u}rgen K{\"u}bler.
\newblock \emph{Theory of Itinerant Electron Magnetism}, volume 106.
\newblock Oxford University Press, 2017.
\newblock \doi{10.1093/oso/9780192895639.001.0001}.

\bibitem[Kukreja et~al.(2018)Kukreja, Hua, Ruby, Barbour, Hu, Mazzoli, Wilkins,
  Fullerton, and Shpyrko]{magnetite3}
Roopali Kukreja, Nelson Hua, Joshua Ruby, Andi Barbour, Wen Hu, Claudio
  Mazzoli, Stuart Wilkins, Eric~E. Fullerton, and Oleg~G. Shpyrko.
\newblock {Orbital Domain Dynamics in Magnetite below the {Verwey} Transition}.
\newblock \emph{Phys. Rev. Lett.}, 121:\penalty0 177601, 2018.
\newblock \doi{10.1103/PhysRevLett.121.177601}.
\newblock URL \url{https://link.aps.org/doi/10.1103/PhysRevLett.121.177601}.

\bibitem[Kung et~al.(2015)Kung, Baumbach, Bauer, Thorsm{\o}lle, Zhang, Haule,
  Mydosh, and Blumberg]{kung2015chirality}
H-H Kung, RE~Baumbach, ED~Bauer, VK~Thorsm{\o}lle, W-L Zhang, Kristjan Haule,
  JA~Mydosh, and Girsh Blumberg.
\newblock {Chirality density wave of the “hidden order” phase in
  URu$_2$Si$_2$}.
\newblock \emph{Science}, 347\penalty0 (6228):\penalty0 1339--1342, 2015.
\newblock \doi{10.1103/PhysRevB.91.155106}.

\bibitem[Kunisada et~al.(2020)Kunisada, Isono, Kohama, Sakai, Bareille,
  Sakuragi, Noguchi, Kurokawa, Kuroda, Ishida, et~al.]{kunisada2020observation}
So~Kunisada, Shunsuke Isono, Yoshimitsu Kohama, Shiro Sakai, C{\'e}dric
  Bareille, Shunsuke Sakuragi, Ryo Noguchi, Kifu Kurokawa, Kenta Kuroda,
  Yukiaki Ishida, et~al.
\newblock Observation of small {Fermi} pockets protected by clean {CuO$_2$}
  sheets of a high-$t_c$ superconductor.
\newblock \emph{Science}, 369\penalty0 (6505):\penalty0 833--838, 2020.
\newblock \doi{10.1126/science.aay7311}.

\bibitem[Kurosaki et~al.(2005)Kurosaki, Shimizu, Miyagawa, Kanoda, and
  Saito]{kurosaki2005mott}
Y~Kurosaki, Y~Shimizu, K~Miyagawa, K~Kanoda, and G~Saito.
\newblock Mott transition from a spin liquid to a {Fermi} liquid in the
  spin-frustrated organic conductor {$\kappa$-(ET)$_2$Cu$_2$(CN)$_3$}.
\newblock \emph{Phys. Rev. Lett.}, 95\penalty0 (17):\penalty0 177001, 2005.
\newblock \doi{10.1103/PhysRevLett.95.177001}.

\bibitem[Laanait et~al.(2019)Laanait, Romero, Yin, Young, Treichler,
  Starchenko, Borisevich, Sergeev, and Matheson]{laanait2019exascale}
Nouamane Laanait, Joshua Romero, Junqi Yin, M~Todd Young, Sean Treichler,
  Vitalii Starchenko, Albina Borisevich, Alex Sergeev, and Michael Matheson.
\newblock Exascale deep learning for scientific inverse problems.
\newblock \emph{arXiv}, page arXiv:1909.11150, 2019.
\newblock \doi{10.48550/arXiv.1909.11150}.

\bibitem[Laflorencie(2016)]{laflorencie2016quantum}
Nicolas Laflorencie.
\newblock Quantum entanglement in condensed matter systems.
\newblock \emph{Physics Reports}, 646:\penalty0 1--59, 2016.
\newblock \doi{10.1016/j.physrep.2016.06.008}.

\bibitem[Lai et~al.(2018)Lai, Grefe, Paschen, and Si]{lai2018weyl}
Hsin-Hua Lai, Sarah~E Grefe, Silke Paschen, and Qimiao Si.
\newblock {Weyl--Kondo semimetal in heavy-fermion systems}.
\newblock \emph{Proceedings of the National Academy of Sciences}, 115\penalty0
  (1):\penalty0 93--97, 2018.
\newblock \doi{10.1073/pnas.1715851115}.

\bibitem[Laughlin(1999)]{laughlin2000fractional}
Robert~B Laughlin.
\newblock Nobel lecture: Fractional quantization.
\newblock \emph{Reviews of Modern Physics}, 71\penalty0 (4):\penalty0 863,
  1999.
\newblock \doi{10.1103/RevModPhys.71.863}.

\bibitem[Laurell and Okamoto(2020)]{laurell2020dynamical}
Pontus Laurell and Satoshi Okamoto.
\newblock {Dynamical and thermal magnetic properties of the Kitaev spin liquid
  candidate $\alpha$-RuCl$_3$}.
\newblock \emph{npj Quantum Materials}, 5\penalty0 (1):\penalty0 1--10, 2020.
\newblock \doi{10.1038/s41535-019-0203-y}.

\bibitem[Laurell et~al.(2021)Laurell, Scheie, Mukherjee, Koza, Enderle,
  Tylczynski, Okamoto, Coldea, Tennant, and Alvarez]{laurell2020dynamics}
Pontus Laurell, Allen Scheie, Chiron~J. Mukherjee, Michael~M. Koza, Mechtild
  Enderle, Zbigniew Tylczynski, Satoshi Okamoto, Radu Coldea, D.~Alan Tennant,
  and Gonzalo Alvarez.
\newblock {Quantifying and Controlling Entanglement in the Quantum Magnet
  ${\mathrm{Cs}}_{2}{\mathrm{CoCl}}_{4}$}.
\newblock \emph{Phys. Rev. Lett.}, 127:\penalty0 037201, 2021.
\newblock \doi{10.1103/PhysRevLett.127.037201}.

\bibitem[Laurita et~al.(2016)Laurita, Morris, Koohpayeh, Rosa, Phelan, Fisk,
  McQueen, and Armitage]{laurita2016anomalous}
NJ~Laurita, CM~Morris, SM~Koohpayeh, PFS Rosa, WA~Phelan, Z~Fisk, TM~McQueen,
  and NP~Armitage.
\newblock {Anomalous three-dimensional bulk ac conduction within the {Kondo}
  gap of SmB$_6$ single crystals}.
\newblock \emph{Phys. Rev. B}, 94\penalty0 (16):\penalty0 165154, 2016.
\newblock \doi{10.1103/PhysRevB.94.165154}.

\bibitem[LeBlanc et~al.(2015)LeBlanc, Antipov, Becca, Bulik, Chan, Chung, Deng,
  Ferrero, Henderson, Jim\'enez-Hoyos, Kozik, Liu, Millis, Prokof'ev, Qin,
  Scuseria, Shi, Svistunov, Tocchio, Tupitsyn, White, Zhang, Zheng, Zhu, and
  Gull]{leblanc15}
J.~P.~F. LeBlanc, Andrey~E. Antipov, Federico Becca, Ireneusz~W. Bulik, Garnet
  Kin-Lic Chan, Chia-Min Chung, Youjin Deng, Michel Ferrero, Thomas~M.
  Henderson, Carlos~A. Jim\'enez-Hoyos, E.~Kozik, Xuan-Wen Liu, Andrew~J.
  Millis, N.~V. Prokof'ev, Mingpu Qin, Gustavo~E. Scuseria, Hao Shi, B.~V.
  Svistunov, Luca~F. Tocchio, I.~S. Tupitsyn, Steven~R. White, Shiwei Zhang,
  Bo-Xiao Zheng, Zhenyue Zhu, and Emanuel Gull.
\newblock Solutions of the two-dimensional hubbard model: Benchmarks and
  results from a wide range of numerical algorithms.
\newblock \emph{Phys. Rev. X}, 5:\penalty0 041041, 2015.
\newblock \doi{10.1103/PhysRevX.5.041041}.
\newblock URL \url{https://link.aps.org/doi/10.1103/PhysRevX.5.041041}.

\bibitem[LeCun et~al.(2015)LeCun, Bengio, and Hinton]{LeCun2015}
Y.~LeCun, Y.~Bengio, and G.~Hinton.
\newblock {Deep Learning}.
\newblock \emph{Nature}, 521:\penalty0 436, 2015.
\newblock \doi{10.1038/nature14539}.

\bibitem[Lee et~al.(2018)Lee, Wang, Zaletel, Vishwanath, and He]{LeeQED2018}
Jong~Yeon Lee, Chong Wang, Michael~P. Zaletel, Ashvin Vishwanath, and Yin-Chen
  He.
\newblock {Emergent Multi-Flavor ${\mathrm{QED}}_{3}$ at the Plateau Transition
  between Fractional Chern Insulators: Applications to Graphene
  Heterostructures}.
\newblock \emph{Phys. Rev. X}, 8:\penalty0 031015, 2018.
\newblock \doi{10.1103/PhysRevX.8.031015}.
\newblock URL \url{https://link.aps.org/doi/10.1103/PhysRevX.8.031015}.

\bibitem[Lee et~al.(2019{\natexlab{a}})Lee, You, Sachdev, and
  Vishwanath]{LeeShastry2019}
Jong~Yeon Lee, Yi-Zhuang You, Subir Sachdev, and Ashvin Vishwanath.
\newblock {Signatures of a Deconfined Phase Transition on the
  Shastry-Sutherland Lattice: Applications to Quantum Critical
  ${\mathrm{SrCu}}_{2}({\mathrm{BO}}_{3}{)}_{2}$}.
\newblock \emph{Phys. Rev. X}, 9:\penalty0 041037, 2019{\natexlab{a}}.
\newblock \doi{10.1103/PhysRevX.9.041037}.
\newblock URL \url{https://link.aps.org/doi/10.1103/PhysRevX.9.041037}.

\bibitem[Lee et~al.(2019{\natexlab{b}})Lee, You, Sachdev, and
  Vishwanath]{lee2019signatures}
Jong~Yeon Lee, Yi-Zhuang You, Subir Sachdev, and Ashvin Vishwanath.
\newblock Signatures of a deconfined phase transition on the shastry-sutherland
  lattice: Applications to quantum critical srcu 2 (bo 3) 2.
\newblock \emph{Physical Review X}, 9\penalty0 (4):\penalty0 041037,
  2019{\natexlab{b}}.
\newblock \doi{10.1103/PhysRevX.9.041037}.

\bibitem[Lee and Ramakrishnan(1985)]{MIT-theoryreview1}
Patrick~A. Lee and T.~V. Ramakrishnan.
\newblock Disordered electronic systems.
\newblock \emph{Rev. Mod. Phys.}, 57:\penalty0 287--337, 1985.
\newblock \doi{10.1103/RevModPhys.57.287}.
\newblock URL \url{https://link.aps.org/doi/10.1103/RevModPhys.57.287}.

\bibitem[Lee et~al.(2006)Lee, Nagaosa, and Wen]{lee2006doping}
Patrick~A Lee, Naoto Nagaosa, and Xiao-Gang Wen.
\newblock {Doping a Mott insulator: Physics of high-temperature
  superconductivity}.
\newblock \emph{Reviews of modern physics}, 78\penalty0 (1):\penalty0 17, 2006.
\newblock \doi{10.1103/RevModPhys.78.17}.

\bibitem[Lee(2018)]{lee2018recent}
Sung-Sik Lee.
\newblock Recent developments in non-fermi liquid theory.
\newblock \emph{Annual Review of Condensed Matter Physics}, 9:\penalty0
  227--244, 2018.
\newblock \doi{10.1146/annurev-conmatphys-031016-025531}.

\bibitem[Leedahl et~al.(2019)Leedahl, Sundermann, Amorese, Severing,
  Gretarsson, Zhang, Komarek, Maignan, Haverkort, and Tjeng]{leedahl2019origin}
Brett Leedahl, Martin Sundermann, Andrea Amorese, Andrea Severing, Hlynur
  Gretarsson, Lunyong Zhang, Alexander~C Komarek, Antoine Maignan, Maurits~W
  Haverkort, and Liu~Hao Tjeng.
\newblock {Origin of Ising magnetism in Ca$_3$Co$_2$O$_6$ unveiled by orbital
  imaging}.
\newblock \emph{Nature Communications}, 10\penalty0 (1):\penalty0 1--7, 2019.
\newblock \doi{10.1038/s41467-019-13273-4}.

\bibitem[Lefler et~al.(2019)Lefler, Ducho{\v{n}}, Karapetrov, Wang, Schneider,
  and May]{lefler2019reconfigurable}
Benjamin~M Lefler, Tom{\'a}{\v{s}} Ducho{\v{n}}, Goran Karapetrov, Jiayi Wang,
  Claus~M Schneider, and Steven~J May.
\newblock Reconfigurable lateral anionic heterostructures in oxide thin films
  via lithographically defined topochemistry.
\newblock \emph{Physical Review Materials}, 3\penalty0 (7):\penalty0 073802,
  2019.
\newblock \doi{10.1103/PhysRevMaterials.3.073802}.

\bibitem[Leggett(1998)]{LEGGETT19981729}
A.J. Leggett.
\newblock Where is the energy saved in cuprate superconductivity?
\newblock \emph{Journal of Physics and Chemistry of Solids}, 59\penalty0
  (10):\penalty0 1729 -- 1732, 1998.
\newblock \doi{10.1016/S0022-3697(98)00091-2}.

\bibitem[Leggett(1975)]{leggett1975theoretical}
Anthony~J Leggett.
\newblock A theoretical description of the new phases of liquid $^3$he.
\newblock \emph{Reviews of Modern Physics}, 47\penalty0 (2):\penalty0 331,
  1975.
\newblock \doi{10.1103/RevModPhys.47.331}.

\bibitem[Leggett(1996)]{leggett1996interlayer}
Anthony~J Leggett.
\newblock Interlayer tunneling models of cuprate superconductivity:
  Implications of a recent experiment.
\newblock \emph{Science}, 274\penalty0 (5287):\penalty0 587--589, 1996.
\newblock \doi{10.1126/science.274.5287.587}.

\bibitem[Leggett(1999)]{leggett1999midinfrared}
Anthony~J Leggett.
\newblock A “midinfrared” scenario for cuprate superconductivity.
\newblock \emph{Proceedings of the National Academy of Sciences}, 96\penalty0
  (15):\penalty0 8365--8372, 1999.
\newblock \doi{10.1073/pnas.96.15.8365}.

\bibitem[Leggett(2006{\natexlab{a}})]{leggett2006some}
Anthony~J Leggett.
\newblock Some thoughts about two dimensionality and cuprate superconductivity.
\newblock \emph{Journal of Superconductivity and Novel Magnetism}, 19\penalty0
  (3-5):\penalty0 187--192, 2006{\natexlab{a}}.
\newblock \doi{10.1007/s10948-006-0154-y}.

\bibitem[Leggett(2006{\natexlab{b}})]{leggett2006we}
Anthony~J Leggett.
\newblock {What DO we know about high T$_c$?}
\newblock \emph{Nature Physics}, 2\penalty0 (3):\penalty0 134--136,
  2006{\natexlab{b}}.
\newblock \doi{10.1038/nphys254}.

\bibitem[Legros et~al.(2019)Legros, Benhabib, Tabis, Laliberté, Dion, Lizaire,
  Vignolle, Vignolles, Raffy, Li, Auban-Senzier, Doiron-Leyraud, Fournier,
  Colson, Taillefer, and Proust]{legros_universal_2019}
A.~Legros, S.~Benhabib, W.~Tabis, F.~Laliberté, M.~Dion, M.~Lizaire,
  B.~Vignolle, D.~Vignolles, H.~Raffy, Z.~Z. Li, P.~Auban-Senzier,
  N.~Doiron-Leyraud, P.~Fournier, D.~Colson, L.~Taillefer, and C.~Proust.
\newblock Universal {T} -linear resistivity and {Planckian} dissipation in
  overdoped cuprates.
\newblock \emph{Nature Physics}, 15\penalty0 (2):\penalty0 142--147, 2019.
\newblock \doi{10.1038/s41567-018-0334-2}.
\newblock URL \url{https://www.nature.com/articles/s41567-018-0334-2}.

\bibitem[Lepori and Dell’Anna(2017)]{DellAnnaNJP2017}
Luca Lepori and L~Dell’Anna.
\newblock Long-range topological insulators and weakened bulk-boundary
  correspondence.
\newblock \emph{New Journal of Physics}, 19\penalty0 (10):\penalty0 103030,
  2017.
\newblock \doi{10.1088/1367-2630/aa84d0}.

\bibitem[Levallois et~al.(2016{\natexlab{a}})Levallois, Tran, Pouliot, Presura,
  Greene, Eckstein, Uccelli, Giannini, Gu, Leggett, and van~der
  Marel]{PhysRevX.6.031027}
J.~Levallois, M.~K. Tran, D.~Pouliot, C.~N. Presura, L.~H. Greene, J.~N.
  Eckstein, J.~Uccelli, E.~Giannini, G.~D. Gu, A.~J. Leggett, and D.~van~der
  Marel.
\newblock Temperature-dependent ellipsometry measurements of partial coulomb
  energy in superconducting cuprates.
\newblock \emph{Phys. Rev. X}, 6:\penalty0 031027, 2016{\natexlab{a}}.
\newblock \doi{10.1103/PhysRevX.6.031027}.
\newblock URL \url{https://link.aps.org/doi/10.1103/PhysRevX.6.031027}.

\bibitem[Levallois et~al.(2016{\natexlab{b}})Levallois, Tran, Pouliot, Presura,
  Greene, Eckstein, Uccelli, Giannini, Gu, Leggett,
  et~al.]{levallois2016temperature}
Julien Levallois, MK~Tran, D~Pouliot, CN~Presura, LH~Greene, James~N Eckstein,
  J~Uccelli, Enrico Giannini, GD~Gu, Anthony~J Leggett, et~al.
\newblock Temperature-dependent ellipsometry measurements of partial coulomb
  energy in superconducting cuprates.
\newblock \emph{Physical Review X}, 6\penalty0 (3):\penalty0 031027,
  2016{\natexlab{b}}.
\newblock \doi{10.1103/PhysRevX.6.031027}.

\bibitem[Levine et~al.(2019)Levine, Turner, Kehayias, Hart, Langellier, Trubko,
  Glenn, Fu, and Walsworth]{levine2019principles}
Edlyn~V. Levine, Matthew~J. Turner, Pauli Kehayias, Connor~A. Hart, Nicholas
  Langellier, Raisa Trubko, David~R. Glenn, Roger~R. Fu, and Ronald~L.
  Walsworth.
\newblock Principles and techniques of the quantum diamond microscope.
\newblock \emph{Nanophotonics}, 8\penalty0 (11):\penalty0 1945--1973, 2019.
\newblock \doi{10.1515/nanoph-2019-0209}.

\bibitem[Lewenstein et~al.(2007)Lewenstein, Sanpera, Ahufinger, Damski,
  Sen(De), and Sen]{doi:10.1080/00018730701223200}
Maciej Lewenstein, Anna Sanpera, Veronica Ahufinger, Bogdan Damski, Aditi
  Sen(De), and Ujjwal Sen.
\newblock Ultracold atomic gases in optical lattices: mimicking condensed
  matter physics and beyond.
\newblock \emph{Advances in Physics}, 56\penalty0 (2):\penalty0 243--379, 2007.
\newblock \doi{10.1080/00018730701223200}.

\bibitem[Li et~al.(2019{\natexlab{a}})Li, Lee, Wang, Osada, Crossley, Lee, Cui,
  Hikita, and Hwang]{li2019superconductivity}
Danfeng Li, Kyuho Lee, Bai~Yang Wang, Motoki Osada, Samuel Crossley, Hye~Ryoung
  Lee, Yi~Cui, Yasuyuki Hikita, and Harold~Y Hwang.
\newblock Superconductivity in an infinite-layer nickelate.
\newblock \emph{Nature}, 572\penalty0 (7771):\penalty0 624--627,
  2019{\natexlab{a}}.
\newblock \doi{10.1038/s41586-019-1496-5}.

\bibitem[Li et~al.(2018)Li, Zhou, Parham, Reber, Berger, Arnold, and
  Dessau]{li2018coherent}
Haoxiang Li, Xiaoqing Zhou, Stephen Parham, Theodore~J Reber, Helmuth Berger,
  Gerald~B Arnold, and Daniel~S Dessau.
\newblock {Coherent organization of electronic correlations as a mechanism to
  enhance and stabilize high-$T_c$ cuprate superconductivity}.
\newblock \emph{Nature Communications}, 9\penalty0 (1):\penalty0 1--9, 2018.
\newblock \doi{10.1038/s41467-017-02422-2}.

\bibitem[Li et~al.(2013)Li, Pollanen, Zimmerman, Collett, Gannon, and
  Halperin]{li2013superfluid}
JIA Li, J~Pollanen, AM~Zimmerman, CA~Collett, WJ~Gannon, and William~P
  Halperin.
\newblock {The superfluid glass phase of $^3$He-A}.
\newblock \emph{Nature Physics}, 9\penalty0 (12):\penalty0 775--779, 2013.
\newblock \doi{10.1038/nphys2806}.

\bibitem[Li et~al.(2008)Li, Checkelsky, Hor, Uher, Hebard, Cava, and
  Ong]{li2008phase}
Lu~Li, Joseph~G Checkelsky, Yew~San Hor, Ctirad Uher, Arthur~F Hebard,
  Robert~Joseph Cava, and Nai~Phuan Ong.
\newblock {Phase transitions of {Dirac} electrons in bismuth}.
\newblock \emph{Science}, 321\penalty0 (5888):\penalty0 547--550, 2008.
\newblock \doi{10.1126/science.1158908}.

\bibitem[Li et~al.(2020)Li, Sun, Kurdak, and Allen]{li2020emergent}
Lu~Li, Kai Sun, Cagliyan Kurdak, and JW~Allen.
\newblock {Emergent mystery in the {Kondo} insulator samarium hexaboride}.
\newblock \emph{Nature Reviews Physics}, pages 1--17, 2020.
\newblock \doi{10.1038/s42254-020-0210-8}.

\bibitem[Li et~al.(2019{\natexlab{b}})Li, Qiu, Zhang, Baldini, Lu, Rappe, and
  Nelson]{li2019terahertz}
Xian Li, Tian Qiu, Jiahao Zhang, Edoardo Baldini, Jian Lu, Andrew~M. Rappe, and
  Keith~A. Nelson.
\newblock {Terahertz field–induced ferroelectricity in quantum paraelectric
  SrTiO$_3$}.
\newblock \emph{Science}, 364\penalty0 (6445):\penalty0 1079--1082,
  2019{\natexlab{b}}.
\newblock \doi{10.1126/science.aaw4913}.

\bibitem[Li and Haldane(2018)]{li2018topological}
Yi~Li and FDM Haldane.
\newblock {Topological nodal Cooper pairing in doped {Weyl} metals}.
\newblock \emph{Phys. Rev. Lett.}, 120\penalty0 (6):\penalty0 067003, 2018.
\newblock \doi{10.1103/PhysRevLett.120.067003}.

\bibitem[Li and Yao(2019)]{Li_Yao}
Zi-Xiang Li and Hong Yao.
\newblock Sign-problem-free fermionic quantum {Monte} carlo: Developments and
  applications.
\newblock \emph{Annual Review of Condensed Matter Physics}, 10\penalty0
  (1):\penalty0 337--356, 2019.
\newblock \doi{10.1146/annurev-conmatphys-033117-054307}.
\newblock URL \url{https://doi.org/10.1146/annurev-conmatphys-033117-054307}.

\bibitem[Liang and Wang(2018)]{liang2018singleshot}
Jinyang Liang and Lihong~V. Wang.
\newblock {Single-shot ultrafast optical imaging}.
\newblock \emph{Optica}, 5\penalty0 (9):\penalty0 1113, 2018.
\newblock \doi{10.1364/OPTICA.5.001113}.

\bibitem[Liao et~al.(2014)Liao, Huijben, Koster, and Rijnders]{Liao2014}
Zhaoliang Liao, Mark Huijben, Gertjan Koster, and Guus Rijnders.
\newblock Uniaxial magnetic anisotropy induced low field anomalous anisotropic
  magnetoresistance in manganite thin films.
\newblock \emph{APL Materials}, 2\penalty0 (9):\penalty0 096112, 2014.
\newblock \doi{10.1063/1.4895956}.

\bibitem[Lieb et~al.(1961)Lieb, Schultz, and Mattis]{LSM_original}
Elliott Lieb, Theodore Schultz, and Daniel Mattis.
\newblock Two soluble models of an antiferromagnetic chain.
\newblock \emph{Annals of Physics}, 16\penalty0 (3):\penalty0 407 -- 466, 1961.
\newblock \doi{10.1016/0003-4916(61)90115-4}.
\newblock URL
  \url{http://www.sciencedirect.com/science/article/pii/0003491661901154}.

\bibitem[Lim et~al.(2020)Lim, Chrysler, Kumar, Mauthe, Kumah, Richardson,
  LeBeau, and Ngai]{lim2020suspended}
Zheng~Hui Lim, Matthew Chrysler, Abinash Kumar, Jacob~P Mauthe, Divine~P Kumah,
  Chris Richardson, James~M LeBeau, and Joseph~H Ngai.
\newblock Suspended single-crystalline oxide structures on silicon through
  wet-etch techniques: Effects of oxygen vacancies and dislocations on etch
  rates.
\newblock \emph{Journal of Vacuum Science \& Technology A: Vacuum, Surfaces,
  and Films}, 38\penalty0 (1):\penalty0 013406, 2020.
\newblock \doi{10.1116/1.5135035}.

\bibitem[Lin et~al.(2013)Lin, Zhu, Fauqu\'e, and Behnia]{STO-SC2}
Xiao Lin, Zengwei Zhu, Beno\^{\i}t Fauqu\'e, and Kamran Behnia.
\newblock Fermi surface of the most dilute superconductor.
\newblock \emph{Phys. Rev. X}, 3:\penalty0 021002, 2013.
\newblock \doi{10.1103/PhysRevX.3.021002}.
\newblock URL \url{https://link.aps.org/doi/10.1103/PhysRevX.3.021002}.

\bibitem[Lin et~al.(2017)Lin, Rischau, Buchauer, Jaoui, Fauqu{\'e}, and
  Behnia]{STOIoffeRegel}
Xiao Lin, Carl~Willem Rischau, Lisa Buchauer, Alexandre Jaoui, Beno{\^\i}t
  Fauqu{\'e}, and Kamran Behnia.
\newblock Metallicity without quasi-particles in room-temperature strontium
  titanate.
\newblock \emph{npj Quantum Materials}, 2\penalty0 (1):\penalty0 1--8, 2017.
\newblock \doi{10.1038/s41535-017-0044-5}.

\bibitem[Lindner et~al.(2011{\natexlab{a}})Lindner, Refael, and
  Galitski]{LindnerNature2011}
Netanel~H. Lindner, Gil Refael, and Victor Galitski.
\newblock Floquet topological insulator in semiconductor quantum wells.
\newblock \emph{Nature Physics}, 7\penalty0 (6):\penalty0 490--495,
  2011{\natexlab{a}}.
\newblock \doi{10.1038/nphys1926}.
\newblock URL \url{https://doi.org/10.1038/nphys1926}.

\bibitem[Lindner et~al.(2011{\natexlab{b}})Lindner, Refael, and
  Galitski]{lindner2011floquet}
Netanel~H Lindner, Gil Refael, and Victor Galitski.
\newblock Floquet topological insulator in semiconductor quantum wells.
\newblock \emph{Nature Physics}, 7\penalty0 (6):\penalty0 490--495,
  2011{\natexlab{b}}.
\newblock \doi{10.1038/nphys1926}.

\bibitem[Liu et~al.(2019)Liu, Hal{\'a}sz, and Balents]{liu2019competing}
Chunxiao Liu, G{\'a}bor~B Hal{\'a}sz, and Leon Balents.
\newblock {Competing orders in pyrochlore magnets from a Z$_2$ spin liquid
  perspective}.
\newblock \emph{Phys. Rev. B}, 100\penalty0 (7):\penalty0 075125, 2019.
\newblock \doi{10.1103/PhysRevB.100.075125}.

\bibitem[Liu et~al.(2020)Liu, Hao, Khalaf, Lee, Ronen, Yoo, Najafabadi,
  Watanabe, Taniguchi, Vishwanath, et~al.]{Liu2019Spin}
Xiaomeng Liu, Zeyu Hao, Eslam Khalaf, Jong~Yeon Lee, Yuval Ronen, Hyobin Yoo,
  Danial~Haei Najafabadi, Kenji Watanabe, Takashi Taniguchi, Ashvin Vishwanath,
  et~al.
\newblock Tunable spin-polarized correlated states in twisted double bilayer
  graphene.
\newblock \emph{Nature}, 583\penalty0 (7815):\penalty0 221--225, 2020.
\newblock \doi{10.1038/s41586-020-2458-7}.

\bibitem[Logvenov et~al.(2009)Logvenov, Gozar, and Bozovic]{logvenov2009high}
G~Logvenov, A~Gozar, and I~Bozovic.
\newblock High-temperature superconductivity in a single copper-oxygen plane.
\newblock \emph{Science}, 326\penalty0 (5953):\penalty0 699--702, 2009.
\newblock \doi{10.1126/science.1178863}.

\bibitem[L{\"o}hneysen et~al.(1994)L{\"o}hneysen, Pietrus, Portisch, Schlager,
  Schr{\"o}der, Sieck, and Trappmann]{lohneysen1994non}
H~v L{\"o}hneysen, T~Pietrus, G~Portisch, HG~Schlager, A~Schr{\"o}der, M~Sieck,
  and T~Trappmann.
\newblock Non-fermi-liquid behavior in a heavy-fermion alloy at a magnetic
  instability.
\newblock \emph{Phys. Rev. Lett.}, 72\penalty0 (20):\penalty0 3262, 1994.
\newblock \doi{10.1103/PhysRevLett.72.3262}.

\bibitem[L{\"o}hneysen et~al.(2007{\natexlab{a}})L{\"o}hneysen, Rosch, Vojta,
  and W{\"o}lfle]{Loehneysen07}
Hilbert~v L{\"o}hneysen, Achim Rosch, Matthias Vojta, and Peter W{\"o}lfle.
\newblock Fermi-liquid instabilities at magnetic quantum phase transitions.
\newblock \emph{Reviews of Modern Physics}, 79\penalty0 (3):\penalty0 1015,
  2007{\natexlab{a}}.
\newblock \doi{10.1103/RevModPhys.79.1015}.

\bibitem[L{\"o}hneysen et~al.(2007{\natexlab{b}})L{\"o}hneysen, Rosch, Vojta,
  and W{\"o}lfle]{lohneysen2007fermi}
Hilbert~v L{\"o}hneysen, Achim Rosch, Matthias Vojta, and Peter W{\"o}lfle.
\newblock Fermi-liquid instabilities at magnetic quantum phase transitions.
\newblock \emph{Reviews of Modern Physics}, 79\penalty0 (3):\penalty0 1015,
  2007{\natexlab{b}}.
\newblock \doi{10.1103/RevModPhys.79.1015}.

\bibitem[Loubeyre et~al.(2020)Loubeyre, Occelli, and Dumas]{loubeyre2020highp}
Paul Loubeyre, Florent Occelli, and Paul Dumas.
\newblock {Synchrotron infrared spectroscopic evidence of the probable
  transition to metal hydrogen}.
\newblock \emph{Nature}, 577\penalty0 (7792):\penalty0 631--635, 2020.
\newblock \doi{10.1038/s41586-019-1927-3}.

\bibitem[Lu et~al.(2016)Lu, Baek, Hong, Kourkoutis, Hikita, and
  Hwang]{lu2016synthesis}
Di~Lu, David~J Baek, Seung~Sae Hong, Lena~F Kourkoutis, Yasuyuki Hikita, and
  Harold~Y Hwang.
\newblock Synthesis of freestanding single-crystal perovskite films and
  heterostructures by etching of sacrificial water-soluble layers.
\newblock \emph{Nature materials}, 15\penalty0 (12):\penalty0 1255--1260, 2016.
\newblock \doi{10.1038/nmat4749}.

\bibitem[Lu et~al.(2017)Lu, Li, Hwang, Ofori-Okai, Kurihara, Suemoto, and
  Nelson]{lu2017coherent}
Jian Lu, Xian Li, Harold~Y Hwang, Benjamin~K Ofori-Okai, Takayuki Kurihara,
  Tohru Suemoto, and Keith~A Nelson.
\newblock Coherent two-dimensional terahertz magnetic resonance spectroscopy of
  collective spin waves.
\newblock \emph{Phys. Rev. Lett.}, 118\penalty0 (20):\penalty0 207204, 2017.
\newblock \doi{10.1103/PhysRevLett.118.207204}.

\bibitem[Lu et~al.(2019)Lu, Stepanov, Yang, Xie, Aamir, Das, Urgell, Watanabe,
  Taniguchi, Zhang, et~al.]{lu2019superconductors}
Xiaobo Lu, Petr Stepanov, Wei Yang, Ming Xie, Mohammed~Ali Aamir, Ipsita Das,
  Carles Urgell, Kenji Watanabe, Takashi Taniguchi, Guangyu Zhang, et~al.
\newblock Superconductors, orbital magnets and correlated states in magic-angle
  bilayer graphene.
\newblock \emph{Nature}, 574\penalty0 (7780):\penalty0 653--657, 2019.
\newblock \doi{10.1038/s41586-019-1695-0}.

\bibitem[Lucas and Hartnoll(2017{\natexlab{a}})]{Lucas11344}
Andrew Lucas and Sean~A. Hartnoll.
\newblock Resistivity bound for hydrodynamic bad metals.
\newblock \emph{Proceedings of the National Academy of Sciences}, 114\penalty0
  (43):\penalty0 11344--11349, 2017{\natexlab{a}}.
\newblock \doi{10.1073/pnas.1711414114}.
\newblock URL \url{https://www.pnas.org/content/114/43/11344}.

\bibitem[Lucas and Hartnoll(2017{\natexlab{b}})]{lucas2017resistivity}
Andrew Lucas and Sean~A Hartnoll.
\newblock Resistivity bound for hydrodynamic bad metals.
\newblock \emph{Proceedings of the National Academy of Sciences}, 114\penalty0
  (43):\penalty0 11344--11349, 2017{\natexlab{b}}.
\newblock \doi{10.1103/PhysRevLett.118.096603}.

\bibitem[Luther(1979)]{PhysRevB.19.320}
A.~Luther.
\newblock Tomonaga fermions and the {Dirac} equation in three dimensions.
\newblock \emph{Phys. Rev. B}, 19:\penalty0 320--330, 1979.
\newblock \doi{10.1103/PhysRevB.19.320}.
\newblock URL \url{https://link.aps.org/doi/10.1103/PhysRevB.19.320}.

\bibitem[Luttinger(1960)]{luttingerFS}
J.~M. Luttinger.
\newblock Fermi surface and some simple equilibrium properties of a system of
  interacting fermions.
\newblock \emph{Phys. Rev.}, 119:\penalty0 1153--1163, 1960.
\newblock \doi{10.1103/PhysRev.119.1153}.
\newblock URL \url{https://link.aps.org/doi/10.1103/PhysRev.119.1153}.

\bibitem[Lv et~al.(2015)Lv, Wang, Peng, Ding, Wang, Wang, He, Ji, Zhong,
  Schneeloch, et~al.]{lv2015mapping}
Yan-Feng Lv, Wen-Lin Wang, Jun-Ping Peng, Hao Ding, Yang Wang, Lili Wang,
  Ke~He, Shuai-Hua Ji, Ruidan Zhong, John Schneeloch, et~al.
\newblock {Mapping the electronic structure of each ingredient oxide layer of
  high-$T_c$ cuprate superconductor {Bi}$_2${Sr}$_2${CaCu}$_2${O}$_{8+\delta}$
  }.
\newblock \emph{Phys. Rev. Lett.}, 115\penalty0 (23):\penalty0 237002, 2015.
\newblock \doi{10.1103/PhysRevLett.115.237002}.

\bibitem[Lüscher(1989)]{LUSCHER1989557}
M.~Lüscher.
\newblock Bosonization in 2 + 1 dimensions.
\newblock \emph{Nuclear Physics B}, 326\penalty0 (3):\penalty0 557 -- 582,
  1989.
\newblock \doi{10.1016/0550-3213(89)90544-0}.
\newblock URL
  \url{http://www.sciencedirect.com/science/article/pii/0550321389905440}.

\bibitem[Ma et~al.(2015)Ma, Cui, Ueda, Tang, Chen, Tamura, Wu, Fujioka, Tokura,
  and Shen]{ma2015mim}
E.~Y. Ma, Y.-T. Cui, Kentaro Ueda, Shujie Tang, Kai Chen, Nobumichi Tamura,
  Phillip~M Wu, Jun Fujioka, Yoshinori Tokura, and Z.-X. Shen.
\newblock {Mobile metallic domain walls in an all-in-all-out magnetic
  insulator}.
\newblock \emph{Science}, 350\penalty0 (6260):\penalty0 538--541, 2015.
\newblock \doi{10.1126/science.aac8289}.

\bibitem[Ma(2018)]{ma2018modern}
Shang-Keng Ma.
\newblock \emph{Modern Theory of Critical Phenomena}.
\newblock Routledge, 2018.
\newblock \doi{10.4324/9780429498886}.

\bibitem[Machado et~al.(2019)Machado, Kahanamoku-Meyer, Else, Nayak, and
  Yao]{PhysRevResearch.1.033202}
Francisco Machado, Gregory~D. Kahanamoku-Meyer, Dominic~V. Else, Chetan Nayak,
  and Norman~Y. Yao.
\newblock Exponentially slow heating in short and long-range interacting
  {Floquet} systems.
\newblock \emph{Phys. Rev. Research}, 1:\penalty0 033202, 2019.
\newblock \doi{10.1103/PhysRevResearch.1.033202}.
\newblock URL \url{https://link.aps.org/doi/10.1103/PhysRevResearch.1.033202}.

\bibitem[Maciejko and Fiete(2015)]{maciejko2015fractionalized}
Joseph Maciejko and Gregory~A Fiete.
\newblock Fractionalized topological insulators.
\newblock \emph{Nature Physics}, 11\penalty0 (5):\penalty0 385--388, 2015.
\newblock \doi{10.1038/nphys3311}.

\bibitem[Maciejko et~al.(2010)Maciejko, Qi, Karch, and
  Zhang]{maciejko2010fractional}
Joseph Maciejko, Xiao-Liang Qi, Andreas Karch, and Shou-Cheng Zhang.
\newblock Fractional topological insulators in three dimensions.
\newblock \emph{Physical Review Letters}, 105\penalty0 (24):\penalty0 246809,
  2010.
\newblock \doi{10.1103/PhysRevLett.105.246809}.

\bibitem[Mackenzie et~al.(2017)Mackenzie, Scaffidi, Hicks, and
  Maeno]{mackenzie2017even}
Andrew~P Mackenzie, Thomas Scaffidi, Clifford~W Hicks, and Yoshiteru Maeno.
\newblock {Even odder after twenty-three years: the superconducting order
  parameter puzzle of Sr$_2$ RuO$_4$}.
\newblock \emph{npj Quantum Materials}, 2\penalty0 (1):\penalty0 1--9, 2017.
\newblock \doi{10.1038/s41535-017-0045-4}.

\bibitem[Maeno et~al.(1994)Maeno, Hashimoto, Yoshida, Nishizaki, Fujita,
  Bednorz, and Lichtenberg]{maeno1994superconductivity}
Y~Maeno, H~Hashimoto, K~Yoshida, S~Nishizaki, T~Fujita, JG~Bednorz, and
  F~Lichtenberg.
\newblock Superconductivity in a layered perovskite without copper.
\newblock \emph{Nature}, 372\penalty0 (6506):\penalty0 532--534, 1994.
\newblock \doi{10.1038/372532a0}.

\bibitem[Maier et~al.(2004)Maier, Jarrell, Macridin, and
  Slezak]{maier2004kinetic}
Th~A Maier, M~Jarrell, A~Macridin, and C~Slezak.
\newblock Kinetic energy driven pairing in cuprate superconductors.
\newblock \emph{Phys. Rev. Lett.}, 92\penalty0 (2):\penalty0 027005, 2004.
\newblock \doi{10.1103/PhysRevLett.92.027005}.

\bibitem[Maisuradze et~al.(2010)Maisuradze, Schnelle, Khasanov, Gumeniuk,
  Nicklas, Rosner, Leithe-Jasper, Grin, Amato, and Thalmeier]{Maisuradze10}
A~Maisuradze, W~Schnelle, R~Khasanov, R~Gumeniuk, M~Nicklas, H~Rosner,
  A~Leithe-Jasper, Yu~Grin, A~Amato, and P~Thalmeier.
\newblock {Evidence for time-reversal symmetry breaking in superconducting
  PrPt$_4$Ge$_{12}$ }.
\newblock \emph{Phys. Rev. B}, 82\penalty0 (2):\penalty0 024524, 2010.
\newblock \doi{10.1103/PhysRevB.82.024524}.

\bibitem[Mankowsky et~al.(2014)Mankowsky, Subedi, F\"{o}rst, Mariager, Chollet,
  Lemke, Robinson, Glownia, Minitti, Frano, Fechner, Spaldin, Loew, Keimer,
  Georges, and Cavalleri]{Mankowsky2014}
R.~Mankowsky, A.~Subedi, M.~F\"{o}rst, S.~O. Mariager, M.~Chollet, H.~T. Lemke,
  J.~S. Robinson, J.~M. Glownia, M.~P. Minitti, A.~Frano, M.~Fechner, N.~A.
  Spaldin, T.~Loew, B.~Keimer, A.~Georges, and A.~Cavalleri.
\newblock {Nonlinear lattice dynamics as a basis for enhanced superconductivity
  in {YBa}$_2$Cu$_3$O$_{6.5}$}.
\newblock \emph{Nature}, 516\penalty0 (7529):\penalty0 71--73, 2014.
\newblock \doi{10.1038/nature13875}.
\newblock URL \url{https://doi.org/10.1038/nature13875}.

\bibitem[Manousakis(2002)]{manousakis2002quantum}
Efstratios Manousakis.
\newblock A quantum-dot array as model for copper-oxide superconductors: A
  dedicated quantum simulator for the many-fermion problem.
\newblock \emph{Journal of Low Temperature Physics}, 126\penalty0
  (5-6):\penalty0 1501--1513, 2002.
\newblock \doi{10.1023/A:1014295416763}.

\bibitem[Maple et~al.(2010)Maple, Baumbach, Butch, Hamlin, and
  Janoschek]{Maple10}
M~Brian Maple, Ryan~E Baumbach, Nicholas~P Butch, James~J Hamlin, and Marc
  Janoschek.
\newblock {Non-Fermi liquid regimes and superconductivity in the low
  temperature phase diagrams of strongly correlated d-and f-electron
  materials}.
\newblock \emph{Journal of Low Temperature Physics}, 161\penalty0
  (1-2):\penalty0 4--54, 2010.
\newblock \doi{10.1007/s10909-010-0212-5}.

\bibitem[Maple et~al.(1994)Maple, Seaman, Gajewski, Dalichaouch, Barbetta,
  De~Andrade, Mook, Lukefahr, Bernal, and MacLaughlin]{Maple94}
MB~Maple, CL~Seaman, DA~Gajewski, Y~Dalichaouch, VB~Barbetta, MC~De~Andrade,
  HA~Mook, HG~Lukefahr, OO~Bernal, and DE~MacLaughlin.
\newblock Non {Fermi} liquid behavior in strongly correlated f-electron
  materials.
\newblock \emph{Journal of Low Temperature Physics}, 95\penalty0
  (1-2):\penalty0 225--243, 1994.
\newblock \doi{10.1007/BF00754938}.

\bibitem[Maple et~al.(1995)Maple, De~Andrade, Herrmann, Dalichaouch, Gajewski,
  Seaman, Chau, Movshovich, Aronson, and Osborn]{Maple95}
MB~Maple, MC~De~Andrade, J~Herrmann, Y~Dalichaouch, DA~Gajewski, CL~Seaman,
  R~Chau, R~Movshovich, MC~Aronson, and R~Osborn.
\newblock Non {Fermi} liquid ground states in strongly correlated f-electron
  materials.
\newblock \emph{Journal of Low Temperature Physics}, 99\penalty0
  (3-4):\penalty0 223--249, 1995.
\newblock \doi{10.1007/BF00752290}.

\bibitem[Maple et~al.(2006)Maple, Frederick, Ho, Yuhasz, and
  Yanagisawa]{Maple06}
MB~Maple, NA~Frederick, P-C Ho, WM~Yuhasz, and T~Yanagisawa.
\newblock {Unconventional Superconductivity and Heavy Fermion Behavior in
  PrOs$_4$Sb$_{12}$ }.
\newblock \emph{Journal of superconductivity and novel magnetism}, 19\penalty0
  (3-5):\penalty0 299--315, 2006.
\newblock \doi{10.1007/s10948-006-0165-8}.

\bibitem[Maricq(1982)]{PhysRevB.25.6622}
M.~Matti Maricq.
\newblock {Application of average Hamiltonian theory to the NMR of solids}.
\newblock \emph{Phys. Rev. B}, 25:\penalty0 6622--6632, 1982.
\newblock \doi{10.1103/PhysRevB.25.6622}.
\newblock URL \url{https://link.aps.org/doi/10.1103/PhysRevB.25.6622}.

\bibitem[Mart{\'\i}nez-P{\'e}rez and Koelle(2017)]{martinez2017nanosquids}
Maria~Jos{\'e} Mart{\'\i}nez-P{\'e}rez and Dieter Koelle.
\newblock Nanosquids: Basics \& recent advances.
\newblock \emph{Physical Sciences Reviews}, 2\penalty0 (8), 2017.
\newblock \doi{10.1515/9783110456806-012}.

\bibitem[Mata-Pinz{\'o}n et~al.(2016)Mata-Pinz{\'o}n, Valladares, Valladares,
  and Valladares]{mata2016superconductivity}
Zaahel Mata-Pinz{\'o}n, Ariel~A Valladares, Renela~M Valladares, and Alexander
  Valladares.
\newblock Superconductivity in bismuth. a new look at an old problem.
\newblock \emph{PLoS One}, 11\penalty0 (1):\penalty0 e0147645, 2016.
\newblock \doi{10.1371/journal.pone.0147645}.

\bibitem[Matsumoto et~al.(2016)Matsumoto, Tsujimoto, Tomita, Sakai, and
  Nakatsuji]{Matsumoto16}
Yosuke Matsumoto, Masaki Tsujimoto, Takahiro Tomita, Akito Sakai, and Satoru
  Nakatsuji.
\newblock {Heavy Fermion Superconductivity in Non-magnetic Cage Compound
  PrV$_2$Al$_{20}$ }.
\newblock In \emph{Journal of Physics: Conference Series}, volume 683, page
  012013, 2016.
\newblock \doi{10.1088/1742-6596/683/1/012013}.

\bibitem[Matt et~al.(2015)Matt, Fatuzzo, Sassa, M{\aa}nsson, Fatale, Bitetta,
  Shi, Pailh{\`e}s, Berntsen, Kurosawa, et~al.]{matt2015electron}
Christian~E Matt, Claudia~G Fatuzzo, Yasmine Sassa, Martin M{\aa}nsson,
  S~Fatale, V~Bitetta, X~Shi, St{\'e}phane Pailh{\`e}s, MH~Berntsen, Tohru
  Kurosawa, et~al.
\newblock {Electron scattering, charge order, and pseudogap physics in
  La$_{1.6-x}$Nd$_{0.4}$Sr$_x$CuO$_4$: an angle-resolved photoemission
  spectroscopy study}.
\newblock \emph{Phys. Rev. B}, 92\penalty0 (13):\penalty0 134524, 2015.
\newblock \doi{10.1103/PhysRevB.92.134524}.

\bibitem[Matthias et~al.(1962)Matthias, Geballe, Compton, Corenzwit, and
  Hull]{dopedCr1}
B.~T. Matthias, T.~H. Geballe, V.~B. Compton, E.~Corenzwit, and G.~W. Hull.
\newblock Superconductivity of chromium alloys.
\newblock \emph{Phys. Rev.}, 128:\penalty0 588--590, 1962.
\newblock \doi{10.1103/PhysRev.128.588}.
\newblock URL \url{https://link.aps.org/doi/10.1103/PhysRev.128.588}.

\bibitem[McIver et~al.(2020)McIver, Schulte, Stein, Matsuyama, Jotzu, Meier,
  and Cavalleri]{mciver2020ahe}
J.~W. McIver, B.~Schulte, F.-U. Stein, T.~Matsuyama, G.~Jotzu, G.~Meier, and
  A.~Cavalleri.
\newblock {Light-induced anomalous Hall effect in graphene}.
\newblock \emph{Nature Physics}, 16\penalty0 (1):\penalty0 38--41, 2020.
\newblock \doi{10.1038/s41567-019-0698-y}.

\bibitem[McLeod et~al.(2020)McLeod, Zhang, Gu, Jin, Zhang, Post, Zhao, Millis,
  Wu, Rondinelli, et~al.]{mcleod2019multi}
Alexander~S McLeod, Jingdi Zhang, MQ~Gu, Feng Jin, G~Zhang, Kirk~W Post,
  XG~Zhao, Andrew~J Millis, WB~Wu, James~M Rondinelli, et~al.
\newblock Multi-messenger nanoprobes of hidden magnetism in a strained
  manganite.
\newblock \emph{Nature materials}, 19\penalty0 (4):\penalty0 397--404, 2020.
\newblock \doi{10.1038/s41563-019-0533-y}.

\bibitem[McLeod et~al.(2017)McLeod, Van~Heumen, Ramirez, Wang, Saerbeck,
  Guenon, Goldflam, Anderegg, Kelly, Mueller, et~al.]{mcleod2017nanotextured}
AS~McLeod, E~Van~Heumen, JG~Ramirez, S~Wang, T~Saerbeck, S~Guenon, M~Goldflam,
  L~Anderegg, P~Kelly, A~Mueller, et~al.
\newblock {Nanotextured phase coexistence in the correlated insulator
  V$_2$O$_3$}.
\newblock \emph{Nature Physics}, 13\penalty0 (1):\penalty0 80--86, 2017.
\newblock \doi{10.1038/nphys3882}.

\bibitem[Meier et~al.(2018)Meier, Khandarkhaeva, Petitgirard, K{\"o}rber,
  Lauerer, R{\"o}ssler, and Dubrovinsky]{meier2018nmr}
Thomas Meier, Saiana Khandarkhaeva, Sylvain Petitgirard, Thomas K{\"o}rber,
  Alexander Lauerer, Ernst R{\"o}ssler, and Leonid Dubrovinsky.
\newblock Nmr at pressures up to 90 gpa.
\newblock \emph{Journal of Magnetic Resonance}, 292:\penalty0 44--47, 2018.
\newblock \doi{10.1016/j.jmr.2018.05.002}.

\bibitem[Menth et~al.(1969)Menth, Buehler, and Geballe]{SmB61}
A.~Menth, E.~Buehler, and T.~H. Geballe.
\newblock {Magnetic and Semiconducting Properties of Sm${\mathrm{B}}_{6}$}.
\newblock \emph{Phys. Rev. Lett.}, 22:\penalty0 295--297, 1969.
\newblock \doi{10.1103/PhysRevLett.22.295}.
\newblock URL \url{https://link.aps.org/doi/10.1103/PhysRevLett.22.295}.

\bibitem[Meregalli and Savrasov(1998)]{meregalli1998electron}
V~Meregalli and S~Yu Savrasov.
\newblock {Electron-phonon coupling and properties of doped BaBiO$_3$}.
\newblock \emph{Phys. Rev. B}, 57\penalty0 (22):\penalty0 14453, 1998.
\newblock \doi{10.1103/PhysRevB.57.14453}.

\bibitem[Merritt et~al.(2019)Merritt, Castellan, Keller, Park, Fernandez-Baca,
  Gu, and Reznik]{merritt2019low}
AM~Merritt, J-P Castellan, T~Keller, SR~Park, JA~Fernandez-Baca, GD~Gu, and
  D~Reznik.
\newblock {Low-energy phonons in {Bi}$_2${Sr}$_2${CaCu}$_2${O}$_{8+\delta}$ and
  their possible interaction with electrons measured by inelastic neutron
  scattering}.
\newblock \emph{Phys. Rev. B}, 100\penalty0 (14):\penalty0 144502, 2019.
\newblock \doi{10.1103/PhysRevB.100.144502}.

\bibitem[Mesaros et~al.(2011)Mesaros, Fujita, Eisaki, Uchida, Davis, Sachdev,
  Zaanen, Lawler, and Kim]{mesaros2011topological}
A~Mesaros, K~Fujita, H~Eisaki, S~Uchida, JC~Davis, S~Sachdev, J~Zaanen,
  MJ~Lawler, and Eun-Ah Kim.
\newblock Topological defects coupling smectic modulations to intra-unit-cell
  nematicity in cuprates.
\newblock \emph{Science}, 333\penalty0 (6041):\penalty0 426--430, 2011.
\newblock \doi{10.1126/science.1201082}.

\bibitem[Michon et~al.(2018)Michon, Ataei, Bourgeois-Hope, Collignon, Li,
  Badoux, Gourgout, Lalibert{\'e}, Zhou, Doiron-Leyraud,
  et~al.]{michon2018wiedemann}
B~Michon, A~Ataei, P~Bourgeois-Hope, C~Collignon, SY~Li, S~Badoux, A~Gourgout,
  F~Lalibert{\'e}, J-S Zhou, Nicolas Doiron-Leyraud, et~al.
\newblock Wiedemann-franz law and abrupt change in conductivity across the
  pseudogap critical point of a cuprate superconductor.
\newblock \emph{Physical Review X}, 8\penalty0 (4):\penalty0 041010, 2018.
\newblock \doi{10.1103/PhysRevX.8.041010}.

\bibitem[Mielke(1998)]{mielke1998flow}
Andreas Mielke.
\newblock Flow equations for band-matrices.
\newblock \emph{The European Physical Journal B-Condensed Matter and Complex
  Systems}, 5\penalty0 (3):\penalty0 605--611, 1998.
\newblock \doi{10.1007/s100510050485}.

\bibitem[Mikami et~al.(2016)Mikami, Kitamura, Yasuda, Tsuji, Oka, and
  Aoki]{PhysRevB.93.144307}
Takahiro Mikami, Sota Kitamura, Kenji Yasuda, Naoto Tsuji, Takashi Oka, and
  Hideo Aoki.
\newblock Brillouin-wigner theory for high-frequency expansion in periodically
  driven systems: Application to {Floquet} topological insulators.
\newblock \emph{Phys. Rev. B}, 93:\penalty0 144307, 2016.
\newblock \doi{10.1103/PhysRevB.93.144307}.
\newblock URL \url{https://link.aps.org/doi/10.1103/PhysRevB.93.144307}.

\bibitem[Mirebeau et~al.(2002)Mirebeau, Goncharenko, Cadavez-Peres, Bramwell,
  Gingras, and Gardner]{Mirebeau2002}
I.~Mirebeau, I.~N. Goncharenko, P.~Cadavez-Peres, S.~T. Bramwell, M.~J.~P.
  Gingras, and J.~S. Gardner.
\newblock Pressure-induced crystallization of a spin liquid.
\newblock \emph{Nature}, 420\penalty0 (6911):\penalty0 54--57, 2002.
\newblock \doi{10.1038/nature01157}.
\newblock URL \url{https://doi.org/10.1038/nature01157}.

\bibitem[Missell(1985)]{amorphousSC2}
Frank~P Missell.
\newblock {Proceedings of the Latin American Symposium of Physics of Amorphous
  Systems-v.1}.
\newblock pages 161--175, 1985.
\newblock \doi{not available}.

\bibitem[Misumi et~al.(2020)Misumi, Yamaguchi, Zhang, Matsushita, Wada,
  Tsuchiizu, and Awaga]{misumi2020quantum}
Yuki Misumi, Akira Yamaguchi, Zhongyue Zhang, Taku Matsushita, Nobuo Wada,
  Masahisa Tsuchiizu, and Kunio Awaga.
\newblock Quantum spin liquid state in a two-dimensional semiconductive
  metal--organic framework.
\newblock \emph{Journal of the American Chemical Society}, 142:\penalty0 16513,
  2020.
\newblock \doi{10.1021/jacs.0c05472}.

\bibitem[Mito et~al.(2012)Mito, Imakyurei, Deguchi, Matsumoto, Tajiri, Hara,
  Ozaki, Takeya, and Takano]{Masaki2012}
Masaki Mito, Takumi Imakyurei, Hiroyuki Deguchi, Kaname Matsumoto, Takayuki
  Tajiri, Hiroshi Hara, Toshinori Ozaki, Hiroyuki Takeya, and Yoshihiko Takano.
\newblock {Uniaxial Strain Effects on Cuprate Superconductor
  YBa$_2$Cu$_4$O$_8$}.
\newblock \emph{Journal of the Physical Society of Japan}, 81\penalty0
  (11):\penalty0 113709, 2012.
\newblock \doi{10.1143/JPSJ.81.113709}.

\bibitem[Mitrano et~al.(2018)Mitrano, Husain, Vig, Kogar, Rak, Rubeck,
  Schmalian, Uchoa, Schneeloch, Zhong, et~al.]{mitrano2018anomalous}
M~Mitrano, AA~Husain, S~Vig, A~Kogar, MS~Rak, SI~Rubeck, J~Schmalian, B~Uchoa,
  J~Schneeloch, R~Zhong, et~al.
\newblock Anomalous density fluctuations in a strange metal.
\newblock \emph{Proceedings of the National Academy of Sciences}, 115\penalty0
  (21):\penalty0 5392--5396, 2018.
\newblock \doi{10.1073/pnas.1721495115}.

\bibitem[Mizuguchi et~al.(2008)Mizuguchi, Tomioka, Tsuda, Yamaguchi, and
  Takano]{mizuguchi2008superconductivity}
Yoshikazu Mizuguchi, Fumiaki Tomioka, Shunsuke Tsuda, Takahide Yamaguchi, and
  Yoshihiko Takano.
\newblock {Superconductivity at 27 K in tetragonal FeSe under high pressure}.
\newblock \emph{Applied Physics Letters}, 93\penalty0 (15):\penalty0 152505,
  2008.
\newblock \doi{10.1063/1.3000616}.

\bibitem[Mohan et~al.(2016)Mohan, Saxena, Kundu, and Rao]{PhysRevB.94.235419}
Priyanka Mohan, Ruchi Saxena, Arijit Kundu, and Sumathi Rao.
\newblock Brillouin-wigner theory for {Floquet} topological phase transitions
  in spin-orbit-coupled materials.
\newblock \emph{Phys. Rev. B}, 94:\penalty0 235419, 2016.
\newblock \doi{10.1103/PhysRevB.94.235419}.
\newblock URL \url{https://link.aps.org/doi/10.1103/PhysRevB.94.235419}.

\bibitem[Moll(2018)]{moll2018focused}
Philip~JW Moll.
\newblock Focused ion beam microstructuring of quantum matter.
\newblock \emph{Annual Review of Condensed Matter Physics}, 9:\penalty0
  147--162, 2018.
\newblock \doi{10.1146/annurev-conmatphys-033117-054021}.

\bibitem[Moll et~al.(2016{\natexlab{a}})Moll, Nair, Helm, Potter, Kimchi,
  Vishwanath, and Analytis]{moll2016transport}
Philip~JW Moll, Nityan~L Nair, Toni Helm, Andrew~C Potter, Itamar Kimchi,
  Ashvin Vishwanath, and James~G Analytis.
\newblock {Transport evidence for Fermi-arc-mediated chirality transfer in the
  {Dirac} semimetal Cd$_3$ As$_2$}.
\newblock \emph{Nature}, 535\penalty0 (7611):\penalty0 266--270,
  2016{\natexlab{a}}.
\newblock \doi{10.1038/nature18276}.

\bibitem[Moll et~al.(2016{\natexlab{b}})Moll, Potter, Nair, Ramshaw, Modic,
  Riggs, Zeng, Ghimire, Bauer, Kealhofer, et~al.]{moll2016magnetic}
Philip~JW Moll, Andrew~C Potter, Nityan~L Nair, BJ~Ramshaw, KA~Modic, Scott
  Riggs, Bin Zeng, Nirmal~J Ghimire, Eric~D Bauer, Robert Kealhofer, et~al.
\newblock {Magnetic torque anomaly in the quantum limit of {Weyl} semimetals}.
\newblock \emph{Nature Communications}, 7\penalty0 (1):\penalty0 1--7,
  2016{\natexlab{b}}.
\newblock \doi{10.1038/ncomms12492}.

\bibitem[Morampudi et~al.(2017)Morampudi, Turner, Pollmann, and
  Wilczek]{Morampudi_PRL_2017}
Siddhardh~C. Morampudi, Ari~M. Turner, Frank Pollmann, and Frank Wilczek.
\newblock Statistics of fractionalized excitations through threshold
  spectroscopy.
\newblock \emph{Phys. Rev. Lett.}, 118:\penalty0 227201, 2017.
\newblock \doi{10.1103/PhysRevLett.118.227201}.
\newblock URL \url{https://link.aps.org/doi/10.1103/PhysRevLett.118.227201}.

\bibitem[Mori et~al.(2016)Mori, Kuwahara, and Saito]{PhysRevLett.116.120401}
Takashi Mori, Tomotaka Kuwahara, and Keiji Saito.
\newblock Rigorous bound on energy absorption and generic relaxation in
  periodically driven quantum systems.
\newblock \emph{Phys. Rev. Lett.}, 116:\penalty0 120401, 2016.
\newblock \doi{10.1103/PhysRevLett.116.120401}.
\newblock URL \url{https://link.aps.org/doi/10.1103/PhysRevLett.116.120401}.

\bibitem[Morice et~al.(2017)Morice, Montiel, and
  P{\'e}pin]{morice2017evolution}
C~Morice, X~Montiel, and C~P{\'e}pin.
\newblock Evolution of {Hall} resistivity and spectral function with doping in
  the {SU}(2) theory of cuprates.
\newblock \emph{Phys. Rev. B}, 96\penalty0 (13):\penalty0 134511, 2017.
\newblock \doi{10.1103/PhysRevB.96.134511}.

\bibitem[Morimoto and Nagaosa(2016)]{morimoto2016topological}
Takahiro Morimoto and Naoto Nagaosa.
\newblock Topological nature of nonlinear optical effects in solids.
\newblock \emph{Science Advances}, 2\penalty0 (5):\penalty0 e1501524, 2016.
\newblock \doi{10.1126/sciadv.1501524}.

\bibitem[Moriya and Takahashi(1984)]{moriya1984itinerant}
Toru Moriya and Yoshinori Takahashi.
\newblock Itinerant electron magnetism.
\newblock \emph{Annual Review of Materials Science}, 14\penalty0 (1):\penalty0
  1--25, 1984.
\newblock \doi{10.1146/annurev.ms.14.080184.000245}.

\bibitem[Mourigal et~al.(2013)Mourigal, Enderle, Kl{\"o}pperpieper, Caux,
  Stunault, and R{\o}nnow]{mourigal2013fractional}
Martin Mourigal, Mechthild Enderle, Axel Kl{\"o}pperpieper, Jean-S{\'e}bastien
  Caux, Anne Stunault, and Henrik~M R{\o}nnow.
\newblock Fractional spinon excitations in the quantum {Heisenberg}
  antiferromagnetic chain.
\newblock \emph{Nature Physics}, 9\penalty0 (7):\penalty0 435--441, 2013.
\newblock \doi{10.1038/nphys2652}.

\bibitem[Mourik et~al.(2012)Mourik, Zuo, Frolov, Plissard, Bakkers, and
  Kouwenhoven]{mourik2012signatures}
Vincent Mourik, Kun Zuo, Sergey~M Frolov, SR~Plissard, Erik~PAM Bakkers, and
  Leo~P Kouwenhoven.
\newblock Signatures of {Majorana} fermions in hybrid
  superconductor-semiconductor nanowire devices.
\newblock \emph{Science}, 336\penalty0 (6084):\penalty0 1003--1007, 2012.
\newblock \doi{10.1126/science.1222360}.

\bibitem[Mousatov and Hartnoll(2020)]{mousatov2019planckian}
Connie~H Mousatov and Sean~A Hartnoll.
\newblock On the planckian bound for heat diffusion in insulators.
\newblock \emph{Nature Physics}, 16\penalty0 (5):\penalty0 579--584, 2020.
\newblock \doi{10.1073/pnas.1915224117}.

\bibitem[Mousatov et~al.(2019)Mousatov, Esterlis, and
  Hartnoll]{mousatov2019bad}
Connie~H Mousatov, Ilya Esterlis, and Sean~A Hartnoll.
\newblock Bad metallic transport in a modified hubbard model.
\newblock \emph{Phys. Rev. Lett.}, 122\penalty0 (18):\penalty0 186601, 2019.
\newblock \doi{10.1103/PhysRevLett.122.186601}.

\bibitem[Mousatov et~al.(2020)Mousatov, Berg, and Hartnoll]{berg2018theory}
Connie~H Mousatov, Erez Berg, and Sean~A Hartnoll.
\newblock {Theory of the strange metal Sr$_3$Ru$_2$O$_7$}.
\newblock \emph{Proceedings of the National Academy of Sciences}, 117\penalty0
  (6):\penalty0 2852--2857, 2020.
\newblock \doi{10.1073/pnas.1915224117}.

\bibitem[Moussa and Cohen(2006)]{moussa2006two}
Jonathan~E Moussa and Marvin~L Cohen.
\newblock Two bounds on the maximum phonon-mediated superconducting transition
  temperature.
\newblock \emph{Phys. Rev. B}, 74\penalty0 (9):\penalty0 094520, 2006.
\newblock \doi{10.1103/PhysRevB.74.094520}.

\bibitem[M\"uhlbacher and Rabani(2008)]{dynamical_sign}
Lothar M\"uhlbacher and Eran Rabani.
\newblock Real-time path integral approach to nonequilibrium many-body quantum
  systems.
\newblock \emph{Phys. Rev. Lett.}, 100:\penalty0 176403, 2008.
\newblock \doi{10.1103/PhysRevLett.100.176403}.
\newblock URL \url{https://link.aps.org/doi/10.1103/PhysRevLett.100.176403}.

\bibitem[Muller et~al.(2008)Muller, Kourkoutis, Murfitt, Song, Hwang, Silcox,
  Dellby, and Krivanek]{muller_atomic-scale_2008}
D~A Muller, L~Fitting Kourkoutis, M~Murfitt, J~H Song, H~Y Hwang, J~Silcox,
  N~Dellby, and O~L Krivanek.
\newblock Atomic-scale chemical imaging of composition and bonding by
  aberration-corrected microscopy.
\newblock \emph{Science (New York, N.Y.)}, 319\penalty0 (5866):\penalty0
  1073--6, 2008.
\newblock \doi{10.1126/science.1148820}.
\newblock URL \url{http://www.ncbi.nlm.nih.gov/pubmed/18292338}.

\bibitem[Muller et~al.(2012)Muller, Mundy, Kourkoutis, Warusawithana, Ludwig,
  Roy, Pawlicki, Heeg, Richter, Paetel, et~al.]{muller2012atomic}
DA~Muller, JA~Mundy, L~Fitting Kourkoutis, MP~Warusawithana, J~Ludwig, P~Roy,
  AA~Pawlicki, T~Heeg, C~Richter, S~Paetel, et~al.
\newblock Atomic-resolution electron spectroscopy of interfaces and defects in
  complex oxides.
\newblock In \emph{Frontiers in Electronic Materials: A Collection of Extended
  Abstracts of the Nature Conference Frontiers in Electronic Materials, June
  17th to 20th 2012, Aachen, Germany}, pages 32--32. Wiley Online Library,
  2012.
\newblock \doi{10.1002/9783527667703.ch2}.

\bibitem[Mundy et~al.(2014)Mundy, Hikita, Hidaka, Yajima, Higuchi, Hwang,
  Muller, and Kourkoutis]{mundy2014visualizing}
Julia~A Mundy, Yasuyuki Hikita, Takeaki Hidaka, Takeaki Yajima, Takuya Higuchi,
  Harold~Y Hwang, David~A Muller, and Lena~F Kourkoutis.
\newblock Visualizing the interfacial evolution from charge compensation to
  metallic screening across the manganite metal--insulator transition.
\newblock \emph{Nature Communications}, 5\penalty0 (1):\penalty0 1--6, 2014.
\newblock \doi{10.1038/ncomms4464}.

\bibitem[Murakami and Nagaosa(2003)]{murakami2003berry}
Shuichi Murakami and Naoto Nagaosa.
\newblock Berry phase in magnetic superconductors.
\newblock \emph{Phys. Rev. Lett.}, 90\penalty0 (5):\penalty0 057002, 2003.
\newblock \doi{10.1126/science.1087128}.

\bibitem[Mydosh and Oppeneer(2014)]{mydosh2014hidden}
John~A Mydosh and Peter~M Oppeneer.
\newblock {Hidden order behaviour in URu$_2$Si$_2$ (A critical review of the
  status of hidden order in 2014)}.
\newblock \emph{Philosophical Magazine}, 94\penalty0 (32-33):\penalty0
  3642--3662, 2014.
\newblock \doi{10.1080/14786435.2014.916428}.

\bibitem[Nag et~al.(2019)Nag, Slager, Higuchi, and Oka]{RobertPRB2019}
Tanay Nag, Robert-Jan Slager, Takuya Higuchi, and Takashi Oka.
\newblock Dynamical synchronization transition in interacting electron systems.
\newblock \emph{Phys. Rev. B}, 100:\penalty0 134301, 2019.
\newblock \doi{10.1103/PhysRevB.100.134301}.
\newblock URL \url{https://link.aps.org/doi/10.1103/PhysRevB.100.134301}.

\bibitem[Nagata et~al.(1998)Nagata, Uehara, Goto, Akimitsu, Motoyama, Eisaki,
  Uchida, Takahashi, Nakanishi, and M\^ori]{Nagata1998}
T.~Nagata, M.~Uehara, J.~Goto, J.~Akimitsu, N.~Motoyama, H.~Eisaki, S.~Uchida,
  H.~Takahashi, T.~Nakanishi, and N.~M\^ori.
\newblock {Pressure-Induced Dimensional Crossover and Superconductivity in the
  Hole-Doped Two-Leg Ladder Compound
  ${\mathrm{Sr}}_{14\ensuremath{-}\mathit{x}}{\mathrm{Ca}}_{\mathit{x}}{\mathrm{Cu}}_{24}{O}_{41}$}.
\newblock \emph{Phys. Rev. Lett.}, 81:\penalty0 1090--1093, 1998.
\newblock \doi{10.1103/PhysRevLett.81.1090}.
\newblock URL \url{https://link.aps.org/doi/10.1103/PhysRevLett.81.1090}.

\bibitem[Nakamura et~al.(2018)Nakamura, Ikeda, Sawabe, Matsuda, and
  Takeyama]{nakamura2018record}
Daisuke Nakamura, A~Ikeda, H~Sawabe, YH~Matsuda, and S~Takeyama.
\newblock {Record indoor magnetic field of 1200 T generated by electromagnetic
  flux-compression}.
\newblock \emph{Review of Scientific Instruments}, 89\penalty0 (9):\penalty0
  095106, 2018.
\newblock \doi{10.1063/1.5044557}.

\bibitem[Narumi et~al.(2012)Narumi, Nakamura, Kinoshita, Matsuda, and
  Nojiri]{narumi2012x}
Y~Narumi, T~Nakamura, T~Kinoshita, YH~Matsuda, and Hiroyuki Nojiri.
\newblock {X-ray Spectroscopies in Pulsed High Magnetic Fields: New Frontier
  with Flying Magnets and Rolling Capacitor Banks}.
\newblock \emph{Synchrotron Radiation News}, 25\penalty0 (6):\penalty0 12--17,
  2012.
\newblock \doi{10.1080/08940886.2012.736833}.

\bibitem[Nasu and Motome(2019)]{nasu2019nonequilibrium}
Joji Nasu and Yukitoshi Motome.
\newblock {Nonequilibrium Majorana dynamics by quenching a magnetic field in
  Kitaev spin liquids}.
\newblock \emph{Physical Review Research}, 1\penalty0 (3):\penalty0 033007,
  2019.
\newblock \doi{10.1103/PhysRevResearch.1.033007}.

\bibitem[Neacsu et~al.(2009)Neacsu, van Aken, Fiebig, and
  Raschke]{neacsu2009second}
Catalin~C Neacsu, Bas~B van Aken, Manfred Fiebig, and Markus~B Raschke.
\newblock {Second-harmonic near-field imaging of ferroelectric domain structure
  of YMnO$_3$}.
\newblock \emph{Phys. Rev. B}, 79\penalty0 (10):\penalty0 100107, 2009.
\newblock \doi{10.1103/PhysRevB.79.100107}.

\bibitem[Nelson et~al.(2004)Nelson, Mao, Maeno, and Liu]{nelson2004odd}
KD~Nelson, ZQ~Mao, Y~Maeno, and Ying Liu.
\newblock {Odd-parity superconductivity in Sr$_2$RuO$_4$}.
\newblock \emph{Science}, 306\penalty0 (5699):\penalty0 1151--1154, 2004.
\newblock \doi{10.1126/science.1103881}.

\bibitem[Neville et~al.(1972)Neville, Hoeneisen, and Mead]{NevilleSTO}
RC~Neville, B~Hoeneisen, and CA~Mead.
\newblock Permittivity of strontium titanate.
\newblock \emph{Journal of Applied Physics}, 43\penalty0 (5):\penalty0
  2124--2131, 1972.
\newblock \doi{10.1063/1.1661463}.

\bibitem[Nickerson et~al.(1971)Nickerson, White, Lee, Bachmann, Geballe, and
  Hull]{SmB62}
J.~C. Nickerson, R.~M. White, K.~N. Lee, R.~Bachmann, T.~H. Geballe, and G.~W.
  Hull.
\newblock {Physical Properties of Sm${\mathrm{B}}_{6}$}.
\newblock \emph{Phys. Rev. B}, 3:\penalty0 2030--2042, 1971.
\newblock \doi{10.1103/PhysRevB.3.2030}.
\newblock URL \url{https://link.aps.org/doi/10.1103/PhysRevB.3.2030}.

\bibitem[Norman(2011)]{norman2011challenge}
Michael~R Norman.
\newblock The challenge of unconventional superconductivity.
\newblock \emph{Science}, 332\penalty0 (6026):\penalty0 196--200, 2011.
\newblock \doi{10.1126/science.1200181}.

\bibitem[Norman and P{\'e}pin(2003)]{norman2003electronic}
MR~Norman and C~P{\'e}pin.
\newblock The electronic nature of high temperature cuprate superconductors.
\newblock \emph{Reports on Progress in Physics}, 66\penalty0 (10):\penalty0
  1547, 2003.
\newblock \doi{10.1088/0034-4885/66/10/R01}.

\bibitem[Norman et~al.(2000)Norman, Randeria, Janko, and
  Campuzano]{norman2000condensation}
MR~Norman, M~Randeria, B~Janko, and JC~Campuzano.
\newblock Condensation energy and spectral functions in high-temperature
  superconductors.
\newblock \emph{Phys. Rev. B}, 61\penalty0 (21):\penalty0 14742, 2000.
\newblock \doi{10.1103/PhysRevB.61.14742}.

\bibitem[Nova et~al.(2019)Nova, Disa, Fechner, and
  Cavalleri]{nova2019metastable}
T.~F. Nova, A.~S. Disa, M.~Fechner, and A.~Cavalleri.
\newblock {Metastable ferroelectricity in optically strained SrTiO$_3$}.
\newblock \emph{Science}, 364\penalty0 (6445):\penalty0 1075--1079, 2019.
\newblock \doi{10.1126/science.aaw4911}.

\bibitem[Novoselov et~al.(2016)Novoselov, Mishchenko, Carvalho, and
  Neto]{novoselov20162d}
KS~Novoselov, A~Mishchenko, A~Carvalho, and AH~Castro Neto.
\newblock {2{D} materials and van der Waals heterostructures}.
\newblock \emph{Science}, 353\penalty0 (6298):\penalty0 aac9439, 2016.
\newblock \doi{10.1126/science.aac9439}.

\bibitem[O'Brien et~al.(2016)O'Brien, Hermanns, and Trebst]{PhysRevB.93.085101}
Kevin O'Brien, Maria Hermanns, and Simon Trebst.
\newblock {Classification of gapless ${\mathbb{Z}}_{2}$ spin liquids in
  three-dimensional Kitaev models}.
\newblock \emph{Phys. Rev. B}, 93:\penalty0 085101, 2016.
\newblock \doi{10.1103/PhysRevB.93.085101}.
\newblock URL \url{https://link.aps.org/doi/10.1103/PhysRevB.93.085101}.

\bibitem[Oeschler et~al.(2003)Oeschler, Gegenwart, Lang, Movshovich, Sarrao,
  Thompson, and Steglich]{PhysRevLett.91.076402}
N.~Oeschler, P.~Gegenwart, M.~Lang, R.~Movshovich, J.~L. Sarrao, J.~D.
  Thompson, and F.~Steglich.
\newblock {Uniaxial Pressure Effects on
  ${\mathrm{C}\mathrm{e}\mathrm{I}\mathrm{r}\mathrm{I}\mathrm{n}}_{5}$ and
  ${\mathrm{C}\mathrm{e}\mathrm{C}\mathrm{o}\mathrm{I}\mathrm{n}}_{5}$ Studied
  by Low-Temperature Thermal Expansion}.
\newblock \emph{Phys. Rev. Lett.}, 91:\penalty0 076402, 2003.
\newblock \doi{10.1103/PhysRevLett.91.076402}.
\newblock URL \url{https://link.aps.org/doi/10.1103/PhysRevLett.91.076402}.

\bibitem[Oka and Kitamura(2019)]{OkaReview2019}
Takashi Oka and Sota Kitamura.
\newblock Floquet engineering of quantum materials.
\newblock \emph{Annual Review of Condensed Matter Physics}, 10\penalty0
  (1):\penalty0 387--408, 2019.
\newblock \doi{10.1146/annurev-conmatphys-031218-013423}.

\bibitem[Okamoto(1971)]{Cu2Sethermopower1}
Kimihiko Okamoto.
\newblock {Thermoelectric Power and Phase Transition of Cu$_2$Se}.
\newblock \emph{Japanese Journal of Applied Physics}, 10\penalty0 (4):\penalty0
  508--508, 1971.
\newblock \doi{10.1143/jjap.10.508}.
\newblock URL \url{https://doi.org/10.1143%2Fjjap.10.508}.

\bibitem[Osheroff et~al.(1972)Osheroff, Richardson, and
  Lee]{osheroff1972evidence}
DD~Osheroff, RC~Richardson, and DM~Lee.
\newblock Evidence for a new phase of solid he$^3$.
\newblock \emph{Phys. Rev. Lett.}, 28\penalty0 (14):\penalty0 885, 1972.
\newblock \doi{10.1103/PhysRevLett.28.885}.

\bibitem[Oshikawa(2000)]{oshikawa2000topological}
Masaki Oshikawa.
\newblock {Topological approach to Luttinger's theorem and the {Fermi} surface
  of a {Kondo} lattice}.
\newblock \emph{Phys. Rev. Lett.}, 84\penalty0 (15):\penalty0 3370, 2000.
\newblock \doi{10.1103/PhysRevLett.84.3370}.

\bibitem[Paddison et~al.(2017)Paddison, Daum, Dun, Ehlers, Liu, Stone, Zhou,
  and Mourigal]{paddison2017continuous}
Joseph~AM Paddison, Marcus Daum, Zhiling Dun, Georg Ehlers, Yaohua Liu,
  Matthew~B Stone, Haidong Zhou, and Martin Mourigal.
\newblock {Continuous excitations of the triangular-lattice quantum spin liquid
  YbMgGaO$_4$}.
\newblock \emph{Nature Physics}, 13\penalty0 (2):\penalty0 117--122, 2017.
\newblock \doi{10.1038/nphys3971}.

\bibitem[Paeckel et~al.(2019)Paeckel, Köhler, Swoboda, Manmana, Schollwöck,
  and Hubig]{tMPS}
Sebastian Paeckel, Thomas Köhler, Andreas Swoboda, Salvatore~R. Manmana,
  Ulrich Schollwöck, and Claudius Hubig.
\newblock Time-evolution methods for matrix-product states.
\newblock \emph{Annals of Physics}, 411:\penalty0 167998, 2019.
\newblock \doi{10.1016/j.aop.2019.167998}.
\newblock URL
  \url{http://www.sciencedirect.com/science/article/pii/S0003491619302532}.

\bibitem[Pan et~al.(2008)Pan, Xia, Stormer, Tsui, Vicente, Adams, Sullivan,
  Pfeiffer, Baldwin, and West]{pan2008experimental}
Wei Pan, JS~Xia, HL~Stormer, DC~Tsui, C~Vicente, ED~Adams, NS~Sullivan,
  LN~Pfeiffer, KW~Baldwin, and KW~West.
\newblock Experimental studies of the fractional quantum hall effect in the
  first excited landau level.
\newblock \emph{Phys. Rev. B}, 77\penalty0 (7):\penalty0 075307, 2008.
\newblock \doi{10.1103/PhysRevB.77.075307}.

\bibitem[Paolasini et~al.(2007)Paolasini, Detlefs, Mazzoli, Wilkins, Deen,
  Bombardi, Kernavanois, De~Bergevin, Yakhou, Valade,
  et~al.]{paolasini2007id20}
L~Paolasini, C~Detlefs, Claudio Mazzoli, S~Wilkins, PP~Deen, A~Bombardi,
  N~Kernavanois, F~De~Bergevin, F~Yakhou, JP~Valade, et~al.
\newblock {ID20: a beamline for magnetic and resonant X-ray scattering
  investigations under extreme conditions}.
\newblock \emph{Journal of Synchrotron Radiation}, 14\penalty0 (4):\penalty0
  301--312, 2007.
\newblock \doi{10.1107/S0909049507024879}.

\bibitem[Pashkin et~al.(2010)Pashkin, Dressel, Hanfland, and
  Kuntscher]{Pashkin2010}
A.~Pashkin, M.~Dressel, M.~Hanfland, and C.~A. Kuntscher.
\newblock {Deconfinement transition and dimensional crossover in the
  Bechgaard-Fabre salts: Pressure- and temperature-dependent optical
  investigations}.
\newblock \emph{Phys. Rev. B}, 81:\penalty0 125109, 2010.
\newblock \doi{10.1103/PhysRevB.81.125109}.
\newblock URL \url{https://link.aps.org/doi/10.1103/PhysRevB.81.125109}.

\bibitem[Patel and Sachdev(2019)]{patel2019theory}
Aavishkar~A Patel and Subir Sachdev.
\newblock Theory of a planckian metal.
\newblock \emph{Phys. Rev. Lett.}, 123\penalty0 (6):\penalty0 066601, 2019.
\newblock \doi{10.1103/PhysRevLett.123.066601}.

\bibitem[Patel et~al.(2018)Patel, McGreevy, Arovas, and
  Sachdev]{patel2018magnetotransport}
Aavishkar~A Patel, John McGreevy, Daniel~P Arovas, and Subir Sachdev.
\newblock Magnetotransport in a model of a disordered strange metal.
\newblock \emph{Physical Review X}, 8\penalty0 (2):\penalty0 021049, 2018.
\newblock \doi{10.1103/PhysRevX.8.021049}.

\bibitem[Patri et~al.(2019)Patri, Sakai, Lee, Paramekanti, Nakatsuji, and
  Kim]{patri2019unveiling}
Adarsh~S Patri, Akito Sakai, SungBin Lee, Arun Paramekanti, Satoru Nakatsuji,
  and Yong~Baek Kim.
\newblock Unveiling hidden multipolar orders with magnetostriction.
\newblock \emph{Nature Communications}, 10\penalty0 (1):\penalty0 1--8, 2019.
\newblock \doi{10.1038/s41467-019-11913-3}.

\bibitem[Paul et~al.(2020)Paul, Chung, Birol, and Changlani]{Paul_2020}
Arpita Paul, Chia-Min Chung, Turan Birol, and Hitesh~J. Changlani.
\newblock Spin-lattice coupling and the emergence of the trimerized phase in
  the $s=1$ kagome antiferromagnet
  ${\mathrm{na}}_{2}{\mathrm{ti}}_{3}{\mathrm{cl}}_{8}$.
\newblock \emph{Phys. Rev. Lett.}, 124:\penalty0 167203, 2020.
\newblock \doi{10.1103/PhysRevLett.124.167203}.
\newblock URL \url{https://link.aps.org/doi/10.1103/PhysRevLett.124.167203}.

\bibitem[Paul et~al.(2008)Paul, P{\'e}pin, and Norman]{paul2008multiscale}
I~Paul, C~P{\'e}pin, and MR~Norman.
\newblock Multiscale fluctuations near a {Kondo} breakdown quantum critical
  point.
\newblock \emph{Phys. Rev. B}, 78\penalty0 (3):\penalty0 035109, 2008.
\newblock \doi{10.1103/PhysRevB.78.035109}.

\bibitem[Pavarini et~al.(2001)Pavarini, Dasgupta, Saha-Dasgupta, Jepsen, and
  Andersen]{pavirini}
E.~Pavarini, I.~Dasgupta, T.~Saha-Dasgupta, O.~Jepsen, and O.~K. Andersen.
\newblock {Band-Structure Trend in Hole-Doped Cuprates and Correlation with
  ${\mathit{T}}_{\mathit{c}\mathrm{max}}$}.
\newblock \emph{Phys. Rev. Lett.}, 87:\penalty0 047003, 2001.
\newblock \doi{10.1103/PhysRevLett.87.047003}.

\bibitem[Penz and Bowers(1968)]{PotassiumMR2}
P.~A. Penz and R.~Bowers.
\newblock Strain-dependent magnetoresistance of potassium.
\newblock \emph{Phys. Rev.}, 172:\penalty0 991--1001, 1968.
\newblock \doi{10.1103/PhysRev.172.991}.
\newblock URL \url{https://link.aps.org/doi/10.1103/PhysRev.172.991}.

\bibitem[Pepperhoff and Acet(2013)]{ironbook}
Werner Pepperhoff and Mehmet Acet.
\newblock \emph{Constitution and Magnetism of Iron and its Alloys}.
\newblock Springer Science \& {BCS} iness Media, 2013.
\newblock \doi{10.1007/978-3-662-04345-5}.

\bibitem[Peruzzo et~al.(2013)Peruzzo, Mcclean, Shadbolt, Yung, Zhou, Love,
  Aspuru-Guzik, and O'Brien]{Peruzzo13}
Alberto Peruzzo, Jarrod Mcclean, Peter Shadbolt, Man~Hong Yung, Xiaoqi Zhou,
  Peter Love, Alán Aspuru-Guzik, and Jeremy O'Brien.
\newblock A variational eigenvalue solver on a quantum processor.
\newblock \emph{Nature Communications}, 5, 2013.
\newblock \doi{10.1038/ncomms5213}.

\bibitem[Petruzielo et~al.(2012)Petruzielo, Holmes, Changlani, Nightingale, and
  Umrigar]{SQMC}
F.~R. Petruzielo, A.~A. Holmes, Hitesh~J. Changlani, M.~P. Nightingale, and
  C.~J. Umrigar.
\newblock Semistochastic projector {Monte} {Carlo} method.
\newblock \emph{Phys. Rev. Lett.}, 109:\penalty0 230201, 2012.
\newblock \doi{10.1103/PhysRevLett.109.230201}.

\bibitem[Pfau et~al.(2019)Pfau, Rotundu, Palmstrom, Chen, Hashimoto, Lu,
  Kemper, Fisher, and Shen]{pfau2019detailed}
H~Pfau, CR~Rotundu, JC~Palmstrom, SD~Chen, M~Hashimoto, D~Lu, AF~Kemper,
  IR~Fisher, and Z-X Shen.
\newblock {Detailed band structure of twinned and detwinned BaFe$_2$As$_2$
  studied with angle-resolved photoemission spectroscopy}.
\newblock \emph{Phys. Rev. B}, 99\penalty0 (3):\penalty0 035118, 2019.
\newblock \doi{10.1103/PhysRevB.99.035118}.

\bibitem[Pfeiffer(2018)]{pfeiffer2018x}
Franz Pfeiffer.
\newblock X-ray ptychography.
\newblock \emph{Nature Photonics}, 12\penalty0 (1):\penalty0 9--17, 2018.
\newblock \doi{10.1038/s41566-017-0072-5}.

\bibitem[Pham et~al.(2011)Pham, Le~Sage, Stanwix, Yeung, Glenn, Trifonov,
  Cappellaro, Hemmer, Lukin, Park, et~al.]{pham2011magnetic}
Linh~My Pham, David Le~Sage, Paul~L Stanwix, Tsun~Kwan Yeung, D~Glenn, Alexei
  Trifonov, Paola Cappellaro, Philip~R Hemmer, Mikhail~D Lukin, Hongkun Park,
  et~al.
\newblock Magnetic field imaging with nitrogen-vacancy ensembles.
\newblock \emph{New Journal of Physics}, 13\penalty0 (4):\penalty0 045021,
  2011.
\newblock \doi{10.1088/1367-2630/13/4/045021}.

\bibitem[Phelan et~al.(2014)Phelan, Koohpayeh, Cottingham, Freeland, Leiner,
  Broholm, and McQueen]{Phelan2014}
W.~A. Phelan, S.~M. Koohpayeh, P.~Cottingham, J.~W. Freeland, J.~C. Leiner,
  C.~L. Broholm, and T.~M. McQueen.
\newblock Correlation between bulk thermodynamic measurements and the
  low-temperature-resistance plateau in ${\mathrm{smb}}_{6}$.
\newblock \emph{Phys. Rev. X}, 4:\penalty0 031012, 2014.
\newblock \doi{10.1103/PhysRevX.4.031012}.

\bibitem[Phillips and Chamon(2005)]{phillips2005breakdown}
Philip Phillips and Claudio Chamon.
\newblock Breakdown of one-parameter scaling in quantum critical scenarios for
  high-temperature copper-oxide superconductors.
\newblock \emph{Phys. Rev. Lett.}, 95\penalty0 (10):\penalty0 107002, 2005.
\newblock \doi{10.1103/PhysRevB.72.115119}.

\bibitem[Piccinini et~al.(2005)Piccinini, Cestelli~Guidi, Marcelli, Calvani,
  Burattini, Nucara, Postorino, Sacchetti, Arcangeletti, Sheregii,
  et~al.]{piccinini2005far}
M~Piccinini, M~Cestelli~Guidi, A~Marcelli, P~Calvani, E~Burattini, A~Nucara,
  P~Postorino, A~Sacchetti, E~Arcangeletti, E~Sheregii, et~al.
\newblock Far-infrared synchrotron radiation spectroscopy of solids in normal
  and extreme conditions.
\newblock \emph{physica status solidi (c)}, 2\penalty0 (1):\penalty0 236--239,
  2005.
\newblock \doi{10.1002/pssc.200460154}.

\bibitem[Pippard and Pippard(1989)]{pippard1989magnetoresistance}
Alfred~Brian Pippard and Pippard Alfred~Brian Pippard.
\newblock \emph{Magnetoresistance in metals}, volume~2.
\newblock Cambridge university press, 1989.
\newblock \doi{10.1017/CBO9780511629020}.

\bibitem[Po et~al.(2018)Po, Watanabe, and Vishwanath]{PhysRevLett.121.126402}
Hoi~Chun Po, Haruki Watanabe, and Ashvin Vishwanath.
\newblock Fragile topology and {Wannier} obstructions.
\newblock \emph{Phys. Rev. Lett.}, 121:\penalty0 126402, 2018.
\newblock \doi{10.1103/PhysRevLett.121.126402}.
\newblock URL \url{https://link.aps.org/doi/10.1103/PhysRevLett.121.126402}.

\bibitem[Polchinski(1994)]{POLCHINSKI1994617}
Joseph Polchinski.
\newblock Low-energy dynamics of the spinon-gauge system.
\newblock \emph{Nuclear Physics B}, 422\penalty0 (3):\penalty0 617 -- 633,
  1994.
\newblock \doi{10.1016/0550-3213(94)90449-9}.
\newblock URL
  \url{http://www.sciencedirect.com/science/article/pii/0550321394904499}.

\bibitem[Pollanen et~al.(2012)Pollanen, Li, Collett, Gannon, Halperin, and
  Sauls]{pollanen2012new}
J~Pollanen, JIA Li, CA~Collett, WJ~Gannon, William~P Halperin, and James~A
  Sauls.
\newblock {New chiral phases of superfluid $^3$He stabilized by anisotropic
  silica aerogel}.
\newblock \emph{Nature Physics}, 8\penalty0 (4):\penalty0 317--320, 2012.
\newblock \doi{10.1038/nphys2220}.

\bibitem[Polshyn et~al.(2019)Polshyn, Yankowitz, Chen, Zhang, Watanabe,
  Taniguchi, Dean, and Young]{polshyn2019large}
Hryhoriy Polshyn, Matthew Yankowitz, Shaowen Chen, Yuxuan Zhang, K~Watanabe,
  T~Taniguchi, Cory~R Dean, and Andrea~F Young.
\newblock Large linear-in-temperature resistivity in twisted bilayer graphene.
\newblock \emph{Nature Physics}, 15\penalty0 (10):\penalty0 1011--1016, 2019.
\newblock \doi{10.1038/s41567-019-0596-3}.

\bibitem[Portugall et~al.(1999)Portugall, Puhlmann, Müller, Barczewski,
  Stolpe, and von Ortenberg]{Portugall_1999}
O~Portugall, N~Puhlmann, H~U Müller, M~Barczewski, I~Stolpe, and M~von
  Ortenberg.
\newblock Megagauss magnetic field generation in single-turn coils: new
  frontiers for scientific experiments.
\newblock \emph{Journal of Physics D: Applied Physics}, 32\penalty0
  (18):\penalty0 2354--2366, 1999.
\newblock \doi{10.1088/0022-3727/32/18/306}.
\newblock URL \url{https://doi.org/10.1088%2F0022-3727%2F32%2F18%2F306}.

\bibitem[Powell and McKenzie(2011)]{Powell_2011}
B~J Powell and Ross~H McKenzie.
\newblock Quantum frustration in organic mott insulators: from spin liquids to
  unconventional superconductors.
\newblock \emph{Reports on Progress in Physics}, 74\penalty0 (5):\penalty0
  056501, 2011.
\newblock \doi{10.1088/0034-4885/74/5/056501}.
\newblock URL \url{https://doi.org/10.1088%2F0034-4885%2F74%2F5%2F056501}.

\bibitem[Prakash et~al.(2017)Prakash, Kumar, Thamizhavel, and
  Ramakrishnan]{bismuthSC}
Om~Prakash, Anil Kumar, A~Thamizhavel, and S~Ramakrishnan.
\newblock Evidence for bulk superconductivity in pure bismuth single crystals
  at ambient pressure.
\newblock \emph{Science}, 355\penalty0 (6320):\penalty0 52--55, 2017.
\newblock \doi{10.1126/science.aaf8227}.

\bibitem[Preskill(2018)]{Preskill2018}
John Preskill.
\newblock Quantum {C}omputing in the {NISQ} era and beyond.
\newblock \emph{{Quantum}}, 2:\penalty0 79, 2018.
\newblock \doi{10.22331/q-2018-08-06-79}.
\newblock URL \url{https://doi.org/10.22331/q-2018-08-06-79}.

\bibitem[Proke{\v{s}} et~al.(2017)Proke{\v{s}}, Bartkowiak, Rivin, Prokhnenko,
  F{\"o}rster, Gerischer, Wahle, Huang, and Mydosh]{prokevs2017magnetic}
K~Proke{\v{s}}, M~Bartkowiak, O~Rivin, O~Prokhnenko, T~F{\"o}rster,
  S~Gerischer, R~Wahle, Y-K Huang, and JA~Mydosh.
\newblock {Magnetic structure in a U(Ru$_{0.92}$Rh$_{0.08}$)$_2$Si$_2$ single
  crystal studied by neutron diffraction in static magnetic fields up to 24 T}.
\newblock \emph{Phys. Rev. B}, 96\penalty0 (12):\penalty0 121117, 2017.
\newblock \doi{10.1103/PhysRevB.96.121117}.

\bibitem[Prokof'ev et~al.(1998)Prokof'ev, Svistunov, and
  Tupitsyn]{Prokofev1998}
N.~V. Prokof'ev, B.~V. Svistunov, and I.~S. Tupitsyn.
\newblock Exact, complete, and universal continuous-time worldline {Monte}
  {Carlo} approach to the statistics of discrete quantum systems.
\newblock \emph{Journal of Experimental and Theoretical Physics}, 87\penalty0
  (2):\penalty0 310--321, 1998.
\newblock \doi{10.1134/1.558661}.
\newblock URL \url{https://doi.org/10.1134/1.558661}.

\bibitem[Prokof'ev and Svistunov(2007)]{prokofev07}
Nikolay Prokof'ev and Boris Svistunov.
\newblock Bold diagrammatic {Monte} {Carlo} technique: When the sign problem is
  welcome.
\newblock \emph{Phys. Rev. Lett.}, 99:\penalty0 250201, 2007.
\newblock \doi{10.1103/PhysRevLett.99.250201}.
\newblock URL \url{https://link.aps.org/doi/10.1103/PhysRevLett.99.250201}.

\bibitem[Proust and Taillefer(2019)]{proust2019remarkable}
Cyril Proust and Louis Taillefer.
\newblock The remarkable underlying ground states of cuprate superconductors.
\newblock \emph{Annual Review of Condensed Matter Physics}, 10:\penalty0
  409--429, 2019.
\newblock \doi{10.1103/PhysRevLett.117.167001}.

\bibitem[Pustogow et~al.(2019)Pustogow, Luo, Chronister, Su, Sokolov,
  Jerzembeck, Mackenzie, Hicks, Kikugawa, Raghu, Bauer, and
  Brown]{pustogow2019constraints}
A.~Pustogow, Yongkang Luo, A.~Chronister, Y.-S. Su, D.~A. Sokolov,
  F.~Jerzembeck, A.~P. Mackenzie, C.~W. Hicks, N.~Kikugawa, S.~Raghu, E.~D.
  Bauer, and S.~E. Brown.
\newblock {Constraints on the superconducting order parameter in Sr$_2$RuO$_4$
  from oxygen-17 nuclear magnetic resonance}.
\newblock \emph{Nature}, 574\penalty0 (7776):\penalty0 72--75, 2019.
\newblock \doi{10.1038/s41586-019-1596-2}.

\bibitem[Qi and Zhang(2011)]{qi2011topological}
Xiao-Liang Qi and Shou-Cheng Zhang.
\newblock Topological insulators and superconductors.
\newblock \emph{Reviews of Modern Physics}, 83\penalty0 (4):\penalty0 1057,
  2011.
\newblock \doi{10.1103/RevModPhys.83.1057}.

\bibitem[Quito and Flint(2020)]{quito2020floquet}
V.~L. Quito and Rebecca Flint.
\newblock Floquet engineering magnetic materials with polarized and unpolarized
  light.
\newblock \emph{arXiv preprint arXiv:2003.04272}, 2020.
\newblock \doi{10.1103/PhysRevLett.126.177201}.

\bibitem[Raghu et~al.(2015)Raghu, Torroba, and Wang]{raghu2015metallic}
Srinivas Raghu, Gonzalo Torroba, and Huajia Wang.
\newblock Metallic quantum critical points with finite {BCS} couplings.
\newblock \emph{Phys. Rev. B}, 92\penalty0 (20):\penalty0 205104, 2015.
\newblock \doi{10.1103/PhysRevB.92.205104}.

\bibitem[Rahav et~al.(2003)Rahav, Gilary, and Fishman]{PhysRevA.68.013820}
Saar Rahav, Ido Gilary, and Shmuel Fishman.
\newblock Effective hamiltonians for periodically driven systems.
\newblock \emph{Phys. Rev. A}, 68:\penalty0 013820, 2003.
\newblock \doi{10.1103/PhysRevA.68.013820}.
\newblock URL \url{https://link.aps.org/doi/10.1103/PhysRevA.68.013820}.

\bibitem[Ramazanoglu et~al.(2018)Ramazanoglu, Ueland, Pratt, Harriger, Lynn,
  Ehlers, Granroth, Bud'ko, Canfield, Schlagel, Goldman, Lograsso, and
  McQueeney]{dopedcr2}
M.~Ramazanoglu, B.~G. Ueland, D.~K. Pratt, L.~W. Harriger, J.~W. Lynn,
  G.~Ehlers, G.~E. Granroth, S.~L. Bud'ko, P.~C. Canfield, D.~L. Schlagel,
  A.~I. Goldman, T.~A. Lograsso, and R.~J. McQueeney.
\newblock {Suppression of antiferromagnetic spin fluctuations in
  superconducting ${\mathrm{Cr}}_{0.8}{\mathrm{Ru}}_{0.2}$}.
\newblock \emph{Phys. Rev. B}, 98:\penalty0 134512, 2018.
\newblock \doi{10.1103/PhysRevB.98.134512}.
\newblock URL \url{https://link.aps.org/doi/10.1103/PhysRevB.98.134512}.

\bibitem[Ramshaw et~al.(2018)Ramshaw, Modic, Shekhter, Zhang, Kim, Moll,
  Bachmann, Chan, Betts, Balakirev, et~al.]{ramshaw2018quantum}
BJ~Ramshaw, KA~Modic, Arkady Shekhter, Yi~Zhang, Eun-Ah Kim, Philip~JW Moll,
  Maja~D Bachmann, MK~Chan, JB~Betts, F~Balakirev, et~al.
\newblock {Quantum limit transport and destruction of the {Weyl} nodes in
  TaAs}.
\newblock \emph{Nature communications}, 9\penalty0 (1):\penalty0 1--9, 2018.
\newblock \doi{10.1038/s41467-018-04542-9}.

\bibitem[Ran et~al.(2019{\natexlab{a}})Ran, Eckberg, Ding, Furukawa, Metz,
  Saha, Liu, Zic, Kim, Paglione, et~al.]{ran2019nearly}
Sheng Ran, Chris Eckberg, Qing-Ping Ding, Yuji Furukawa, Tristin Metz, Shanta~R
  Saha, I-Lin Liu, Mark Zic, Hyunsoo Kim, Johnpierre Paglione, et~al.
\newblock Nearly ferromagnetic spin-triplet superconductivity.
\newblock \emph{Science}, 365\penalty0 (6454):\penalty0 684--687,
  2019{\natexlab{a}}.
\newblock \doi{10.1126/science.aav8645}.

\bibitem[Ran et~al.(2019{\natexlab{b}})Ran, Liu, Eo, Campbell, Neves, Fuhrman,
  Saha, Eckberg, Kim, Graf, et~al.]{ran2019extreme}
Sheng Ran, I-Lin Liu, Yun~Suk Eo, Daniel~J Campbell, Paul~M Neves, Wesley~T
  Fuhrman, Shanta~R Saha, Christopher Eckberg, Hyunsoo Kim, David Graf, et~al.
\newblock Extreme magnetic field-boosted superconductivity.
\newblock \emph{Nature Physics}, 15\penalty0 (12):\penalty0 1250--1254,
  2019{\natexlab{b}}.
\newblock \doi{10.1038/s41567-019-0670-x}.

\bibitem[Reimann et~al.(2018)Reimann, Schlauderer, Schmid, Langer, Baierl,
  Kokh, Tereshchenko, Kimura, Lange, G{\"{u}}dde, H{\"{o}}fer, and
  Huber]{reimann2018arpes}
J.~Reimann, S.~Schlauderer, C.~P. Schmid, F.~Langer, S.~Baierl, K.~A. Kokh,
  O.~E. Tereshchenko, A.~Kimura, C.~Lange, J.~G{\"{u}}dde, U.~H{\"{o}}fer, and
  R.~Huber.
\newblock {Subcycle observation of lightwave-driven {Dirac} currents in a
  topological surface band}.
\newblock \emph{Nature}, 562\penalty0 (7727):\penalty0 396--400, 2018.
\newblock \doi{10.1038/s41586-018-0544-x}.

\bibitem[Reitz and Overhauser(1968)]{PotassiumMR1}
John~R. Reitz and A.~W. Overhauser.
\newblock Magnetoresistance of potassium.
\newblock \emph{Phys. Rev.}, 171:\penalty0 749--753, 1968.
\newblock \doi{10.1103/PhysRev.171.749}.
\newblock URL \url{https://link.aps.org/doi/10.1103/PhysRev.171.749}.

\bibitem[Requist and Gross(2019)]{Requist19}
Ryan Requist and E.~K.~U. Gross.
\newblock Model hamiltonian for strongly correlated systems: Systematic,
  self-consistent, and unique construction.
\newblock \emph{Phys. Rev. B}, 99:\penalty0 125114, 2019.
\newblock \doi{10.1103/PhysRevB.99.125114}.
\newblock URL \url{https://link.aps.org/doi/10.1103/PhysRevB.99.125114}.

\bibitem[Rhodes et~al.(2019)Rhodes, Chae, Ribeiro-Palau, and
  Hone]{rhodes2019disorder}
Daniel Rhodes, Sang~Hoon Chae, Rebeca Ribeiro-Palau, and James Hone.
\newblock {Disorder in van der Waals heterostructures of 2{D} materials}.
\newblock \emph{Nature materials}, 18\penalty0 (6):\penalty0 541, 2019.
\newblock \doi{10.1038/s41563-019-0366-8}.

\bibitem[Ricc{\`{o}} et~al.(2018)Ricc{\`{o}}, Kim, Tamai, {McKeown Walker},
  Bruno, Cucchi, Cappelli, Besnard, Kim, Dudin, Hoesch, Gutmann, Georges,
  Perry, and Baumberger]{ricco2018cro}
S.~Ricc{\`{o}}, M.~Kim, A.~Tamai, S.~{McKeown Walker}, F.~Y. Bruno, I.~Cucchi,
  E.~Cappelli, C.~Besnard, T.~K. Kim, P.~Dudin, M.~Hoesch, M.~J. Gutmann,
  A.~Georges, R.~S. Perry, and F.~Baumberger.
\newblock {In situ strain tuning of the metal-insulator-transition of
  Ca$_2$RuO$_4$ in angle-resolved photoemission experiments}.
\newblock \emph{Nature Communications}, 9\penalty0 (1):\penalty0 4535, 2018.
\newblock \doi{10.1038/s41467-018-06945-0}.

\bibitem[Riggs et~al.(2011)Riggs, Vafek, Kemper, Betts, Migliori, Balakirev,
  Hardy, Liang, Bonn, and Boebinger]{riggs2011heat}
Scott~C Riggs, O~Vafek, JB~Kemper, JB~Betts, A~Migliori, FF~Balakirev,
  WN~Hardy, Ruixing Liang, DA~Bonn, and GS~Boebinger.
\newblock Heat capacity through the magnetic-field-induced resistive transition
  in an underdoped high-temperature superconductor.
\newblock \emph{Nature Physics}, 7\penalty0 (4):\penalty0 332--335, 2011.
\newblock \doi{10.1038/nphys1921}.

\bibitem[Robinson and Harder(2009)]{robinson2009coherent}
Ian Robinson and Ross Harder.
\newblock Coherent x-ray diffraction imaging of strain at the nanoscale.
\newblock \emph{Nature materials}, 8\penalty0 (4):\penalty0 291--298, 2009.
\newblock \doi{10.1038/nmat2400}.

\bibitem[Robinson et~al.(2020)Robinson, Assefa, Cao, Gu, Harder, Maxey, and
  Dean]{robinson2020domain}
Ian Robinson, Tadesse~A Assefa, Yue Cao, Genda Gu, Ross Harder, Evan Maxey, and
  Mark~PM Dean.
\newblock {Domain Texture of the Orthorhombic Phase of La$_{2-x}$ Ba$_x$
  CuO$_4$}.
\newblock \emph{Journal of Superconductivity and Novel Magnetism}, 33\penalty0
  (1):\penalty0 99--106, 2020.
\newblock \doi{10.1007/s10948-019-05252-z}.

\bibitem[Rodriguez-Vega and Seradjeh(2018)]{PhysRevLett.121.036402}
M.~Rodriguez-Vega and B.~Seradjeh.
\newblock Universal fluctuations of {Floquet} topological invariants at low
  frequencies.
\newblock \emph{Phys. Rev. Lett.}, 121:\penalty0 036402, 2018.
\newblock \doi{10.1103/PhysRevLett.121.036402}.
\newblock URL \url{https://link.aps.org/doi/10.1103/PhysRevLett.121.036402}.

\bibitem[Rodriguez-Vega et~al.(2018)Rodriguez-Vega, Fertig, and
  Seradjeh]{PhysRevB.98.041113}
M.~Rodriguez-Vega, H.~A. Fertig, and B.~Seradjeh.
\newblock Quantum noise detects {Floquet} topological phases.
\newblock \emph{Phys. Rev. B}, 98:\penalty0 041113, 2018.
\newblock \doi{10.1103/PhysRevB.98.041113}.
\newblock URL \url{https://link.aps.org/doi/10.1103/PhysRevB.98.041113}.

\bibitem[Rodriguez-Vega et~al.(2019)Rodriguez-Vega, Kumar, and
  Seradjeh]{PhysRevB.100.085138}
Martin Rodriguez-Vega, Abhishek Kumar, and Babak Seradjeh.
\newblock Higher-order {Floquet} topological phases with corner and bulk bound
  states.
\newblock \emph{Phys. Rev. B}, 100:\penalty0 085138, 2019.
\newblock \doi{10.1103/PhysRevB.100.085138}.
\newblock URL \url{https://link.aps.org/doi/10.1103/PhysRevB.100.085138}.

\bibitem[Ronning et~al.(2017)Ronning, Helm, Shirer, Bachmann, Balicas, Chan,
  Ramshaw, Mcdonald, Balakirev, Jaime, et~al.]{ronning2017electronic}
Filip Ronning, Toni Helm, KR~Shirer, MD~Bachmann, Luis Balicas, Mun~Keat Chan,
  BJ~Ramshaw, Ross~David Mcdonald, Fedor~Fedorovich Balakirev, Marcelo Jaime,
  et~al.
\newblock {Electronic in-plane symmetry breaking at field-tuned quantum
  criticality in CeRhIn$_5$}.
\newblock \emph{Nature}, 548\penalty0 (7667):\penalty0 313--317, 2017.
\newblock \doi{10.1038/nature23315}.

\bibitem[Rosenbaum et~al.(1980)Rosenbaum, Andres, Thomas, and Bhatt]{Si-MIT1}
T.~F. Rosenbaum, K.~Andres, G.~A. Thomas, and R.~N. Bhatt.
\newblock Sharp metal-insulator transition in a random solid.
\newblock \emph{Phys. Rev. Lett.}, 45:\penalty0 1723--1726, 1980.
\newblock \doi{10.1103/PhysRevLett.45.1723}.
\newblock URL \url{https://link.aps.org/doi/10.1103/PhysRevLett.45.1723}.

\bibitem[Rosenbaum et~al.(1983)Rosenbaum, Milligan, Paalanen, Thomas, Bhatt,
  and Lin]{Si-MIT4}
T.~F. Rosenbaum, R.~F. Milligan, M.~A. Paalanen, G.~A. Thomas, R.~N. Bhatt, and
  W.~Lin.
\newblock Metal-insulator transition in a doped semiconductor.
\newblock \emph{Phys. Rev. B}, 27:\penalty0 7509--7523, 1983.
\newblock \doi{10.1103/PhysRevB.27.7509}.
\newblock URL \url{https://link.aps.org/doi/10.1103/PhysRevB.27.7509}.

\bibitem[Ross et~al.(2011)Ross, Savary, Gaulin, and Balents]{ross2011qsi}
Kate~A. Ross, Lucile Savary, Bruce~D. Gaulin, and Leon Balents.
\newblock {Quantum Excitations in Quantum Spin Ice}.
\newblock \emph{Physical Review X}, 1\penalty0 (2):\penalty0 021002, 2011.
\newblock \doi{10.1103/PhysRevX.1.021002}.

\bibitem[Rossi et~al.(2017)Rossi, Prokof'ev, Svistunov, Houcke, and
  Werner]{rossi17b}
R.~Rossi, N.~Prokof'ev, B.~Svistunov, K.~Van Houcke, and F.~Werner.
\newblock Polynomial complexity despite the fermionic sign.
\newblock \emph{{EPL} (Europhysics Letters)}, 118\penalty0 (1):\penalty0 10004,
  2017.
\newblock \doi{10.1209/0295-5075/118/10004}.
\newblock URL \url{https://doi.org/10.1209%2F0295-5075%2F118%2F10004}.

\bibitem[Rossi(2017)]{rossi17a}
Riccardo Rossi.
\newblock Determinant diagrammatic {Monte} {Carlo} algorithm in the
  thermodynamic limit.
\newblock \emph{Phys. Rev. Lett.}, 119:\penalty0 045701, 2017.
\newblock \doi{10.1103/PhysRevLett.119.045701}.
\newblock URL \url{https://link.aps.org/doi/10.1103/PhysRevLett.119.045701}.

\bibitem[Rozenberg et~al.(2006)Rozenberg, Pasternak, Xu, Amiel, Hanfland,
  Amboage, Taylor, and Jeanloz]{magnetite2}
G.~Kh. Rozenberg, M.~P. Pasternak, W.~M. Xu, Y.~Amiel, M.~Hanfland, M.~Amboage,
  R.~D. Taylor, and R.~Jeanloz.
\newblock Origin of the {Verwey} transition in magnetite.
\newblock \emph{Phys. Rev. Lett.}, 96:\penalty0 045705, 2006.
\newblock \doi{10.1103/PhysRevLett.96.045705}.
\newblock URL \url{https://link.aps.org/doi/10.1103/PhysRevLett.96.045705}.

\bibitem[Rudner and Lindner(2020)]{rudner2019floquet}
Mark~S Rudner and Netanel~H Lindner.
\newblock Band structure engineering and non-equilibrium dynamics in {Floquet}
  topological insulators.
\newblock \emph{Nature Reviews Physics}, pages 1--16, 2020.
\newblock \doi{10.1038/s42254-020-0170-z}.

\bibitem[Rudner et~al.(2013)Rudner, Lindner, Berg, and
  Levin]{PhysRevX.3.031005}
Mark~S. Rudner, Netanel~H. Lindner, Erez Berg, and Michael Levin.
\newblock Anomalous edge states and the bulk-edge correspondence for
  periodically driven two-dimensional systems.
\newblock \emph{Phys. Rev. X}, 3:\penalty0 031005, 2013.
\newblock \doi{10.1103/PhysRevX.3.031005}.
\newblock URL \url{https://link.aps.org/doi/10.1103/PhysRevX.3.031005}.

\bibitem[Rusakov et~al.(2014)Rusakov, Phillips, and Zgid]{Rusakov14}
Alexander~A. Rusakov, Jordan~J. Phillips, and Dominika Zgid.
\newblock Local hamiltonians for quantitative {Green}'s function embedding
  methods.
\newblock \emph{The Journal of Chemical Physics}, 141\penalty0 (19):\penalty0
  194105, 2014.
\newblock \doi{10.1063/1.4901432}.
\newblock URL \url{, https://doi.org/10.1063/1.4901432}.

\bibitem[Ryu et~al.(2010)Ryu, Schnyder, Furusaki, and
  Ludwig]{ryu2010topological}
Shinsei Ryu, Andreas~P Schnyder, Akira Furusaki, and Andreas~WW Ludwig.
\newblock Topological insulators and superconductors: tenfold way and
  dimensional hierarchy.
\newblock \emph{New Journal of Physics}, 12\penalty0 (6):\penalty0 065010,
  2010.
\newblock \doi{10.1088/1367-2630/12/6/065010}.

\bibitem[Saal et~al.(2013)Saal, Kirklin, Aykol, Meredig, and
  Wolverton]{Saal2013}
James~E. Saal, Scott Kirklin, Muratahan Aykol, Bryce Meredig, and C.~Wolverton.
\newblock Materials design and discovery with high-throughput density
  functional theory: The open quantum materials database (oqmd).
\newblock \emph{JOM}, 65\penalty0 (11), 2013.
\newblock \doi{10.1007/s11837-013-0755-4}.
\newblock URL \url{https://doi.org/10.1007/s11837-013-0755-4}.

\bibitem[Sachdev(1999)]{sachdev1999quantum}
Subir Sachdev.
\newblock Quantum phase transitions.
\newblock \emph{Physics world}, 12\penalty0 (4):\penalty0 33, 1999.
\newblock \doi{10.1017/CBO9780511973765}.

\bibitem[Sachdev(2007)]{sachdev2007quantum}
Subir Sachdev.
\newblock Quantum phase transitions.
\newblock \emph{Handbook of Magnetism and Advanced Magnetic Materials}, 2007.
\newblock \doi{10.1017/CBO9780511973765}.

\bibitem[Sachdev(2008)]{sachdev.2008}
Subir Sachdev.
\newblock Quantum magnetism and criticality.
\newblock \emph{Nature Physics}, 4:\penalty0 173, 2008.
\newblock \doi{10.1038/nphys894}.
\newblock URL \url{https://doi.org/10.1038/nphys894}.

\bibitem[Sachdev and Ye(1993)]{sachdev1993gapless}
Subir Sachdev and Jinwu Ye.
\newblock Gapless spin-fluid ground state in a random quantum {Heisenberg}
  magnet.
\newblock \emph{Phys. Rev. Lett.}, 70\penalty0 (21):\penalty0 3339, 1993.
\newblock \doi{10.1103/PhysRevLett.70.3339}.

\bibitem[Sakai et~al.(2012)Sakai, Kuga, and Nakatsuji]{Sakai12}
Akito Sakai, Kentaro Kuga, and Satoru Nakatsuji.
\newblock {Superconductivity in the ferroquadrupolar state in the quadrupolar
  {Kondo} lattice PrTi$_2$Al$_{20}$}.
\newblock \emph{Journal of the Physical Society of Japan}, 81\penalty0
  (8):\penalty0 083702, 2012.
\newblock \doi{10.1143/JPSJ.81.083702}.

\bibitem[Sakurai et~al.(2015)Sakurai, Fujimoto, Matsui, Kawasaki, Okubo, Ohta,
  Matsubayashi, Uwatoko, and Tanaka]{sakurai2015development}
T~Sakurai, K~Fujimoto, R~Matsui, K~Kawasaki, S~Okubo, H~Ohta, K~Matsubayashi,
  Y~Uwatoko, and H~Tanaka.
\newblock Development of multi-frequency esr system for high-pressure
  measurements up to 2.5 gpa.
\newblock \emph{Journal of Magnetic Resonance}, 259:\penalty0 108--113, 2015.
\newblock \doi{10.1016/j.jmr.2015.08.005}.

\bibitem[Sal{\'{e}}n et~al.(2019)Sal{\'{e}}n, Basini, Bonetti, Hebling,
  Krasilnikov, Nikitin, Shamuilov, Tibai, Zhaunerchyk, and
  Goryashko]{salen2019thz}
Peter Sal{\'{e}}n, Martina Basini, Stefano Bonetti, J{\'{a}}nos Hebling,
  Mikhail Krasilnikov, Alexey~Y. Nikitin, Georgii Shamuilov, Zolt{\'{a}}n
  Tibai, Vitali Zhaunerchyk, and Vitaliy Goryashko.
\newblock {Matter manipulation with extreme terahertz light: Progress in the
  enabling {THz} technology}.
\newblock \emph{Physics Reports}, 836--837:\penalty0 1--74, 2019.
\newblock \doi{10.1016/j.physrep.2019.09.002}.
\newblock URL \url{https://doi.org/10.1016/j.physrep.2019.09.002}.

\bibitem[Sandvik(1997)]{Sandvik_Heisenberg}
Anders~W. Sandvik.
\newblock Finite-size scaling of the ground-state parameters of the
  two-dimensional {Heisenberg} model.
\newblock \emph{Phys. Rev. B}, 56:\penalty0 11678--11690, 1997.
\newblock \doi{10.1103/PhysRevB.56.11678}.
\newblock URL \url{https://link.aps.org/doi/10.1103/PhysRevB.56.11678}.

\bibitem[Sandvik(1999)]{Sandvik_SSE}
Anders~W. Sandvik.
\newblock Stochastic series expansion method with operator-loop update.
\newblock \emph{Phys. Rev. B}, 59:\penalty0 R14157--R14160, 1999.
\newblock \doi{10.1103/PhysRevB.59.R14157}.
\newblock URL \url{https://link.aps.org/doi/10.1103/PhysRevB.59.R14157}.

\bibitem[Sarkar et~al.(2019)Sarkar, Mandal, Poniatowski, Chan, and
  Greene]{sarkarsciadv}
Tarapada Sarkar, P.~R. Mandal, N.~R. Poniatowski, M.~K. Chan, and Richard~L.
  Greene.
\newblock Correlation between scale-invariant normal-state resistivity and
  superconductivity in an electron-doped cuprate.
\newblock \emph{Science Advances}, 5\penalty0 (5), 2019.
\newblock \doi{10.1126/sciadv.aav6753}.
\newblock URL \url{https://advances.sciencemag.org/content/5/5/eaav6753}.

\bibitem[Sato and Ando(2017)]{sato2017topological}
Masatoshi Sato and Yoichi Ando.
\newblock Topological superconductors: a review.
\newblock \emph{Reports on Progress in Physics}, 80\penalty0 (7):\penalty0
  076501, 2017.
\newblock \doi{10.1088/1361-6633/aa6ac7}.

\bibitem[Savary and Balents(2016)]{savary2016quantum}
Lucile Savary and Leon Balents.
\newblock Quantum spin liquids: a review.
\newblock \emph{Reports on Progress in Physics}, 80\penalty0 (1):\penalty0
  016502, 2016.
\newblock \doi{10.1088/0034-4885/80/1/016502}.

\bibitem[Savary and Balents(2017)]{savary2017disorder}
Lucile Savary and Leon Balents.
\newblock Disorder-induced quantum spin liquid in spin ice pyrochlores.
\newblock \emph{Phys. Rev. Lett.}, 118\penalty0 (8):\penalty0 087203, 2017.
\newblock \doi{10.1103/PhysRevLett.118.087203}.

\bibitem[Savary et~al.(2014)Savary, Moon, and Balents]{savary2014new}
Lucile Savary, Eun-Gook Moon, and Leon Balents.
\newblock New type of quantum criticality in the pyrochlore iridates.
\newblock \emph{Physical Review X}, 4\penalty0 (4):\penalty0 041027, 2014.
\newblock \doi{10.1103/PhysRevX.4.041027}.

\bibitem[Scalapino(2012)]{scalapino.2012}
D.~J. Scalapino.
\newblock A common thread: The pairing interaction for unconventional
  superconductors.
\newblock \emph{Rev. Mod. Phys.}, 84:\penalty0 1383--1417, 2012.
\newblock \doi{10.1103/RevModPhys.84.1383}.
\newblock URL \url{https://link.aps.org/doi/10.1103/RevModPhys.84.1383}.

\bibitem[Scalapino and White(1998)]{scalapino1998superconducting}
DJ~Scalapino and Steven~R White.
\newblock Superconducting condensation energy and an antiferromagnetic
  exchange-based pairing mechanism.
\newblock \emph{Phys. Rev. B}, 58\penalty0 (13):\penalty0 8222, 1998.
\newblock \doi{10.1103/PhysRevB.58.8222}.

\bibitem[Scheie et~al.(2021)Scheie, Laurell, Samarakoon, Lake, Nagler,
  Granroth, Okamoto, Alvarez, and Tennant]{scheie2021witnessing}
A.~Scheie, Pontus Laurell, A.~M. Samarakoon, B.~Lake, S.~E. Nagler, G.~E.
  Granroth, S.~Okamoto, G.~Alvarez, and D.~A. Tennant.
\newblock Witnessing entanglement in quantum magnets using neutron scattering.
\newblock \emph{Phys. Rev. B}, 103:\penalty0 224434, 2021.
\newblock \doi{10.1103/PhysRevB.103.224434}.

\bibitem[Schlauderer et~al.(2019)Schlauderer, Lange, Baierl, Ebnet, Schmid,
  Valovcin, Zvezdin, Kimel, Mikhaylovskiy, and Huber]{schlauderer2019temporal}
S.~Schlauderer, C.~Lange, S.~Baierl, T.~Ebnet, C.~P. Schmid, D.~C. Valovcin,
  A.~K. Zvezdin, A.~V. Kimel, R.~V. Mikhaylovskiy, and R.~Huber.
\newblock {Temporal and spectral fingerprints of ultrafast all-coherent spin
  switching}.
\newblock \emph{Nature}, 569\penalty0 (7756):\penalty0 383--387, 2019.
\newblock \doi{10.1038/s41586-019-1174-7}.

\bibitem[Schmehr et~al.(2019)Schmehr, Aling, Zoghlin, and
  Wilson]{schmehr2019high}
Julian~L Schmehr, Michael Aling, Eli Zoghlin, and Stephen~D Wilson.
\newblock High-pressure laser floating zone furnace.
\newblock \emph{Review of Scientific Instruments}, 90\penalty0 (4):\penalty0
  043906, 2019.
\newblock \doi{10.1063/1.5085327}.

\bibitem[Schmidt(2012)]{SchmidtPRB2012}
Manuel~J. Schmidt.
\newblock Strong correlations at topological insulator surfaces and the
  breakdown of the bulk-boundary correspondence.
\newblock \emph{Phys. Rev. B}, 86:\penalty0 161110, 2012.
\newblock \doi{10.1103/PhysRevB.86.161110}.
\newblock URL \url{https://link.aps.org/doi/10.1103/PhysRevB.86.161110}.

\bibitem[Schofield(1999)]{schofield1999non}
Andrew~J Schofield.
\newblock Non-fermi liquids.
\newblock \emph{Contemporary Physics}, 40\penalty0 (2):\penalty0 95--115, 1999.
\newblock \doi{10.1080/001075199181342}.

\bibitem[Schollw{\"o}ck(2005)]{schollwock2005density}
Ulrich Schollw{\"o}ck.
\newblock The density-matrix renormalization group.
\newblock \emph{Reviews of modern physics}, 77\penalty0 (1):\penalty0 259,
  2005.
\newblock \doi{10.1103/RevModPhys.77.259}.

\bibitem[Schr{\"o}der et~al.(1998)Schr{\"o}der, Aeppli, Bucher, Ramazashvili,
  and Coleman]{schroder1998scaling}
A~Schr{\"o}der, G~Aeppli, E~Bucher, R~Ramazashvili, and Piers Coleman.
\newblock Scaling of magnetic fluctuations near a quantum phase transition.
\newblock \emph{Physical Review Letters}, 80\penalty0 (25):\penalty0 5623,
  1998.
\newblock \doi{10.1103/PhysRevLett.80.5623}.

\bibitem[Schubert et~al.(2014)Schubert, Hohenleutner, Langer, Urbanek, Lange,
  Huttner, Golde, Meier, Kira, Koch, and Huber]{schubert2014thz}
O~Schubert, M~Hohenleutner, F~Langer, B~Urbanek, C~Lange, U~Huttner, D~Golde,
  T~Meier, M~Kira, S~W Koch, and R~Huber.
\newblock {Sub-cycle control of terahertz high-harmonic generation by dynamical
  Bloch oscillations}.
\newblock \emph{Nature Photonics}, 8\penalty0 (2):\penalty0 119--123, 2014.
\newblock \doi{10.1038/nphoton.2013.349}.

\bibitem[Schuberth et~al.(2016)Schuberth, Tippmann, Steinke, Lausberg, Steppke,
  Brando, Krellner, Geibel, Yu, Si, et~al.]{schuberth2016emergence}
Erwin Schuberth, Marc Tippmann, Lucia Steinke, Stefan Lausberg, Alexander
  Steppke, Manuel Brando, Cornelius Krellner, Christoph Geibel, Rong Yu, Qimiao
  Si, et~al.
\newblock {Emergence of superconductivity in the canonical heavy-electron metal
  YbRh$_2$Si$_2$}.
\newblock \emph{Science}, 351\penalty0 (6272):\penalty0 485--488, 2016.
\newblock \doi{10.1126/science.aaa9733}.

\bibitem[Schumann et~al.(2007)Schumann, Winkler, and
  Kirschner]{schumann2007mapping}
FO~Schumann, C~Winkler, and J~Kirschner.
\newblock Mapping out electron--electron interactions in angular space.
\newblock \emph{New Journal of Physics}, 9\penalty0 (10):\penalty0 372, 2007.
\newblock \doi{10.1088/1367-2630/9/10/372}.

\bibitem[Schweizer et~al.(2019)Schweizer, Grusdt, Berngruber, Barbiero, Demler,
  Goldman, Bloch, and Aidelsburger]{schweizer2019floquet}
Christian Schweizer, Fabian Grusdt, Moritz Berngruber, Luca Barbiero, Eugene
  Demler, Nathan Goldman, Immanuel Bloch, and Monika Aidelsburger.
\newblock Floquet approach to $\mathbb{Z}$ 2 lattice gauge theories with
  ultracold atoms in optical lattices.
\newblock \emph{Nature Physics}, 15\penalty0 (11):\penalty0 1168--1173, 2019.
\newblock \doi{10.1038/s41567-019-0649-7}.

\bibitem[Seaman et~al.(1991)Seaman, Maple, Lee, Ghamaty, Torikachvili, Kang,
  Liu, Allen, and Cox]{Seaman91}
CL~Seaman, MB~Maple, BW~Lee, S~Ghamaty, MS~Torikachvili, J-S Kang, LZ~Liu,
  JW~Allen, and DL~Cox.
\newblock {Evidence for non-Fermi liquid behavior in the {Kondo} alloy Y$_{1-
  x}$U$_x$Pd$_3$}.
\newblock \emph{Phys. Rev. Lett.}, 67\penalty0 (20):\penalty0 2882, 1991.
\newblock \doi{10.1103/PhysRevLett.67.2882}.

\bibitem[Sebastian and Proust(2015)]{sebastian2015quantum}
Suchitra~E Sebastian and Cyril Proust.
\newblock Quantum oscillations in hole-doped cuprates.
\newblock \emph{Annu. Rev. Condens. Matter Phys.}, 6\penalty0 (1):\penalty0
  411--430, 2015.
\newblock \doi{10.1146/annurev-conmatphys-030212-184305}.

\bibitem[Sekihara et~al.(2013)Sekihara, Masutomi, and Okamoto]{PbSC1}
Takayuki Sekihara, Ryuichi Masutomi, and Tohru Okamoto.
\newblock {Two-Dimensional Superconducting State of Monolayer Pb Films Grown on
  GaAs(110) in a Strong Parallel Magnetic Field}.
\newblock \emph{Phys. Rev. Lett.}, 111:\penalty0 057005, 2013.
\newblock \doi{10.1103/PhysRevLett.111.057005}.
\newblock URL \url{https://link.aps.org/doi/10.1103/PhysRevLett.111.057005}.

\bibitem[Senga et~al.(2019)Senga, Suenaga, Barone, Morishita, Mauri, and
  Pichler]{senga2019position}
Ryosuke Senga, Kazu Suenaga, Paolo Barone, Shigeyuki Morishita, Francesco
  Mauri, and Thomas Pichler.
\newblock Position and momentum mapping of vibrations in graphene
  nanostructures.
\newblock \emph{Nature}, 573\penalty0 (7773):\penalty0 247--250, 2019.
\newblock \doi{10.1038/s41586-019-1477-8}.

\bibitem[Senthil(2008)]{senthil2008theory}
T~Senthil.
\newblock Theory of a continuous mott transition in two dimensions.
\newblock \emph{Phys. Rev. B}, 78\penalty0 (4):\penalty0 045109, 2008.
\newblock \doi{10.1103/PhysRevB.78.045109}.

\bibitem[Senthil et~al.(2004{\natexlab{a}})Senthil, Vojta, and
  Sachdev]{senthil2004weak}
T~Senthil, Matthias Vojta, and Subir Sachdev.
\newblock Weak magnetism and non-fermi liquids near heavy-fermion critical
  points.
\newblock \emph{Phys. Rev. B}, 69\penalty0 (3):\penalty0 035111,
  2004{\natexlab{a}}.
\newblock \doi{10.1103/PhysRevB.69.035111}.

\bibitem[Senthil et~al.(2004{\natexlab{b}})Senthil, Vishwanath, Balents,
  Sachdev, and Fisher]{senthil2004deconfined}
Todadri Senthil, Ashvin Vishwanath, Leon Balents, Subir Sachdev, and Matthew~PA
  Fisher.
\newblock Deconfined quantum critical points.
\newblock \emph{Science}, 303\penalty0 (5663):\penalty0 1490--1494,
  2004{\natexlab{b}}.
\newblock \doi{10.1126/science.1091806}.

\bibitem[Serlin et~al.(2020)Serlin, Tschirhart, Polshyn, Zhang, Zhu, Watanabe,
  Taniguchi, Balents, and Young]{serlin2020intrinsic}
M~Serlin, CL~Tschirhart, H~Polshyn, Y~Zhang, J~Zhu, K~Watanabe, T~Taniguchi,
  L~Balents, and AF~Young.
\newblock Intrinsic quantized anomalous hall effect in a moir{\'e}
  heterostructure.
\newblock \emph{Science}, 367\penalty0 (6480):\penalty0 900--903, 2020.
\newblock \doi{10.1126/science.aay5533}.

\bibitem[Shao et~al.(2015)Shao, Gan, Epifanovsky, Gilbert, Wormit, Kussmann,
  Lange, Behn, Deng, Feng, Ghosh, Goldey, Horn, Jacobson, Kaliman, Khaliullin,
  Kuś, Landau, Liu, Proynov, Rhee, Richard, Rohrdanz, Steele, Sundstrom, III,
  Zimmerman, Zuev, Albrecht, Alguire, Austin, Beran, Bernard, Berquist,
  Brandhorst, Bravaya, Brown, Casanova, Chang, Chen, Chien, Closser,
  Crittenden, Diedenhofen, Jr., Do, Dutoi, Edgar, Fatehi, Fusti-Molnar,
  Ghysels, Golubeva-Zadorozhnaya, Gomes, Hanson-Heine, Harbach, Hauser,
  Hohenstein, Holden, Jagau, Ji, Kaduk, Khistyaev, Kim, Kim, King, Klunzinger,
  Kosenkov, Kowalczyk, Krauter, Lao, Laurent, Lawler, Levchenko, Lin, Liu,
  Livshits, Lochan, Luenser, Manohar, Manzer, Mao, Mardirossian, Marenich,
  Maurer, Mayhall, Neuscamman, Oana, Olivares-Amaya, O’Neill, Parkhill,
  Perrine, Peverati, Prociuk, Rehn, Rosta, Russ, Sharada, Sharma, Small, Sodt,
  Stein, Stück, Su, Thom, Tsuchimochi, Vanovschi, Vogt, Vydrov, Wang, Watson,
  Wenzel, White, Williams, Yang, Yeganeh, Yost, You, Zhang, Zhang, Zhao,
  Brooks, Chan, Chipman, Cramer, III, Gordon, Hehre, Klamt, III, Schmidt,
  Sherrill, Truhlar, Warshel, Xu, Aspuru-Guzik, Baer, Bell, Besley, Chai,
  Dreuw, Dunietz, Furlani, Gwaltney, Hsu, Jung, Kong, Lambrecht, Liang,
  Ochsenfeld, Rassolov, Slipchenko, Subotnik, Voorhis, Herbert, Krylov, Gill,
  and Head-Gordon]{Shao_qchem}
Yihan Shao, Zhengting Gan, Evgeny Epifanovsky, Andrew~T.B. Gilbert, Michael
  Wormit, Joerg Kussmann, Adrian~W. Lange, Andrew Behn, Jia Deng, Xintian Feng,
  Debashree Ghosh, Matthew Goldey, Paul~R. Horn, Leif~D. Jacobson, Ilya
  Kaliman, Rustam~Z. Khaliullin, Tomasz Kuś, Arie Landau, Jie Liu, Emil~I.
  Proynov, Young~Min Rhee, Ryan~M. Richard, Mary~A. Rohrdanz, Ryan~P. Steele,
  Eric~J. Sundstrom, H.~Lee~Woodcock III, Paul~M. Zimmerman, Dmitry Zuev, Ben
  Albrecht, Ethan Alguire, Brian Austin, Gregory J.~O. Beran, Yves~A. Bernard,
  Eric Berquist, Kai Brandhorst, Ksenia~B. Bravaya, Shawn~T. Brown, David
  Casanova, Chun-Min Chang, Yunqing Chen, Siu~Hung Chien, Kristina~D. Closser,
  Deborah~L. Crittenden, Michael Diedenhofen, Robert A.~DiStasio Jr., Hainam
  Do, Anthony~D. Dutoi, Richard~G. Edgar, Shervin Fatehi, Laszlo Fusti-Molnar,
  An~Ghysels, Anna Golubeva-Zadorozhnaya, Joseph Gomes, Magnus~W.D.
  Hanson-Heine, Philipp~H.P. Harbach, Andreas~W. Hauser, Edward~G. Hohenstein,
  Zachary~C. Holden, Thomas-C. Jagau, Hyunjun Ji, Benjamin Kaduk, Kirill
  Khistyaev, Jaehoon Kim, Jihan Kim, Rollin~A. King, Phil Klunzinger, Dmytro
  Kosenkov, Tim Kowalczyk, Caroline~M. Krauter, Ka~Un Lao, Adèle~D. Laurent,
  Keith~V. Lawler, Sergey~V. Levchenko, Ching~Yeh Lin, Fenglai Liu, Ester
  Livshits, Rohini~C. Lochan, Arne Luenser, Prashant Manohar, Samuel~F. Manzer,
  Shan-Ping Mao, Narbe Mardirossian, Aleksandr~V. Marenich, Simon~A. Maurer,
  Nicholas~J. Mayhall, Eric Neuscamman, C.~Melania Oana, Roberto
  Olivares-Amaya, Darragh~P. O’Neill, John~A. Parkhill, Trilisa~M. Perrine,
  Roberto Peverati, Alexander Prociuk, Dirk~R. Rehn, Edina Rosta, Nicholas~J.
  Russ, Shaama~M. Sharada, Sandeep Sharma, David~W. Small, Alexander Sodt,
  Tamar Stein, David Stück, Yu-Chuan Su, Alex~J.W. Thom, Takashi Tsuchimochi,
  Vitalii Vanovschi, Leslie Vogt, Oleg Vydrov, Tao Wang, Mark~A. Watson, Jan
  Wenzel, Alec White, Christopher~F. Williams, Jun Yang, Sina Yeganeh, Shane~R.
  Yost, Zhi-Qiang You, Igor~Ying Zhang, Xing Zhang, Yan Zhao, Bernard~R.
  Brooks, Garnet~K.L. Chan, Daniel~M. Chipman, Christopher~J. Cramer, William
  A.~Goddard III, Mark~S. Gordon, Warren~J. Hehre, Andreas Klamt, Henry
  F.~Schaefer III, Michael~W. Schmidt, C.~David Sherrill, Donald~G. Truhlar,
  Arieh Warshel, Xin Xu, Alán Aspuru-Guzik, Roi Baer, Alexis~T. Bell,
  Nicholas~A. Besley, Jeng-Da Chai, Andreas Dreuw, Barry~D. Dunietz, Thomas~R.
  Furlani, Steven~R. Gwaltney, Chao-Ping Hsu, Yousung Jung, Jing Kong,
  Daniel~S. Lambrecht, WanZhen Liang, Christian Ochsenfeld, Vitaly~A. Rassolov,
  Lyudmila~V. Slipchenko, Joseph~E. Subotnik, Troy~Van Voorhis, John~M.
  Herbert, Anna~I. Krylov, Peter~M.W. Gill, and Martin Head-Gordon.
\newblock Advances in molecular quantum chemistry contained in the q-chem 4
  program package.
\newblock \emph{Molecular Physics}, 113\penalty0 (2):\penalty0 184--215, 2015.
\newblock \doi{10.1080/00268976.2014.952696}.
\newblock URL \url{, https://doi.org/10.1080/00268976.2014.952696}.

\bibitem[Sheckelton et~al.(2014)Sheckelton, Foronda, Pan, Moir, McDonald,
  Lancaster, Baker, Armitage, Imai, Blundell, et~al.]{sheckelton2014local}
JP~Sheckelton, FR~Foronda, LiDong Pan, C~Moir, RD~McDonald, T~Lancaster,
  PJ~Baker, NP~Armitage, T~Imai, SJ~Blundell, et~al.
\newblock {Local magnetism and spin correlations in the geometrically
  frustrated cluster magnet LiZn$_2$Mo$_3$O$_8$}.
\newblock \emph{Phys. Rev. B}, 89\penalty0 (6):\penalty0 064407, 2014.
\newblock \doi{10.1103/PhysRevB.89.064407}.

\bibitem[Shen et~al.(2020{\natexlab{a}})Shen, Chu, Wu, Li, Wang, Zhao, Tang,
  Liu, Tian, Watanabe, et~al.]{Shen2019Observation}
Cheng Shen, Yanbang Chu, QuanSheng Wu, Na~Li, Shuopei Wang, Yanchong Zhao, Jian
  Tang, Jieying Liu, Jinpeng Tian, Kenji Watanabe, et~al.
\newblock Correlated states in twisted double bilayer graphene.
\newblock \emph{Nature Physics}, 16\penalty0 (5):\penalty0 520--525,
  2020{\natexlab{a}}.
\newblock \doi{10.1038/s41567-020-0825-9}.

\bibitem[Shen et~al.(2017)Shen, Yang, and Liu]{shen2017situ}
Dawei Shen, Haifeng Yang, and Zhengtai Liu.
\newblock In situ engineering and characterization of correlated materials with
  integrated ombe--arpes.
\newblock \emph{Modern Technologies for Creating the Thin-film Systems and
  Coatings}, page~59, 2017.
\newblock \doi{10.1016/j.elspec.2015.06.002}.

\bibitem[Shen and Mao(2016)]{Shen2017}
Guoyin Shen and Ho~Kwang Mao.
\newblock High-pressure studies with x-rays using diamond anvil cells.
\newblock \emph{Reports on Progress in Physics}, 80\penalty0 (1):\penalty0
  016101, 2016.
\newblock \doi{10.1088/1361-6633/80/1/016101}.
\newblock URL \url{https://doi.org/10.1088%2F1361-6633%2F80%2F1%2F016101}.

\bibitem[Shen et~al.(2020{\natexlab{b}})Shen, Kuhn, Dalgliesh, de~Haan,
  Geerits, Irfan, Li, Lu, Parnell, Plomp, et~al.]{shen2019unveiling}
Jiazhou Shen, Steven~J Kuhn, Robert~M Dalgliesh, VO~de~Haan, N~Geerits,
  Abdul~AM Irfan, Fankang Li, Shufan Lu, Stephen~R Parnell, Jerome Plomp,
  et~al.
\newblock Unveiling contextual realities by microscopically entangling a
  neutron.
\newblock \emph{Nature Communications}, 11\penalty0 (1):\penalty0 1--6,
  2020{\natexlab{b}}.
\newblock \doi{10.1038/s41467-020-14741-y}.

\bibitem[Shi et~al.(2020)Shi, Baity, Sasagawa, and Popovi{\'c}]{shi2020vortex}
Zhenzhong Shi, PG~Baity, T~Sasagawa, and Dragana Popovi{\'c}.
\newblock Vortex phase diagram and the normal state of cuprates with charge and
  spin orders.
\newblock \emph{Science Advances}, 6\penalty0 (7):\penalty0 eaay8946, 2020.
\newblock \doi{10.1126/sciadv.aay8946}.

\bibitem[Shibauchi et~al.(2014)Shibauchi, Carrington, and
  Matsuda]{shibauchi2014quantum}
T~Shibauchi, A~Carrington, and Y~Matsuda.
\newblock A quantum critical point lying beneath the superconducting dome in
  iron pnictides.
\newblock \emph{Annu. Rev. Condens. Matter Phys.}, 5\penalty0 (1):\penalty0
  113--135, 2014.
\newblock \doi{10.1146/annurev-conmatphys-031113-133921}.

\bibitem[Shier and Ginsberg(1966)]{amorphousbismuthSC}
J.~S. Shier and D.~M. Ginsberg.
\newblock Superconducting transitions of amorphous bismuth alloys.
\newblock \emph{Phys. Rev.}, 147:\penalty0 384--391, 1966.
\newblock \doi{10.1103/PhysRev.147.384}.
\newblock URL \url{https://link.aps.org/doi/10.1103/PhysRev.147.384}.

\bibitem[Shimizu et~al.(2001)Shimizu, Kimura, Furomoto, Takeda, Kontani, Onuki,
  and Amaya]{Katsuya2001}
Katsuya Shimizu, Tomohiro Kimura, Shigeyuki Furomoto, Keiki Takeda, Kazuyoshi
  Kontani, Yoshichika Onuki, and Kiichi Amaya.
\newblock {Superconductivity in the non-magnetic state of iron under pressure}.
\newblock \emph{Nature}, 412\penalty0 (6844):\penalty0 316--318, 2001.
\newblock \doi{10.1038/35085536}.
\newblock URL \url{http://www.nature.com/articles/35085536}.

\bibitem[Shishido et~al.(2005)Shishido, Settai, Harima, and
  {\=O}nuki]{shishido2005drastic}
Hiroaki Shishido, Rikio Settai, Hisatomo Harima, and Yoshichika {\=O}nuki.
\newblock {A drastic change of the {Fermi} surface at a critical pressure in
  CeRhIn$_5$: dHvA study under pressure}.
\newblock \emph{Journal of the Physical Society of Japan}, 74\penalty0
  (4):\penalty0 1103--1106, 2005.
\newblock \doi{10.1143/JPSJ.74.1103}.

\bibitem[Si(2010)]{si.2010}
Qimiao Si.
\newblock Quantum criticality and global phase diagram of magnetic heavy
  fermions.
\newblock \emph{Physica Status Solidi (b)}, 247\penalty0 (3):\penalty0
  476--484, 2010.
\newblock \doi{10.1002/pssb.200983082}.
\newblock URL
  \url{https://onlinelibrary.wiley.com/doi/abs/10.1002/pssb.200983082}.

\bibitem[Si and Paschen(2013)]{si2013quantum}
Qimiao Si and Silke Paschen.
\newblock Quantum phase transitions in heavy fermion metals and {Kondo}
  insulators.
\newblock \emph{physica status solidi (b)}, 250\penalty0 (3):\penalty0
  425--438, 2013.
\newblock \doi{10.1002/pssb.201300005}.

\bibitem[Si and Steglich(2010)]{si2010heavy}
Qimiao Si and Frank Steglich.
\newblock Heavy fermions and quantum phase transitions.
\newblock \emph{Science}, 329\penalty0 (5996):\penalty0 1161--1166, 2010.
\newblock \doi{10.1126/science.1191195}.

\bibitem[Si et~al.(2001)Si, Rabello, Ingersent, and Smith]{si2001locally}
Qimiao Si, Silvio Rabello, Kevin Ingersent, and J~Lleweilun Smith.
\newblock Locally critical quantum phase transitions in strongly correlated
  metals.
\newblock \emph{Nature}, 413\penalty0 (6858):\penalty0 804--808, 2001.
\newblock \doi{10.1038/35101507}.

\bibitem[Silva et~al.(2018)Silva, Blinov, Rubtsov, Smirnova, and
  Ivanov]{silva2018hhg}
R.~E.~F. Silva, Igor~V. Blinov, Alexey~N. Rubtsov, O.~Smirnova, and M.~Ivanov.
\newblock {High-harmonic spectroscopy of ultrafast many-body dynamics in
  strongly correlated systems}.
\newblock \emph{Nature Photonics}, 12\penalty0 (5):\penalty0 266--270, 2018.
\newblock \doi{10.1038/s41566-018-0129-0}.
\newblock URL \url{http://www.nature.com/articles/s41566-018-0129-0}.

\bibitem[Sipe and Ghahramani(1993)]{sipe1993nonlinear}
JE~Sipe and Ed~Ghahramani.
\newblock Nonlinear optical response of semiconductors in the
  independent-particle approximation.
\newblock \emph{Phys. Rev. B}, 48\penalty0 (16):\penalty0 11705, 1993.
\newblock \doi{10.1103/PhysRevB.48.11705}.

\bibitem[Skinner(2019)]{skinner2019properties}
Brian Skinner.
\newblock Properties of the donor impurity band in mixed valence insulators.
\newblock \emph{Physical Review Materials}, 3\penalty0 (10):\penalty0 104601,
  2019.
\newblock \doi{10.1103/PhysRevMaterials.3.104601}.

\bibitem[Sleight(2015)]{bismuthates}
Arthur~W Sleight.
\newblock Bismuthates: {BaBiO}$_3$ and related superconducting phases.
\newblock \emph{Physica C: Superconductivity and Its Applications},
  514:\penalty0 152--165, 2015.
\newblock \doi{10.1016/j.physc.2015.02.012}.

\bibitem[Slonczewski and Weiss(1958)]{graphitebands2}
J.~C. Slonczewski and P.~R. Weiss.
\newblock Band structure of graphite.
\newblock \emph{Phys. Rev.}, 109:\penalty0 272--279, 1958.
\newblock \doi{10.1103/PhysRev.109.272}.
\newblock URL \url{https://link.aps.org/doi/10.1103/PhysRev.109.272}.

\bibitem[Sodemann and Fu(2015)]{sodemann2015nhe}
Inti Sodemann and Liang Fu.
\newblock Quantum nonlinear hall effect induced by {Berry} curvature dipole in
  time-reversal invariant materials.
\newblock \emph{Phys. Rev. Lett.}, 115:\penalty0 216806, 2015.
\newblock \doi{10.1103/PhysRevLett.115.216806}.
\newblock URL \url{https://link.aps.org/doi/10.1103/PhysRevLett.115.216806}.

\bibitem[Sondhi et~al.(1997)Sondhi, Girvin, Carini, and
  Shahar]{sondhi1997continuous}
Shivaji~Lal Sondhi, SM~Girvin, JP~Carini, and D~Shahar.
\newblock Continuous quantum phase transitions.
\newblock \emph{Reviews of modern physics}, 69\penalty0 (1):\penalty0 315,
  1997.
\newblock \doi{10.1103/RevModPhys.69.315}.

\bibitem[Song et~al.(2012)Song, Rachel, Flindt, Klich, Laflorencie, and
  Le~Hur]{song2012bipartite}
H~Francis Song, Stephan Rachel, Christian Flindt, Israel Klich, Nicolas
  Laflorencie, and Karyn Le~Hur.
\newblock Bipartite fluctuations as a probe of many-body entanglement.
\newblock \emph{Phys. Rev. B}, 85\penalty0 (3):\penalty0 035409, 2012.
\newblock \doi{10.1103/PhysRevB.85.035409}.

\bibitem[Song and Gabor(2018)]{song2018electron}
Justin~CW Song and Nathaniel~M Gabor.
\newblock {Electron quantum metamaterials in van der Waals heterostructures}.
\newblock \emph{Nature nanotechnology}, 13\penalty0 (11):\penalty0 986--993,
  2018.
\newblock \doi{10.1103/PhysRevLett.118.137001}.

\bibitem[Song et~al.(2017)Song, Jian, and Balents]{song2017strongly}
Xue-Yang Song, Chao-Ming Jian, and Leon Balents.
\newblock {Strongly correlated metal built from Sachdev-Ye-Kitaev models}.
\newblock \emph{Phys. Rev. Lett.}, 119\penalty0 (21):\penalty0 216601, 2017.
\newblock \doi{10.1103/PhysRevLett.119.216601}.

\bibitem[Song et~al.(2010)Song, Otte, Shvarts, Zhao, Kuk, Blankenship, Band,
  Hess, and Stroscio]{song2010stm}
Young~Jae Song, Alexander~F. Otte, Vladimir Shvarts, Zuyu Zhao, Young Kuk,
  Steven~R. Blankenship, Alan Band, Frank~M. Hess, and Joseph~A. Stroscio.
\newblock {Invited Review Article: A 10 mK scanning probe microscopy facility}.
\newblock \emph{Review of Scientific Instruments}, 81\penalty0 (12):\penalty0
  121101, 2010.
\newblock \doi{10.1063/1.3520482}.

\bibitem[Spanton et~al.(2018)Spanton, Zibrov, Zhou, Taniguchi, Watanabe,
  Zaletel, and Young]{spanton2018observation}
Eric~M Spanton, Alexander~A Zibrov, Haoxin Zhou, Takashi Taniguchi, Kenji
  Watanabe, Michael~P Zaletel, and Andrea~F Young.
\newblock {Observation of fractional Chern insulators in a van der Waals
  heterostructure}.
\newblock \emph{Science}, 360\penalty0 (6384):\penalty0 62--66, 2018.
\newblock \doi{10.1126/science.aan8458}.

\bibitem[Sprau et~al.(2017)Sprau, Kostin, Kreisel, B{\"o}hmer, Taufour,
  Canfield, Mukherjee, Hirschfeld, Andersen, and Davis]{sprau2017discovery}
Peter~O Sprau, Andrey Kostin, Andreas Kreisel, Anna~E B{\"o}hmer, Valentin
  Taufour, Paul~C Canfield, Shantanu Mukherjee, Peter~J Hirschfeld,
  Brian~M{\o}ller Andersen, and JC~S{\'e}amus Davis.
\newblock {Discovery of orbital-selective Cooper pairing in FeSe}.
\newblock \emph{Science}, 357\penalty0 (6346):\penalty0 75--80, 2017.
\newblock \doi{10.1126/science.aal1575}.

\bibitem[Stahl and Eckstein(2019)]{stahl2019noise}
Christopher Stahl and Martin Eckstein.
\newblock Noise correlations in time-and angle-resolved photoemission
  spectroscopy.
\newblock \emph{Phys. Rev. B}, 99\penalty0 (24):\penalty0 241111, 2019.
\newblock \doi{10.1103/PhysRevB.99.241111}.

\bibitem[Stearns(1978)]{Stearns1978}
Mary~Beth Stearns.
\newblock Observation of an even-denominator quantum number in the fractional
  quantum hall effect.
\newblock \emph{Physics Today}, 31\penalty0 (4):\penalty0 34, 1978.
\newblock \doi{10.1103/PhysRevLett.59.1776}.

\bibitem[Stefani et~al.(2002)Stefani, Iacobucci, Ruocco, and
  Gotter]{stefani2002electron}
G~Stefani, S~Iacobucci, A~Ruocco, and R~Gotter.
\newblock Electron--electron coincidence spectroscopies at surfaces.
\newblock \emph{Journal of electron spectroscopy and related phenomena},
  127\penalty0 (1-2):\penalty0 1--10, 2002.
\newblock \doi{10.1016/S0368-2048(02)00166-4}.

\bibitem[Steglich et~al.(1979)Steglich, Aarts, Bredl, Lieke, Meschede, Franz,
  and Sch{\"a}fer]{steglich1979superconductivity}
Frank Steglich, J~Aarts, CD~Bredl, W~Lieke, D~Meschede, W~Franz, and
  H~Sch{\"a}fer.
\newblock {Superconductivity in the presence of strong pauli paramagnetism:
  CeCu$_2$Si$_2$}.
\newblock \emph{Phys. Rev. Lett.}, 43\penalty0 (25):\penalty0 1892, 1979.
\newblock \doi{10.1103/PhysRevLett.43.1892}.

\bibitem[Stepanov et~al.(2020)Stepanov, Das, Lu, Fahimniya, Watanabe,
  Taniguchi, Koppens, Lischner, Levitov, and Efetov]{stepanov2019interplay}
Petr Stepanov, Ipsita Das, Xiaobo Lu, Ali Fahimniya, Kenji Watanabe, Takashi
  Taniguchi, Frank~HL Koppens, Johannes Lischner, Leonid Levitov, and Dmitri~K
  Efetov.
\newblock Untying the insulating and superconducting orders in magic-angle
  graphene.
\newblock \emph{Nature}, 583\penalty0 (7816):\penalty0 375--378, 2020.
\newblock \doi{10.1038/s41586-020-2459-6}.

\bibitem[Sterpetti et~al.(2017)Sterpetti, Biscaras, Erb, and
  Shukla]{sterpetti2017comprehensive}
Edoardo Sterpetti, Johan Biscaras, Andreas Erb, and Abhay Shukla.
\newblock {Comprehensive phase diagram of two-dimensional space charge doped
  {Bi}$_2${Sr}$_2${CaCu}$_2${O}$_{8+x}$ }.
\newblock \emph{Nature Communications}, 8\penalty0 (1):\penalty0 1--8, 2017.
\newblock \doi{10.1038/s41467-017-02104-z}.

\bibitem[Stewart(1984)]{stewart1984heavy}
GR~Stewart.
\newblock Heavy-fermion systems.
\newblock \emph{Reviews of Modern Physics}, 56\penalty0 (4):\penalty0 755,
  1984.
\newblock \doi{10.1103/RevModPhys.56.755}.

\bibitem[Stewart(2001)]{stewart2001non}
GR~Stewart.
\newblock Non-fermi-liquid behavior in d-and f-electron metals.
\newblock \emph{Reviews of modern Physics}, 73\penalty0 (4):\penalty0 797,
  2001.
\newblock \doi{10.1103/RevModPhys.73.797}.

\bibitem[Stewart(1964)]{Stewart}
P.~Stewart.
\newblock \url{https://en.wikipedia.org/wiki/I_know_it_when_I_see_it}, 1964.

\bibitem[Storey(2016)]{storey2016hall}
JG~Storey.
\newblock Hall effect and {Fermi} surface reconstruction via electron pockets
  in the high-{T$_c$} cuprates.
\newblock \emph{EPL (Europhysics Letters)}, 113\penalty0 (2):\penalty0 27003,
  2016.
\newblock \doi{10.1209/0295-5075/113/27003}.

\bibitem[Stoudenmire and White(2012)]{2DDMRG}
E.M. Stoudenmire and Steven~R. White.
\newblock Studying two-dimensional systems with the density matrix
  renormalization group.
\newblock \emph{Annual Review of Condensed Matter Physics}, 3\penalty0
  (1):\penalty0 111--128, 2012.
\newblock \doi{10.1146/annurev-conmatphys-020911-125018}.

\bibitem[Strand et~al.(2009)Strand, Van~Harlingen, Kycia, and
  Halperin]{strand2009evidence}
JD~Strand, Dale~J Van~Harlingen, JB~Kycia, and William~P Halperin.
\newblock {Evidence for complex superconducting order parameter symmetry in the
  low-temperature phase of {UPt$_3$} from {Josephson} interferometry}.
\newblock \emph{Phys. Rev. Lett.}, 103\penalty0 (19):\penalty0 197002, 2009.
\newblock \doi{10.1103/PhysRevLett.103.197002}.

\bibitem[Su and Zhang(2020)]{su2020coincidence}
Yuehua Su and Chao Zhang.
\newblock Coincidence angle-resolved photoemission spectroscopy: Proposal for
  detection of two-particle correlations.
\newblock \emph{Physical Review B}, 101\penalty0 (20):\penalty0 205110, 2020.
\newblock \doi{10.1103/PhysRevB.101.205110}.

\bibitem[Sulpizio et~al.(2019)Sulpizio, Ella, Rozen, Birkbeck, Perello, Dutta,
  Ben-Shalom, Taniguchi, Watanabe, Holder, et~al.]{sulpizio2019visualizing}
Joseph~A. Sulpizio, Lior Ella, Asaf Rozen, John Birkbeck, David~J. Perello,
  Debarghya Dutta, Moshe Ben-Shalom, Takashi Taniguchi, Kenji Watanabe, Tobias
  Holder, et~al.
\newblock Visualizing {Poiseuille} flow of hydrodynamic electrons.
\newblock \emph{Nature}, 576\penalty0 (7785):\penalty0 75--79, 2019.
\newblock \doi{10.1038/s41586-019-1788-9}.

\bibitem[Sun et~al.(2018)Sun, Basov, and Fogler]{sun2018universal}
Zhiyuan Sun, Dmitry~N Basov, and Michael~M Fogler.
\newblock Universal linear and nonlinear electrodynamics of a {Dirac} fluid.
\newblock \emph{Proceedings of the National Academy of Sciences}, 115\penalty0
  (13):\penalty0 3285--3289, 2018.
\newblock \doi{10.1073/pnas.1717010115}.

\bibitem[Sunku et~al.(2018)Sunku, Ni, Jiang, Yoo, Sternbach, McLeod, Stauber,
  Xiong, Taniguchi, Watanabe, et~al.]{sunku2018photonic}
SS~Sunku, GuangXin Ni, Bor-Yuan Jiang, Hyobin Yoo, Aaron Sternbach, AS~McLeod,
  T~Stauber, Lin Xiong, Takashi Taniguchi, Kenji Watanabe, et~al.
\newblock Photonic crystals for nano-light in moir{\'e} graphene superlattices.
\newblock \emph{Science}, 362\penalty0 (6419):\penalty0 1153--1156, 2018.
\newblock \doi{10.1126/science.aau5144}.

\bibitem[Suzuki et~al.(2002)Suzuki, Naher, Shimoguchi, Mizuno, Ryu, and
  Fujishita]{Suzuki2002}
H.~Suzuki, S.~Naher, T.~Shimoguchi, M.~Mizuno, A.~Ryu, and H.~Fujishita.
\newblock X-ray diffraction measurement below 1 {K}.
\newblock \emph{Journal of Low Temperature Physics}, 128\penalty0 (1):\penalty0
  1--7, 2002.
\newblock URL \url{https://doi.org/10.1023/A:1015729021796}.

\bibitem[Suzuki et~al.(2004)Suzuki, Xue, Naher, Yamauchi, Asada, Mizuno, Abe,
  Kunii, and Kasaya]{suzuki2004}
H.~Suzuki, Y.~Xue, S.~Naher, R.~Yamauchi, S.~Asada, M.~Mizuno, S.~Abe,
  S.~Kunii, and M.~Kasaya.
\newblock Phase transitions studied by ultra-low-temperature x-ray diffraction.
\newblock \emph{Physica B: Condensed Matter}, 345\penalty0 (1):\penalty0
  239--242, 2004.
\newblock \doi{10.1016/j.physb.2003.11.063}.
\newblock URL
  \url{http://www.sciencedirect.com/science/article/pii/S0921452603010676}.

\bibitem[Swingle et~al.(2011)Swingle, Barkeshli, McGreevy, and
  Senthil]{swingle2011correlated}
Brian Swingle, Maissam Barkeshli, John McGreevy, and Todadri Senthil.
\newblock Correlated topological insulators and the fractional magnetoelectric
  effect.
\newblock \emph{Physical Review B}, 83\penalty0 (19):\penalty0 195139, 2011.
\newblock \doi{10.1103/PhysRevB.83.195139}.

\bibitem[Takagi et~al.(1997)Takagi, Nohara, and Cava]{takagi1997borocarbide}
H~Takagi, M~Nohara, and RJ~Cava.
\newblock Borocarbide superconductors: Materials and physical properties.
\newblock \emph{Physica B: Condensed Matter}, 237:\penalty0 292--295, 1997.
\newblock \doi{10.1016/S0921-4526(97)00175-0}.

\bibitem[Takagi et~al.(2019)Takagi, Takayama, Jackeli, Khaliullin, and
  Nagler]{kitaevspinliquid}
Hidenori Takagi, Tomohiro Takayama, George Jackeli, Giniyat Khaliullin, and
  Stephen~E. Nagler.
\newblock {Concept and realization of Kitaev quantum spin liquids}.
\newblock \emph{Nature Reviews Physics}, 1:\penalty0 264--280, 2019.
\newblock \doi{10.1038/s42254-019-0038-2}.

\bibitem[Takahashi and Hill(2005)]{takahashi2005rotating}
Susumu Takahashi and Stephen Hill.
\newblock Rotating cavity for high-field angle-dependent microwave spectroscopy
  of low-dimensional conductors and magnets.
\newblock \emph{Review of scientific instruments}, 76\penalty0 (2):\penalty0
  023114, 2005.
\newblock \doi{10.1063/1.1852859}.

\bibitem[Takenaka et~al.(2018)Takenaka, Ishihara, Miao, Huang, Xu, Zhu, Su,
  Cheng, and Shibauchi]{takenaka2018signature}
T~Takenaka, K~Ishihara, Y~Miao, X~Huang, W~Xu, D~Zhu, N~Su, J-G Cheng, and
  T~Shibauchi.
\newblock {Signature of Unconventional Superconductivity in a Copper-based
  Metal-Organic Framework with Perfect Kagome Structure}.
\newblock \emph{arXiv preprint arXiv:1810.00569}, 2018.
\newblock \doi{10.48550/arXiv.1810.00569}.

\bibitem[Tan et~al.(2015)Tan, Hsu, Zeng, Hatnean, Harrison, Zhu, Hartstein,
  Kiourlappou, Srivastava, Johannes, et~al.]{SebastianSmB6}
BS~Tan, Y-T Hsu, B~Zeng, M~Ciomaga Hatnean, N~Harrison, Z~Zhu, M~Hartstein,
  M~Kiourlappou, A~Srivastava, MD~Johannes, et~al.
\newblock Unconventional {Fermi} surface in an insulating state.
\newblock \emph{Science}, 349\penalty0 (6245):\penalty0 287--290, 2015.
\newblock \doi{10.1126/science.aaa7974}.

\bibitem[Tang et~al.(2011)Tang, Liang, Qiu, and Gao]{MR-Bi2Se3}
Hao Tang, Dong Liang, Richard~LJ Qiu, and Xuan~PA Gao.
\newblock Two-dimensional transport-induced linear magneto-resistance in
  topological insulator {Bi$_2$Se$_3$} nanoribbons.
\newblock \emph{ACS Nano}, 5\penalty0 (9):\penalty0 7510--7516, 2011.
\newblock \doi{10.1021/nn2024607}.

\bibitem[Tao et~al.(2017)Tao, Singh, Rossi, Gerritsen, Hendriksen,
  Khajetoorians, Christianen, Maan, Zeitler, and Bryant]{tao2017stm}
W.~Tao, S.~Singh, L.~Rossi, J.~W. Gerritsen, B.~L.~M. Hendriksen, A.~A.
  Khajetoorians, P.~C.~M. Christianen, J.~C. Maan, U.~Zeitler, and B.~Bryant.
\newblock {A low-temperature scanning tunneling microscope capable of
  microscopy and spectroscopy in a Bitter magnet at up to 34 T}.
\newblock \emph{Review of Scientific Instruments}, 88\penalty0 (9):\penalty0
  093706, 2017.
\newblock \doi{10.1063/1.4995372}.

\bibitem[Tarantini et~al.(2011)Tarantini, Gurevich, Jaroszynski, Balakirev,
  Bellingeri, Pallecchi, Ferdeghini, Shen, Wen, and
  Larbalestier]{Tarantini2011}
C.~Tarantini, A.~Gurevich, J.~Jaroszynski, F.~Balakirev, E.~Bellingeri,
  I.~Pallecchi, C.~Ferdeghini, B.~Shen, H.~H. Wen, and D.~C. Larbalestier.
\newblock Significant enhancement of upper critical fields by doping and strain
  in iron-based superconductors.
\newblock \emph{Phys. Rev. B}, 84:\penalty0 184522, 2011.
\newblock \doi{10.1103/PhysRevB.84.184522}.
\newblock URL \url{https://link.aps.org/doi/10.1103/PhysRevB.84.184522}.

\bibitem[Tatarova et~al.(2017)Tatarova, Dias, Henriques, Abrashev, Bundaleska,
  Kovacevic, Bundaleski, Cvelbar, Valcheva, Arnaudov,
  et~al.]{tatarova2017towards}
E~Tatarova, A~Dias, J~Henriques, M~Abrashev, N~Bundaleska, E~Kovacevic,
  N~Bundaleski, U~Cvelbar, E~Valcheva, B~Arnaudov, et~al.
\newblock Towards large-scale in free-standing graphene and $n$-graphene
  sheets.
\newblock \emph{Scientific Reports}, 7\penalty0 (1):\penalty0 1--16, 2017.
\newblock \doi{10.1038/s41598-017-10810-3}.

\bibitem[Tediosi et~al.(2007)Tediosi, Armitage, Giannini, and Van
  Der~Marel]{tediosi2007charge}
Riccardo Tediosi, NP~Armitage, Enrico Giannini, and Dirk Van Der~Marel.
\newblock Charge carrier interaction with a purely electronic collective mode:
  plasmarons and the infrared response of elemental bismuth.
\newblock \emph{Phys. Rev. Lett.}, 99\penalty0 (1):\penalty0 016406, 2007.
\newblock \doi{10.1103/PhysRevLett.99.016406}.

\bibitem[Terashima et~al.(2018)Terashima, Matsuda, Kohama, Ikeda, {Kondo},
  Kindo, and Iga]{terashima2018magnetic}
Taku~T Terashima, Yasuhiro~H Matsuda, Yoshimitsu Kohama, Akihiko Ikeda, Akihiro
  {Kondo}, Koichi Kindo, and Fumitoshi Iga.
\newblock {Magnetic-field-induced {Kondo} metal realized in YbB$_{12}$}.
\newblock \emph{Physical Review Letters}, 120\penalty0 (25):\penalty0 257206,
  2018.
\newblock \doi{10.1103/PhysRevLett.120.257206}.

\bibitem[Tetienne et~al.(2014)Tetienne, Hingant, Kim, Diez, Adam, Garcia, Roch,
  Rohart, Thiaville, Ravelosona, et~al.]{tetienne2014nanoscale}
J-P Tetienne, T~Hingant, J-V Kim, L~Herrera Diez, J-P Adam, K~Garcia, J-F Roch,
  S~Rohart, A~Thiaville, D~Ravelosona, et~al.
\newblock Nanoscale imaging and control of domain-wall hopping with a
  nitrogen-vacancy center microscope.
\newblock \emph{Science}, 344\penalty0 (6190):\penalty0 1366--1369, 2014.
\newblock \doi{10.1126/science.1250113}.

\bibitem[Thiel et~al.(2019)Thiel, Wang, Tschudin, Rohner, Guti{\'e}rrez-Lezama,
  Ubrig, Gibertini, Giannini, Morpurgo, and Maletinsky]{thiel2019probing}
Lucas Thiel, Zhe Wang, M{\"a}rta~A Tschudin, Dominik Rohner, Ignacio
  Guti{\'e}rrez-Lezama, Nicolas Ubrig, Marco Gibertini, Enrico Giannini,
  Alberto~F Morpurgo, and Patrick Maletinsky.
\newblock Probing magnetism in 2{D} materials at the nanoscale with single-spin
  microscopy.
\newblock \emph{Science}, 364\penalty0 (6444):\penalty0 973--976, 2019.
\newblock \doi{10.1126/science.aav6926}.

\bibitem[Thomas et~al.(1983)Thomas, Paalanen, and Rosenbaum]{Si-MIT2}
G.~A. Thomas, M.~Paalanen, and T.~F. Rosenbaum.
\newblock Measurements of conductivity near the metal-insulator critical point.
\newblock \emph{Phys. Rev. B}, 27:\penalty0 3897--3900, 1983.
\newblock \doi{10.1103/PhysRevB.27.3897}.
\newblock URL \url{https://link.aps.org/doi/10.1103/PhysRevB.27.3897}.

\bibitem[Thomas et~al.(2016)Thomas, Kim, Chung, Grant, Fisk, and Xia]{MR-SmB6}
S.~Thomas, D.~J. Kim, S.~B. Chung, T.~Grant, Z.~Fisk, and Jing Xia.
\newblock {Weak antilocalization and linear magnetoresistance in the surface
  state of ${\mathrm{SmB}}_{6}$}.
\newblock \emph{Phys. Rev. B}, 94:\penalty0 205114, 2016.
\newblock \doi{10.1103/PhysRevB.94.205114}.
\newblock URL \url{https://link.aps.org/doi/10.1103/PhysRevB.94.205114}.

\bibitem[Thorsm{\o}lle and Armitage(2010)]{thorsmolle2010ultrafast}
VK~Thorsm{\o}lle and NP~Armitage.
\newblock Ultrafast (but many-body) relaxation in a low-density electron glass.
\newblock \emph{Phys. Rev. Lett.}, 105\penalty0 (8):\penalty0 086601, 2010.
\newblock \doi{10.1103/PhysRevLett.105.086601}.

\bibitem[Thuneberg et~al.(1998)Thuneberg, Yip, Fogelstr{\"o}m, and
  Sauls]{thuneberg1998models}
EV~Thuneberg, SK~Yip, M~Fogelstr{\"o}m, and James~A Sauls.
\newblock {Models for Superfluid $^3$He in Aerogel}.
\newblock \emph{Phys. Rev. Lett.}, 80\penalty0 (13):\penalty0 2861, 1998.
\newblock \doi{10.1103/PhysRevLett.80.2861}.

\bibitem[Tian et~al.(2014)Tian, Chang, Cao, He, Ma, Xue, and Chen]{MR-TIs}
Jifa Tian, Cuizu Chang, Helin Cao, Ke~He, Xucun Ma, Qikun Xue, and Yong~P Chen.
\newblock Quantum and classical magnetoresistance in ambipolar topological
  insulator transistors with gate-tunable bulk and surface conduction.
\newblock \emph{Scientific reports}, 4:\penalty0 4859, 2014.
\newblock \doi{10.1038/srep04859}.

\bibitem[Titum et~al.(2016)Titum, Berg, Rudner, Refael, and
  Lindner]{PhysRevX.6.021013}
Paraj Titum, Erez Berg, Mark~S. Rudner, Gil Refael, and Netanel~H. Lindner.
\newblock Anomalous {Floquet}-{Anderson} insulator as a nonadiabatic quantized
  charge pump.
\newblock \emph{Phys. Rev. X}, 6:\penalty0 021013, 2016.
\newblock \doi{10.1103/PhysRevX.6.021013}.
\newblock URL \url{https://link.aps.org/doi/10.1103/PhysRevX.6.021013}.

\bibitem[Toews et~al.(2013)Toews, Zhang, Ross, Dabkowska, Gaulin, and
  Hill]{toews2013thermal}
WH~Toews, Songtian~S Zhang, KA~Ross, HA~Dabkowska, BD~Gaulin, and RW~Hill.
\newblock {Thermal Conductivity of Ho$_2$Ti$_2$O$_7$ along the [111]
  Direction}.
\newblock \emph{Phys. Rev. Lett.}, 110\penalty0 (21):\penalty0 217209, 2013.
\newblock \doi{10.1103/PhysRevLett.110.217209}.

\bibitem[Tokunaga et~al.(2008)Tokunaga, Kumai, Takeshita, Kaneko, He, Arima,
  and Tokura]{Tokunaga2008}
Y.~Tokunaga, R.~Kumai, N.~Takeshita, Y.~Kaneko, J.~P. He, T.~Arima, and
  Y.~Tokura.
\newblock {Effects of uniaxial stress on orbital stripe direction in half-doped
  layered manganites: ${\text{Eu}}_{0.5}{\text{Ca}}_{1.5}{\text{MnO}}_{4}$ and
  $\text{Pr}{(\text{Sr},\text{Ca})}_{2}{\text{Mn}}_{2}{\text{O}}_{7}$}.
\newblock \emph{Phys. Rev. B}, 78:\penalty0 155105, 2008.
\newblock \doi{10.1103/PhysRevB.78.155105}.
\newblock URL \url{https://link.aps.org/doi/10.1103/PhysRevB.78.155105}.

\bibitem[Tokunaga et~al.(2019)Tokunaga, Orlova, Bruyant, Aoki, Mayaffre,
  Kr{\"a}mer, Julien, Berthier, Horvati{\'c}, Higa, et~al.]{tokunaga2019high}
Y~Tokunaga, A~Orlova, N~Bruyant, D~Aoki, H~Mayaffre, S~Kr{\"a}mer, M-H Julien,
  C~Berthier, M~Horvati{\'c}, N~Higa, et~al.
\newblock {High-field phase diagram of the heavy-fermion metal {CeIn$_3$}:
  Pulsed-field {NMR} study on single crystals up to 56 T}.
\newblock \emph{Phys. Rev. B}, 99\penalty0 (8):\penalty0 085142, 2019.
\newblock \doi{10.1103/PhysRevB.99.085142}.

\bibitem[Torchinsky and Hsieh(2017)]{torchinsky2017shg}
Darius~H. Torchinsky and David Hsieh.
\newblock {Rotational Anisotropy Nonlinear Harmonic Generation}.
\newblock In Challa~S.S.R. Kumar, editor, \emph{Magnetic Characterization
  Techniques for Nanomaterials}, pages 1--49. Springer Berlin Heidelberg, 2017.
\newblock \doi{10.1007/978-3-662-52780-1_1}.

\bibitem[Toyama et~al.(2018)Toyama, Huang, Nakamura, Bondarenko, Tupchaya,
  Gruznev, Takayama, Zotov, Saranin, and Hasegawa]{PbSC2}
Haruko Toyama, Hongrui Huang, Tomonori Nakamura, Leonid~V. Bondarenko,
  Alexandra~Y. Tupchaya, Dimitry~V. Gruznev, Akari Takayama, Andrey~V. Zotov,
  Alexander~A. Saranin, and Shuji Hasegawa.
\newblock {Thickness Dependence of Surface Structure and Superconductivity in
  Pb Atomic Layers}.
\newblock \emph{Journal of the Physical Society of Japan}, 87\penalty0
  (11):\penalty0 113601, 2018.
\newblock \doi{10.7566/JPSJ.87.113601}.
\newblock URL \url{, https://doi.org/10.7566/JPSJ.87.113601}.

\bibitem[Trainer et~al.(2019)Trainer, Zhang, Bobba, Xi, Hla, and
  Iavarone]{trainer2019stm}
Daniel~J. Trainer, Yuan Zhang, Fabrizio Bobba, Xiaoxing Xi, Saw-Wai Hla, and
  Maria Iavarone.
\newblock {The Effects of Atomic-Scale Strain Relaxation on the Electronic
  Properties of Monolayer MoS$_2$}.
\newblock \emph{ACS Nano}, 13\penalty0 (7):\penalty0 8284--8291, 2019.
\newblock \doi{10.1021/acsnano.9b03652}.

\bibitem[Travaglini and Wachter(1984)]{Travaglini1984}
G.~Travaglini and P.~Wachter.
\newblock Intermediate-valent sm${\mathrm{b}}_{6}$ and the hybridization model:
  An optical study.
\newblock \emph{Phys. Rev. B}, 29:\penalty0 893--898, 1984.
\newblock \doi{10.1103/PhysRevB.29.893}.

\bibitem[Tripathi et~al.(2007)Tripathi, Chandra, and
  Coleman]{tripathi2007sleuthing}
V~Tripathi, Premala Chandra, and Piers Coleman.
\newblock Sleuthing hidden order.
\newblock \emph{Nature Physics}, 3\penalty0 (2):\penalty0 78--80, 2007.
\newblock \doi{10.1038/nphys524}.

\bibitem[Trivedi and Ceperley(1990)]{Trivedi_Ceperley}
Nandini Trivedi and D.~M. Ceperley.
\newblock Ground-state correlations of quantum antiferromagnets: A
  green-function {Monte} {Carlo} study.
\newblock \emph{Phys. Rev. B}, 41:\penalty0 4552--4569, 1990.
\newblock \doi{10.1103/PhysRevB.41.4552}.

\bibitem[Tsuei and Kirtley(2000)]{tsuei2000pairing}
CC~Tsuei and JR~Kirtley.
\newblock Pairing symmetry in cuprate superconductors.
\newblock \emph{Reviews of Modern Physics}, 72\penalty0 (4):\penalty0 969,
  2000.
\newblock \doi{10.1103/RevModPhys.72.969}.

\bibitem[Tsuei et~al.(1977)Tsuei, Johnson, Laibowitz, and
  Viggiano]{amorphousSC1}
CC~Tsuei, WL~Johnson, RB~Laibowitz, and JM~Viggiano.
\newblock The ratio of energy gap to transition temperature in amorphous
  superconductors.
\newblock \emph{Solid State Communications}, 24\penalty0 (9):\penalty0
  615--618, 1977.
\newblock \doi{10.1016/0038-1098(77)90374-X}.

\bibitem[Tsvetkov et~al.(1998)Tsvetkov, Van~der Marel, Moler, Kirtley, De~Boer,
  Meetsma, Ren, Koleshnikov, Dulic, Damascelli, et~al.]{tsvetkov1998global}
AA~Tsvetkov, D~Van~der Marel, KA~Moler, JR~Kirtley, JL~De~Boer, A~Meetsma,
  ZF~Ren, N~Koleshnikov, D~Dulic, A~Damascelli, et~al.
\newblock {Global and local measures of the intrinsic Josephson coupling in
  Tl$_2$Ba$_2$CuO$_6$ as a test of the interlayer tunnelling model}.
\newblock \emph{Nature}, 395\penalty0 (6700):\penalty0 360--362, 1998.
\newblock \doi{10.1038/26439}.

\bibitem[Tusche et~al.(2015)Tusche, Krasyuk, and Kirschner]{tusche2015spin}
Christian Tusche, Alexander Krasyuk, and J{\"u}rgen Kirschner.
\newblock Spin resolved bandstructure imaging with a high resolution momentum
  microscope.
\newblock \emph{Ultramicroscopy}, 159:\penalty0 520--529, 2015.
\newblock \doi{10.1016/j.ultramic.2015.03.020}.

\bibitem[Uehara et~al.(1999)Uehara, Mori, Chen, and
  Cheong]{uehara1999percolative}
M~Uehara, S~Mori, CH~Chen, and S-W Cheong.
\newblock Percolative phase separation underlies colossal magnetoresistance in
  mixed-valent manganites.
\newblock \emph{nature}, 399\penalty0 (6736):\penalty0 560--563, 1999.
\newblock \doi{10.1038/21142}.

\bibitem[Valla et~al.(2002)Valla, Johnson, Yusof, Wells, Li, Loureiro, Cava,
  Mikami, Mori, Yoshimura, and Sasaki]{Valla2002}
T.~Valla, P.~D. Johnson, Z.~Yusof, B.~Wells, Q.~Li, S.~M. Loureiro, R.~J. Cava,
  M.~Mikami, Y.~Mori, M.~Yoshimura, and T.~Sasaki.
\newblock Coherence-incoherence and dimensional crossover in layered strongly
  correlated metals.
\newblock \emph{Nature}, 417\penalty0 (6889):\penalty0 627--630, 2002.
\newblock \doi{10.1038/nature00774}.
\newblock URL \url{https://doi.org/10.1038/nature00774}.

\bibitem[Van~Harlingen(1995)]{van1995phase}
Dale~J Van~Harlingen.
\newblock {Phase-sensitive tests of the symmetry of the pairing state in the
  high-temperature superconductors—evidence for d$_{x^2- y^2}$ symmetry}.
\newblock \emph{Reviews of Modern Physics}, 67\penalty0 (2):\penalty0 515,
  1995.
\newblock \doi{10.1103/RevModPhys.67.515}.

\bibitem[Varma(1997)]{varma1997non}
CM~Varma.
\newblock Non-fermi-liquid states and pairing instability of a general model of
  copper oxide metals.
\newblock \emph{Phys. Rev. B}, 55\penalty0 (21):\penalty0 14554, 1997.
\newblock \doi{10.1103/PhysRevB.55.14554}.

\bibitem[Varma et~al.(1989)Varma, Littlewood, Schmitt-Rink, Abrahams, and
  Ruckenstein]{varma1989phenomenology}
CM~Varma, P~Be Littlewood, S~Schmitt-Rink, E~Abrahams, and AE~Ruckenstein.
\newblock {Phenomenology of the normal state of Cu-O high-temperature
  superconductors}.
\newblock \emph{Phys. Rev. Lett.}, 63\penalty0 (18):\penalty0 1996, 1989.
\newblock \doi{10.1103/PhysRevLett.63.1996}.

\bibitem[Vasyukov et~al.(2013)Vasyukov, Anahory, Embon, Halbertal, Cuppens,
  Neeman, Finkler, Segev, Myasoedov, Rappaport, et~al.]{vasyukov2013scanning}
Denis Vasyukov, Yonathan Anahory, Lior Embon, Dorri Halbertal, Jo~Cuppens, Lior
  Neeman, Amit Finkler, Yehonathan Segev, Yuri Myasoedov, Michael~L Rappaport,
  et~al.
\newblock A scanning superconducting quantum interference device with single
  electron spin sensitivity.
\newblock \emph{Nature nanotechnology}, 8\penalty0 (9):\penalty0 639, 2013.
\newblock \doi{10.1038/nnano.2013.169}.

\bibitem[Verret et~al.(2017{\natexlab{a}})Verret, Charlebois, S{\'e}n{\'e}chal,
  and Tremblay]{verret2017subgap}
S~Verret, M~Charlebois, D~S{\'e}n{\'e}chal, and A-MS Tremblay.
\newblock Subgap structures and pseudogap in cuprate superconductors: Role of
  density waves.
\newblock \emph{Phys. Rev. B}, 95\penalty0 (5):\penalty0 054518,
  2017{\natexlab{a}}.
\newblock \doi{10.1103/PhysRevB.95.054518}.

\bibitem[Verret et~al.(2017{\natexlab{b}})Verret, Simard, Charlebois,
  S{\'e}n{\'e}chal, and Tremblay]{verret2017phenomenological}
S~Verret, O~Simard, M~Charlebois, D~S{\'e}n{\'e}chal, and A-MS Tremblay.
\newblock Phenomenological theories of the low-temperature pseudogap: Hall
  number, specific heat, and seebeck coefficient.
\newblock \emph{Phys. Rev. B}, 96\penalty0 (12):\penalty0 125139,
  2017{\natexlab{b}}.
\newblock \doi{10.1103/PhysRevB.96.125139}.

\bibitem[Verstraete et~al.(2008)Verstraete, Murg, and Cirac]{TPS_review}
F.~Verstraete, V.~Murg, and J.I. Cirac.
\newblock Matrix product states, projected entangled pair states, and
  variational renormalization group methods for quantum spin systems.
\newblock \emph{Advances in Physics}, 57\penalty0 (2):\penalty0 143--224, 2008.
\newblock \doi{10.1080/14789940801912366}.

\bibitem[Vidal(2008)]{vidal_mera}
G.~Vidal.
\newblock Class of quantum many-body states that can be efficiently simulated.
\newblock \emph{Phys. Rev. Lett.}, 101:\penalty0 110501, 2008.
\newblock \doi{10.1103/PhysRevLett.101.110501}.
\newblock URL \url{https://link.aps.org/doi/10.1103/PhysRevLett.101.110501}.

\bibitem[Vig et~al.(2017)Vig, Kogar, Mitrano, Husain, Mishra, Rak, Venema,
  Johnson, Gu, Fradkin, et~al.]{vig2017measurement}
Sean Vig, Anshul Kogar, Matteo Mitrano, Ali~A Husain, Vivek Mishra, Melinda~S
  Rak, Luc Venema, Peter~D Johnson, Genda~D Gu, Eduardo Fradkin, et~al.
\newblock Measurement of the dynamic charge response of materials using
  low-energy, momentum-resolved electron energy-loss spectroscopy (m-eels).
\newblock \emph{SciPost Phys}, 3:\penalty0 026, 2017.
\newblock \doi{10.21468/SciPostPhys.3.2.021}.

\bibitem[Villaume et~al.(2008)Villaume, Bourdarot, Hassinger, Raymond, Taufour,
  Aoki, and Flouquet]{villaume2008signature}
A~Villaume, F~Bourdarot, E~Hassinger, S~Raymond, V~Taufour, Dai Aoki, and
  Jacques Flouquet.
\newblock {Signature of hidden order in heavy fermion superconductor
  URu$_2$Si$_2$: resonance at the wave vector Q$_0$=(1, 0, 0)}.
\newblock \emph{Phys. Rev. B}, 78\penalty0 (1):\penalty0 012504, 2008.
\newblock \doi{10.1103/PhysRevB.78.012504}.

\bibitem[Villeneuve et~al.(2017)Villeneuve, Hockett, Vrakking, and
  Niikura]{villeneuve2017coherent}
D.~M. Villeneuve, Paul Hockett, M.~J.~J. Vrakking, and Hiromichi Niikura.
\newblock {Coherent imaging of an attosecond electron wave packet}.
\newblock \emph{Science}, 356\penalty0 (6343):\penalty0 1150--1153, 2017.
\newblock \doi{10.1126/science.aam8393}.

\bibitem[Virk and Sipe(2011)]{virk2011optical}
Kuljit~S Virk and JE~Sipe.
\newblock Optical injection and terahertz detection of the macroscopic
  {Berry}curvature.
\newblock \emph{Phys. Rev. Lett.}, 107\penalty0 (12):\penalty0 120403, 2011.
\newblock \doi{10.1103/PhysRevLett.107.120403}.

\bibitem[Vogl et~al.(2019)Vogl, Laurell, Barr, and Fiete]{PhysRevX.9.021037}
Michael Vogl, Pontus Laurell, Aaron~D. Barr, and Gregory~A. Fiete.
\newblock Flow equation approach to periodically driven quantum systems.
\newblock \emph{Phys. Rev. X}, 9:\penalty0 021037, 2019.
\newblock \doi{10.1103/PhysRevX.9.021037}.
\newblock URL \url{https://link.aps.org/doi/10.1103/PhysRevX.9.021037}.

\bibitem[Vogl et~al.(2020)Vogl, Rodriguez-Vega, and Fiete]{Vogl_2020}
Michael Vogl, Martin Rodriguez-Vega, and Gregory~A. Fiete.
\newblock Effective {Floquet} hamiltonian in the low-frequency regime.
\newblock \emph{Phys. Rev. B}, 101\penalty0 (2), 2020.
\newblock \doi{10.1103/physrevb.101.024303}.
\newblock URL \url{http://dx.doi.org/10.1103/PhysRevB.101.024303}.

\bibitem[Vojta(2019)]{vojta2019disorder}
Thomas Vojta.
\newblock Disorder in quantum many-body systems.
\newblock \emph{Annual Review of Condensed Matter Physics}, 10\penalty0
  (1):\penalty0 233--252, 2019.
\newblock \doi{10.1146/annurev-conmatphys-031218-013433}.
\newblock URL \url{, https://doi.org/10.1146/annurev-conmatphys-031218-013433}.

\bibitem[W.~Hicks et~al.(2008)W.~Hicks, M.~Lippman, A.~Moler, E.~Huber, Ren,
  and Zhao]{w2008scanning}
Clifford W.~Hicks, Thomas M.~Lippman, Kathryn A.~Moler, Martin E.~Huber, Zhi-An
  Ren, and Zhong-Xian Zhao.
\newblock {Scanning SQUID microscopy on polycrystalline SmFeAsO$_{0.85}$ and
  NdFeAsO$_{0.94}$F$_{0.06}$ }.
\newblock \emph{Journal of the Physical Society of Japan}, 77\penalty0 (Suppl.
  C):\penalty0 87--90, 2008.
\newblock \doi{10.1143/JPSJS.77SC.87}.

\bibitem[Wagner and Ceperley(2016)]{Wagner_Ceperley_review}
Lucas~K Wagner and David~M Ceperley.
\newblock Discovering correlated fermions using quantum {Monte} {Carlo}.
\newblock \emph{Reports on Progress in Physics}, 79\penalty0 (9):\penalty0
  094501, 2016.
\newblock \doi{10.1088/0034-4885/79/9/094501}.
\newblock URL \url{https://doi.org/10.1088%2F0034-4885%2F79%2F9%2F094501}.

\bibitem[Wakeham et~al.(2016)Wakeham, Wen, Wang, Fisk, Ronning, and
  Thompson]{Wakeham2016}
N~Wakeham, J~Wen, YQ~Wang, Zachary Fisk, Filip Ronning, and Joe~David Thompson.
\newblock {The effect of magnetic and non-magnetic ion damage on the surface
  state in SmB$_6$}.
\newblock \emph{Journal of Magnetism and Magnetic Materials}, 400:\penalty0
  62--65, 2016.
\newblock \doi{10.1016/j.jmmm.2015.07.045}.

\bibitem[Wallace(1947)]{graphitebands1}
P.~R. Wallace.
\newblock The band theory of graphite.
\newblock \emph{Phys. Rev.}, 71:\penalty0 622--634, 1947.
\newblock \doi{10.1103/PhysRev.71.622}.
\newblock URL \url{https://link.aps.org/doi/10.1103/PhysRev.71.622}.

\bibitem[Wallin et~al.(1994)Wallin, So, Girvin, Young,
  et~al.]{wallin1994superconductor}
Mats Wallin, Erik~S So, SM~Girvin, AP~Young, et~al.
\newblock Superconductor-insulator transition in two-dimensional dirty boson
  systems.
\newblock \emph{Phys. Rev. B}, 49\penalty0 (17):\penalty0 12115, 1994.
\newblock \doi{10.1103/PhysRevB.49.12115}.

\bibitem[Wan and Armitage(2019)]{wan2019resolving}
Yuan Wan and NP~Armitage.
\newblock Resolving continua of fractional excitations by spinon echo in {THz}
  2{D} coherent spectroscopy.
\newblock \emph{Phys. Rev. Lett.}, 122\penalty0 (25):\penalty0 257401, 2019.
\newblock \doi{10.1103/PhysRevLett.122.257401}.

\bibitem[Wandel et~al.(2022)Wandel, Boschini, da~Silva~Neto, Shen, Na, Zohar,
  Wang, Welch, Seaberg, Koralek, et~al.]{wandel2020light}
S.~Wandel, F.~Boschini, E.~H. da~Silva~Neto, L.~Shen, M.~X. Na, S.~Zohar,
  Y.~Wang, S.~B. Welch, M.~H. Seaberg, J.~D. Koralek, et~al.
\newblock Enhanced charge density wave coherence in a light-quenched,
  high-temperature superconductor.
\newblock \emph{Science}, 376\penalty0 (6595):\penalty0 860--864, 2022.
\newblock \doi{10.1126/science.abd7213}.

\bibitem[Wang and Senthil(2014)]{Wang_2014}
Chong Wang and T.~Senthil.
\newblock Interacting fermionic topological insulators/superconductors in three
  dimensions.
\newblock \emph{Phys. Rev. B}, 89\penalty0 (19), 2014.
\newblock \doi{10.1103/physrevb.89.195124}.
\newblock URL \url{http://dx.doi.org/10.1103/PhysRevB.89.195124}.

\bibitem[Wang et~al.(2017)Wang, Ortix, van~den Brink, and
  Efremov]{PhysRevB.96.201104}
Jing Wang, Carmine Ortix, Jeroen van~den Brink, and Dmitry~V. Efremov.
\newblock Fate of interaction-driven topological insulators under disorder.
\newblock \emph{Phys. Rev. B}, 96:\penalty0 201104, 2017.
\newblock \doi{10.1103/PhysRevB.96.201104}.
\newblock URL \url{https://link.aps.org/doi/10.1103/PhysRevB.96.201104}.

\bibitem[Wang et~al.(2020)Wang, Shih, Ghiotto, Xian, Rhodes, Tan, Claassen,
  Kennes, Bai, Kim, et~al.]{wang2019magic}
Lei Wang, En-Min Shih, Augusto Ghiotto, Lede Xian, Daniel~A Rhodes, Cheng Tan,
  Martin Claassen, Dante~M Kennes, Yusong Bai, Bumho Kim, et~al.
\newblock Correlated electronic phases in twisted bilayer transition metal
  dichalcogenides.
\newblock \emph{Nature materials}, pages 1--6, 2020.
\newblock \doi{10.1038/s41563-020-0708-6}.

\bibitem[{Wang} et~al.(2019){Wang}, {Xiang}, {Cortie}, {Yue}, {Li}, {Zhang},
  and {Sang}]{liquidmetalMR}
Xiaolin {Wang}, Feixiang {Xiang}, David {Cortie}, Zengji {Yue}, Zhi {Li},
  Zhidong {Zhang}, and Lina {Sang}.
\newblock {Giant and Linear Magnetoresistance in Liquid Metals at Ambient
  Temperature}.
\newblock \emph{arXiv preprint arXiv:1907.01462}, 2019.
\newblock \doi{10.48550/arXiv.1907.01462}.

\bibitem[Wang et~al.(2018{\natexlab{a}})Wang, Claassen, Pemmaraju, Jia, Moritz,
  and Devereaux]{Wang2018}
Yao Wang, Martin Claassen, Chaitanya~Das Pemmaraju, Chunjing Jia, Brian Moritz,
  and Thomas~P Devereaux.
\newblock Theoretical understanding of photon spectroscopies in correlated
  materials in and out of equilibrium.
\newblock \emph{Nature Reviews Materials}, 3\penalty0 (9):\penalty0 312--323,
  2018{\natexlab{a}}.
\newblock \doi{10.1038/s41578-018-0046-3}.

\bibitem[Wang et~al.(2018{\natexlab{b}})Wang, Rosenbaum, Palmer, Ren, Kim,
  Mandrus, and Feng]{wang2018strongly}
Yishu Wang, TF~Rosenbaum, A~Palmer, Y~Ren, J-W Kim, D~Mandrus, and Yejun Feng.
\newblock Strongly-coupled quantum critical point in an all-in-all-out
  antiferromagnet.
\newblock \emph{Nature Communications}, 9\penalty0 (1):\penalty0 2953,
  2018{\natexlab{b}}.
\newblock \doi{10.1038/s41467-018-05435-7}.

\bibitem[Wang et~al.(2019)Wang, Rosenbaum, and Feng]{wang2019x}
Yishu Wang, TF~Rosenbaum, and Yejun Feng.
\newblock X-ray magnetic diffraction under high pressure.
\newblock \emph{IUCrJ}, 6\penalty0 (4):\penalty0 507--520, 2019.
\newblock \doi{10.1107/S2052252519007061}.

\bibitem[Wang et~al.(2016)Wang, Abanov, Altshuler, Yuzbashyan, and
  Chubukov]{wang2016superconductivity}
Yuxuan Wang, Artem Abanov, Boris~L Altshuler, Emil~A Yuzbashyan, and Andrey~V
  Chubukov.
\newblock Superconductivity near a quantum-critical point: The special role of
  the first matsubara frequency.
\newblock \emph{Phys. Rev. Lett.}, 117\penalty0 (15):\penalty0 157001, 2016.
\newblock \doi{10.1103/PhysRevLett.117.157001}.

\bibitem[Watanabe et~al.(2018)Watanabe, Po, and
  Vishwanath]{watanabe2018structure}
Haruki Watanabe, Hoi~Chun Po, and Ashvin Vishwanath.
\newblock Structure and topology of band structures in the 1651 magnetic space
  groups.
\newblock \emph{Science advances}, 4\penalty0 (8):\penalty0 eaat8685, 2018.
\newblock \doi{10.1126/sciadv.aat8685}.

\bibitem[Watson et~al.(2019)Watson, Sochnikov, Kirtley, Cava, and
  Moler]{watson2019real}
Christopher~A Watson, Ilya Sochnikov, John~R Kirtley, Robert~J Cava, and
  Kathryn~A Moler.
\newblock {Real-space imaging and flux noise spectroscopy of magnetic dynamics
  in Ho$_2 $Ti$_2 $O$_7$}.
\newblock \emph{arXiv preprint arXiv:1903.11465}, 2019.
\newblock \doi{10.48550/arXiv.1903.11465}.

\bibitem[Weber et~al.(2012)Weber, Yee, Haule, and Kotliar]{Weber_2012}
C.~Weber, C.~Yee, K.~Haule, and G.~Kotliar.
\newblock Scaling of the transition temperature of hole-doped cuprate
  superconductors with the charge-transfer energy.
\newblock \emph{{EPL} (Europhysics Letters)}, 100\penalty0 (3):\penalty0 37001,
  2012.
\newblock \doi{10.1209/0295-5075/100/37001}.
\newblock URL \url{https://doi.org/10.1209%2F0295-5075%2F100%2F37001}.

\bibitem[Wegner(1994)]{wegner1994flow}
Franz Wegner.
\newblock Flow-equations for hamiltonians.
\newblock \emph{Annalen der physik}, 506\penalty0 (2):\penalty0 77--91, 1994.
\newblock \doi{10.1002/andp.19945060203}.

\bibitem[Weidinger and Knap(2017)]{weidinger2017floquet}
Simon~A Weidinger and Michael Knap.
\newblock Floquet prethermalization and regimes of heating in a periodically
  driven, interacting quantum system.
\newblock \emph{Scientific reports}, 7\penalty0 (1):\penalty0 1--10, 2017.
\newblock \doi{10.1038/srep45382}.

\bibitem[Wen(2012)]{PhysRevB.85.085103}
Xiao-Gang Wen.
\newblock Symmetry-protected topological phases in noninteracting fermion
  systems.
\newblock \emph{Phys. Rev. B}, 85:\penalty0 085103, 2012.
\newblock \doi{10.1103/PhysRevB.85.085103}.
\newblock URL \url{https://link.aps.org/doi/10.1103/PhysRevB.85.085103}.

\bibitem[Wheatley et~al.(1988)Wheatley, Hsu, and
  Anderson]{wheatley1988interlayer}
JM~Wheatley, TC~Hsu, and PW~Anderson.
\newblock Interlayer effects in high-$t_c$ superconductors.
\newblock \emph{Nature}, 333\penalty0 (6169):\penalty0 121--121, 1988.
\newblock \doi{10.1038/333121a0}.

\bibitem[White(1992)]{white1992}
Steven~R. White.
\newblock Density matrix formulation for quantum renormalization groups.
\newblock \emph{Phys. Rev. Lett.}, 69:\penalty0 2863--2866, 1992.
\newblock \doi{10.1103/PhysRevLett.69.2863}.

\bibitem[White and Feiguin(2004)]{White_Feiguin}
Steven~R. White and Adrian~E. Feiguin.
\newblock Real-time evolution using the density matrix renormalization group.
\newblock \emph{Phys. Rev. Lett.}, 93:\penalty0 076401, 2004.
\newblock \doi{10.1103/PhysRevLett.93.076401}.
\newblock URL \url{https://link.aps.org/doi/10.1103/PhysRevLett.93.076401}.

\bibitem[Wietek and L\"auchli(2018)]{Wietek_Lauchli}
Alexander Wietek and Andreas~M. L\"auchli.
\newblock Sublattice coding algorithm and distributed memory parallelization
  for large-scale exact diagonalizations of quantum many-body systems.
\newblock \emph{Phys. Rev. E}, 98:\penalty0 033309, 2018.
\newblock \doi{10.1103/PhysRevE.98.033309}.
\newblock URL \url{https://link.aps.org/doi/10.1103/PhysRevE.98.033309}.

\bibitem[Willans et~al.(2011)Willans, Chalker, and Moessner]{Willans_PRB_2011}
A.~J. Willans, J.~T. Chalker, and R.~Moessner.
\newblock {Site dilution in the Kitaev honeycomb model}.
\newblock \emph{Phys. Rev. B}, 84:\penalty0 115146, 2011.
\newblock \doi{10.1103/PhysRevB.84.115146}.
\newblock URL \url{https://link.aps.org/doi/10.1103/PhysRevB.84.115146}.

\bibitem[Willett et~al.(1987)Willett, Eisenstein, St{\"o}rmer, Tsui, Gossard,
  and English]{willett1987observation}
R~Willett, JP~Eisenstein, HL~St{\"o}rmer, DC~Tsui, AC~Gossard, and JH~English.
\newblock Observation of an even-denominator quantum number in the fractional
  quantum hall effect.
\newblock \emph{Phys. Rev. Lett.}, 59\penalty0 (15):\penalty0 1776, 1987.
\newblock \doi{10.1103/PhysRevLett.59.1776}.

\bibitem[Wilson(1975)]{NRG_wilson}
Kenneth~G. Wilson.
\newblock The renormalization group: Critical phenomena and the {Kondo}
  problem.
\newblock \emph{Rev. Mod. Phys.}, 47:\penalty0 773--840, 1975.
\newblock \doi{10.1103/RevModPhys.47.773}.
\newblock URL \url{https://link.aps.org/doi/10.1103/RevModPhys.47.773}.

\bibitem[Wolf et~al.(2015)Wolf, Neumann, Nakamura, Sumiya, Ohshima, Isoya, and
  Wrachtrup]{wolf2015subpicotesla}
Thomas Wolf, Philipp Neumann, Kazuo Nakamura, Hitoshi Sumiya, Takeshi Ohshima,
  Junichi Isoya, and J{\"o}rg Wrachtrup.
\newblock Subpicotesla diamond magnetometry.
\newblock \emph{Physical Review X}, 5\penalty0 (4):\penalty0 041001, 2015.
\newblock \doi{10.1103/PhysRevX.5.041001}.

\bibitem[Wu et~al.(2019)Wu, Gong, and Sheng]{Wu_PRB_2019}
Han-Qing Wu, Shou-Shu Gong, and D.~N. Sheng.
\newblock Randomness-induced spin-liquid-like phase in the spin-$\frac{1}{2}$
  ${J}_{1}\ensuremath{-}{J}_{2}$ triangular {Heisenberg} model.
\newblock \emph{Phys. Rev. B}, 99:\penalty0 085141, 2019.
\newblock \doi{10.1103/PhysRevB.99.085141}.
\newblock URL \url{https://link.aps.org/doi/10.1103/PhysRevB.99.085141}.

\bibitem[Xia et~al.(2008)Xia, Schemm, Deutscher, Kivelson, Bonn, Hardy, Liang,
  Siemons, Koster, Fejer, et~al.]{xia2008polar}
Jing Xia, Elizabeth Schemm, G~Deutscher, SA~Kivelson, DA~Bonn, WN~Hardy,
  R~Liang, W~Siemons, Gertjan Koster, MM~Fejer, et~al.
\newblock Polar {Kerr}-effect measurements of the high-temperature
  {YBa$_2$Cu$_3$O$_{6+x}$} superconductor: evidence for broken symmetry near
  the pseudogap temperature.
\newblock \emph{Phys. Rev. Lett.}, 100\penalty0 (12):\penalty0 127002, 2008.
\newblock \doi{10.1103/PhysRevLett.100.127002}.

\bibitem[Xia et~al.(2004)Xia, Pan, Vicente, Adams, Sullivan, Stormer, Tsui,
  Pfeiffer, Baldwin, and West]{xia2004electron}
JS~Xia, W~Pan, CLet Vicente, ED~Adams, NS~Sullivan, HL~Stormer, DC~Tsui,
  LN~Pfeiffer, KW~Baldwin, and KW~West.
\newblock Electron correlation in the second landau level: A competition
  between many nearly degenerate quantum phases.
\newblock \emph{Phys. Rev. Lett.}, 93\penalty0 (17):\penalty0 176809, 2004.
\newblock \doi{10.1103/PhysRevLett.93.176809}.

\bibitem[Xiong(2018)]{XiongJPC2018}
Ye~Xiong.
\newblock Why does bulk boundary correspondence fail in some non-hermitian
  topological models.
\newblock \emph{Journal of Physics Communications}, 2\penalty0 (3):\penalty0
  035043, 2018.
\newblock \doi{10.1088/2399-6528/aab64a}.

\bibitem[Xu and Balents(2018)]{PhysRevLett.121.087001}
Cenke Xu and Leon Balents.
\newblock Topological superconductivity in twisted multilayer graphene.
\newblock \emph{Phys. Rev. Lett.}, 121:\penalty0 087001, 2018.
\newblock \doi{10.1103/PhysRevLett.121.087001}.
\newblock URL \url{https://link.aps.org/doi/10.1103/PhysRevLett.121.087001}.

\bibitem[Xu and Moore(2006)]{PhysRevB.73.045322}
Cenke Xu and J.~E. Moore.
\newblock Stability of the quantum spin hall effect: Effects of interactions,
  disorder, and ${\mathbb{z}}_{2}$ topology.
\newblock \emph{Phys. Rev. B}, 73:\penalty0 045322, 2006.
\newblock \doi{10.1103/PhysRevB.73.045322}.
\newblock URL \url{https://link.aps.org/doi/10.1103/PhysRevB.73.045322}.

\bibitem[Xu and Wu(2018)]{WuPRL2018}
Shenglong Xu and Congjun Wu.
\newblock Space-time crystal and space-time group.
\newblock \emph{Phys. Rev. Lett.}, 120:\penalty0 096401, 2018.
\newblock \doi{10.1103/PhysRevLett.120.096401}.
\newblock URL \url{https://link.aps.org/doi/10.1103/PhysRevLett.120.096401}.

\bibitem[Xu et~al.(2019)Xu, Liu, Pan, Qi, Sun, and Meng]{Xu_2019}
Xiao~Yan Xu, Zi~Hong Liu, Gaopei Pan, Yang Qi, Kai Sun, and Zi~Yang Meng.
\newblock Revealing fermionic quantum criticality from new {Monte} {Carlo}
  techniques.
\newblock \emph{Journal of Physics: Condensed Matter}, 31\penalty0
  (46):\penalty0 463001, 2019.
\newblock \doi{10.1088/1361-648x/ab3295}.
\newblock URL \url{https://doi.org/10.1088%2F1361-648x%2Fab3295}.

\bibitem[Xu et~al.(2016)Xu, Zhang, Li, Yu, Hong, Zhang, and Li]{Xu_YMGO_2016}
Y.~Xu, J.~Zhang, Y.~S. Li, Y.~J. Yu, X.~C. Hong, Q.~M. Zhang, and S.~Y. Li.
\newblock Absence of magnetic thermal conductivity in the quantum spin-liquid
  candidate ${\mathrm{ybmggao}}_{4}$.
\newblock \emph{Phys. Rev. Lett.}, 117:\penalty0 267202, 2016.
\newblock \doi{10.1103/PhysRevLett.117.267202}.
\newblock URL \url{https://link.aps.org/doi/10.1103/PhysRevLett.117.267202}.

\bibitem[Yamada et~al.(2017)Yamada, Fujita, and Oshikawa]{yamada2017designing}
Masahiko~G Yamada, Hiroyuki Fujita, and Masaki Oshikawa.
\newblock {Designing Kitaev spin liquids in metal-organic frameworks}.
\newblock \emph{Phys. Rev. Lett.}, 119\penalty0 (5):\penalty0 057202, 2017.
\newblock \doi{10.1103/PhysRevLett.119.057202}.

\bibitem[Yamada and Shirane(1969)]{YamadaSTO}
Yasusada Yamada and Gen Shirane.
\newblock {Neutron scattering and nature of the soft optical phonon in
  SrTiO$_3$}.
\newblock \emph{Journal of the Physical Society of Japan}, 26\penalty0
  (2):\penalty0 396--403, 1969.
\newblock \doi{10.1143/JPSJ.26.396}.

\bibitem[Yan et~al.(2011)Yan, Huse, and White]{Yan_Huse_White}
Simeng Yan, David~A. Huse, and Steven~R. White.
\newblock Spin-liquid ground state of the s = 1/2 kagome {Heisenberg}
  antiferromagnet.
\newblock \emph{Science}, 332\penalty0 (6034):\penalty0 1173--1176, 2011.
\newblock \doi{10.1126/science.1201080}.
\newblock URL \url{https://science.sciencemag.org/content/332/6034/1173}.

\bibitem[Yang et~al.(2013)Yang, Hebestreit, Josberger, and
  Raschke]{yang2013snom}
Honghua~U. Yang, Erik Hebestreit, Erik~E. Josberger, and Markus~B. Raschke.
\newblock {A cryogenic scattering-type scanning near-field optical microscope}.
\newblock \emph{Review of Scientific Instruments}, 84\penalty0 (2):\penalty0
  023701, 2013.
\newblock \doi{10.1063/1.4789428}.

\bibitem[Yang et~al.(2006)Yang, Rice, and Zhang]{yang2006phenomenological}
Kai-Yu Yang, TM~Rice, and Fu-Chun Zhang.
\newblock Phenomenological theory of the pseudogap state.
\newblock \emph{Phys. Rev. B}, 73\penalty0 (17):\penalty0 174501, 2006.
\newblock \doi{10.1103/PhysRevB.73.174501}.

\bibitem[Yang et~al.(2015)Yang, Sobota, Leuenberger, He, Hashimoto, Lu, Eisaki,
  Kirchmann, and Shen]{yang2015inequivalence}
S-L Yang, Jonathan~A Sobota, Dominik Leuenberger, Yu~He, Makoto Hashimoto,
  DH~Lu, Hiroshi Eisaki, Patrick~S Kirchmann, and Z-X Shen.
\newblock Inequivalence of single-particle and population lifetimes in a
  cuprate superconductor.
\newblock \emph{Phys. Rev. Lett.}, 114\penalty0 (24):\penalty0 247001, 2015.
\newblock \doi{10.1103/PhysRevLett.114.247001}.

\bibitem[Yankowitz et~al.(2019)Yankowitz, Chen, Polshyn, Zhang, Watanabe,
  Taniguchi, Graf, Young, and Dean]{yankowitz2019tuning}
Matthew Yankowitz, Shaowen Chen, Hryhoriy Polshyn, Yuxuan Zhang, K~Watanabe,
  T~Taniguchi, David Graf, Andrea~F Young, and Cory~R Dean.
\newblock Tuning superconductivity in twisted bilayer graphene.
\newblock \emph{Science}, 363\penalty0 (6431):\penalty0 1059--1064, 2019.
\newblock \doi{10.1126/science.aav1910}.

\bibitem[Yao and Wang(2018)]{WangPRL2018}
Shunyu Yao and Zhong Wang.
\newblock Edge states and topological invariants of non-hermitian systems.
\newblock \emph{Phys. Rev. Lett.}, 121:\penalty0 086803, 2018.
\newblock \doi{10.1103/PhysRevLett.121.086803}.
\newblock URL \url{https://link.aps.org/doi/10.1103/PhysRevLett.121.086803}.

\bibitem[Yao et~al.(2019)Yao, Gao, Han, Jain, Moon, Kim, Zhu, Cheong, and
  Oh]{yao2019record}
Xiong Yao, Bin Gao, Myung-Geun Han, Deepti Jain, Jisoo Moon, Jae~Wook Kim,
  Yimei Zhu, Sang-Wook Cheong, and Seongshik Oh.
\newblock {Record High-Proximity-Induced Anomalous Hall Effect in
  (Bi$_x$Sb$_{1-x}$)$_2$Te$_3$ Thin Film Grown on CrGeTe$_3$ Substrate}.
\newblock \emph{Nano letters}, 19\penalty0 (7):\penalty0 4567--4573, 2019.
\newblock \doi{10.1021/acs.nanolett.9b01495}.

\bibitem[Yava{\c{s}} et~al.(2019)Yava{\c{s}}, Sundermann, Chen, Amorese,
  Severing, Gretarsson, Haverkort, and Tjeng]{yavacs2019direct}
Hasan Yava{\c{s}}, Martin Sundermann, Kai Chen, Andrea Amorese, Andrea
  Severing, Hlynur Gretarsson, Maurits~W Haverkort, and Liu~Hao Tjeng.
\newblock Direct imaging of orbitals in quantum materials.
\newblock \emph{Nature Physics}, 15\penalty0 (6):\penalty0 559--562, 2019.
\newblock \doi{10.1038/s41567-019-0471-2}.

\bibitem[Yin et~al.(2013)Yin, Kutepov, and Kotliar]{yin2013correlation}
ZP~Yin, A~Kutepov, and G~Kotliar.
\newblock {Correlation-Enhanced Electron-Phonon Coupling: Applications of GW
  and Screened Hybrid Functional to Bismuthates, Chloronitrides, and Other
  High-$T_c$ Superconductors}.
\newblock \emph{Physical Review X}, 3\penalty0 (2):\penalty0 021011, 2013.
\newblock \doi{10.1103/PhysRevX.3.021011}.

\bibitem[Yu et~al.(2019)Yu, Ma, Cai, Zhong, Ye, Shen, Gu, Chen, and
  Zhang]{yu2019high}
Yijun Yu, Liguo Ma, Peng Cai, Ruidan Zhong, Cun Ye, Jian Shen, Genda~D Gu,
  Xian~Hui Chen, and Yuanbo Zhang.
\newblock {High-temperature superconductivity in monolayer
  {Bi}$_2${Sr}$_2${CaCu}$_2${O}$_{8+\delta}$}.
\newblock \emph{Nature}, 575\penalty0 (7781):\penalty0 156--163, 2019.
\newblock \doi{10.1038/s41586-019-1718-x}.

\bibitem[Zabolotnyy et~al.(2012)Zabolotnyy, Carleschi, Kim, Kordyuk, Trinckauf,
  Geck, Evtushinsky, Doyle, Fittipaldi, Cuoco, Vecchione, B{\"{u}}chner, and
  Borisenko]{zabolotnyy2012sro}
V.~B. Zabolotnyy, E.~Carleschi, T.~K. Kim, A.~A. Kordyuk, J.~Trinckauf,
  J.~Geck, D.~Evtushinsky, B.~P. Doyle, R.~Fittipaldi, M.~Cuoco, A.~Vecchione,
  B.~B{\"{u}}chner, and S.~V. Borisenko.
\newblock {Surface and bulk electronic structure of the unconventional
  superconductor Sr$_2$RuO$_4$: unusual splitting of the $\beta$ band}.
\newblock \emph{New Journal of Physics}, 14\penalty0 (6):\penalty0 063039,
  2012.
\newblock \doi{10.1088/1367-2630/14/6/063039}.

\bibitem[Zaric et~al.(2006)Zaric, Ostojic, Shaver, Kono, Portugall, Frings,
  Rikken, Furis, Crooker, Wei, et~al.]{zaric2006excitons}
S~Zaric, GN~Ostojic, Jonah Shaver, J~Kono, O~Portugall, PH~Frings, GLJA Rikken,
  M~Furis, SA~Crooker, X~Wei, et~al.
\newblock Excitons in carbon nanotubes with broken time-reversal symmetry.
\newblock \emph{Physical Review Letters}, 96\penalty0 (1):\penalty0 016406,
  2006.
\newblock \doi{10.1103/PhysRevLett.96.016406}.

\bibitem[Zayed et~al.(2017)Zayed, R{\"u}egg, Larrea~J., L{\"a}uchli,
  Panagopoulos, Saxena, Ellerby, McMorrow, Str{\"a}ssle, Klotz, Hamel, Sadykov,
  Pomjakushin, Boehm, Jim{\'e}nez-Ruiz, Schneidewind, Pomjakushina, Stingaciu,
  Conder, and R{\o}nnow]{Zayed2017}
M.~E. Zayed, C.~R{\"u}egg, J.~Larrea~J., A.~M. L{\"a}uchli, C.~Panagopoulos,
  S.~S. Saxena, M.~Ellerby, D.~F. McMorrow, Th. Str{\"a}ssle, S.~Klotz,
  G.~Hamel, R.~A. Sadykov, V.~Pomjakushin, M.~Boehm, M.~Jim{\'e}nez-Ruiz,
  A.~Schneidewind, E.~Pomjakushina, M.~Stingaciu, K.~Conder, and H.~M.
  R{\o}nnow.
\newblock {4-spin plaquette singlet state in the Shastry--Sutherland compound
  SrCu$_2$(BO$_3$)$_2$}.
\newblock \emph{Nature Physics}, 13:\penalty0 962 EP --, 2017.
\newblock URL \url{http://dx.doi.org/10.1038/nphys4190}.

\bibitem[Zhang et~al.(2019{\natexlab{a}})Zhang, Liu, Wimmer, and
  Kouwenhoven]{zhang2019next}
Hao Zhang, Dong~E. Liu, Michael Wimmer, and Leo~P. Kouwenhoven.
\newblock Next steps of quantum transport in {Majorana} nanowire devices.
\newblock \emph{Nature Communications}, 10\penalty0 (1):\penalty0 1--7,
  2019{\natexlab{a}}.
\newblock \doi{10.1038/s41467-019-13133-1}.

\bibitem[Zhang et~al.(2014)Zhang, Fritsch, Hao, Bagheri, Gingras, Granroth,
  Jiramongkolchai, Cava, and Gaulin]{zhang2014neutron}
J~Zhang, Katharina Fritsch, Z~Hao, BV~Bagheri, MJP Gingras, Garrett~E Granroth,
  P~Jiramongkolchai, Robert~Joseph Cava, and BD~Gaulin.
\newblock {Neutron spectroscopic study of crystal field excitations in
  Tb$_2$Ti$_2$O$_7$ and Tb$_2$Sn$_2$O$_7$}.
\newblock \emph{Phys. Rev. B}, 89\penalty0 (13):\penalty0 134410, 2014.
\newblock \doi{10.1103/PhysRevB.89.134410}.

\bibitem[Zhang et~al.(2013)Zhang, Jackson, Raghavan, Hwang, and
  Stemmer]{Zhang2013}
Jack~Y. Zhang, Clayton~A. Jackson, Santosh Raghavan, Jinwoo Hwang, and Susanne
  Stemmer.
\newblock {Magnetism and local structure in low-dimensional Mott insulating
  GdTiO${}_{3}$}.
\newblock \emph{Phys. Rev. B}, 88:\penalty0 121104, 2013.
\newblock \doi{10.1103/PhysRevB.88.121104}.
\newblock URL \url{https://link.aps.org/doi/10.1103/PhysRevB.88.121104}.

\bibitem[Zhang et~al.(2017{\natexlab{a}})Zhang, Levenson-Falk, Ramshaw, Bonn,
  Liang, Hardy, Hartnoll, and Kapitulnik]{zhang2017anomalous}
Jiecheng Zhang, Eli~M Levenson-Falk, BJ~Ramshaw, DA~Bonn, Ruixing Liang,
  WN~Hardy, Sean~A Hartnoll, and Aharon Kapitulnik.
\newblock {Anomalous thermal diffusivity in underdoped YBa$_2$Cu$_3$O$_{6+ x}$
  }.
\newblock \emph{Proceedings of the National Academy of Sciences}, 114\penalty0
  (21):\penalty0 5378--5383, 2017{\natexlab{a}}.
\newblock \doi{10.1073/pnas.1703416114}.

\bibitem[Zhang et~al.(2019{\natexlab{b}})Zhang, Kountz, Levenson-Falk, Song,
  Greene, and Kapitulnik]{zhang_thermal_2019}
Jiecheng Zhang, Erik~D. Kountz, Eli~M. Levenson-Falk, Dongjoon Song, Richard~L.
  Greene, and Aharon Kapitulnik.
\newblock {Thermal diffusivity above the Mott-Ioffe-Regel limit}.
\newblock \emph{Phys. Rev. B}, 100:\penalty0 241114, 2019{\natexlab{b}}.
\newblock \doi{10.1103/PhysRevB.100.241114}.
\newblock URL \url{https://link.aps.org/doi/10.1103/PhysRevB.100.241114}.

\bibitem[Zhang et~al.(2016)Zhang, Tan, Liu, Teitelbaum, Post, Jin, Nelson,
  Basov, Wu, and Averitt]{zhang2016lcmo}
Jingdi Zhang, Xuelian Tan, Mengkun Liu, S.~W. Teitelbaum, K.~W. Post, Feng Jin,
  K.A. Nelson, D.~N. Basov, Wenbin Wu, and R.~D. Averitt.
\newblock {Cooperative photoinduced metastable phase control in strained
  manganite films}.
\newblock \emph{Nature Materials}, 15\penalty0 (9):\penalty0 956--960, 2016.
\newblock \doi{10.1038/nmat4695}.

\bibitem[Zhang et~al.(2019{\natexlab{c}})Zhang, Jia, Wang, Li, Duan, Li, Zhao,
  Cao, Dai, Deng, Zhang, Feng, Yu, Liu, Hu, Zhu, and Jin]{Zhang2019}
Jun Zhang, Yating Jia, Xiancheng Wang, Zhi Li, Lei Duan, Wenmin Li, Jianfa
  Zhao, Lipeng Cao, Guangyang Dai, Zheng Deng, Sijia Zhang, Shaomin Feng, Runze
  Yu, Qingqing Liu, Jiangping Hu, Jinlong Zhu, and Changqing Jin.
\newblock {A new quasi-one-dimensional compound Ba$_3$TiTe$_5$ and
  superconductivity induced by pressure}.
\newblock \emph{NPG Asia Materials}, 11\penalty0 (1):\penalty0 60,
  2019{\natexlab{c}}.
\newblock \doi{10.1038/s41427-019-0158-2}.
\newblock URL \url{https://doi.org/10.1038/s41427-019-0158-2}.

\bibitem[Zhang et~al.(2018{\natexlab{a}})Zhang, Yaji, Hashimoto, Ota, {Kondo},
  Okazaki, Wang, Wen, Gu, Ding, et~al.]{zhang2018observation}
Peng Zhang, Koichiro Yaji, Takahiro Hashimoto, Yuichi Ota, Takeshi {Kondo},
  Kozo Okazaki, Zhijun Wang, Jinsheng Wen, GD~Gu, Hong Ding, et~al.
\newblock Observation of topological superconductivity on the surface of an
  iron-based superconductor.
\newblock \emph{Science}, 360\penalty0 (6385):\penalty0 182--186,
  2018{\natexlab{a}}.
\newblock \doi{10.1126/science.aan4596}.

\bibitem[Zhang et~al.(1997)Zhang, Carlson, and Gubernatis]{zhangcpmc}
Shiwei Zhang, J.~Carlson, and J.~E. Gubernatis.
\newblock Constrained path {Monte} {Carlo} method for fermion ground states.
\newblock \emph{Phys. Rev. B}, 55:\penalty0 7464--7477, 1997.
\newblock \doi{10.1103/PhysRevB.55.7464}.
\newblock URL \url{https://link.aps.org/doi/10.1103/PhysRevB.55.7464}.

\bibitem[Zhang et~al.(2019{\natexlab{d}})Zhang, Changlani, Plumb, Tchernyshyov,
  and Moessner]{Zhang_NCNF_2019}
Shu Zhang, Hitesh~J. Changlani, Kemp~W. Plumb, Oleg Tchernyshyov, and Roderich
  Moessner.
\newblock {Dynamical Structure Factor of the Three-Dimensional Quantum Spin
  Liquid Candidate ${\mathrm{NaCaNi}}_{2}{\mathrm{F}}_{7}$}.
\newblock \emph{Phys. Rev. Lett.}, 122:\penalty0 167203, 2019{\natexlab{d}}.
\newblock \doi{10.1103/PhysRevLett.122.167203}.
\newblock URL \url{https://link.aps.org/doi/10.1103/PhysRevLett.122.167203}.

\bibitem[Zhang et~al.(2017{\natexlab{b}})Zhang, Nickel, and
  Mittleman]{zhang2017high}
Wei Zhang, Daniel Nickel, and Daniel Mittleman.
\newblock High-pressure cell for terahertz time-domain spectroscopy.
\newblock \emph{Optics express}, 25\penalty0 (3):\penalty0 2983--2993,
  2017{\natexlab{b}}.
\newblock \doi{10.1364/OE.25.002983}.

\bibitem[Zhang et~al.(2020)Zhang, Gao, Liu, and Chen]{Zhang_2020}
Xiao-Tian Zhang, Yong~Hao Gao, Chunxiao Liu, and Gang Chen.
\newblock Topological thermal hall effect of magnetic monopoles in the
  pyrochlore {U(1)} spin liquid.
\newblock \emph{Phys. Rev. Research}, 2:\penalty0 013066, 2020.
\newblock \doi{10.1103/PhysRevResearch.2.013066}.
\newblock URL \url{https://link.aps.org/doi/10.1103/PhysRevResearch.2.013066}.

\bibitem[Zhang et~al.(2017{\natexlab{c}})Zhang, Zhou, Cui, Zhao, and
  Liu]{zhang2017theoretical}
Xiaoming Zhang, Yinong Zhou, Bin Cui, Mingwen Zhao, and Feng Liu.
\newblock Theoretical discovery of a superconducting two-dimensional
  metal--organic framework.
\newblock \emph{Nano Letters}, 17\penalty0 (10):\penalty0 6166--6170,
  2017{\natexlab{c}}.
\newblock \doi{10.1021/acs.nanolett.7b02795}.

\bibitem[Zhang et~al.(2018{\natexlab{b}})Zhang, Mahmood, Daum, Dun, Paddison,
  Laurita, Hong, Zhou, Armitage, and Mourigal]{zhang2018hierarchy}
Xinshu Zhang, Fahad Mahmood, Marcus Daum, Zhiling Dun, Joseph~AM Paddison,
  Nicholas~J Laurita, Tao Hong, Haidong Zhou, NP~Armitage, and Martin Mourigal.
\newblock {Hierarchy of Exchange Interactions in the Triangular-Lattice Spin
  Liquid YbMgGaO$_4$}.
\newblock \emph{Physical Review X}, 8\penalty0 (3):\penalty0 031001,
  2018{\natexlab{b}}.
\newblock \doi{10.1103/PhysRevX.8.031001}.

\bibitem[Zhang and Satpathy(1991)]{magnetite1}
Ze~Zhang and Sashi Satpathy.
\newblock Electron states, magnetism, and the {Verwey} transition in magnetite.
\newblock \emph{Phys. Rev. B}, 44:\penalty0 13319--13331, 1991.
\newblock \doi{10.1103/PhysRevB.44.13319}.
\newblock URL \url{https://link.aps.org/doi/10.1103/PhysRevB.44.13319}.

\bibitem[Zhao et~al.(2019)Zhao, Hu, Ye, Hoffmann, Kimchi, and
  Cao]{zhao2019nonequilibrium}
Hengdi Zhao, Bing Hu, Feng Ye, Christina Hoffmann, Itamar Kimchi, and Gang Cao.
\newblock {Nonequilibrium orbital transitions via applied electrical current in
  calcium ruthenates}.
\newblock \emph{Phys. Rev. B}, 100\penalty0 (24):\penalty0 241104, 2019.
\newblock \doi{10.1103/PhysRevB.100.241104}.

\bibitem[Zhao et~al.(2017)Zhao, Belvin, Liang, Bonn, Hardy, Armitage, and
  Hsieh]{zhao2017global}
L~Zhao, CA~Belvin, R~Liang, DA~Bonn, WN~Hardy, NP~Armitage, and D~Hsieh.
\newblock {A global inversion-symmetry-broken phase inside the pseudogap region
  of YBa$_2$Cu$_3$O$_y$}.
\newblock \emph{Nature Physics}, 13\penalty0 (3):\penalty0 250--254, 2017.
\newblock \doi{10.1038/nphys3962}.

\bibitem[Zhao et~al.(2018{\natexlab{a}})Zhao, Torchinsky, Harter, de~la Torre,
  and Hsieh]{zhao2018second}
Liuyan Zhao, Darius Torchinsky, John Harter, Alberto de~la Torre, and David
  Hsieh.
\newblock Second harmonic generation spectroscopy of hidden phases.
\newblock \emph{Encyclopedia of Modern Optics}, 2:\penalty0 207--226,
  2018{\natexlab{a}}.
\newblock \doi{10.1016/B978-0-12-803581-8.09533-3}.

\bibitem[Zhao et~al.(2018{\natexlab{b}})Zhao, Li, Chang, Jiang, Wu, Liu,
  Moodera, Zhu, and Chan]{zhao2018direct}
Weiwei Zhao, Mingda Li, Cui-Zu Chang, Jue Jiang, Lijun Wu, Chaoxing Liu,
  Jagadeesh~S Moodera, Yimei Zhu, and Moses~HW Chan.
\newblock {Direct imaging of electron transfer and its influence on
  superconducting pairing at FeSe/SrTiO$_3$ interface}.
\newblock \emph{Science advances}, 4\penalty0 (3):\penalty0 eaao2682,
  2018{\natexlab{b}}.
\newblock \doi{10.1126/sciadv.aao2682}.

\bibitem[Zhao et~al.(2015)Zhao, Liu, Yan, An, Liu, Zhang, Wang, Liu, Jiang, Li,
  Wang, Li, Mandrus, Xie, Pan, and Wang]{MR-WTe2}
Yanfei Zhao, Haiwen Liu, Jiaqiang Yan, Wei An, Jun Liu, Xi~Zhang, Huichao Wang,
  Yi~Liu, Hua Jiang, Qing Li, Yong Wang, Xin-Zheng Li, David Mandrus, X.~C.
  Xie, Minghu Pan, and Jian Wang.
\newblock Anisotropic magnetotransport and exotic longitudinal linear
  magnetoresistance in $\mathrm{WT}{\mathrm{e}}_{2}$ crystals.
\newblock \emph{Phys. Rev. B}, 92:\penalty0 041104, 2015.
\newblock \doi{10.1103/PhysRevB.92.041104}.
\newblock URL \url{https://link.aps.org/doi/10.1103/PhysRevB.92.041104}.

\bibitem[Zhelev et~al.(2016)Zhelev, Reichl, Abhilash, Smith, Nguyen, Mueller,
  and Parpia]{zhelev2016observation}
Nikolay Zhelev, Matthew Reichl, Thanniyil~Sebastian Abhilash, Eric~Nelson
  Smith, KX~Nguyen, EJ~Mueller, and Jeevak~M Parpia.
\newblock {Observation of a new superfluid phase for $^3$He embedded in
  nematically ordered aerogel}.
\newblock \emph{Nature Communications}, 7:\penalty0 12975, 2016.
\newblock \doi{10.1038/ncomms12975}.

\bibitem[Zheng et~al.(2018)Zheng, Changlani, Williams, Busemeyer, and
  Wagner]{Zheng18}
Huihuo Zheng, Hitesh~J. Changlani, Kiel~T. Williams, Brian Busemeyer, and
  Lucas~K. Wagner.
\newblock From real materials to model hamiltonians with density matrix
  downfolding.
\newblock \emph{Frontiers in Physics}, 6:\penalty0 43, 2018.
\newblock \doi{10.3389/fphy.2018.00043}.
\newblock URL
  \url{https://www.frontiersin.org/article/10.3389/fphy.2018.00043}.

\bibitem[Zheng et~al.(2014)Zheng, Wang, Hardy, B{\"o}hmer, Wolf, Meingast, and
  Lortz]{zheng2014high}
Y~Zheng, Y~Wang, F~Hardy, AE~B{\"o}hmer, T~Wolf, C~Meingast, and R~Lortz.
\newblock {High-pressure evolution of the specific heat of a strongly
  underdoped Ba(Fe$_{0.963}$Co$_{0.037}$)As$_2$ iron-based superconductor}.
\newblock \emph{Phys. Rev. B}, 89\penalty0 (5):\penalty0 054514, 2014.
\newblock \doi{10.1103/PhysRevB.89.054514}.

\bibitem[Zhu et~al.(2017{\natexlab{a}})Zhu, McDonald, Shekhter, Ramshaw, Modic,
  Balakirev, and Harrison]{graphitehighfield}
Zengwei Zhu, RD~McDonald, A~Shekhter, BJ~Ramshaw, KA~Modic, FF~Balakirev, and
  N~Harrison.
\newblock Magnetic field tuning of an excitonic insulator between the weak and
  strong coupling regimes in quantum limit graphite.
\newblock \emph{Scientific reports}, 7\penalty0 (1):\penalty0 1--6,
  2017{\natexlab{a}}.
\newblock \doi{10.1038/s41598-017-01693-5}.

\bibitem[Zhu et~al.(2017{\natexlab{b}})Zhu, Maksimov, White, and
  Chernyshev]{Zhu_YMGO_2017}
Zhenyue Zhu, P.~A. Maksimov, Steven~R. White, and A.~L. Chernyshev.
\newblock {Disorder-Induced Mimicry of a Spin Liquid in
  ${\mathrm{YbMgGaO}}_{4}$}.
\newblock \emph{Phys. Rev. Lett.}, 119:\penalty0 157201, 2017{\natexlab{b}}.
\newblock \doi{10.1103/PhysRevLett.119.157201}.
\newblock URL \url{https://link.aps.org/doi/10.1103/PhysRevLett.119.157201}.

\bibitem[Zong et~al.(2019)Zong, Kogar, Bie, Rohwer, Lee, Baldini,
  Erge{\c{c}}en, Yilmaz, Freelon, Sie, et~al.]{zong2019evidence}
Alfred Zong, Anshul Kogar, Ya-Qing Bie, Timm Rohwer, Changmin Lee, Edoardo
  Baldini, Emre Erge{\c{c}}en, Mehmet~B Yilmaz, Byron Freelon, Edbert~J Sie,
  et~al.
\newblock Evidence for topological defects in a photoinduced phase transition.
\newblock \emph{Nature Physics}, 15\penalty0 (1):\penalty0 27--31, 2019.
\newblock \doi{10.1038/s41567-018-0311-9}.

\bibitem[Zou and He(2020)]{zou2020field}
Liujun Zou and Yin-Chen He.
\newblock {Field-induced QCD 3-Chern-Simons quantum criticalities in Kitaev
  materials}.
\newblock \emph{Physical Review Research}, 2\penalty0 (1):\penalty0 013072,
  2020.
\newblock \doi{10.1103/PhysRevResearch.2.013072}.

\bibitem[Zuo et~al.(2015)Zuo, Shi, Xia, Huang, Chen, Jin, Wei, Ouyang, and
  Cheng]{zuo2015magnetic}
Hua-Kun Zuo, Li-Ran Shi, Zheng-Cai Xia, Jun-Wei Huang, Bo-Rong Chen, Zhao Jin,
  Meng Wei, Zhong-Wen Ouyang, and Gang Cheng.
\newblock {The Magnetic Anisotropy and Complete Phase Diagram of CuFeO$_2$
  Measured in a Pulsed High Magnetic Field up to 75 T}.
\newblock \emph{Chinese Physics Letters}, 32\penalty0 (4):\penalty0
  47502--047502, 2015.
\newblock \doi{10.1088/0256-307X/32/4/047502}.

\bibitem[Zvyagin et~al.(2011)Zvyagin, {\v{C}}i{\v{z}}m{\'a}r, Ozerov, Wosnitza,
  Feyerherm, Manmana, and Mila]{zvyagin2011field}
SA~Zvyagin, E~{\v{C}}i{\v{z}}m{\'a}r, M~Ozerov, J~Wosnitza, R~Feyerherm,
  SR~Manmana, and F~Mila.
\newblock Field-induced gap in a quantum spin-1/2 chain in a strong magnetic
  field.
\newblock \emph{Physical Review B}, 83\penalty0 (6):\penalty0 060409, 2011.
\newblock \doi{10.1103/PhysRevB.83.060409}.

\end{thebibliography}

\end{document}